\documentclass{aa}
\usepackage{txfonts}
\usepackage{graphicx}
\usepackage{rotating}
\usepackage{natbib}
\usepackage{lscape}
\usepackage{multirow}
\bibpunct{(}{)}{;}{a}{}{,} 
\newcommand{\ha}{H$\alpha$}
\newcommand{\hb}{H$\beta$}

\newcommand{\pab}{Pa$\beta$}

\newcommand{\hi}{\ion{H}{i}}

\newcommand{\sii}{[\ion{S}{ii}]}

\newcommand{\oi}{[\ion{O}{i}]}

\newcommand{\av}{$A_V$}
\newcommand{\kms}{km\,s$^{-1}$}
\newcommand{\um}{$\mu$m}

\newcommand{\lsun}{L$_{\odot}$}
\newcommand{\msun}{M$_{\odot}$}

\newcommand{\msunyr}{M$_{\odot}$\,yr$^{-1}$}
\newcommand{\rstar}{$R_{\mathrm{*}}$}
\newcommand{\lstar}{$L_{\mathrm{*}}$}
\newcommand{\mstar}{$M_{\mathrm{*}}$}
\newcommand{\teff}{$T_\mathrm{eff}$}
\newcommand{\lacc}{$L_{\mathrm{acc}}$}

\newcommand{\macc}{$\dot{M}_{acc}$}

\begin{document}

\title{X-Shooter spectroscopy of young stellar objects}
\subtitle{VI - \hi\ line decrements\thanks{Based on observations collected at the European Southern Observatory at Paranal, Chile, under  programmes 084.C-0269(A), 085.C-238(A), 086.C-0173(A), 087.C-0244(A), and 089.C-0143(A).}}

\author{S. Antoniucci\inst{1},
        B. Nisini\inst{1},
        T. Giannini\inst{1},
                E. Rigliaco\inst{2,3},
        J.M. Alcal\'a\inst{4},
                A. Natta\inst{5,6},
                B. Stelzer\inst{7}
                }

\institute{ INAF-Osservatorio Astronomico di Roma, Via di Frascati 33, I-00040 Monte Porzio Catone, Italy \and
                        Institute for Astronomy, ETH Z\"urich, Wolfgang-Pauli-Strasse 27, CH-8093 Z\"urich, Switzerland \and
                        INAF-Osservatorio Astronomico di Padova, Vicolo dell'Osservatorio 5, 35122 Padova, Italy \and
                        INAF-Osservatorio Astronomico di Capodimonte, Salita Moiariello, 16 80131, Napoli, Italy \and
                        INAF-Osservatorio Astrofisico of Arcetri, Largo E. Fermi, 5, 50125 Firenze, Italy \and
            School of Cosmic Physics, Dublin Institute for Advanced Studies, 31 Fitzwilliams Place, Dublin 2, Ireland \and
                        INAF-Osservatorio Astronomico di Palermo, Piazza del Parlamento 1, 90134 Palermo, Italy
                        } 

\offprints{Simone Antoniucci, \email{simone.antoniucci@oa-roma.inaf.it}}
\date{Received date / Accepted date}
\titlerunning{\hi\ decrements in T Tauri stars}
\authorrunning{Antoniucci et al.}

\abstract
{Hydrogen recombination emission lines commonly observed in accreting young stellar objects represent a powerful tracer for the gas conditions in the circumstellar structures (accretion columns, and winds or jets).}
{Here we perform
a study of the \hi\ decrements and line profiles, from the Balmer and Paschen \hi\ lines detected in the X-Shooter spectra of a homogeneous sample of 36 T Tauri objects in Lupus, the accretion and stellar properties of which were already derived in a previous work. We aim to obtain information on the \hi\ gas physical conditions to delineate a consistent picture of the \hi\ emission mechanisms in pre-main sequence low-mass stars (\mstar $<$ 2\msun).}
{We have empirically classified the sources based on their \hi\ line profiles and decrements. We identified four Balmer decrement types (which we classified as 1, 2, 3, and 4) and three Paschen decrement types (A, B, and C), characterised by different shapes. 
We first discussed the connection between the decrement types and the source properties and then compared the observed decrements with predictions from recently published local line excitation models.}
{We identify a few groups of sources that display similar \hi\ properties. 
One third of the objects show lines with narrow symmetric profiles, and present similar Balmer and Paschen decrements (straight decrements, types 2 and A). Lines in these sources are consistent with optically thin emission from gas with hydrogen densities of order 10$^9$ cm$^{-3}$ and $5\,000 < T < 15\,000$K. These objects are associated with low mass accretion rates.
Type 4 (L-shaped) Balmer and type B Paschen decrements are found in conjunction with very wide line profiles and are characteristic of strong accretors, with optically thick emission from high-density gas (log $n_H$ $>$ 11 cm$^{-3}$).
Type 1 (curved) Balmer decrements are observed only in three sub-luminous sources viewed edge-on, so we speculate that these are actually type 2 decrements that are reddened because of neglecting a residual amount of extinction in the line emission region.
About 20\% of the objects present type 3 Balmer decrements (bumpy), which, however, cannot be reproduced with current models.
}
{}


\maketitle

\section{Introduction}
\label{sec:intro}

Hydrogen recombination lines (\hi) are the most common features observed in optical and near-infrared spectra of young stellar objects (YSOs) 
and are considered a typical manifestation of accretion activity in these sources. 
These lines are indeed tightly related to the accretion and ejection process, as evidenced by the empirical relationships between \hi\ line 
luminosity and the accretion luminosity that have been found by numerous groups \citep[e.g.][]{muzerolle98a,calvet04,natta06,herczeg08,rigliaco12,alcala14}.
As such, \hi\ lines are commonly used as a proxy to derive mass accretion rates in YSOs 
\citep[e.g.][]{gatti06,antoniucci11,antoniucci14d,biazzo12},
although their actual origin in accretion flows or winds has not been clarified yet.

Observations and modelling of the \hi\ emission lines in Classical T Tauri stars (CTTSs) have been the topic of numerous studies in recent decades.
These lines were first interpreted as originating in circumstellar winds based on detected P Cygni profiles 
\citep[e.g.][]{natta88,hartmann90b}.
Subsequently, new observations at medium and high spectral resolution in fairly large samples of YSOs both at optical and near-infrared wavelengths 
have progressively revealed a great variety of line profiles \citep[e.g.][]{edwards94,fernandez95,reipurth96,muzerolle98b,alencar00,folha01,najita96a},
so that the most common interpretation has gradually shifted towards an origin in magnetospheric accretion flows 
\citep[e.g.][]{calvet92a,muzerolle98b,muzerolle00}.
Both accretion and wind models, however, fail to reproduce all the features of the profiles, 
so that more recent models propose a hybrid scenario in which the lines present contributions from both accretion flows and winds
\citep[e.g.][]{kurosawa08,lima10,kurosawa11}.

Deriving the physical conditions of the \hi\ emitting gas in the circumstellar region of T Tauri stars would provide precious constraints 
to both test the models and identify the excitation mechanism for \hi\ lines, thus helping to derive information on their region of origin. 
Despite the huge observational and modelling effort on \hi\ lines, however, the physical conditions of the gas 
responsible for \hi\ emission are still poorly constrained from an observational point of view.

An effective method to directly derive the temperature and density of the gas is to compare the observed \hi\ line ratios 
and in particular the \hi\ Balmer, Paschen, and Brackett series decrements with predictions of models based on atomic level populations. 
Many works carried out in the past have, however, been limited to the analysis of a few line ratios only 
and were often based on observations performed at different times and/or with different instrumental setups. Additionally, 
the interpretation of the line ratios was done assuming simple standard emission regimes, such as blackbody-like 
emission and classic \citet{baker38} Case B recombination.

Case B recombination has been invoked to explain \hi\ decrements from young objects in several papers \citep{nisini04,bary08,podio08,vacca11,kraus12,kospal11,whelan14}.
The comparison with Case B predictions has suggested a huge diversity of conditions in these works,
with temperatures spanning from T=20\;000 K down to T=1000 K and electron densities in the range $n_e$ = 10$^{7}$-10$^{13}$cm$^{-3}$.
For instance in a sample of 15 CTTSs in Taurus-Auriga \citet{bary08} found that the average Paschen and Brackett decrement 
are compatible with Case B emission with electron density $n_e$ $\sim$ 10$^{10}$ cm$^{-3}$ 
and fairly low temperatures T $<$ 2000 K, in stark contrast with the temperatures expected
for \hi\ emission in accretion columns (T$\sim$6\;000-20\;000 K; e.g. Martin et al., 1996\nocite{martin96}; Muzerolle et al.,2001\nocite{muzerolle01}).

The validity of Case B for circumstellar environments of T Tauri has often been questioned. \citet{edwards13} have 
recently shown that by using the local line excitation calculations developed by Kwan \& Fischer (2011) for 
conditions appropriate for winds and accretion flows, the Paschen decrements of T Tauri stars can 
be reproduced by assuming a narrow range of total \hi\ densities between 2\;10$^{9}$ and 2\;10$^{10}$ cm$^{-3}$,
although they could not provide strong constraints on the temperature.

The wide and simultaneous spectral coverage now provided by an increasing number of new-generation spectrometers 
allows us to observe a great number of optical and near-infrared \hi\ lines that also avoid all intrinsic line variability issues. 
In this paper we use VLT/X-Shooter \citep{vernet11} data covering the 0.3-2.4 \um\ wavelength range to
perform a systematic study of the \hi\ decrements of the Balmer and Paschen series in a large and 
homogeneous sample of T Tauri objects in the Lupus star-forming region,  
whose stellar and accretion properties have already been derived 
from the analysis of the same dataset in a previous work by our group \citep{alcala14}.
Despite their moderate spectral resolution, the X-shooter spectra also provide information on the \hi\ 
line profiles, which are indicative of the kinematic distribution of the \hi\ emitting gas. 

On this basis, we are able on the one hand to investigate the possible relationships between the observed decrements 
and both the source properties and line profile types, on the other hand to compare the decrement to predictions of 
models such as Case B and the Kwan \& Fischer (2011) local line excitation calculations. 
This allows us to test the effective capability of the decrement analysis to provide information on the \hi\ emission 
mechanisms and to draw a picture of the different gas physical conditions existing in the circumstellar environment 
of typical T Tauri stars.

The paper is organised as follows: the sample and data are presented in Sect.~\ref{sec:sample}. Line profiles and 
decrements are presented in Sect.~\ref{sec:lines}, while we empirically analyse their mutual relations in Sect.~\ref{sec:relations}. 
The comparison with the models is treated in Sect.~\ref{sec:models} 
and results are discussed in Sect.\ref{sec:discussion}. Our main conclusions are presented in Sect.~\ref{sec:outro}.

\begin{figure*}[t!]
\centering
\includegraphics[width=6.3cm]{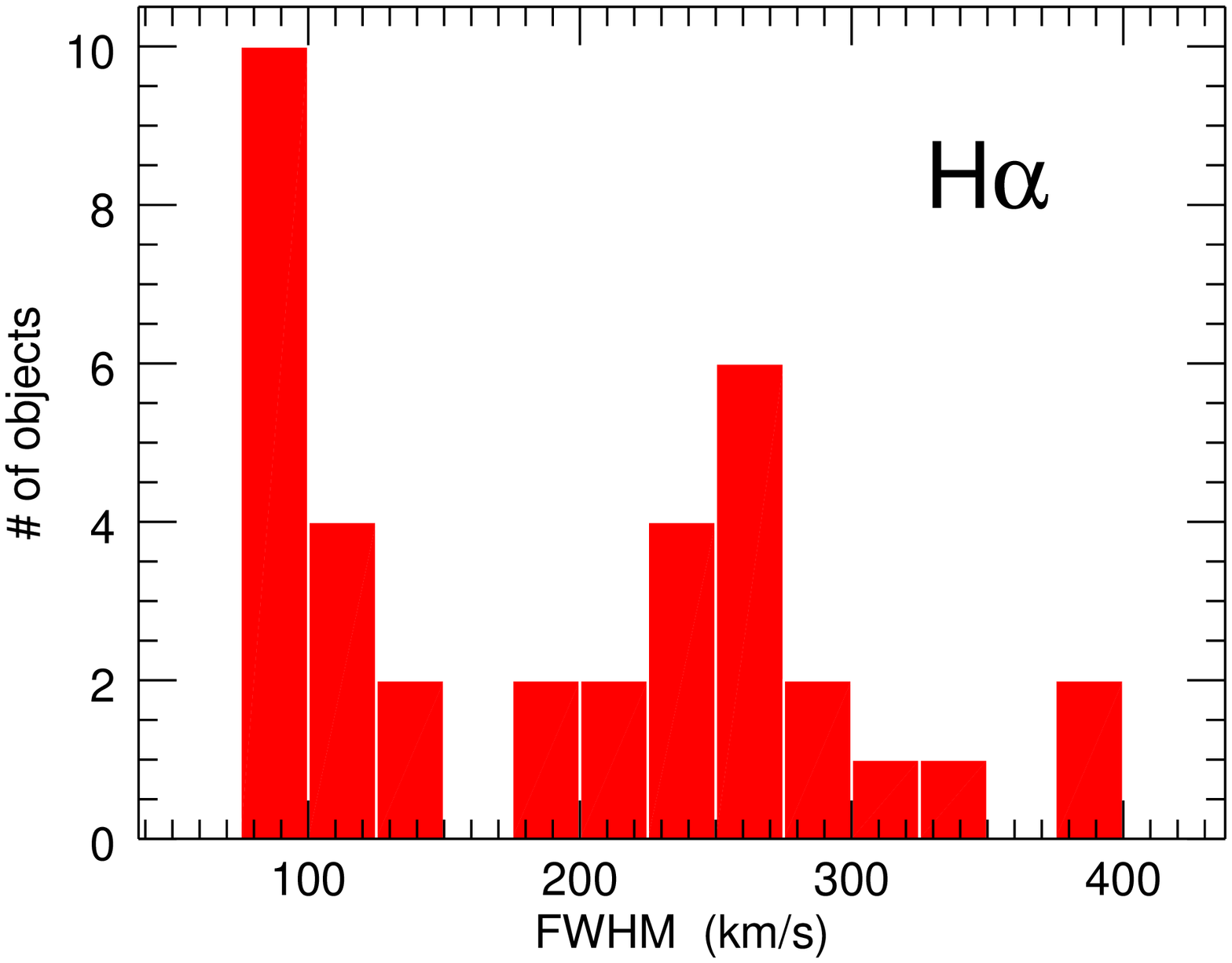}\hspace{-0.5cm}
\includegraphics[width=6.3cm]{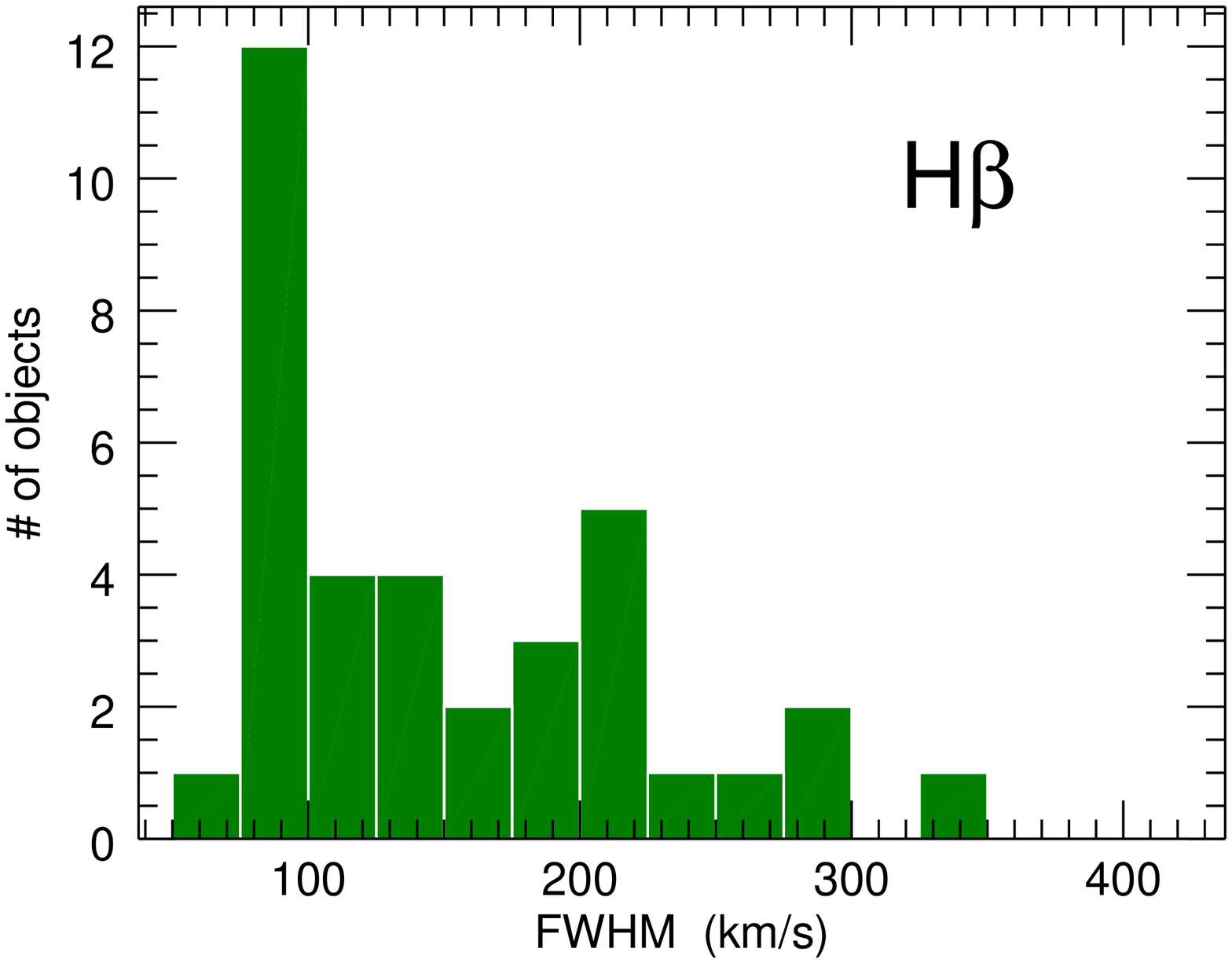}\hspace{-0.5cm}
\includegraphics[width=6.3cm]{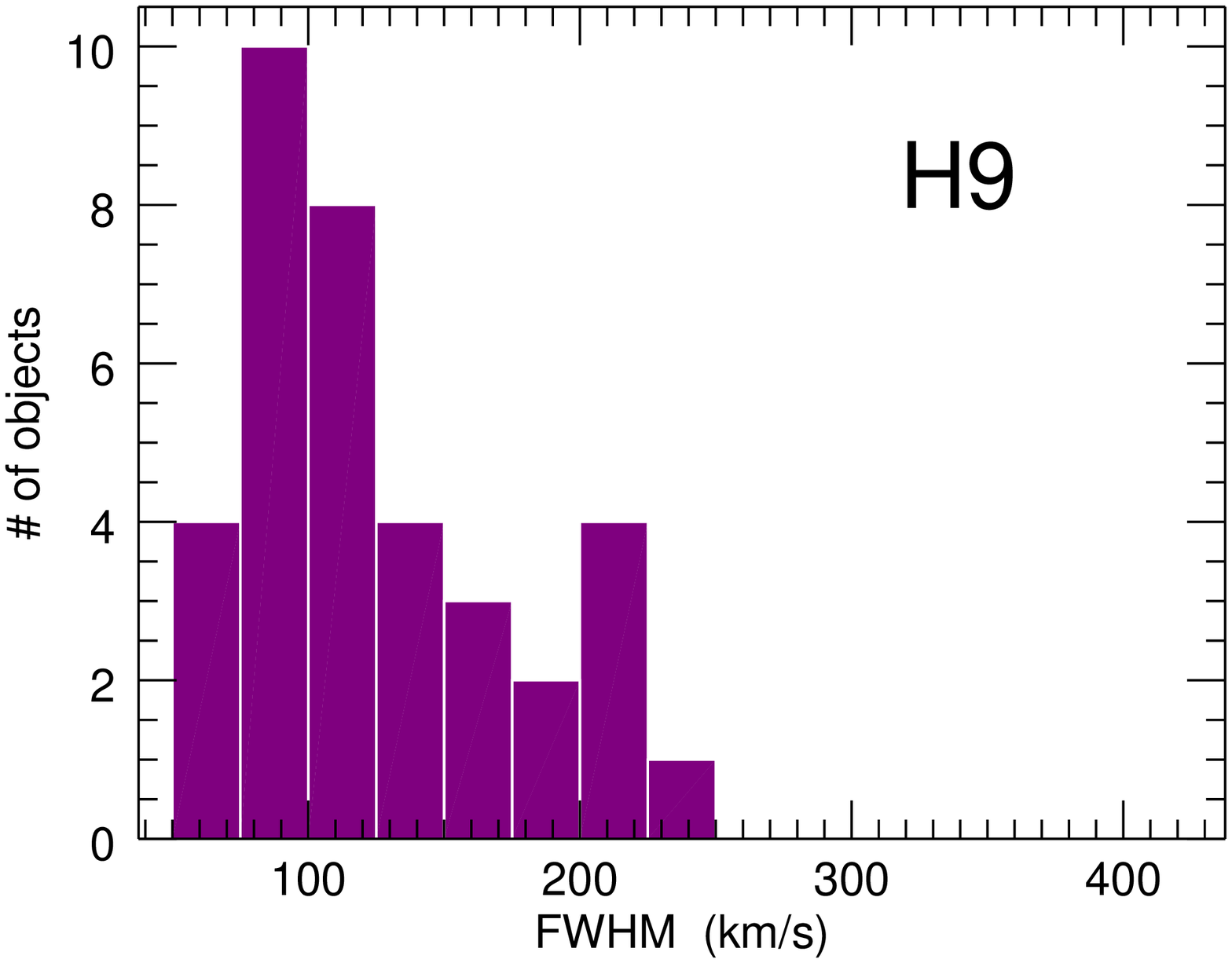}
\caption{\label{fig:histo_fwhm} Distribution of the FWHM of the \ha\ (left), \hb\ (centre), and H9 (right) Balmer lines measured in the sample.} 
\end{figure*}

\section{Sample and dataset}
\label{sec:sample}

The X-Shooter spectra were presented in \citet{alcala14} (hereafter A14) 
and refer to a sample of 36 low-mass Class II sources 
(i.e. with evidence of excess emission from disks) in the Lupus star-forming region, one 
of the closest ($d<200$ pc) low-mass star-forming regions \citep[see][and references therein]{comeron08}.

Most targets were observed with the 1$\arcsec$ slits in the three X-Shooter arms (UVB, VIS, NIR), thus providing a final resolution of about 
60 km/s, 35 km/s, and 50 km/s, respectively. A factor of two higher spectral resolution is available only for Sz74 and Sz83, 
which were observed with the 0.5$\arcsec$ slits.

All details of the observations, the instrument setup, data reduction, and calibration, as well as the description of the procedures for the 
self-consistent determination of the extinction, the stellar parameters, and accretion properties of the sources are given in A14.
The spectra presented in A14 and employed in this work are flux-calibrated and are already corrected for the derived extinction.
The estimated relative error on the flux calibration of the spectra is 10\%.

The properties of the sample are exhaustively described in A14 and \citet{natta14}. 
The sample contains about 50\% of the total Class II YSOs in the Lupus I and Lupus III clouds \citep{merin08}.
The mass of the targets ranges between 0.05\msun\ and 1\msun\ (40\% of the objects have \mstar$<$ 0.2 \msun, 35\% have \mstar\ in the range 0.2-0.5\msun, and 25\% have \mstar between 0.5 and 1.15\msun). 
The inferred spectral types are comprised in the range K7-M8, with a relatively large number of M4-M5 objects.
Derived extinction values are low, with \av\ $<$ 1 mag for all objects except for five that display visual extinctions in the range 1-3.5 mag.
The accretion luminosity (\lacc) of the targets spans 
from $\sim$10$^{-5}$ to $\sim$1 \lsun, with corresponding mass accretion rates (\macc) between 10$^{-12}$ and
10$^{-7}$ \msunyr.

For the convenience of the reader, we report in Table~\ref{tab:main} the main stellar and accretion properties of the objects as determined by A14.

\section{\hi\ lines}
\label{sec:lines}

Hydrogen emission lines are detected in all sources of the sample. All the line fluxes for Balmer lines and Paschen lines up to Pa10 
are taken from Tables C3-C7 of A14. These fluxes were measured by directly integrating the flux-calibrated and extinction-corrected
spectra over the wavelength range of the line, as explained in Sect.~4.4 of A14, without any further correction.
In particular, for each line three independent measurements were performed, considering the lowest, the highest, and the mid position of the local continuum, which depend on the noise level of the spectra around the line. The average and standard deviation of these three measurements were then assumed as the line flux and its associated error.
In this work we have employed the same method to measure the fluxes of a few additional Paschen lines (from Pa11 to Pa14) that were not reported in A14, which we have used to improve the decrement analysis.

The stellar photospheric velocity was measured using the \ion{Li}{i} absorption line at 670.78 nm, located in the VIS arm of X-Shooter, which is detected in all sources.
The \hi\ lines are detected in all three spectrograph arms. Since we want to compare the profile of the different \hi\ lines for each object, we need to align them to the same zero-velocity scale. 
Therefore, in addition to the standard procedures for the wavelength calibration described in A14, which often leave residual offsets between the arms, we paid special attention to re-aligning the wavelength scale of the UVB and NIR arms with that of the VIS arm. To this end, we used the spectral features in emission and/or absorption that were detected in the wavelength interval in common between the arms. 
The typical shifts thus applied are around 10 km/s.

Balmer series lines are observed in all targets, in most cases up to the Balmer jump in the UV region (see A14). 
The continuum-normalised Balmer lines (from H$\alpha$ to H11) are plotted as a function of the rest 
velocity of the stellar photosphere in Fig.~\ref{fig:lines:h} of the appendix for all sources. 

Emission from the Paschen series is detected in 33 sources, but we observe at least four Paschen lines in only 25 objects. 
Lines from \pab\ up to Pa$\delta$ are displayed in Fig.~\ref{fig:lines:pa} of the appendix 
only for those objects where \pab\ is detected with a S/N of greater than or equal to five.

The Br$\gamma$ line is observed in 19 objects, but multiple lines from the the Brackett series are visible only in two objects 
(Sz83 and Sz88A). The Br$\gamma$ for objects where this line is detected with a signal-to-noise (S/N) of greater than or equal to five is reported in Fig.~\ref{fig:lines:br} of the appendix.

The Balmer lines are observed at very high signal-to-noise in almost all cases, while the global quality of the near-IR (Paschen and Brackett) data is much lower, owing to the worse S/N of the X-Shooter spectra in the infrared segment. For this reason, the analysis that we present in this paper is mostly based on the Balmer lines.

\begin{figure*}[t!]
\centering
\includegraphics[width=6.2cm]{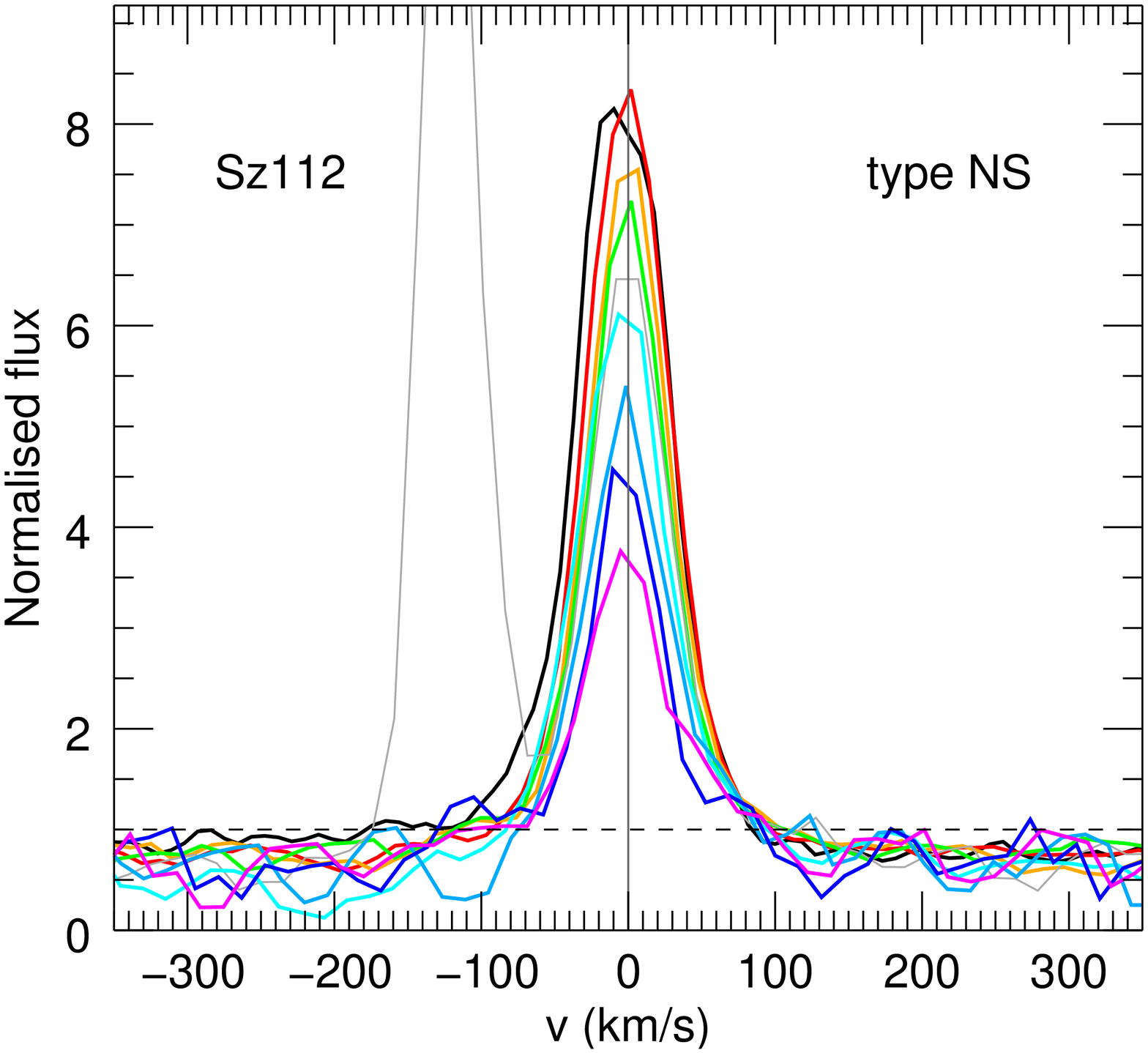}\hspace{-.2cm}
\includegraphics[width=6.2cm]{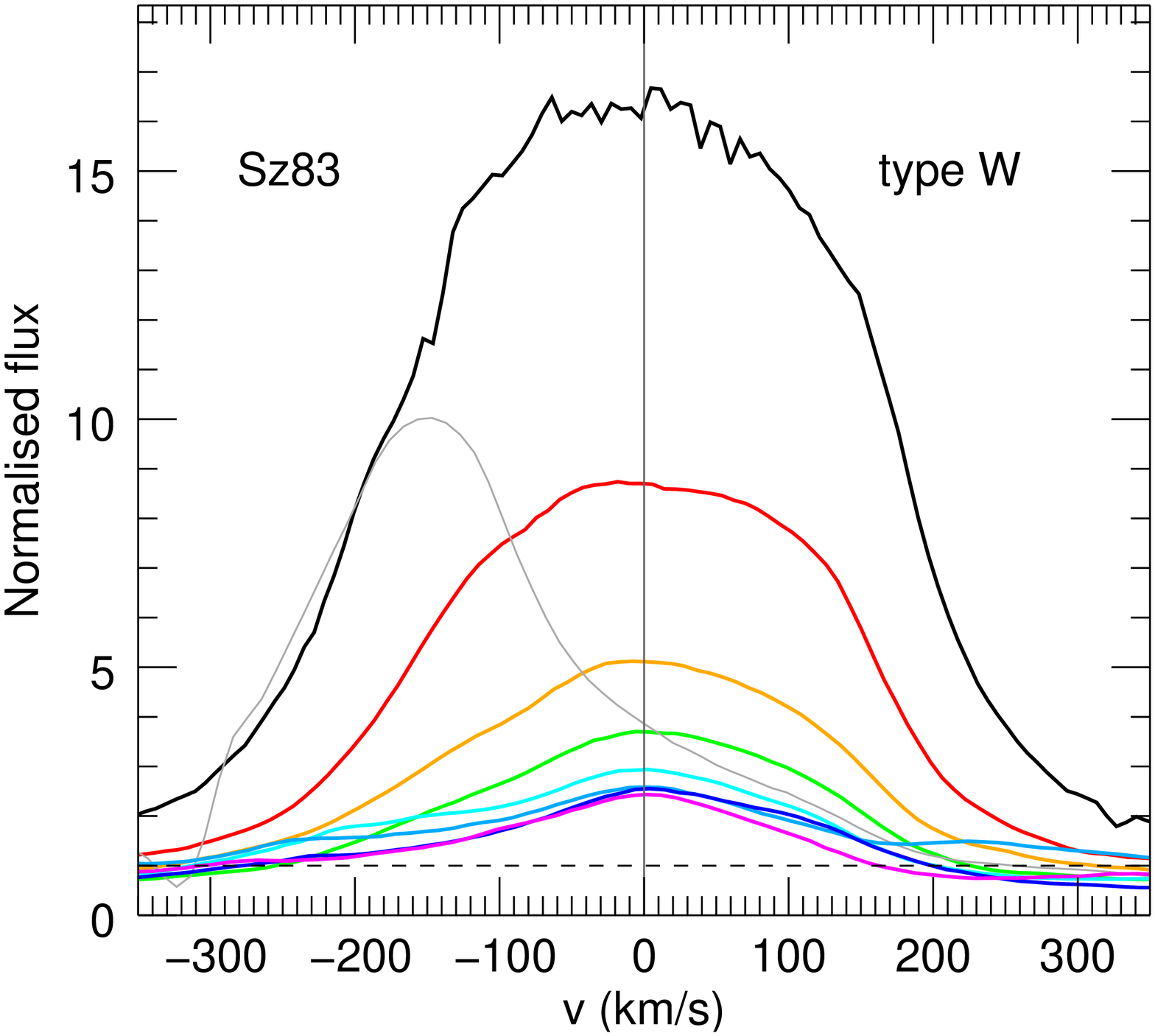}\hspace{-.2cm}
\includegraphics[width=6.2cm]{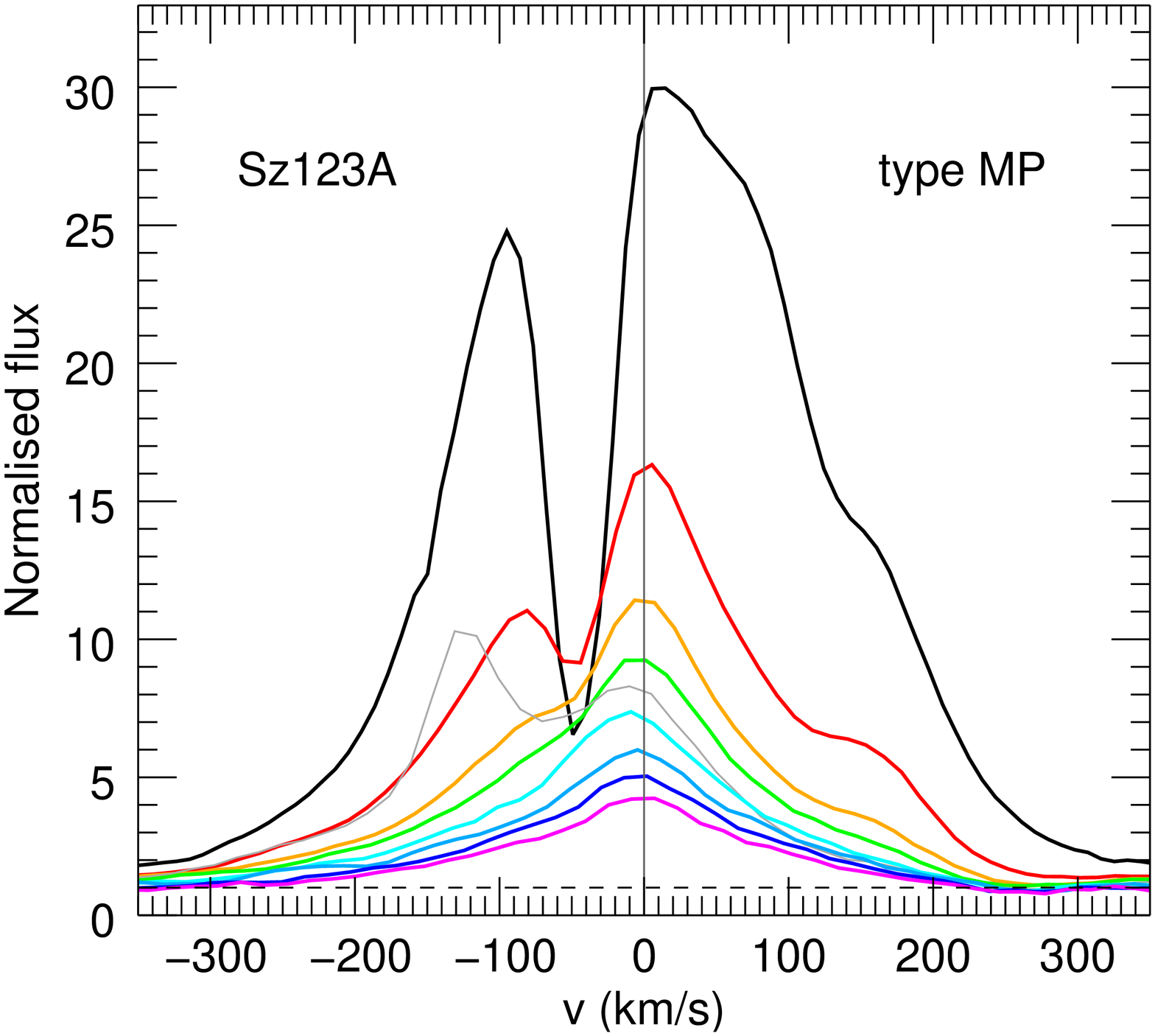}
\caption{\label{fig:lines:h:types} 
Continuum-normalised spectra of three stars that exemplify the main types of \hi\ Balmer line profiles observed in the sample (see text for further details). From left to right: narrow symmetric profile (NS), wide symmetric profile (W), asymmetric or multi-peaked profile (MP). Lines from H$\alpha$ (black) to H11 (purple) are plotted. The profile of the H7 line, which is blended with the \ion{Ca}{ii}-H line, 
is marked in grey.} 
\end{figure*}

\subsection{Line shapes and widths}
\label{sec:lines:shapes}

The moderate resolution of the X-Shooter spectra is sufficient to resolve the wide \hi\ lines, thus revealing a large variety of line profiles, especially for Balmer lines (Fig.~\ref{fig:lines:h}).
These \hi\ profiles can be catalogued using the classification proposed by \citet{reipurth96}, which was originally devised on the basis of the H$\alpha$ profiles observed in a sample of Herbig and T Tauri stars. In this two-dimensional classification profiles are divided depending on the depth (four types: I to IV) and position (blue- or red-shifted) of the absorptions along the profiles with respect to the main peak. 
More specifically, type I profiles show no absorption and are globally symmetric, type II profiles display a weak absorption, type III profiles show a strong absorption (deeper than the half-maximum of the line peak), and type IV profiles display an absorption that goes below the continuum \citep[see][for further details]{reipurth96}.
This should provide information on the distribution and kinematics of the absorbing material in the circumstellar environment.

The H$\alpha$ 
line shows a roughly symmetric profile (type I) in half of the objects, whereas in the 
remaining sources it displays more complex profiles with one or multiple absorptions (type II and III).
Clear evidence for (an inverse) P-Cygni profile (type IV) is found only in Sz91 (3\%). 
Asymmetric lines display secondary peaks that are more or less equally distributed between the red (10, 28\%) and blue (8, 22\%) side. 
These numbers are slightly different from those found by \citet{reipurth96} in their TTS sample, where they observed 
25\% of symmetric H$\alpha$ profiles, 49\% (21\%) of blue (red) absorptions, and 5\% of P Cygni profiles. The different percentage 
of the symmetric profiles might be in part related to the fact that our sample is composed of sources with lower levels of accretion than those of 
\citet{reipurth96}. In our sample the number of blue- and red-shifted absorptions is comparable, indicating no prevalent profile 
shape for asymmetric \hi\ lines.

This type of classification, however, does not take into account the different widths of the lines (i.e. velocity dispersions), 
which may be indicative of a very diverse emission origin. 
Considering the Balmer series, we report in Fig.~\ref{fig:histo_fwhm} the histograms of the measured full-width-at-half-maximum 
(FWHM) of the \ha, \hb, and H9 lines. 
The FWHM distributions clearly reveal two peaks/groups of sources, each containing about half of the objects for \ha\ and \hb: 
the first group display widths around 
or below 100 km/s (narrow line objects), while the second group have larger widths with a peak around 200-250 km/s.

On this basis, considering the \hb\ line (the line we use as reference for computing the decrements, see Sect.~\ref{sec:decs}), in this work we divide objects with roughly symmetric lines (i.e. the lines without two or more separate peaks, which would belong to group I in the classical \citet{reipurth96} classification) in two main groups: sources with narrow symmetric lines (hereafter NS, eleven objects) with FWHM $\sim$ 100 km/s, and sources with wider lines 
(hereafter W, ten objects).
The line profiles in NS sources are highly similar and appear to be slightly blue-shifted in all cases, with an average shift of 
about $-15$ km/s, except for Par-Lup 3-3, whose peak is basically at zero velocity.

The asymmetric \hb\ profiles shown by the remaining targets, which we classify as multi-peaked (hereafter MP, 15 objects), may be interpreted as a combination of different emission and/or absorption components, which testify the complexity of the circumstellar gas distribution and suggest concurrent contributions to the line from accretion and winds.
For instance, clear evidence for blue-shifted absorptions at velocities around 50 km/s is found in eight objects, which is most likely indicative of the presence of strong inner disk winds \citep{lima10}.

The Balmer profiles of three targets that represent the forementioned three profile classes are depicted in Fig.~\ref{fig:lines:h:types}. 
The profiles tend to become more and more symmetric in higher lines of the H series, as expected based on the lower opacities of these lines with respect to \ha\ and \hb.

Paschen lines are detected with a good S/N only in targets that display wide or asymmetric Balmer line profiles. The widths of these \pab\ lines are in the range 50-200 km/s and show in general a more symmetric profile than lower Balmer lines, although they often retain signatures of the same absorption features detected in the Balmer series (Fig.~\ref{fig:lines:h} and \ref{fig:lines:pa}). In that, their profile is similar to the one of higher Balmer lines, as found also in previous works \citep{folha01}.
It is finally worth noting that about one third of the detected \pab\ lines show evidence of an inverse P-Cygni profile indicative of flows of infalling matter.

\begin{figure*}[!t]
\centering
\includegraphics[width=4.4cm]{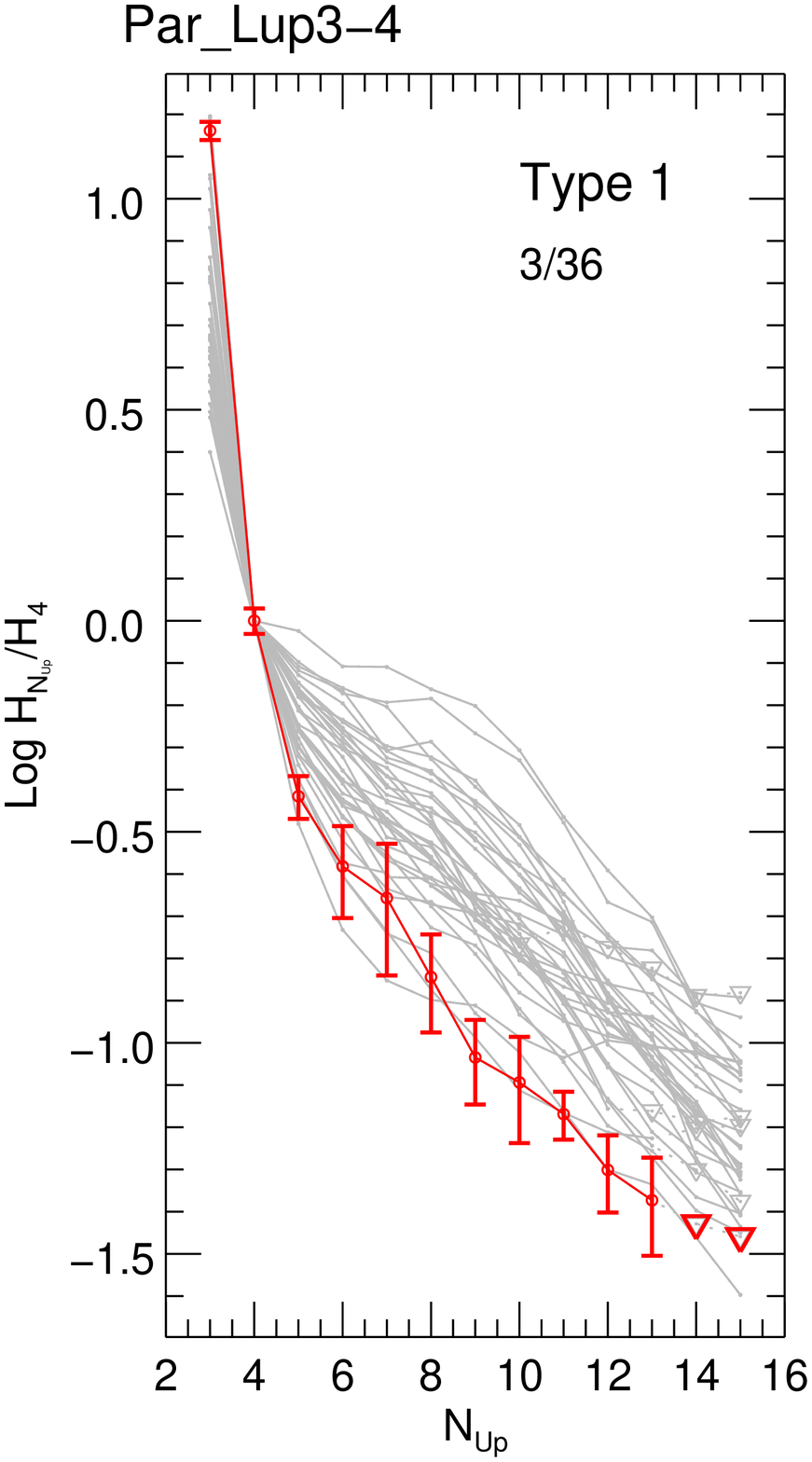}
\includegraphics[width=4.4cm]{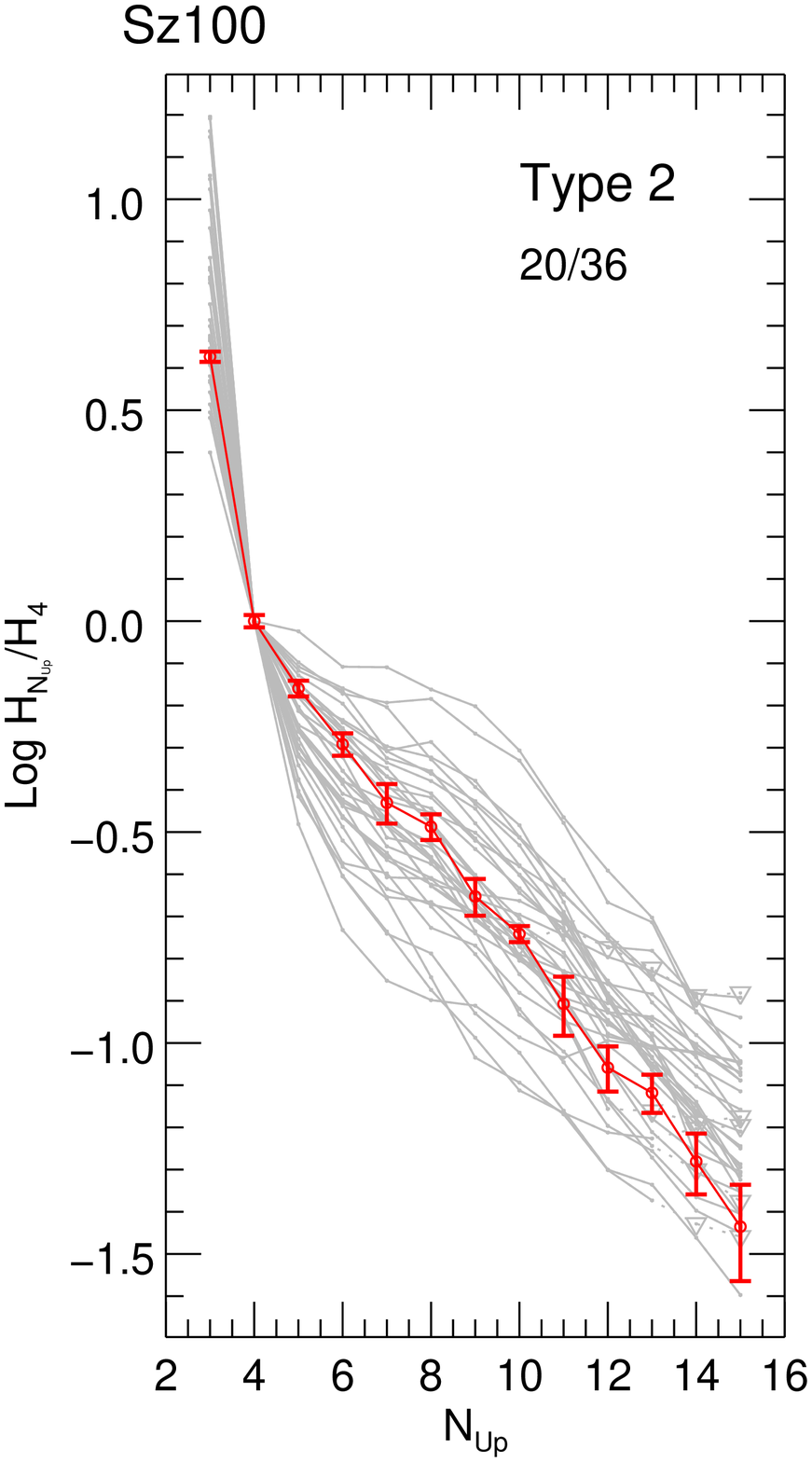}
\includegraphics[width=4.4cm]{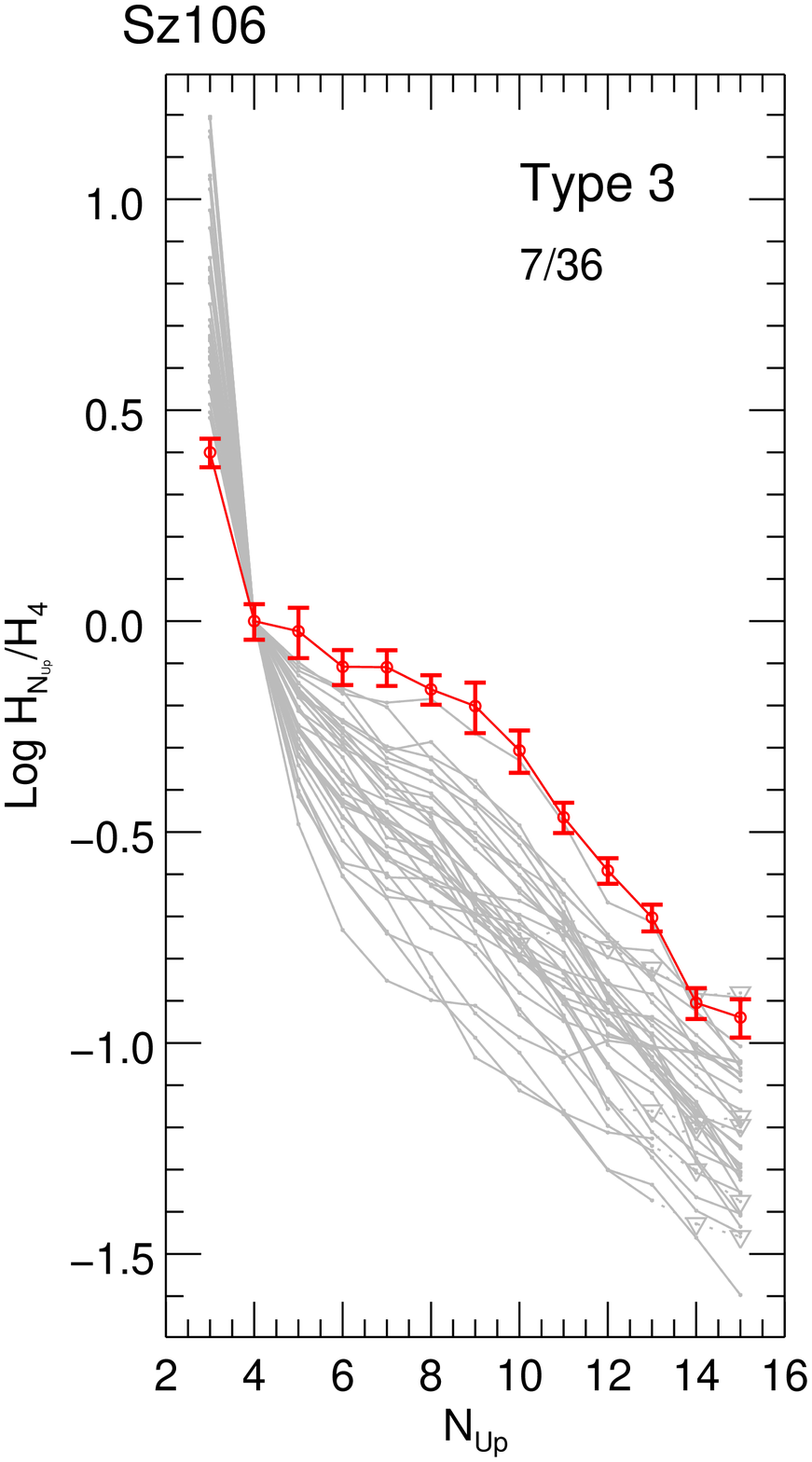}
\includegraphics[width=4.4cm]{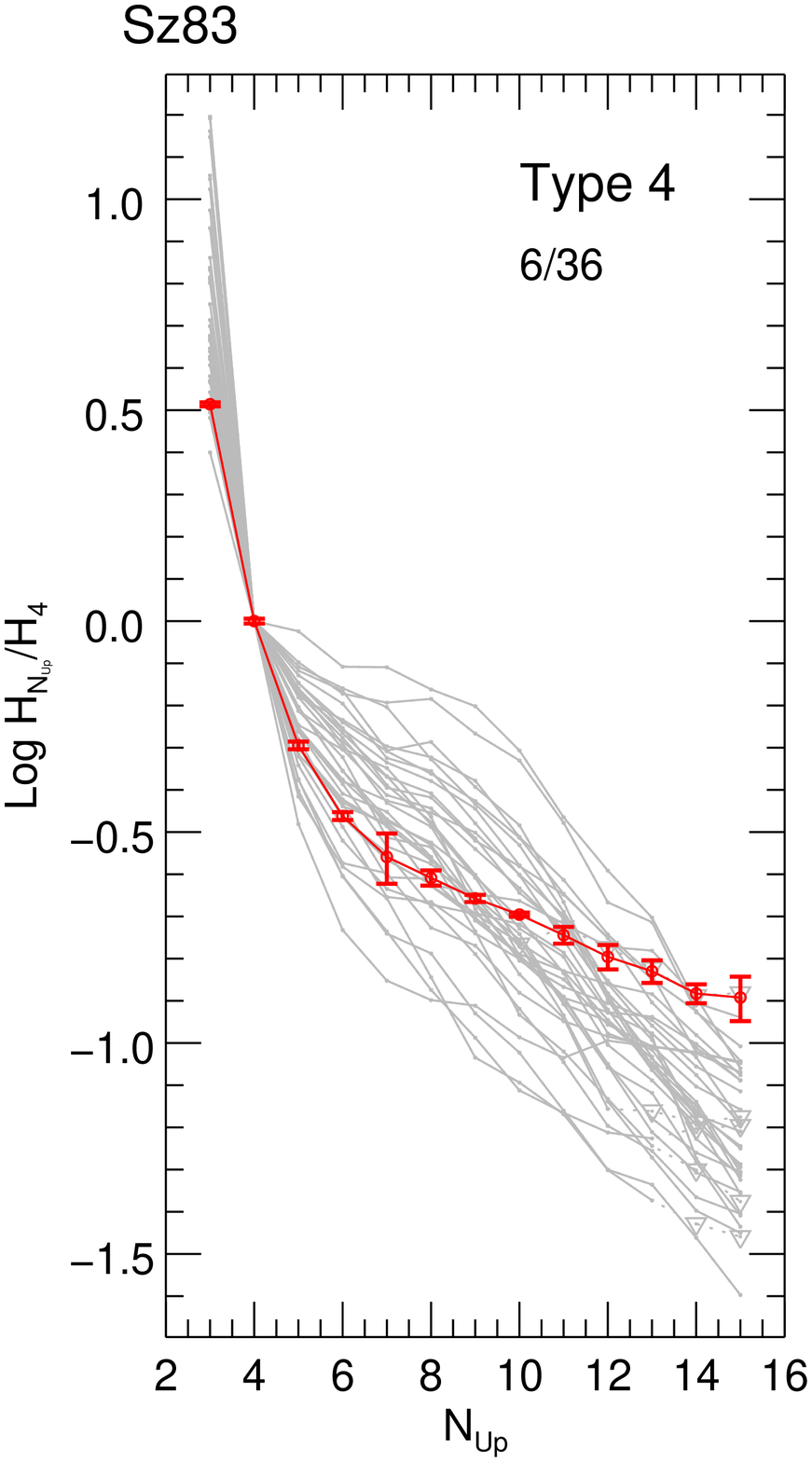}
\caption{\label{fig:decs:types} Four main Balmer decrement shapes observed in our sample. From left to right: type 1 (\textit{curved}, ParLup 3-4), type 2 (\textit{straight}, Sz100), type 3 (\textit{bumpy}, Sz106), and type 4 (\textit{L-shape}, Sz83). Each type is highlighted in red against the decrements observed for all other stars (in grey).} 
\end{figure*}

\begin{figure*}[!t]
\centering
\includegraphics[width=4.4cm]{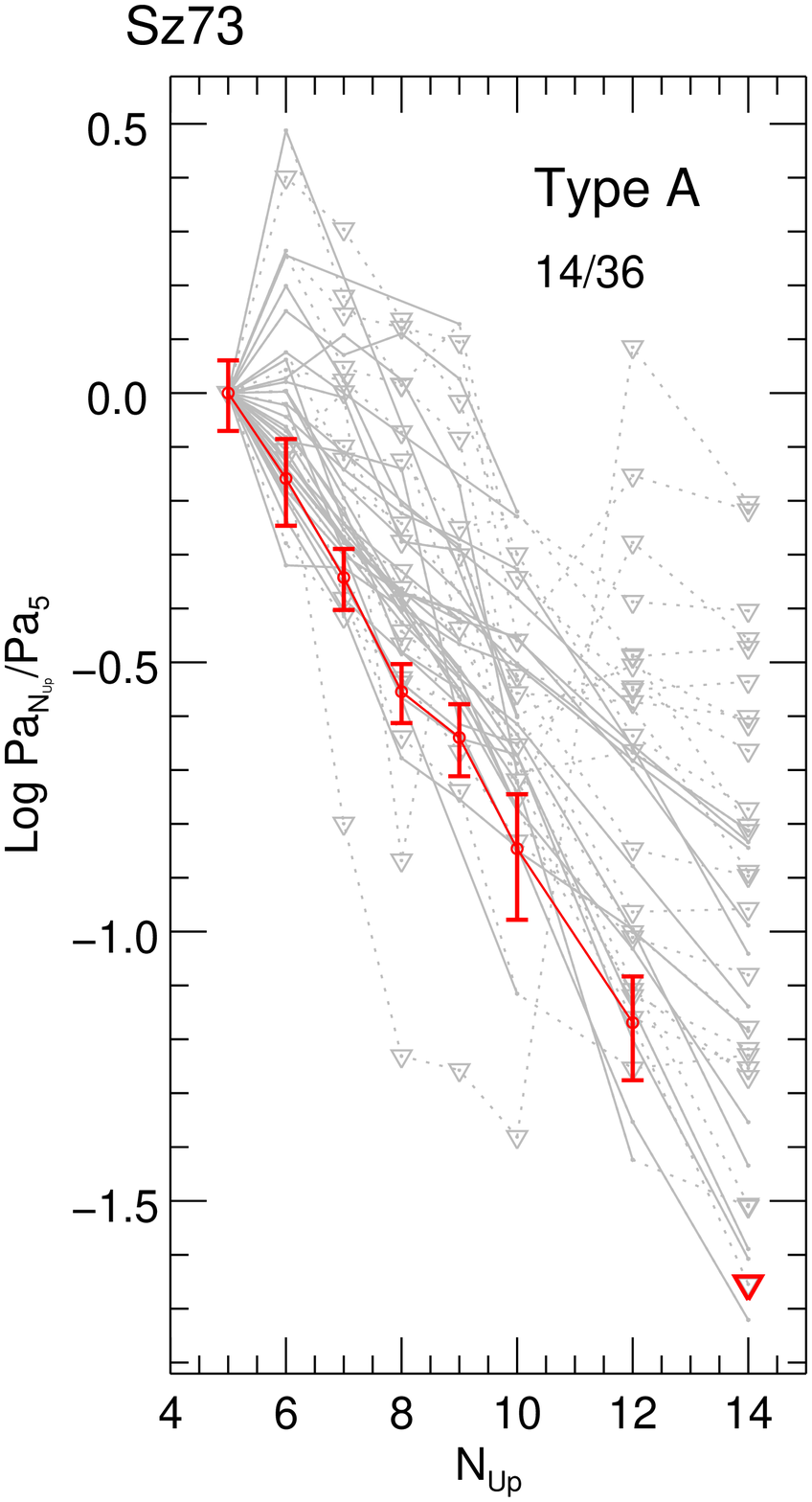}
\includegraphics[width=4.4cm]{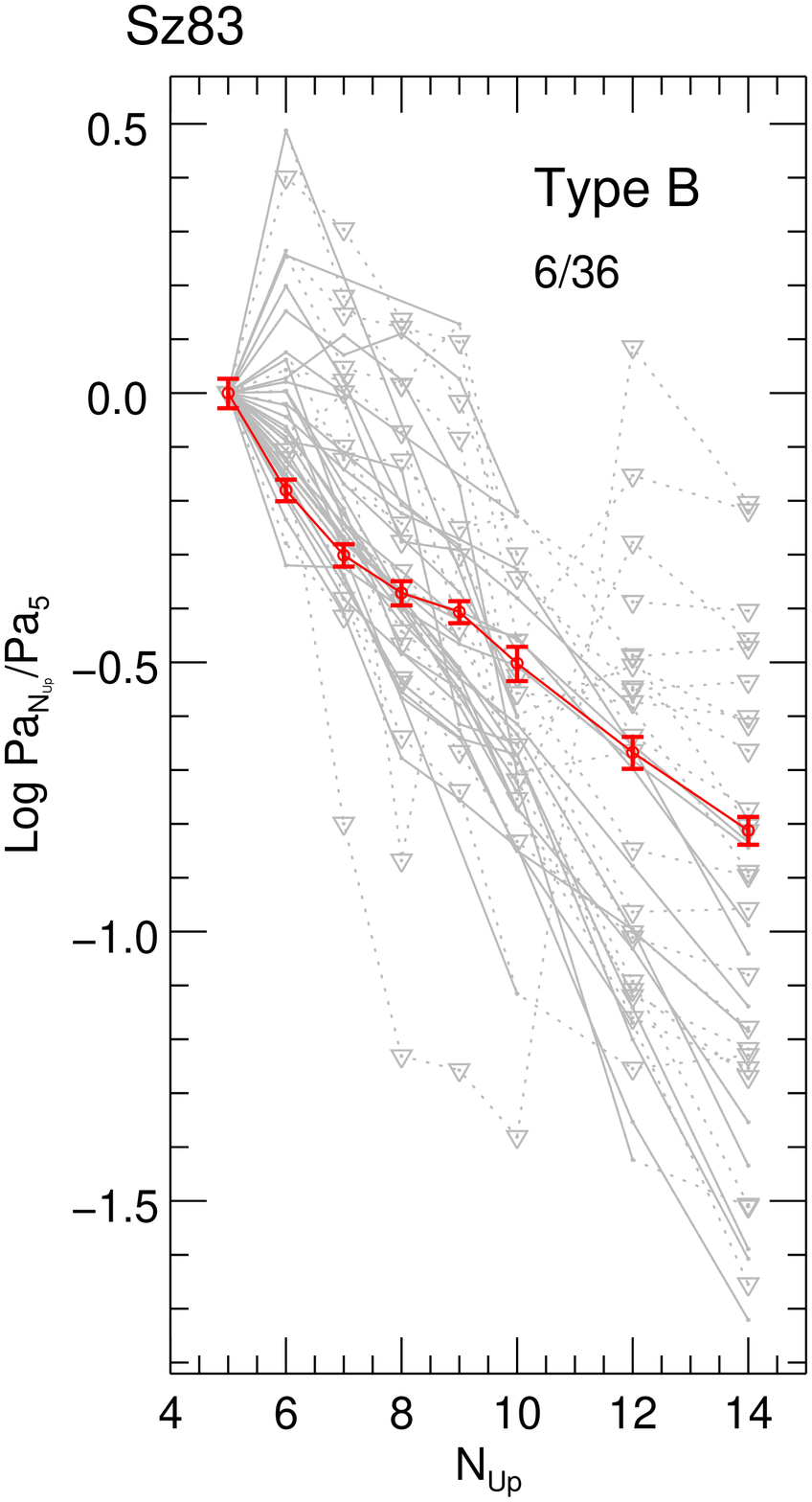}
\includegraphics[width=4.4cm]{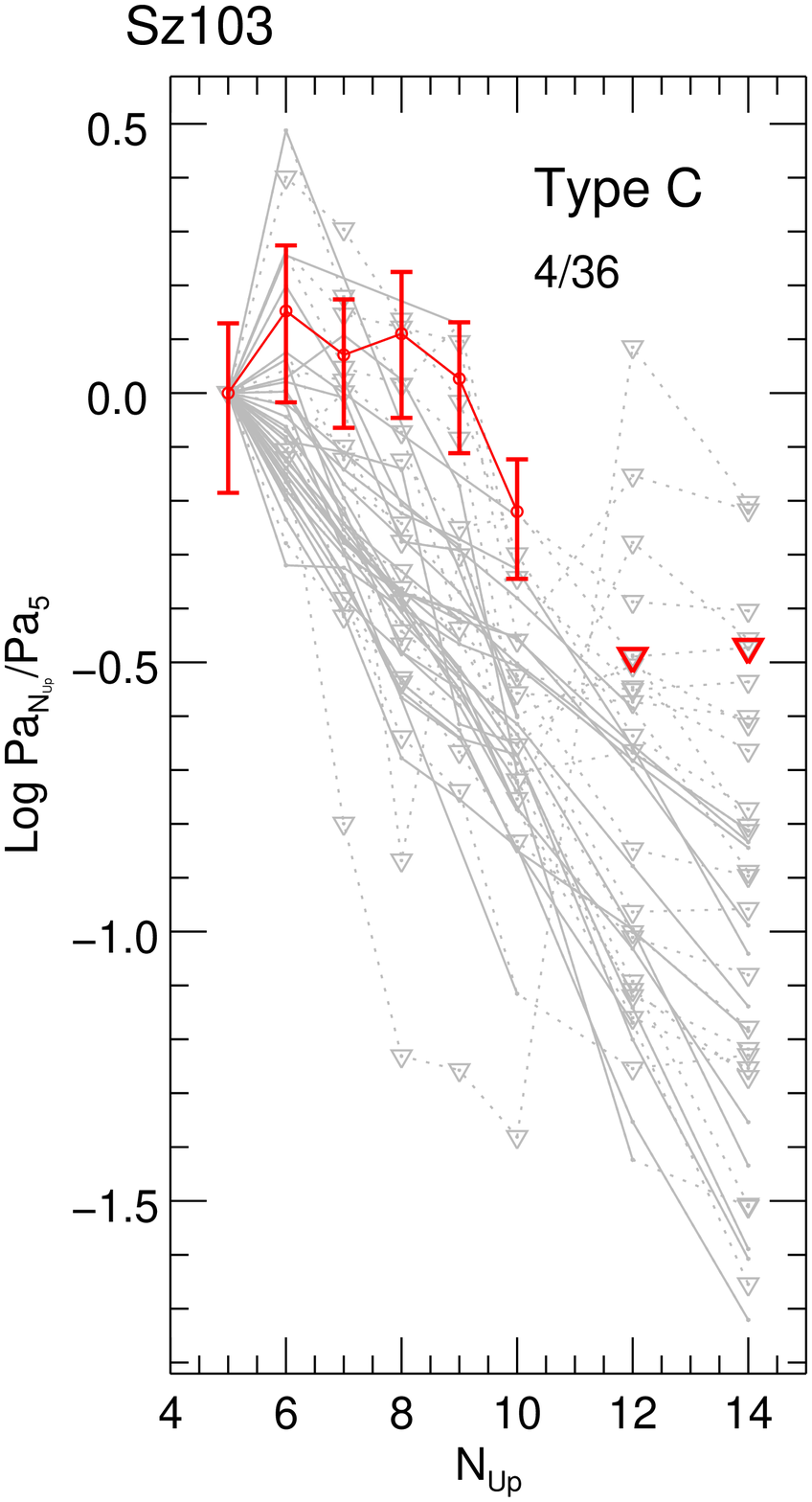}
\caption{\label{fig:decs:pa:types} Three main Paschen decrement shapes observed in the sample. From left to right: type A (Sz73), type B (Sz83), and type C (Sz103).}
\end{figure*}

\begin{table*}[!t]
\begin{small}

\begin{center}
\caption{\label{tab:main} Main source properties and \hi\ emission properties for the sources of the Lupus sample.}
\vspace*{0.2cm}
\begin{tabular}{l|cc|cccc|cc|c}
\hline
\hline
Source & RA(J2000) & DEC(J2000) & \multicolumn{4}{c|}{Main properties} & \multicolumn{2}{c|}{H lines} & Pa lines \\
   & & & $T_\mathrm{eff}$ (K) & $L_\mathrm{*}$ (L$_\odot$) & $M_\mathrm{*}$ (L$_\odot$) & Log $\dot{M}_\mathrm{acc}$ (M$_\odot$/yr) & prof.\tablefootmark{a} & dec.\tablefootmark{b} & dec.\tablefootmark{c}\\

\hline
Sz66                         &15:39:28.28 & $-$34:46:18.0 & 3415. &  0.200 &   0.45 &    -8.73                           &   MP  &   2           &B\\
AKC2006-19                   &15:44:57.90 & $-$34:23:39.5 & 3125. &  0.016 &   0.10 &    -10.85                          &   NS  &   2           &...\\
Sz69                         &15:45:17.42 & $-$34:18:28.5 & 3197. &  0.088 &   0.20 &    -9.50                           &   MP  &   4           &B\\
Sz71                         &15:46:44.73 & $-$34:30:35.5 & 3632. &  0.309 &   0.62 &    -9.23                           &   W   &   2           &A\\
Sz72                         &15:47:50.63 & $-$35:28:35.4 & 3560. &  0.252 &   0.45 &    -8.73                           &   MP  &   4           &B\\
Sz73                         &15:47:56.94 & $-$35:14:34.8 & 4060. &  0.419 &   1.00 &    -8.26                           &   MP  &   2           &A\\
Sz74                         &15:48:05.23 & $-$35:15:52.8 & 3342. &  1.043 &   0.50 &    -8.09                           &   MP  &   2           &?\\
Sz83                         &15:56:42.31 & $-$37:49:15.5 & 4060. &  1.313 &   1.15 &    -7.37                           &   W   &   4           &B\\
Sz84                         &15:58:02.53 & $-$37:36:02.7 & 3125. &  0.122 &   0.17 &    -9.24                           &   MP  &   2           &A\\  
Sz130                        &16:00:31.05 & $-$41:43:37.2 & 3560. &  0.160 &   0.45 &    -9.23                           &   W   &   2           &A\\
Sz88A                        &16:07:00.54 & $-$39:02:19.3 & 3850. &  0.488 &   0.85 &    -8.31                           &   W   &   4           &B\\
Sz88B                        &16:07:00.62 & $-$39:02:18.1 & 3197. &  0.118 &   0.20 &    -9.74                           &   W   &   4           &?\\
Sz91                         &16:07:11.61 & $-$39:03:47.1 & 3705. &  0.311 &   0.62 &    -8.85                           &   MP  &   2           &A\\
Lup713                       &16:07:37.72 & $-$39:21:38.8 & 3057. &  0.020 &   0.08 &    -10.08                          &   MP  &   2           &A\\
Lup604s                      &16:08:00.20 & $-$39:02:59.7 & 3057. &  0.057 &   0.11 &    -10.21                          &   NS  &   2           &A\\
Sz97                         &16:08:21.79 & $-$39:04:21.5 & 3270. &  0.169 &   0.25 &    -9.56                           &   W   &   3           &C\\  
Sz99                         &16:08:24.04 & $-$39:05:49.4 & 3270. &  0.074 &   0.17 &    -9.27                           &   W   &   3           &A\\
Sz100                        &16:08:25.76 & $-$39:06:01.1 & 3057. &  0.169 &   0.17 &    -9.47                           &   NS  &   2           &?\\
Sz103                        &16:08:30.26 & $-$39:06:11.1 & 3270. &  0.188 &   0.25 &    -9.04                           &   MP  &   3           &C\\
Sz104                        &16:08:30.81 & $-$39:05:48.8 & 3125. &  0.102 &   0.15 &    -9.72                           &   NS  &   2           &?\\
Lup706\tablefootmark{d}      &16:08:37.30 & $-$39:23:10.8 & 2795. &  0.003 &   0.06 &    -11.63 (-10.13)\tablefootmark{e}&   MP  &   1           &C\\
Sz106\tablefootmark{d}       &16:08:39.76 & $-$39:06:25.3 & 3777. &  0.098 &   0.62 &    -9.83 (-8.92)\tablefootmark{e}  &   MP  &   3           &C\\
ParLup3-3                    &16:08:49.40 & $-$39:05:39.3 & 3270. &  0.240 &   0.25 &    -9.49                           &   NS  &   2           &...\\
ParLup3-4\tablefootmark{d}   &16:08:51.43 & $-$39:05:30.4 & 3197. &  0.003 &   0.13 &    -11.37 (-9.27)\tablefootmark{e} &   W   &   1           &A\\
Sz110                        &16:08:51.57 & $-$39:03:17.7 & 3270. &  0.276 &   0.35 &    -8.73                           &   W   &   3           &A\\
Sz111                        &16:08:54.69 & $-$39:37:43.1 & 3705. &  0.330 &   0.75 &    -9.32                           &   MP  &   3           &A\\
Sz112                        &16:08:55.52 & $-$39:02:33.9 & 3125. &  0.191 &   0.25 &    -9.81                           &   NS  &   2           &...\\
Sz113                        &16:08:57.80 & $-$39:02:22.7 & 3197. &  0.064 &   0.17 &    -8.80                           &   W   &   4           &B\\
2MASSJ16085953-3856275       &16:08:59.53 & $-$38:56:27.6 & 2600. &  0.009 &   0.03 &    -10.80                          &   NS  &   2           &...\\
SSTc2d160901.4-392512        &16:09:01.40 & $-$39:25:11.9 & 3270. &  0.148 &   0.20 &    -9.59                           &   MP  &   3           &?\\
Sz114                        &16:09:01.84 & $-$39:05:12.5 & 3175. &  0.312 &   0.30 &    -9.11                           &   NS  &   2           &?\\
Sz115                        &16:09:06.21 & $-$39:08:51.8 & 3197. &  0.175 &   0.17 &    -9.19                           &   NS  &   2           &...\\
Lup818s                      &16:09:56.29 & $-$38:59:51.7 & 2990. &  0.025 &   0.08 &    -10.63                          &   NS  &   2           &?\\
Sz123A                       &16:10:51.34 & $-$38:53:14.6 & 3705. &  0.203 &   0.60 &    -8.93                           &   MP  &   2           &A\\
Sz123B\tablefootmark{d}      &16:10:51.31 & $-$38:53:12.8 & 3560. &  0.051 &   0.50 &    -10.03 (-8.86)\tablefootmark{e} &   MP  &   1           &A\\
SST-Lup3-1                   &16:11:59.81 & $-$38:23:38.5 & 3125. &  0.059 &   0.13 &    -10.17                          &   NS  &   2           &A\\
\hline
\end{tabular}
\end{center}
\tablefoot{
Source parameters from \citet{alcala14}.\\
\tablefoottext{a}{Profile type (of the \hb\ line): NS, narrow symmetric (11/36); W, wide (10/36); MP, multi-peaked (15/36).}\\
\tablefoottext{b}{Decrement type: 1-4 as in Fig.~\ref{fig:decs:types}.}\\
\tablefoottext{c}{Decrement type: A-C as in Fig.~\ref{fig:decs:pa:types}. Unclear decrement types are marked with a "?".}\\
\tablefoottext{d}{Sub-luminous source.}\\
\tablefoottext{e}{\macc\ value after correction for the obscuration factor \citep[see][]{alcala14}.}
}
\end{small}
\end{table*}

\subsection{Balmer decrements}
\label{sec:decs}

The very high quality of the Balmer line spectra in conjunction with the simultaneous measurement of all lines of the series provided by X-Shooter allows us to perform a detailed and unprecedented study of the Balmer decrements in a large and homogeneous sample of T Tauri stars.

In computing the decrements, we decided not to use H$\alpha$ as reference, because this line is the most sensitive to opacity effects and most subject to contamination from chromospheric emission \citep[e.g.][]{manara13}. 
The decrements of the targets relative to the H$\beta$ (H4) line and up to the H15 line are shown in the logarithmic plots of Fig.~\ref{fig:decs:h} of the appendix. For every star, the decrement is highlighted in red against the ensemble of all decrements of the sample, which are traced in grey.
The error associated with each decrement point (i.e. line ratio) was obtained by propagating the errors on the line fluxes of the two lines.

Despite the decrement shapes vary with continuity within this ensemble, we can define four major decrement shapes or types, which are summarised in Fig.~\ref{fig:decs:types}, by using a representative source for each type.
Type 1 (curved) decrements are observed in three objects only and present a fairly curved shape outlining the lower part of the decrement ensemble. Type 2 (straight) decrements, which are the most common (20/36), appear as basically straight lines in the logarithmic plot. Type 3 (bumpy) decrements (7/36) present a shape characterised by a wide bump and occupy the upper side of the decrement ensemble, opposite to type 1. Finally, in type 4 (L-shape) curves (6/36) the decrement slope is steep for the first lines of the series, but becomes much flatter for N$_{up} \geq 6$.

We have empirically defined the first three types on the basis of the value of the Log(H9/H4) ratio. This ratio is $< -0.80$ in type 1, between $-$0.80 and $-$0.45 in type 2, and $> -0.45$ in type 3, respectively. Type 4 decrements can instead be defined as those for which the slope of the decrement between the H6 and H14 lines, i.e. the slope of the line connecting the H6 and H14 points in the logarithmic plot of the decrements, 
is greater than -0.07.
The Balmer decrement types of the targets are reported in Table~\ref{tab:main}.

Since the shape of the decrement depends in principle on the chosen reference line, we explored the possibility of using lines with a higher N$_{up}$ as reference, but we obtained no significant variations from the schematic classification we have just detailed. 
The possible dependence of the different decrement shapes on both the line profiles and the source properties is discussed in the following sections.

\begin{figure*}[!t]
\centering
\includegraphics[width=6.cm]{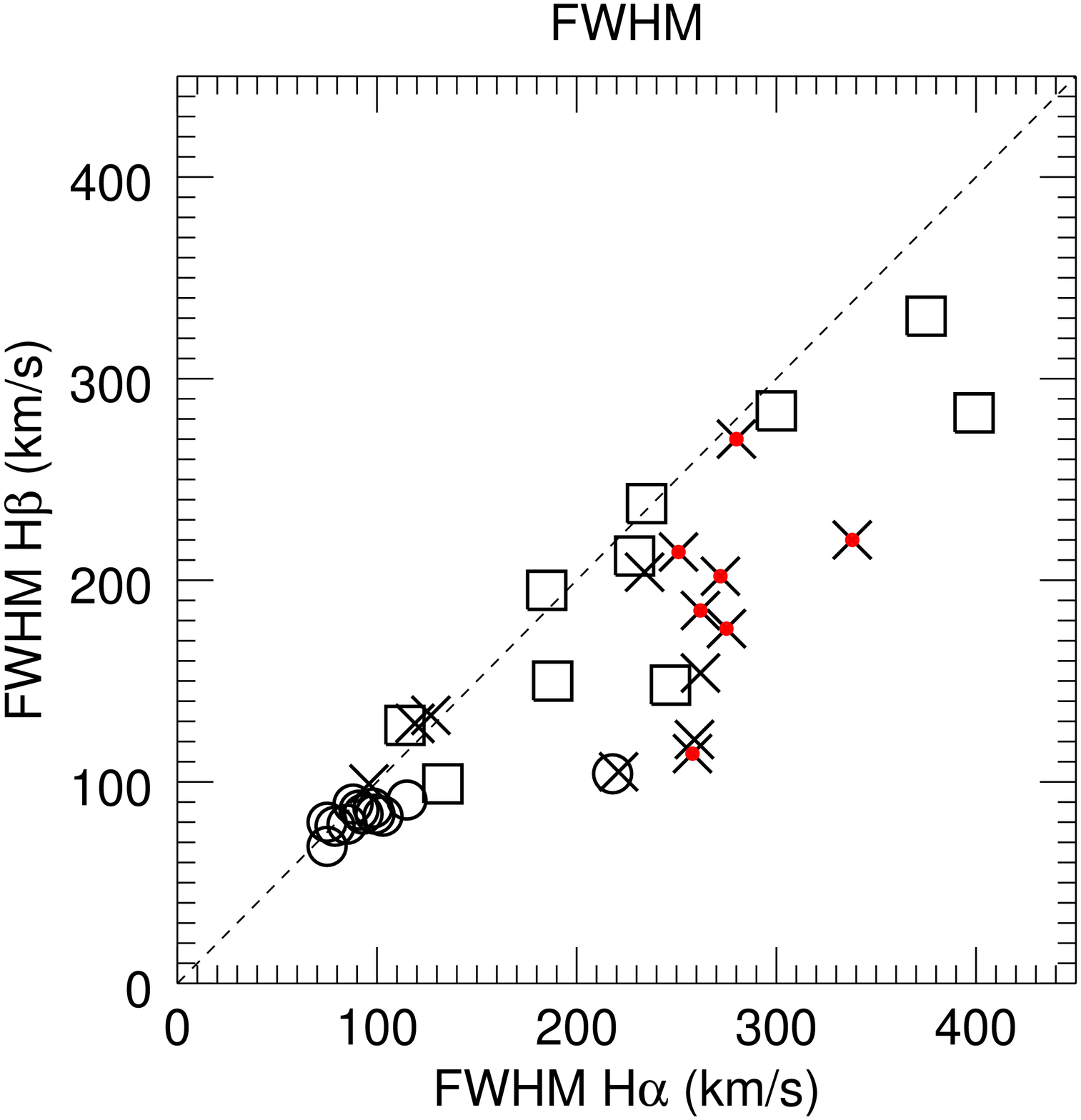}
\includegraphics[width=6.cm]{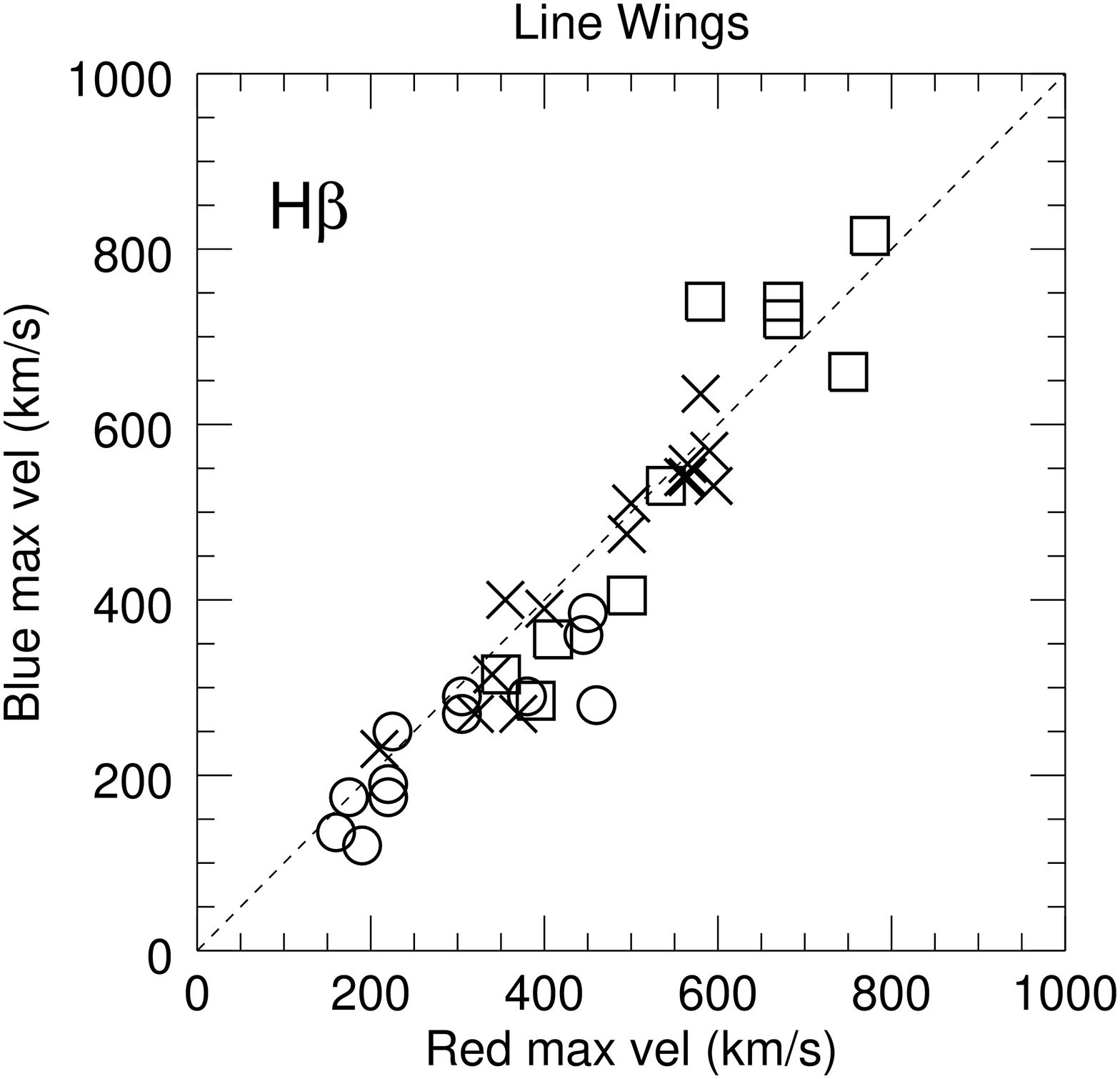}
\includegraphics[width=6.cm]{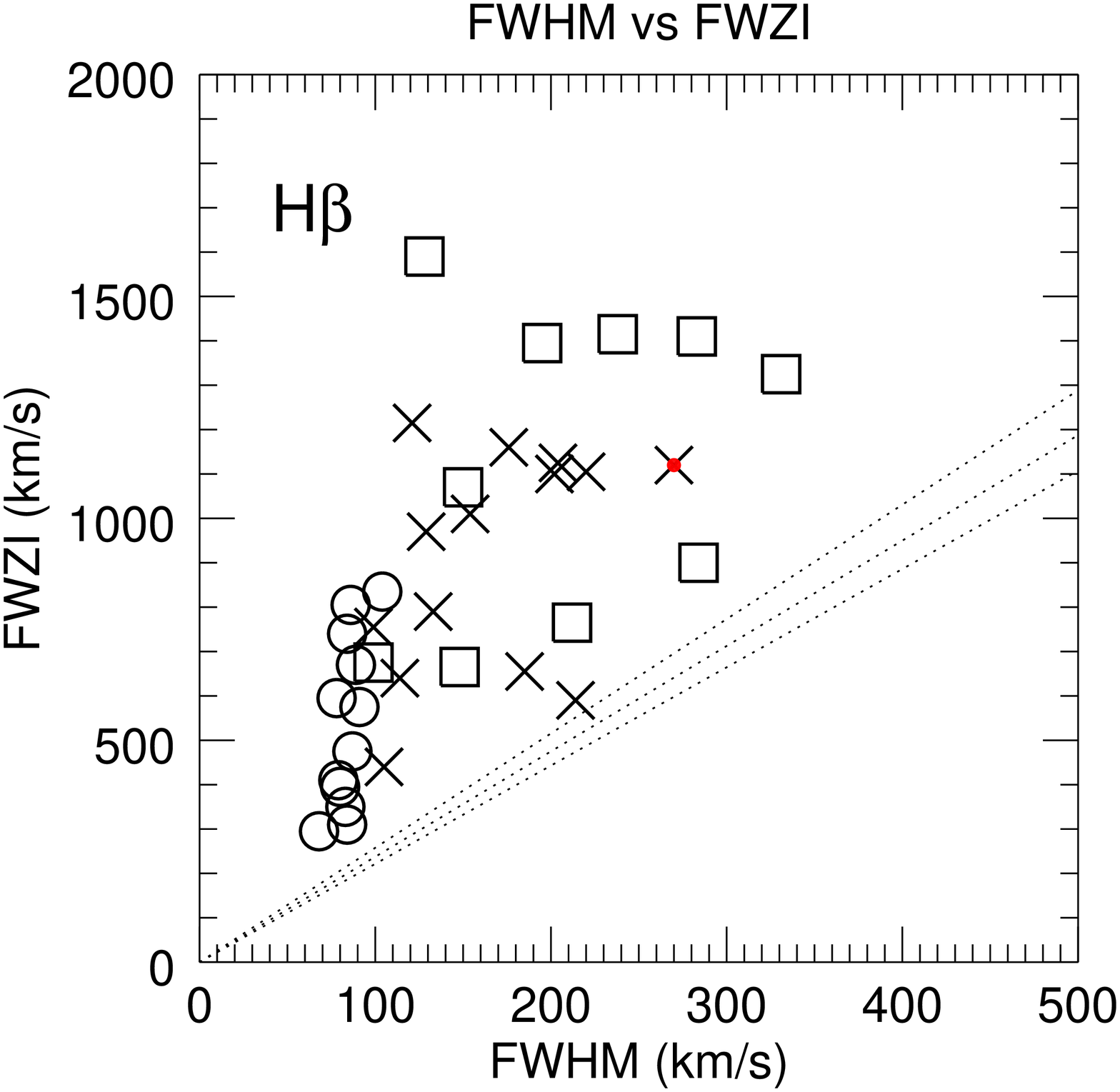}\\
\vspace*{-0.6cm}
\includegraphics[width=6.cm]{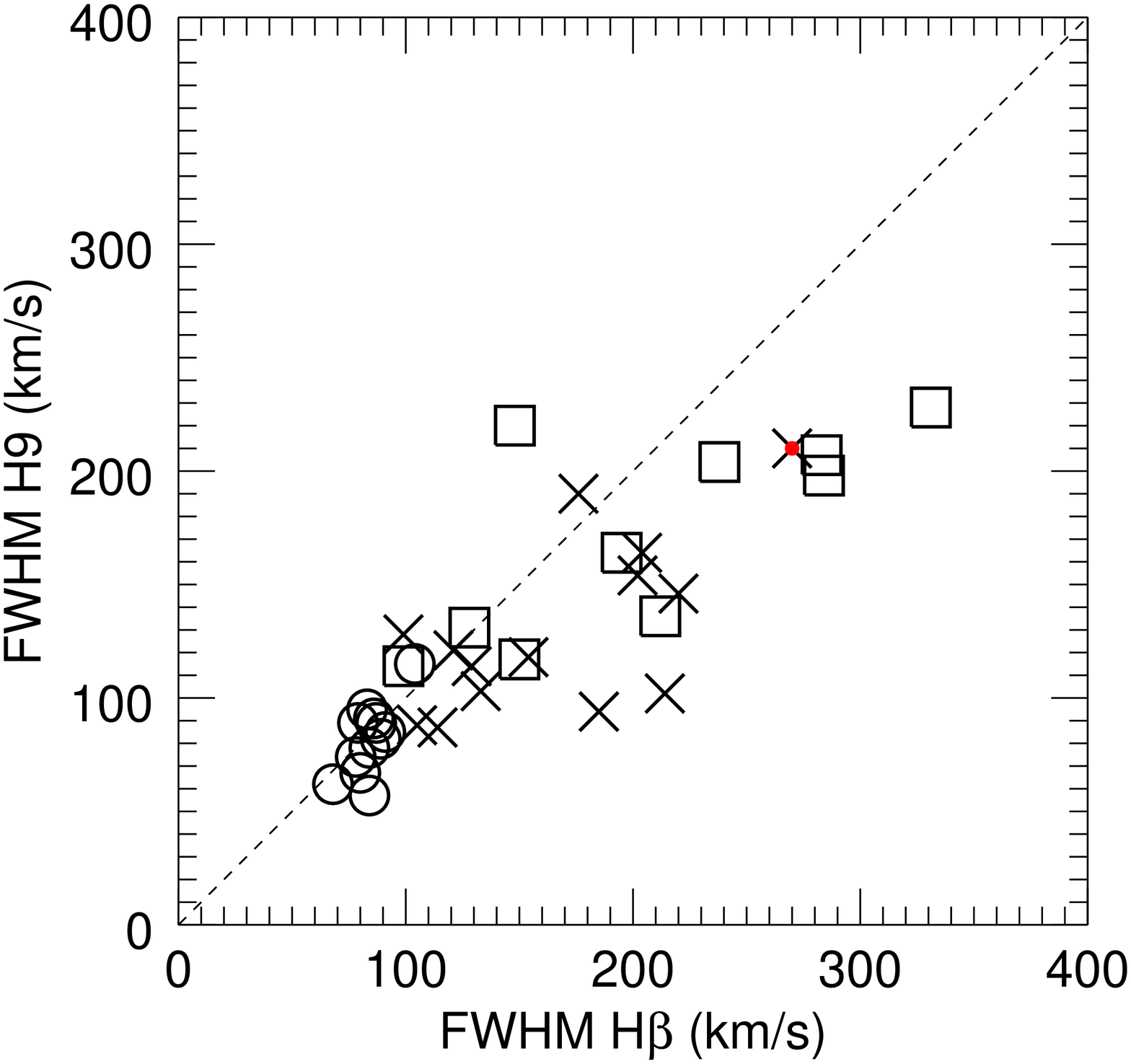}
\includegraphics[width=6.cm]{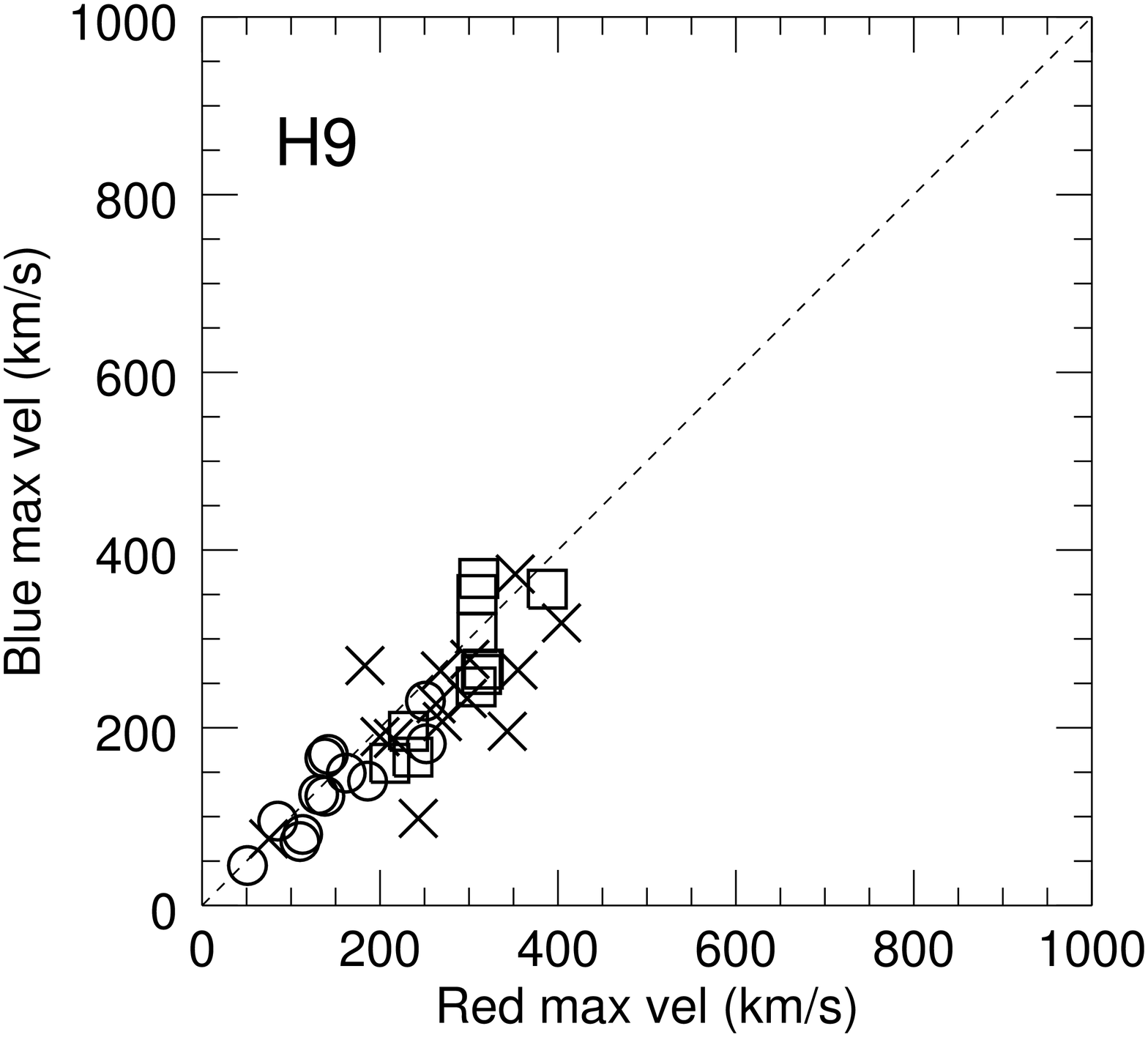}
\includegraphics[width=6.cm]{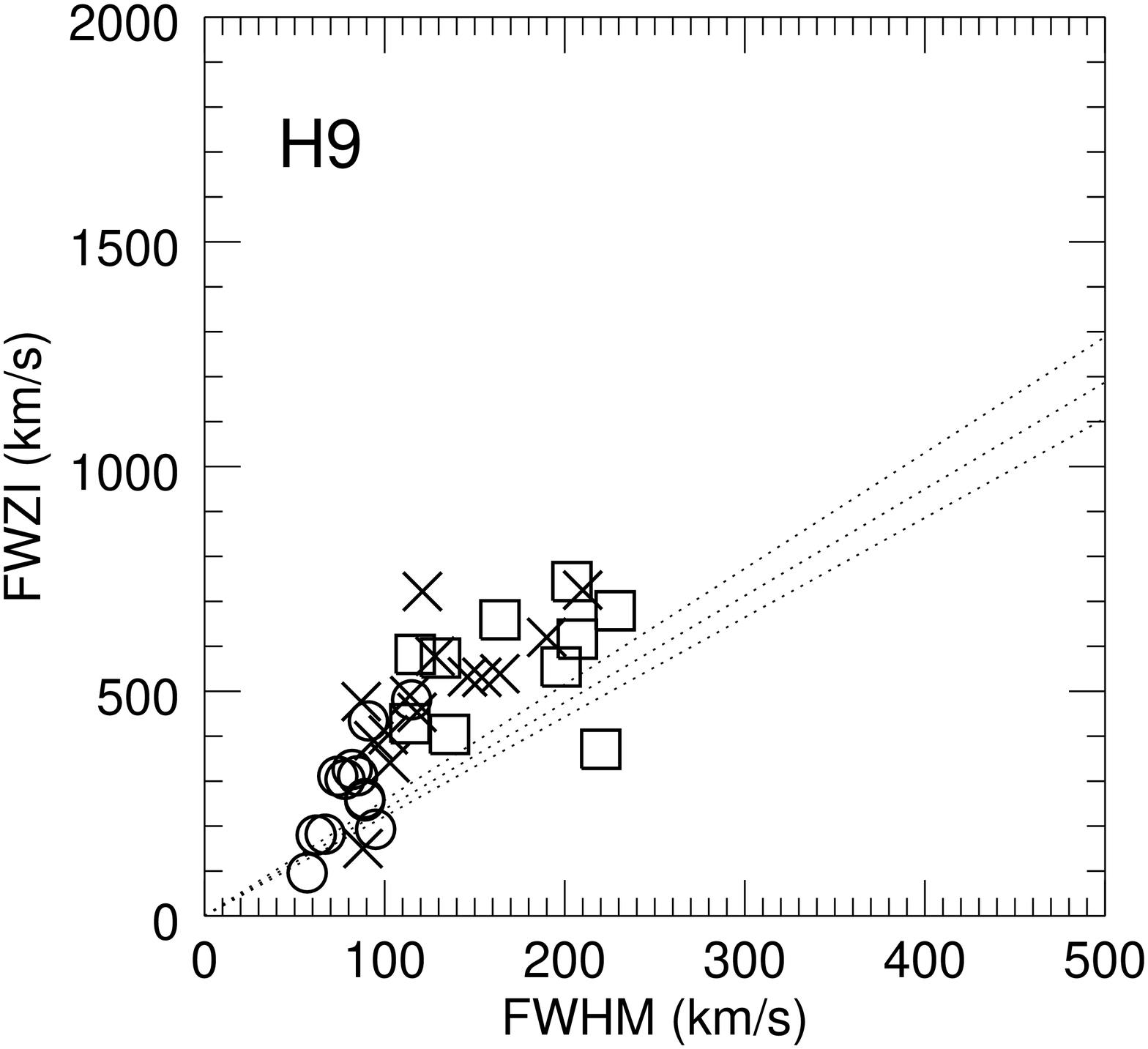}
\caption{\label{fig:lines:vel} Line widths and velocities for Balmer lines of the Lupus sample. \textit{Left panels:} FWHM of \hb\ versus FWHM of \ha\ (\textit{top}); FWHM of H9 versus FWHM of \hb\ (\textit{bottom}). \textit{Central panels:} red/blue wings extension in \hb\ (\textit{top}) and H9 (\textit{bottom}). \textit{Right panels:} FWZI versus FWHM for \hb\ (\textit{top}) and H9 (\textit{bottom}); dashed lines mark the expected correlation between the FWHM and FWZI (measured at 1/30, 1/50, and 1/100 of the peak) in a Gaussian profile.
The different symbols indicate the line profile typology (see Sect.~\ref{sec:lines:shapes} and Table 1: narrow symmetric (circles), wide symmetric (squares), and asymmetric or multi-peaked (crosses). Sources where the \ha\ or \hb\ present a particularly deep central absorption, which may introduce a strong bias in the line FWHM measurement (see text), are marked with a red circle.} 
\end{figure*}

\subsection{Paschen decrements}
\label{sec:paschen}

The Paschen decrements for each target relative to the Pa$\beta$ line and up to the Pa14 line are shown in Fig.~\ref{fig:decs:pa} of the appendix, using the same graphical representation employed for the Balmer decrements. 
The Pa 11 and Pa 13 lines were not considered since these lines are blended with the bright IR triplet \ion{Ca}{ii} lines.
Because of the lower quality of the data of the near-IR arm and the large number of upper limits it is more difficult to identify different decrement classes as done for the Balmer lines. 
We can however distinguish between three major shapes, which we classify as type A, B, and C, and are reported in Fig.~\ref{fig:decs:pa:types}. In the first two types, displayed by the majority of the objects, the decrement is roughly straight, steeper in type A and more gradual in type B. Considering the slope of the line connecting the Pa5 and Pa12 points in the logarithmic plot, we empirically define as type A those in which this slope is lesser or equal than -0.14. 
In type C stars the decrement shows a plateau or small bump at the position of the first lines of the series, which have fluxes equal or higher than Pa$\beta$. This type of decrement is clearly observed only in four sources (Lup706, Sz103, Sz106, Sz97). 
Other objects show possible hints for a similar shape but we cannot derive a firm classification for these targets owing to large error bars or upper limits.
The Paschen decrement class is given in Table~\ref{tab:main}.

\section{Empirical relations involving lines properties and source parameters}
\label{sec:relations}

\subsection{Line widths and velocities}

The full-widths of the \ha, \hb, and H9 lines are directly compared (\ha\ vs \hb\ and \hb\ vs H9) in the left panel of Fig.~\ref{fig:lines:vel}.
The \ha\ line is in general wider than \hb\ and, although this is a real effect in a few objects, we note that the presence of strong 
absorptions in the profile often lowers the intrinsic height of the line peak more for \ha\ than for \hb\ (differential opacity 
effects), thus leading to overestimating the FWHM. 
Using H9 as representative for high Balmer lines, the comparison with the \hb\ line shows that the width 
remains constant throughout the series in sources with narrow lines, whereas in the remaining objects we find a decrease of the FWHM in highest series 
lines, even without considering sources that show clear differential opacity effects.
This suggests an additional \hi\ emission component in objects with wider lines, which is observed mostly in lower Balmer 
lines and tends to disappear with increasing quantum number $N_{up}$.

To estimate the full width at zero intensity of the lines (FWZI) we considered the line width at the level of the fitted local continuum plus 1$\sigma$. In general, the S/N on the local continuum was better for the \hb\ than for \ha\ and H9, so that the determination of the maximum velocities might be slightly underestimated for these latter lines. The blue and red wings of the H lines have on average a similar extent (see central panels of Fig.~\ref{fig:lines:vel}).
The maximum velocities measured on the red and blue side of the \hb\ line span from 200 up to about 800 km/s, which means a FWZI in the range 400-1600 km/s, while for the H9 line the wing velocities go from about 50 to 400 km/s.

The comparison between FWHM and FWZI (right panels of Fig.~\ref{fig:lines:vel} clearly shows that the majority of objects have \ha\ and \hb\ profiles that significantly differ from a Gaussian, having in most cases very extended wings that indicate gas moving at large velocities with respect to the star. A high-velocity wind contribution or broadening from line damping \citep[][]{muzerolle01} might explain these very extended wings, especially in optically thick lines.

\subsection{Line widths and mass accretion rate}

\begin{figure}[t]
\centering
\includegraphics[width=9cm]{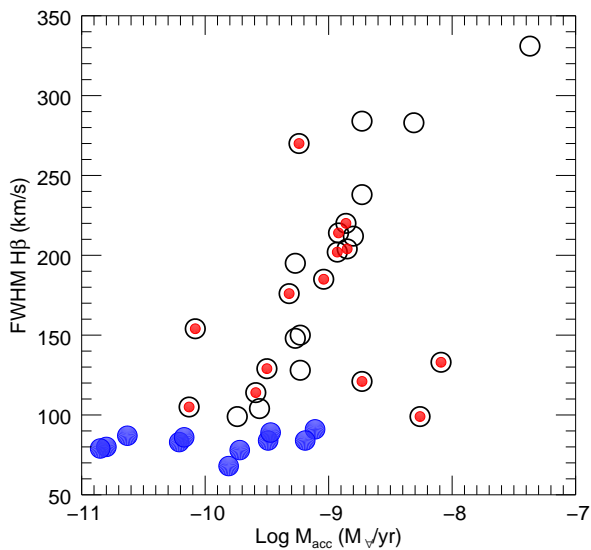}
\caption{\label{fig:macc_width} FWHM of the \hb\ line as a function of the \macc\ of the objects. The blue points indicate the narrow symmetric line sources (NS) while multi-peaked line objects (MP) are flagged with a red point. Accretion rates of the sub-luminous objects have been corrected for their obscuration factor (see Alcal\'a et al. 2014 and Table~\ref{tab:main}).} 
\end{figure}

In Fig.~\ref{fig:macc_width} we plot the FWHM of the \hb\ line versus the mass accretion rate, using blue points to indicate the narrow symmetric line sources (NS) and flagging with a red point the multi-peaked objects (MP). Indeed, in these latter the strong absorptions may alter the line profile in such a way that the measured FWHM is heavily biased compared to the intrinsic width the line would have without optical depth effects.
Taking into consideration this possible effect, the plot confirms that line width is globally related with the accretion activity of the objects, in agreement with known empirical relations connecting for example the width of the \ha\ to the accretion rate \citep[e.g.][]{natta06}.
 
It is also evident that the narrow symmetric line sources are indeed associated with the lower mass accretion rates (Log \macc $< 9$\msunyr), although their line width seems independent on the \macc\ value.

\subsection{Balmer decrements and source parameters}
\label{sec:decs:params}

We searched for possible connections between the observed decrement shapes and the main stellar properties of the targets (namely the stellar mass \mstar, luminosity \lstar, radius \rstar, effective temperature \teff, and visual extinction \av), but we could determine no clear correlation for any of the mentioned parameters. 

Considering the accretion properties of the sources, we find instead a tentative connection of the decrement shape with the mass accretion rate, in the sense that the strongest accretors in the sample show decrement shapes that tend to type 4. Evidence for this can be seen in Fig.~\ref{fig:decs:properties} where the \hi\ decrement curves are displayed separately for each bin of the mass accretion rate distribution histogram, with the strongest accretors indicated (rightmost panel).

\begin{figure}[!t]
\centering
\includegraphics[width=8cm]{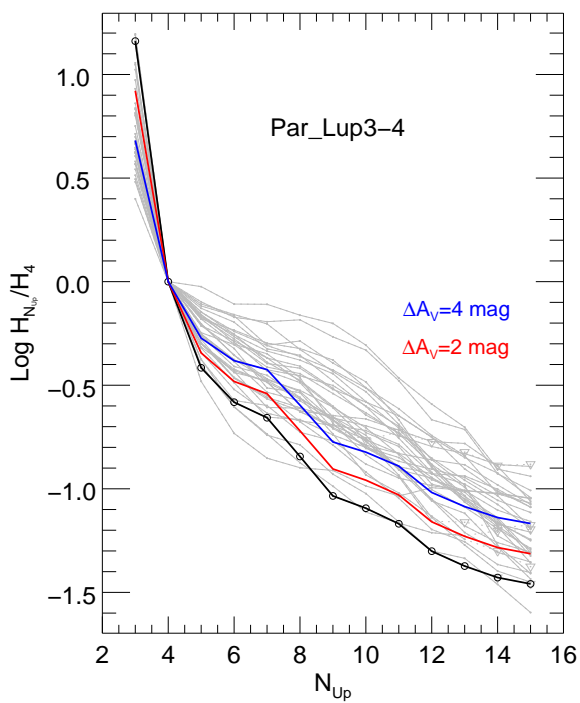}
\caption{\label{fig:effect_extinction} Effect of an additional amount of visual extinction on the type 1 decrement of Par Lup 3-4 (black line). Resulting decrements with $\Delta A_v$=2 mag (red) and $\Delta A_v$=4 mag (blue) are shown.} 
\end{figure}

\begin{figure*}[t]
\centering
\includegraphics[width=4.5cm]{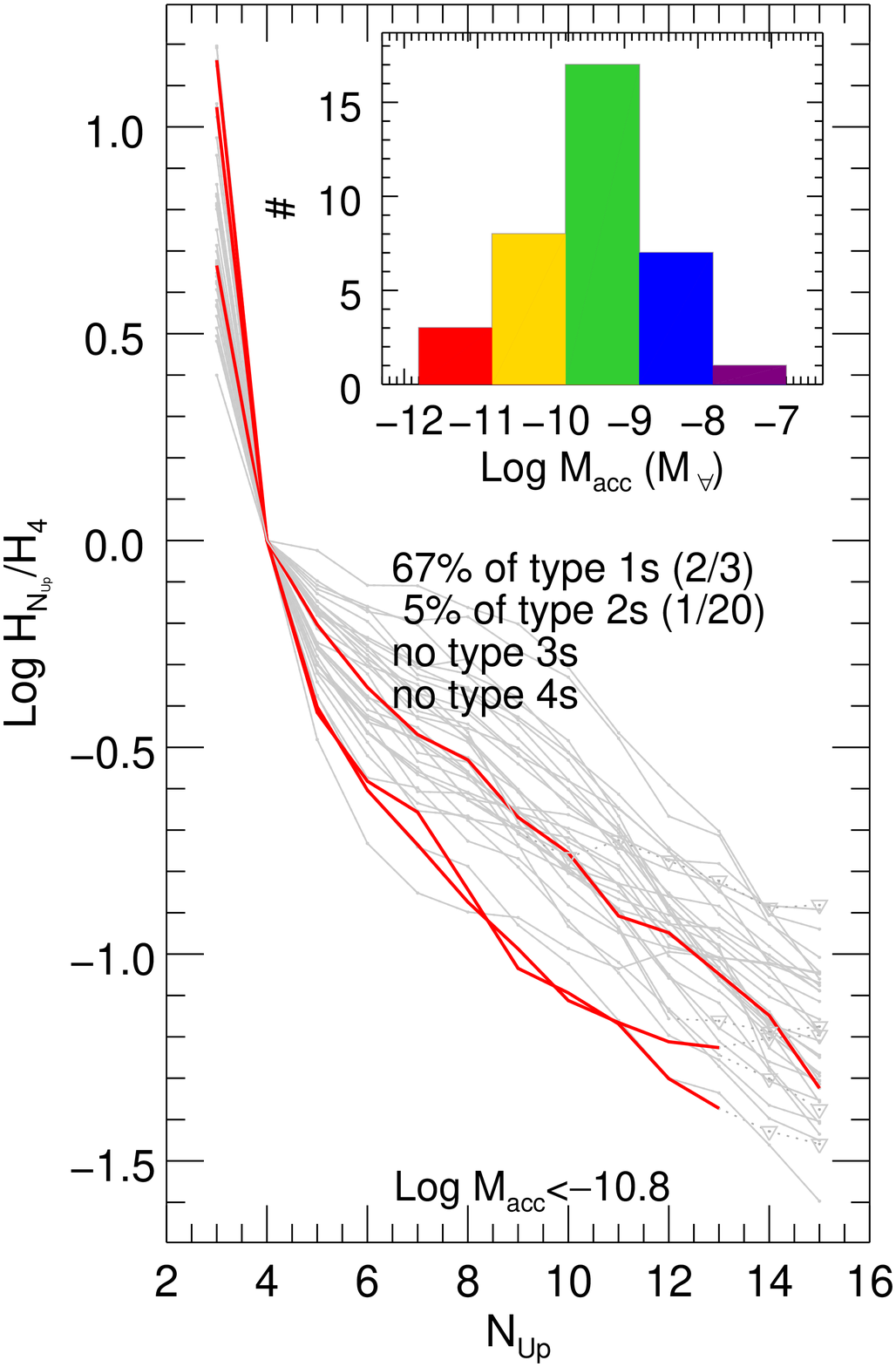}
\includegraphics[width=4.5cm]{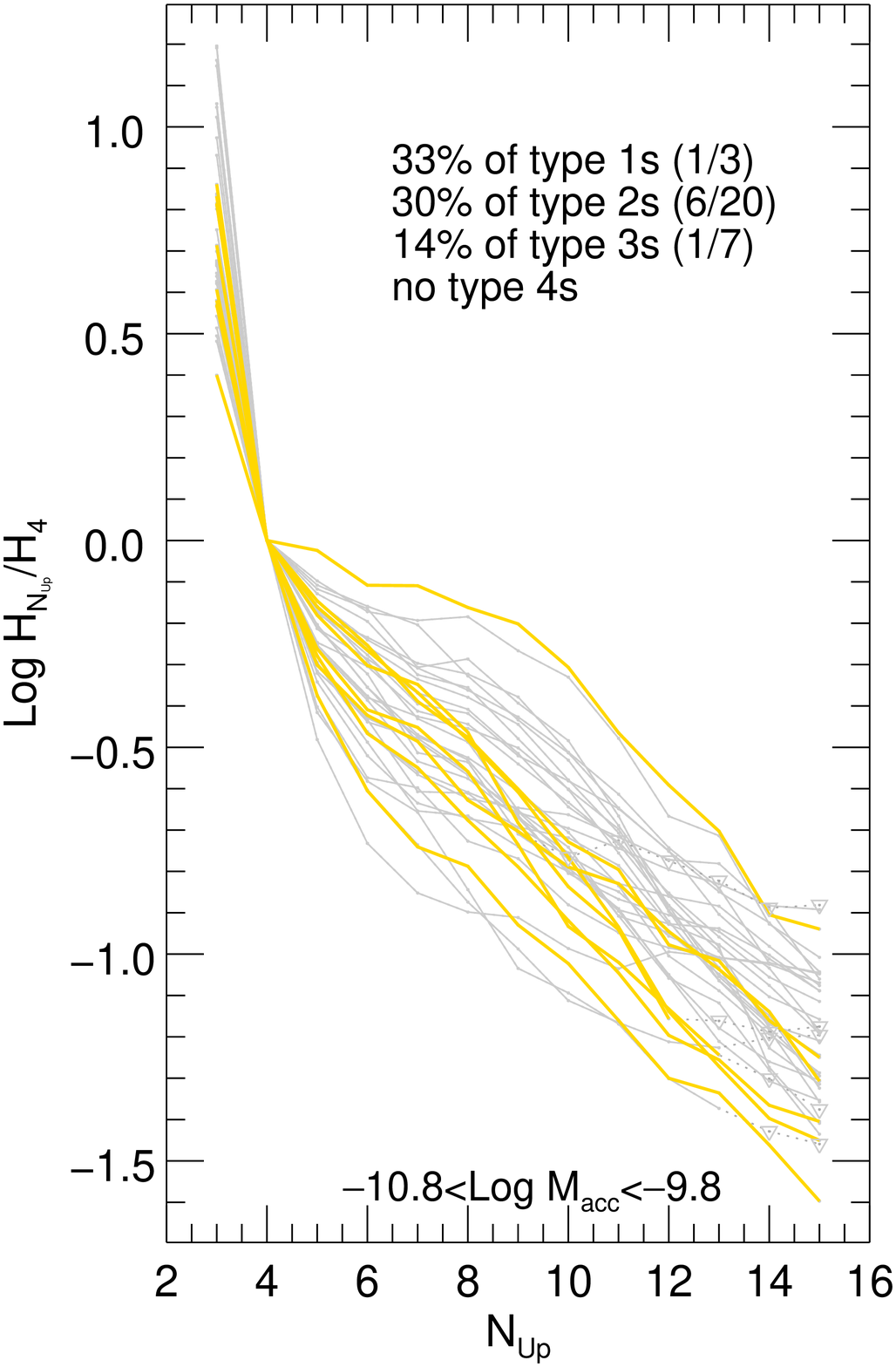}
\includegraphics[width=4.5cm]{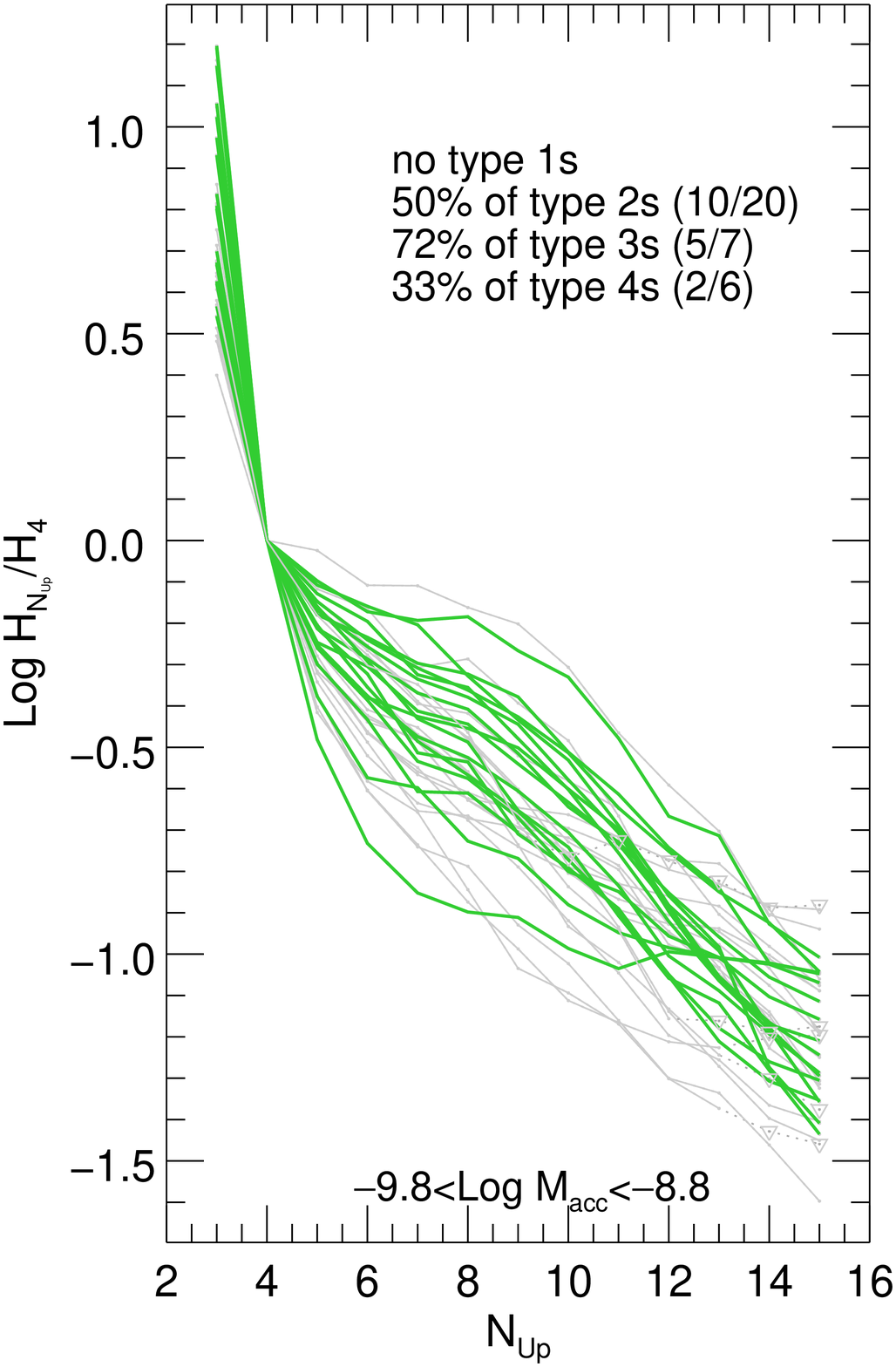}
\includegraphics[width=4.5cm]{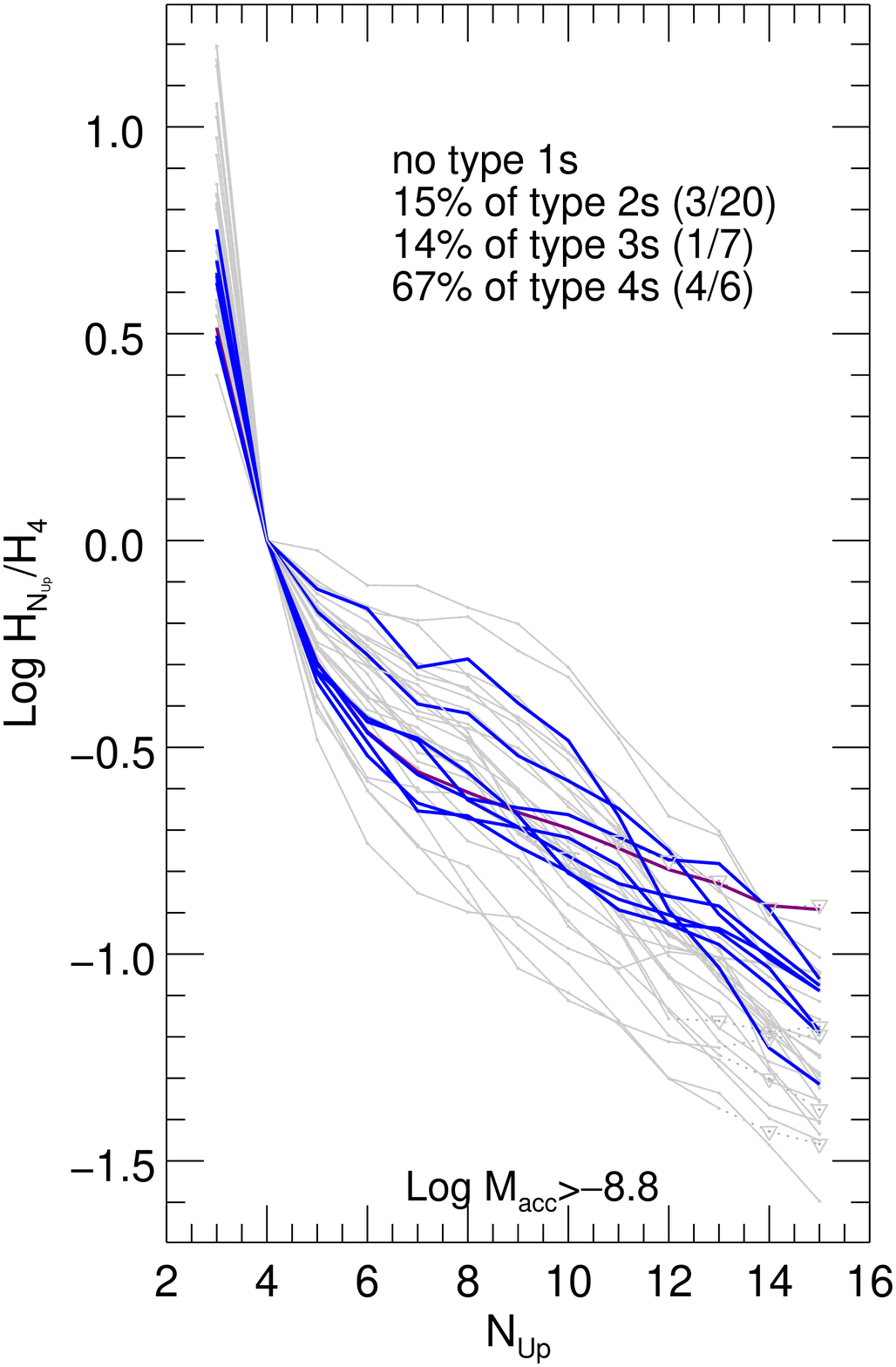}
\caption{\label{fig:decs:properties} Decrements and mass accretion rate. The \macc\ distribution histogram is shown in the first panel; the decrements relative to each bin are displayed using the same colour code as in the histogram.} 
\end{figure*}

An important point to remark is that the three sources displaying a type 1 decrement (Sz106, Par-Lup 3-3, Lup706) have been identified as sub-luminous objects by A14, who provide evidence that the sub-luminosity is due to obscuration of the star by the circumstellar disc, as a result of an almost edge-on view. This finding strongly suggests a direct connection between the edge-on line-of-sight and the observed type 1 decrement.
It is likely that in an edge-on geometry also the \hi\ line emission region is partly obscured by the disc. In this scenario the type 1 decrement could be due to the fact that we are observing only the outer parts of the emission region, which might be characterised by a different extinction. 
The derived value of \av\ for these sources is equal to zero in A14, so that the effect of the disc obscuration has been interpreted by the authors in terms of an obscuration factor (see Sect. 7.4 of A14). This is the factor that one needs to apply to bring the luminosity of the sources into agreement
with that of other objects of the sample with the same spectral type.
This approach corresponds to assuming a grey (i.e. wavelength-independent) extinction.
It is plausible that the approximation of grey extinction might be less accurate for the \hi\ emission region, which can be in principle more extended than the disc of the star from our observing point. If there is an amount of chromatic extinction that is currently not accounted for in the \hi\ emission region, the type 1 decrements observed should be corrected to consider it. Since the net effect of adding the extinction is to raise the decrement curves, if we consider an \av\ of a few magnitudes the observed type 1 decrements are modified in such a way that they appear as type 2 decrements.
This effect is depicted in Fig.~\ref{fig:effect_extinction}, where the type 1 decrement of Par Lup 3-4 is reddened by considering an extinction $A_v$=2 mag and $A_v$=4 mag, using the \citet{weingartner01} exinction law adopted also by A14. 

\subsection{Balmer decrements and line profiles}
\label{sec:decs:lines}

The different combinations of Balmer decrement and profile type that we observe are reported in the top panel of Table~\ref{tab:type_occ}.
The most evident relationship between decrement shape and profiles is that all narrow symmetric line sources (11 objects) display a type 2 (\textit{straight}) decrement, although not all type 2 decrements are associated with objects showing narrow symmetric profiles (e.g. Sz123A, Sz74, Sz71). 
This also means that sources with decrements other than type 2 always present \hi\ lines with wide or asymmetric profiles.

Four objects with a type 4 decrement display wide (W) profiles. In particular, Sz83, Sz88A, and Sz72 are the sources with the widest lines (FWHM(\hb) $\sim$ 300 km/s) in the sample. 
The line profiles in type 4 targets remain basically the same throughout all lines of the series. 

This is not the case in type 3 \textit{bumpy} decrements, 
where we note that in three objects (Sz103, Sz106, Sz111) the profiles of the \hb\ and following lines of the series display optical depth effects (self-absorption in the core of the line) that tend to disappear in higher $N_{up}$ lines (Fig.~\ref{fig:lines:h}), which are optically thinner. For these sources, normalising by the H$\beta$ flux clearly introduces some bias. However, even applying a flux correction to take into account the absorption (i.e. filling the absorption by fitting a Gaussian using only the line wings as reference) does not change significantly the decrement shape. 
We notice that such self-absorption effects are apparent only in the first three or four lines of the series (H3 to H6), while the location of the bump "peak" is around $N_{up}$= 9,10. This further supports the conclusion that the bump is not a bias introduced by disregarding self-absorption. 
Moreover, in the remaining type 3 decrement objects the differential opacity effects are not found and the profiles seem fairly consistent between \hb\ and the remaining lines of the series.

\subsection{Paschen decrements and source parameters}

In the bottom panel of Table~\ref{tab:type_occ} we indicate the occurrence of the different combinations of Balmer and Paschen decrement shapes.
The four objects that display a clear type C Paschen decrement comprise three type 3 and one type 1 object. 
These sources do not seem to show any particular stellar parameter in common.

It must be noted that all targets with a type 4 Balmer decrement display a Paschen decrement of type B.
There is therefore a strong consistency between the Balmer and Paschen \hi\ decrements in these objects, which were shown to be typically associated with strong accretors (see Fig.~\ref{fig:decs:properties} and Sect.~\ref{sec:decs:params}).

The type 2 sources for which we can derive a solid classification of the Paschen decrement display a type A shape, with the only exception of Sz66. 
The remaining type 2 sources, and in particular most narrow symmetric line objects, have uncertain Paschen decrement types. 

Finally, we note that objects that show a \pab\ line with a clear inverse P-Cygni profile display a Paschen A decrement and Balmer decrements of type 2 and 3.

\begin{figure*}[t]
\centering
\includegraphics[width=13cm]{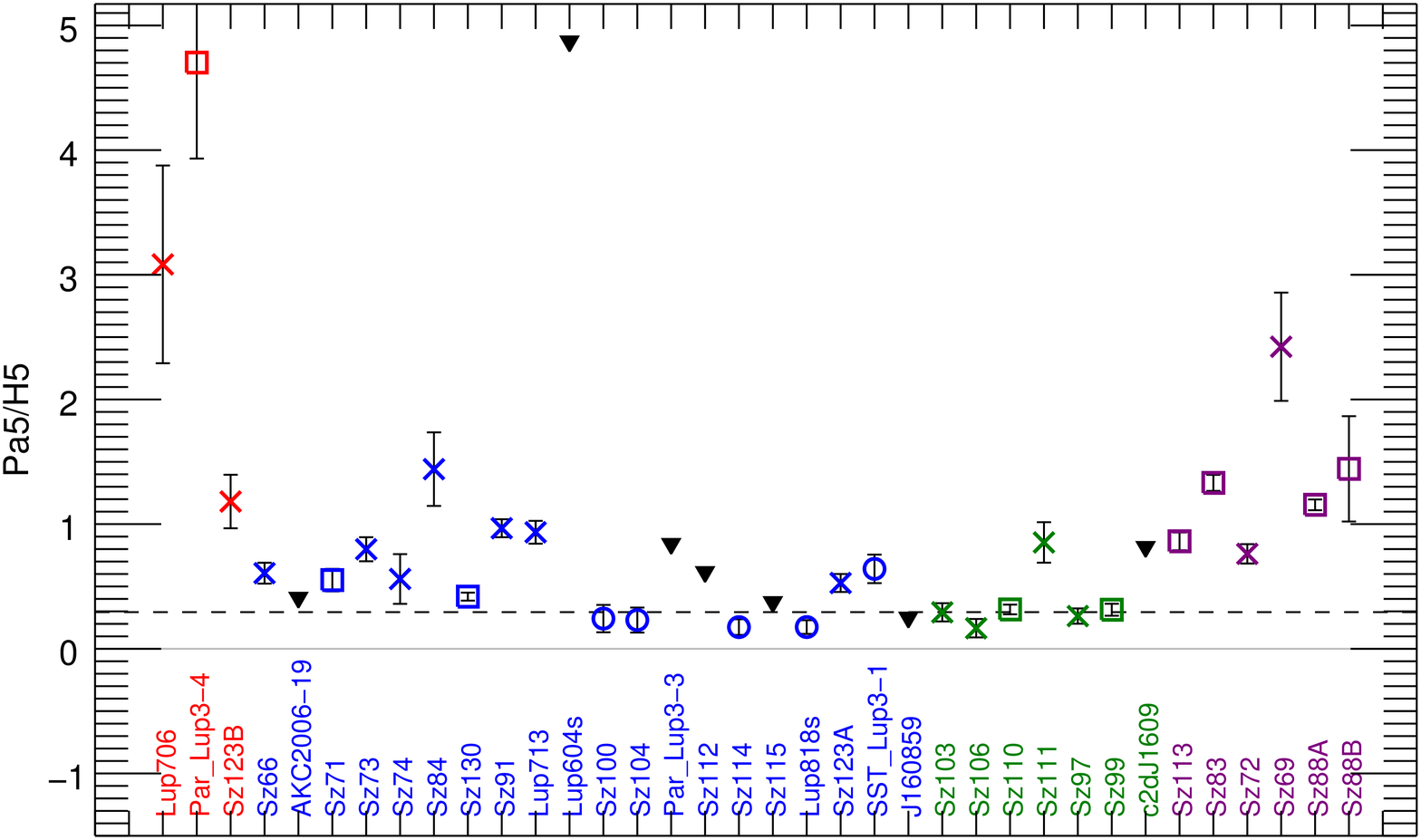}\\
\caption{\label{fig:ratios} 
Values of the Pa5/H5 (Pa$\beta$/H$\gamma$) in the sample. The expected value for optically thin emission is marked with a dashed horizontal line. The colour marks the decrement shape type (red:1, blue:2, green:3, purple:4) while the symbol indicates the line profile typology (circle: narrow symmetric, square: wide, cross: multi-peaked); downward triangles are upper limits.}
\end{figure*}

\subsection{Line ratios and optical depth}
\label{sec:decs:raod}

The large number of \hi\ transitions simultaneously observed with X-Shooter allows us to derive information on the line optical depth, by analysing Paschen and Balmer lines with common upper level. In Fig.~\ref{fig:ratios}, we show the Pa5/H5 ratio (Pa$\beta$/H$\gamma$, that is, the one computed from lines with the highest SNR) measured in the sample along with information on the profile and decrement type of the sources. Once the line fluxes have been corrected for extinction, in the case of optically thin emission, these ratios depend only on the Einstein coefficients of the transitions \citep{goldwire68}.

First, we note that sources with symmetric narrow profiles are compatible with optically thin emission in these lines, while no definite trend can be seen for objects with different profiles types. 
As for the decrement shape, all sources with type 4 decrements do not agree with the expected value, clearly suggesting optically thick emission for these targets. 
Also, type 1 objects display ratios that are significantly greater than the optically thin value. This supports our previous hypothesis that extinction might have been underestimated for these sources (Sect.~\ref{sec:decs:params}), as the net effect of the extinction is to lower the ratios. 
Interestingly, the discrepancy between the measured ratios and the expected optically thin value roughly reflect the different obscuration factors (grey extinction) applied by A14 in these targets (25, 10, 6 for ParLup3-4, Lup706, Sz123B, respectively).

Finally, all type 3 decrement objects, except for Sz111, show ratios that are in principle compatible with optically thin emission in the considered lines. 
However, Sz103 and Sz106 display evident opacity effects in the first lines of the series (see Fig.~\ref{fig:lines:h} and Sect.~\ref{sec:decs:lines}) and assuming that the flux of the Balmer lines needs to be corrected for absorption would even cause a further decrease of the ratio. We will return on this aspect in Sect.~\ref{sec:models:kf:ratios}.

\begin{table}[t]
\begin{center}
\caption{\label{tab:type_occ} Occurrence of different combinations of decrements and profile types. \textit{Top}: Balmer decrement type and \hb\ line profile type. \textit{Bottom}: Balmer decrement type and Paschen decrement type. Uncertain Paschen decrements are marked with a ``?''.}
\begin{tabular}{ c c | c c c }
\hline
\hline
&    &\multicolumn{3}{c}{Profile}\\
&    &  NS  &  W  &  MP \\
\hline
\multirow{4}{*}{\rotatebox{90}{H dec.}}
&1   & 0  & 1  & 2  \\
&2   & 11 & 2  & 7  \\
&3   & 0  & 3  & 4  \\
&4   & 0  & 4  & 2  \\
\hline
\end{tabular}
\end{center}
\hspace{1cm}
\begin{center}
\begin{tabular}{ c c | c c c c }
\hline
\hline
&    &\multicolumn{4}{c}{Pa dec.}\\
&    &  A  &  B  &  C  &  ? \\
\hline
\multirow{4}{*}{\rotatebox{90}{H dec.}}
&1   & 2  & 0  & 1  & 0\\
&2   & 9  & 1  & 0  & 10\\
&3   & 3  & 0  & 3  & 1\\
&4   & 0  & 5  & 0  & 1\\
\hline
\end{tabular}
\end{center}
\hspace{1cm}
\end{table}

\section{Comparison with models}
\label{sec:models}
To derive information on the properties of the gas emitting the \hi\ lines we have compared the observed decrements with predictions of two standard emission models:
the classic Case B recombination \citep{baker38,hummer87}, in which Balmer lines are optically thin and level populations are determined by radiative cascade
from the continuum, and the local line excitation calculations recently developed by \citet{kwan11}, which evaluate self-consistently the line emissivities upon input of the local physical conditions of the gas.

\subsection{Case B}
\label{sec:models:caseb}
\begin{figure}[t]
\centering
\includegraphics[width=4.5cm]{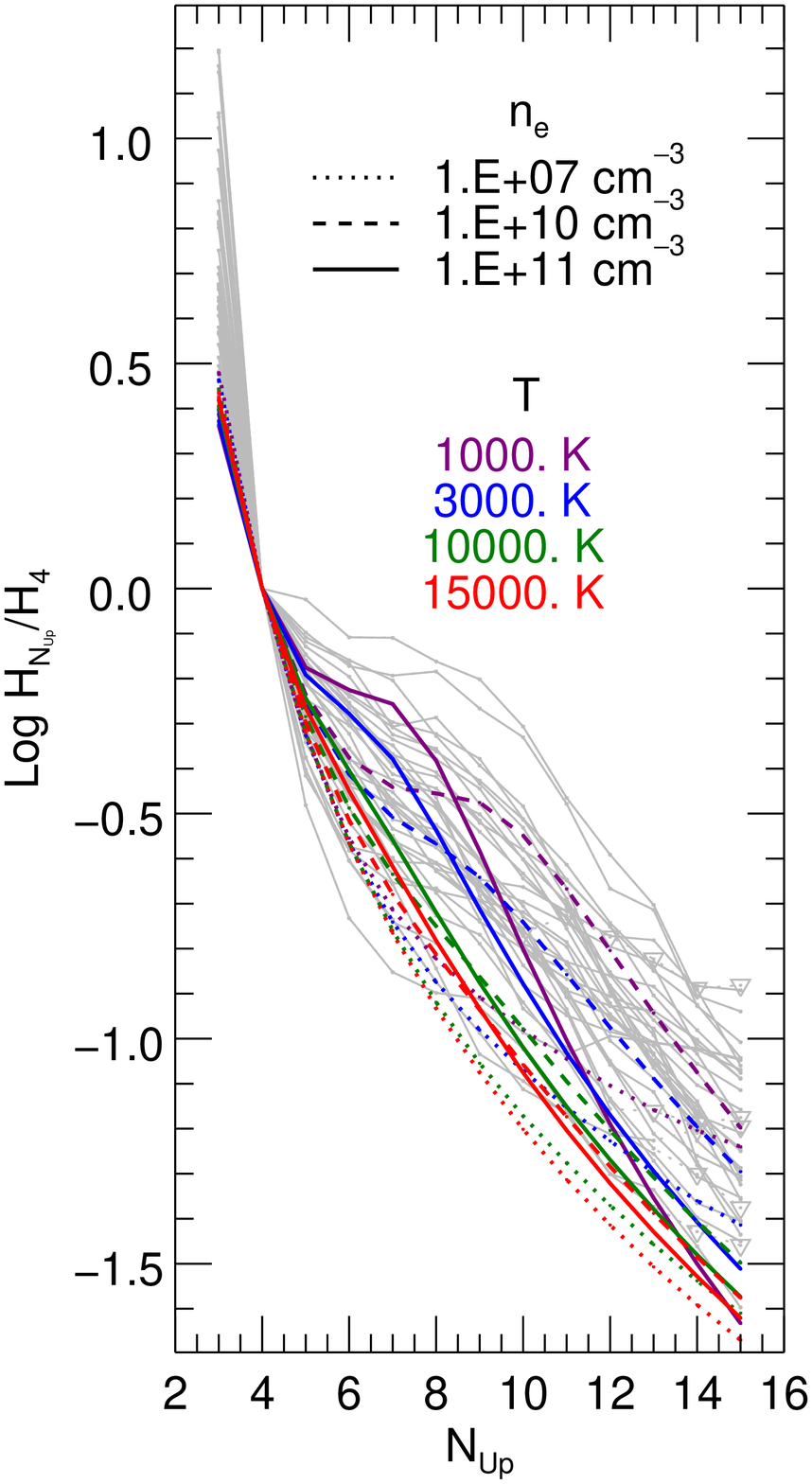}
\hspace{-.3cm}
\includegraphics[width=4.5cm]{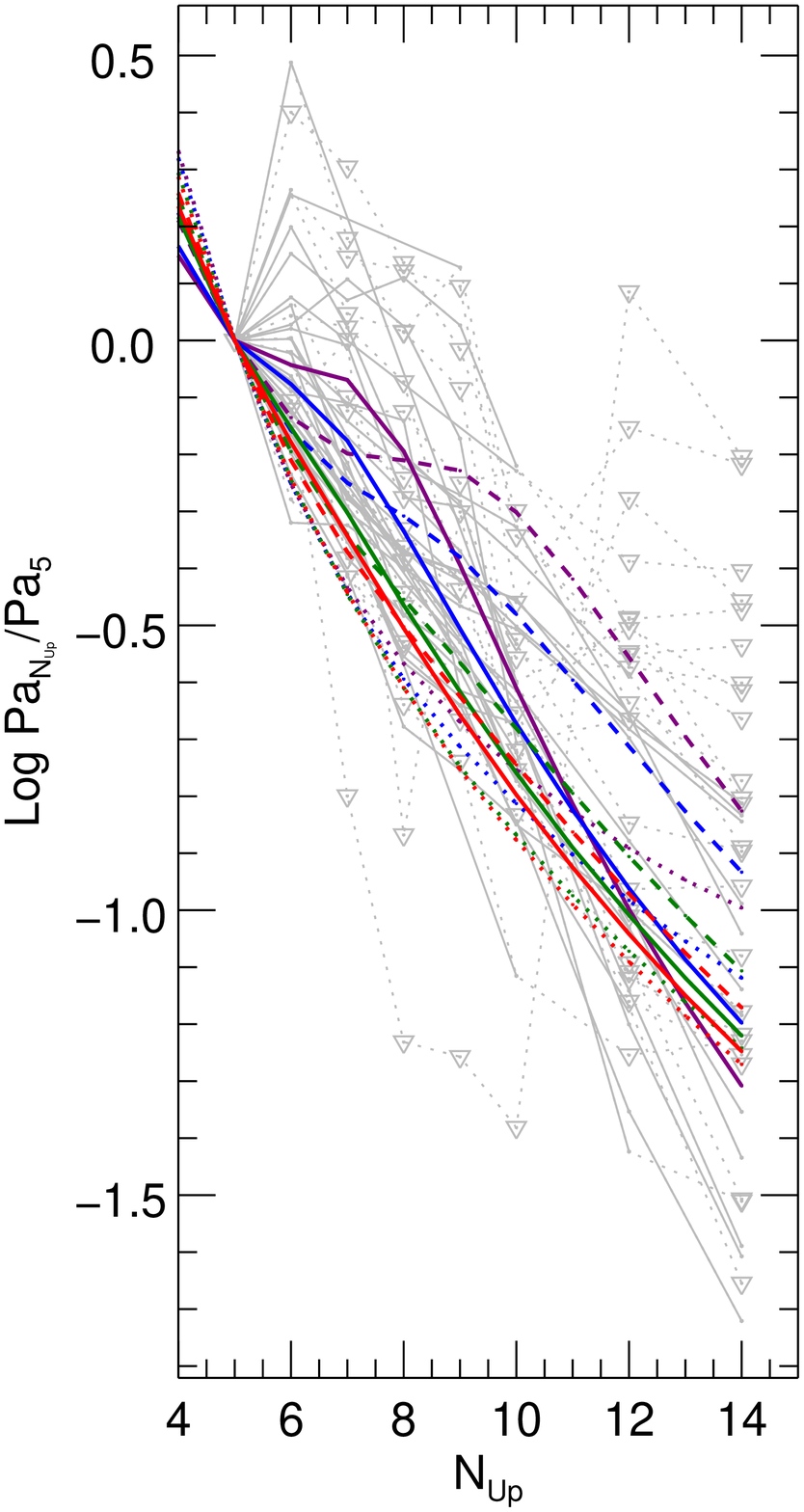}
\caption{\label{fig:caseb} Case B model Balmer (left) and Paschen (right) decrements are superposed to the observed decrements of our sample (grey lines). The various colours and line styles refer to different gas parameters, as indicated in the legend in the left panel.} 
\end{figure}

Case B decrements appear in some cases formally compatible with the observations. This can be seen in Fig.~\ref{fig:caseb}, where Case B predictions 
for different temperatures and electron densities from the online tables provided by \citet{storey95}
are directly compared with our data. 

In Balmer decrements, the Case B curves cover the central and lower part of the Lupus decrement ensemble. 
Low-temperature ($T\sim$3000 K) high-density ($n_e$ $\sim$ 10$^{10}$ cm$^{-3}$) curves are those that fall in the central part of the ensemble, where most of the observed decrements of type 2 are located, which is in line with the results of \citet{bary08}. 
However, the type A Paschen decrements shown by most of the sources with type 2 Balmer curves are more consistent with lower density gas ($n_e$ $\sim$ 10$^{7}$ cm$^{-3}$) at higher temperatures ($T \gtrsim $3000 K).

The characteristic bumpy shape reminiscent of type 3 Balmer decrements appears in Case B curves for very low temperatures ($T\sim1000$ K) and high electron densities, but the Case B predictions remain globally below the observed type 3 decrements.

Although we detect a bump very similar to that predicted by Case B in objects with decrements of type 2 but tending to type 3 (Sz66, Sz84), the line profiles in these sources indicate clear differential optical depth effects that should rule out an optically thin emission regime. 

Finally, Case B completely fails to reproduce the L-shape of the type 4 Balmer decrements, as well as the plateau/bump of the type C Paschen decrements.
This is in agreement with the observation that emission in these objects is not optically thin (see Sect.~\ref{sec:decs:raod} and \ref{sec:models:kf:ratios}).

Even though Case B curves are compatible with some of the observed decrements, the mentioned inconsistencies seem to indicate that the Case B model 
is generally not appropriate for describing the \hi\ emission in T Tauri stars.
Indeed, for Case B assumptions to be valid it is necessary that radiative de-excitation from level $n$=2 is more rapid than excitation by collisions from the same energy level. As evidenced by \citet{edwards13}, in terms of physical conditions this translates in an upper limit for the column density of the neutral hydrogen ($n_{HI}\cdot dl$), $dl$ being the emission length scale. Since in Case B $n_e$ (and so $n_{HI}$) is a free parameter, the model implicitly assumes that $n_{HI}$ and $dl$ have values ensuring that the collisional excitation from $n$=2 remains negligible. 
As the electron density increases ($n_e\gtrsim$10$^{9}$cm$^{-3}$), however, this regime is obtainable only with a very strong photo-ionisation rate, which can be difficult to obtain in the circumstellar environment of standard CTTSs.
Although ionisation by photons from the hot corona of the star or from the accretion shock has been proposed (e.g. Bary et al. 2008), 
the capability of such mechanisms to maintain the Case B regime in T Tauri environments still needs to be proved. 

\subsection{Local line excitation calculations}
\label{sec:models:kf}

The local line excitation calculations by Kwan \& Fischer (2011, hereafter KF; see also Edwards et al. 2013) have been developed to describe line emission under the physical conditions expected for winds and accretion flows in CTTSs. 
Given one set of local physical conditions as input (\hi\ number density $n_H$, temperature $T$, hydrogen ionisation rate $\gamma_{HI}$, and a velocity gradient $dv/dl$) the KF models compute the ionisation fraction $n_e/n_H$, the level populations, and the line optical depths in a self-consistent fashion. These models can manage optically thick emission and thus avoid the implicit limitations of Case B.

In this article, we consider the set of models presented in \citet{kwan11}.
This set was obtained from a 20-level model of the hydrogen atom, by assuming 
a velocity gradient $dv/dl = 150$ km s$^{-1} / 2$\rstar\ and an ionisation rate $\gamma_{HI}$ = 2$\times 10^{-4}$ s$^{-1}$. These
values were originally chosen by KF to approximate the conditions in the region of a wind or an accretion
flow, based on the set of T Tauri spectra analysed by the authors \citep{edwards13}.

\begin{figure*}[!t]
\centering
\includegraphics[width=4.4cm]{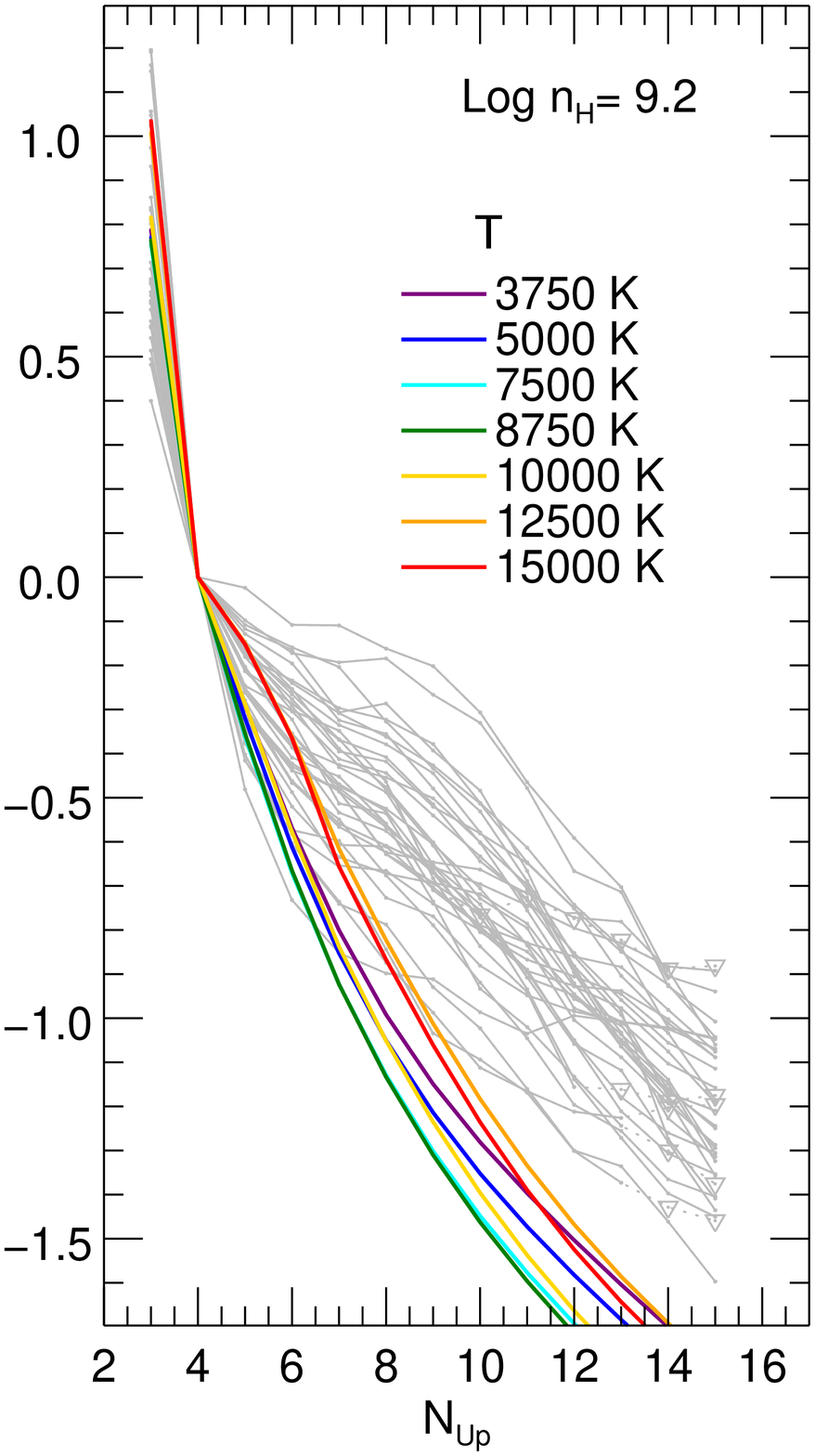}
\includegraphics[width=4.4cm]{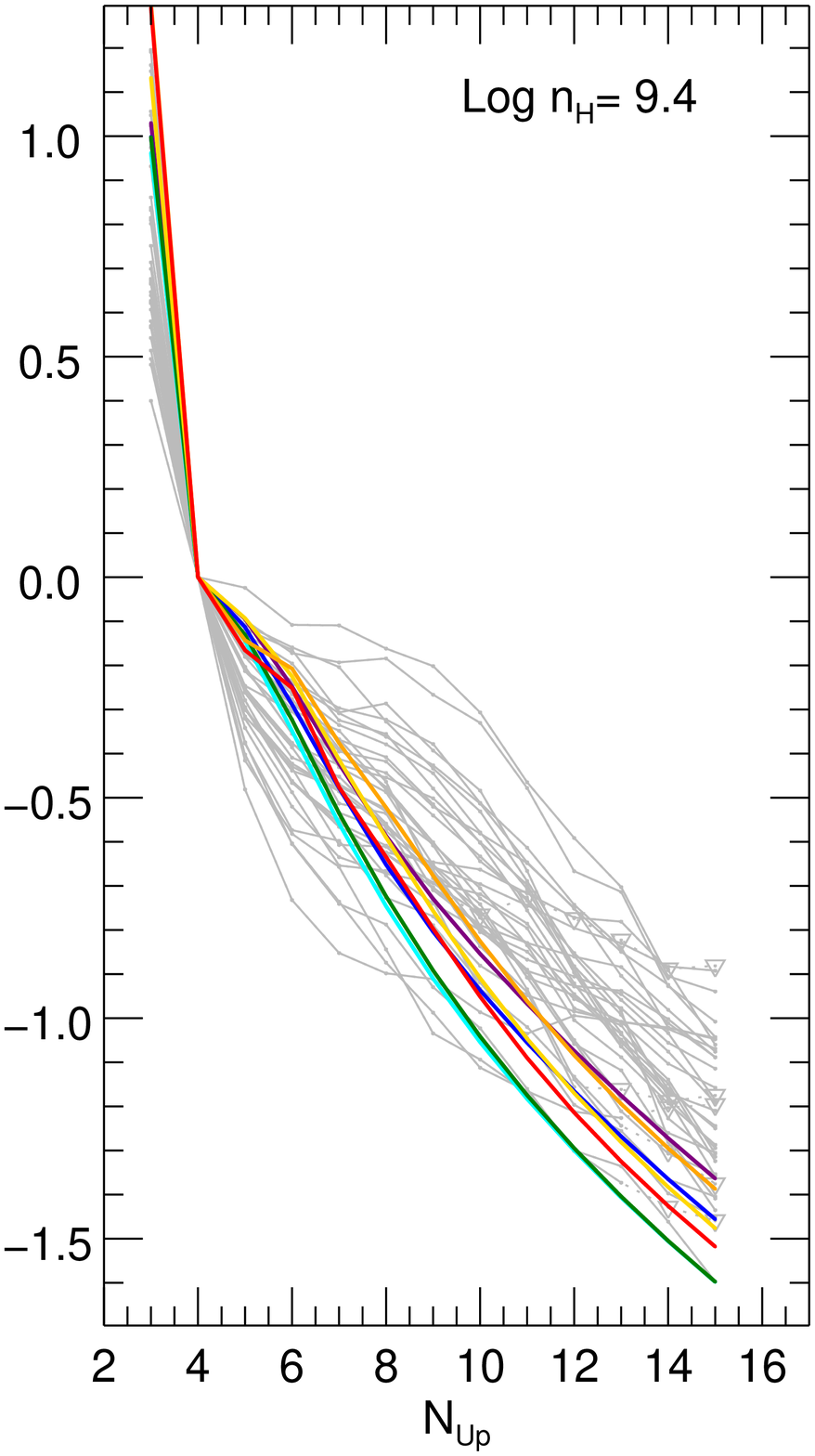}
\includegraphics[width=4.4cm]{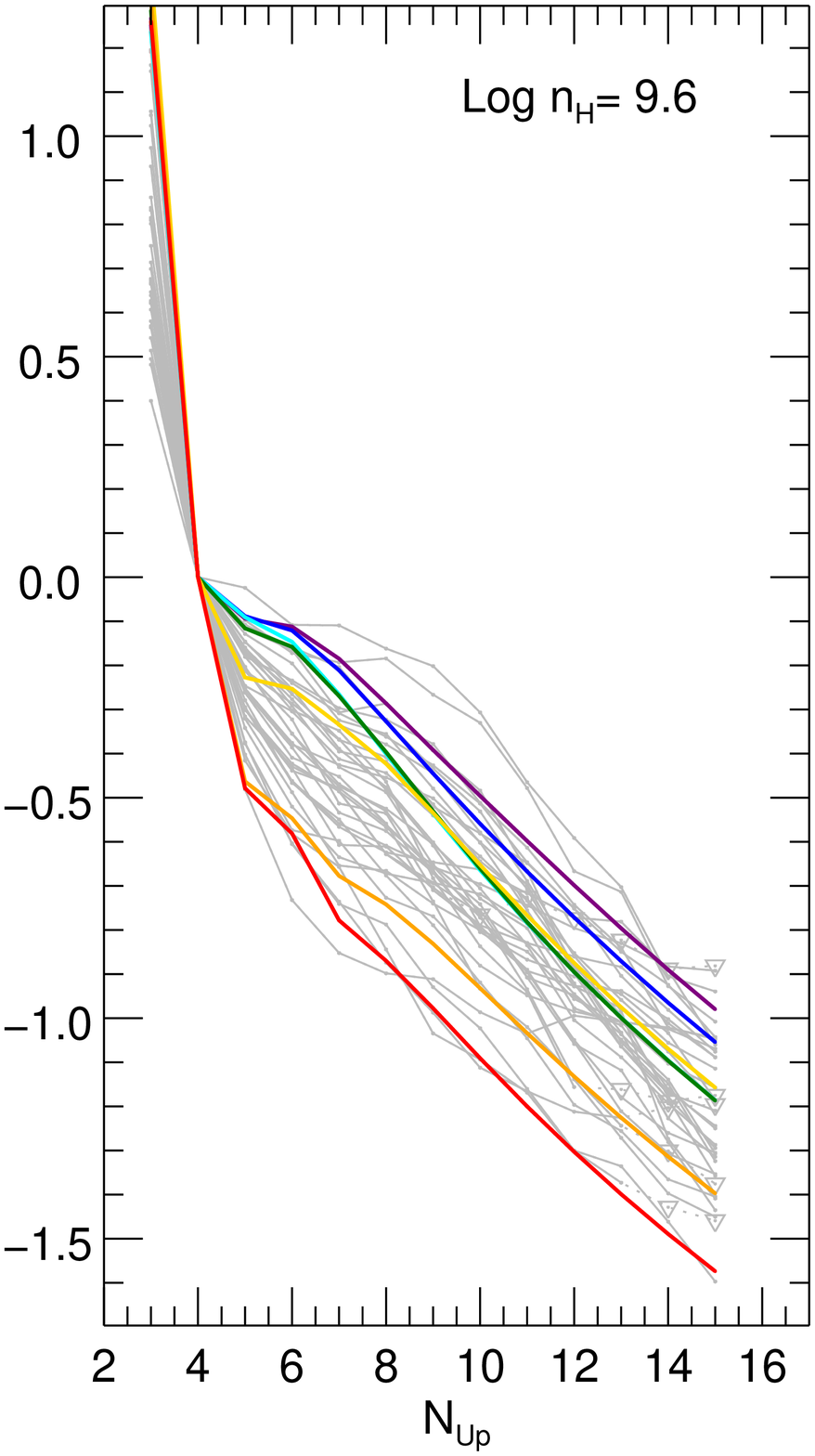}
\includegraphics[width=4.4cm]{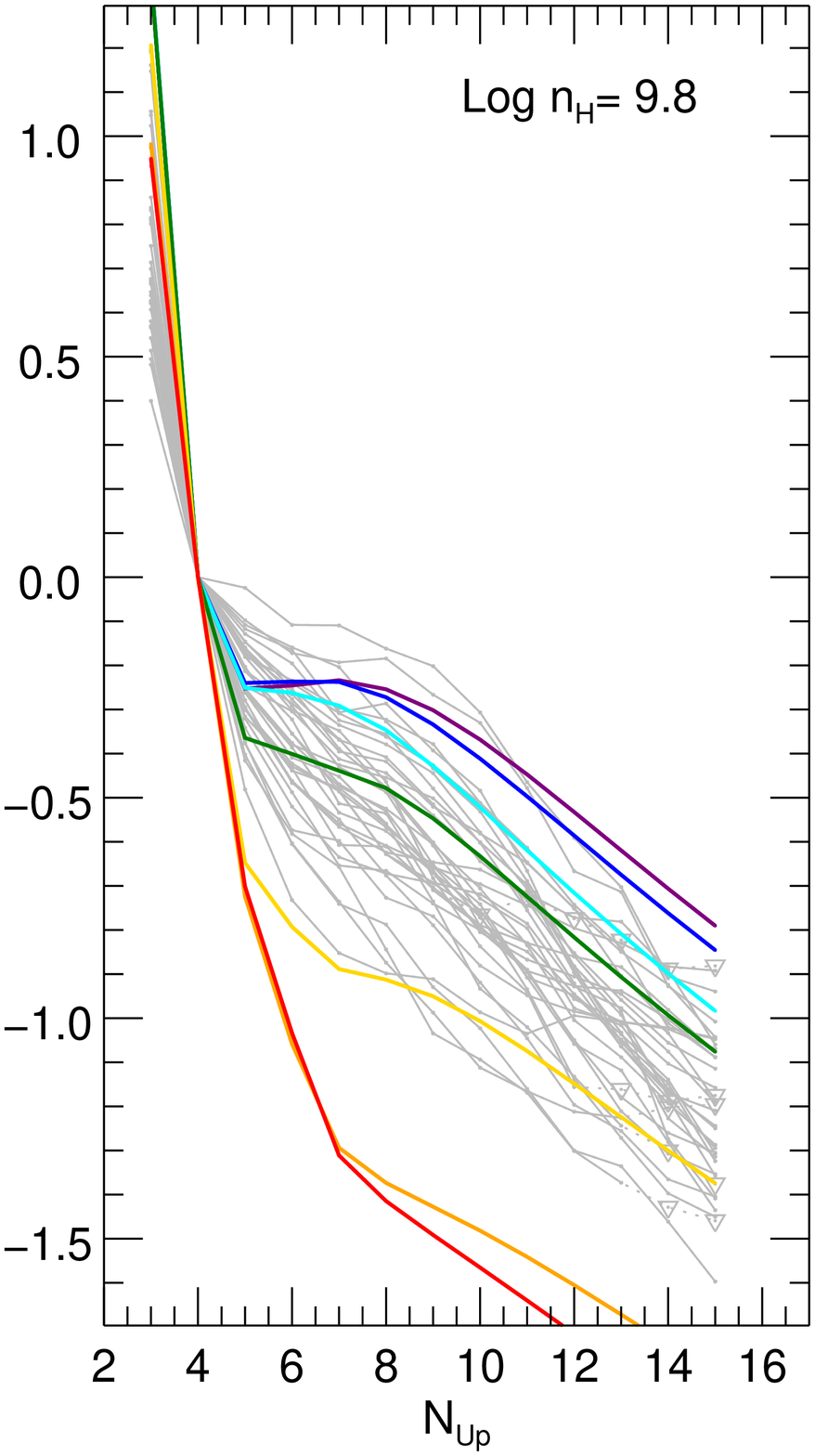}\\
\includegraphics[width=4.4cm]{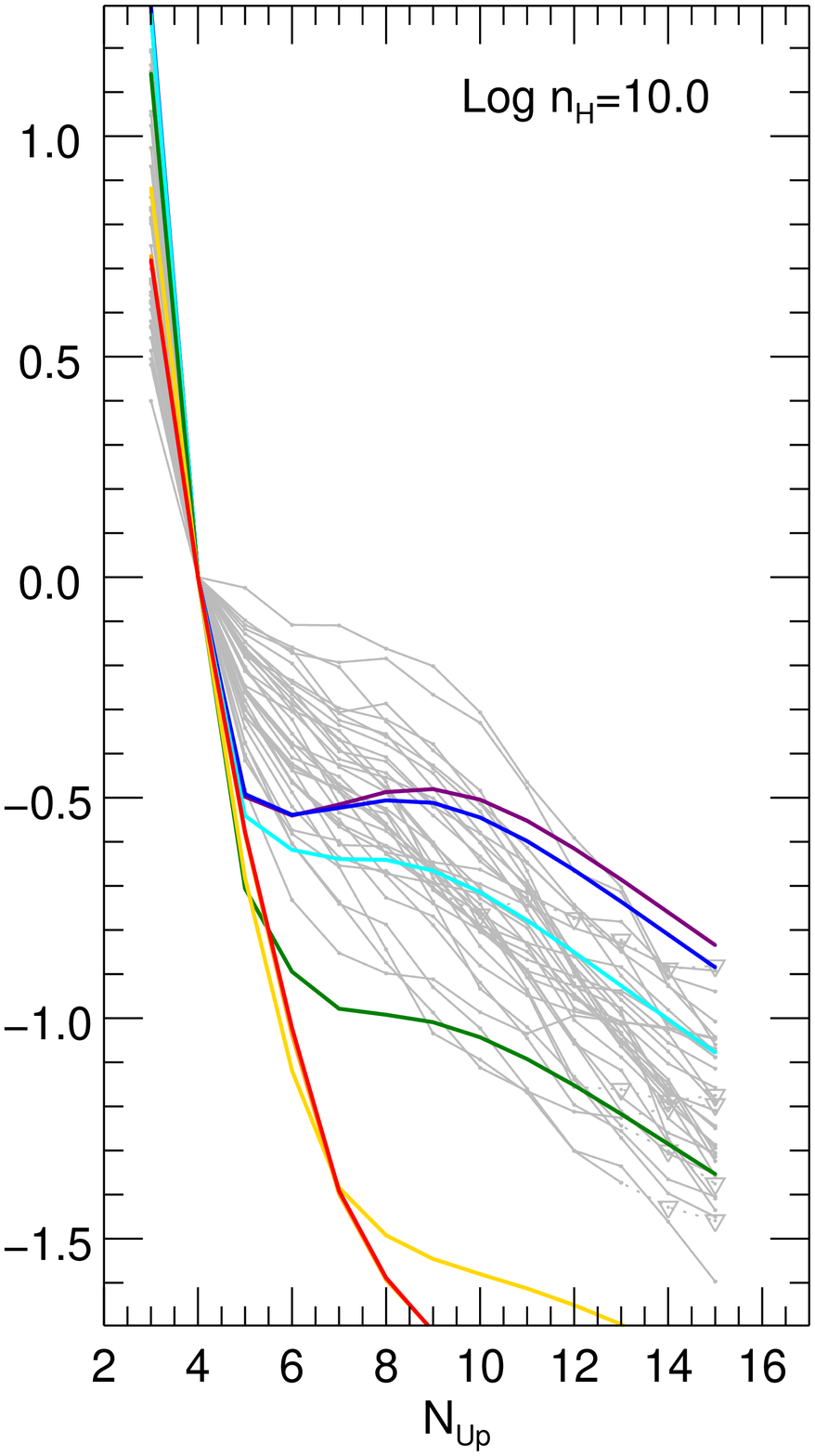}
\includegraphics[width=4.4cm]{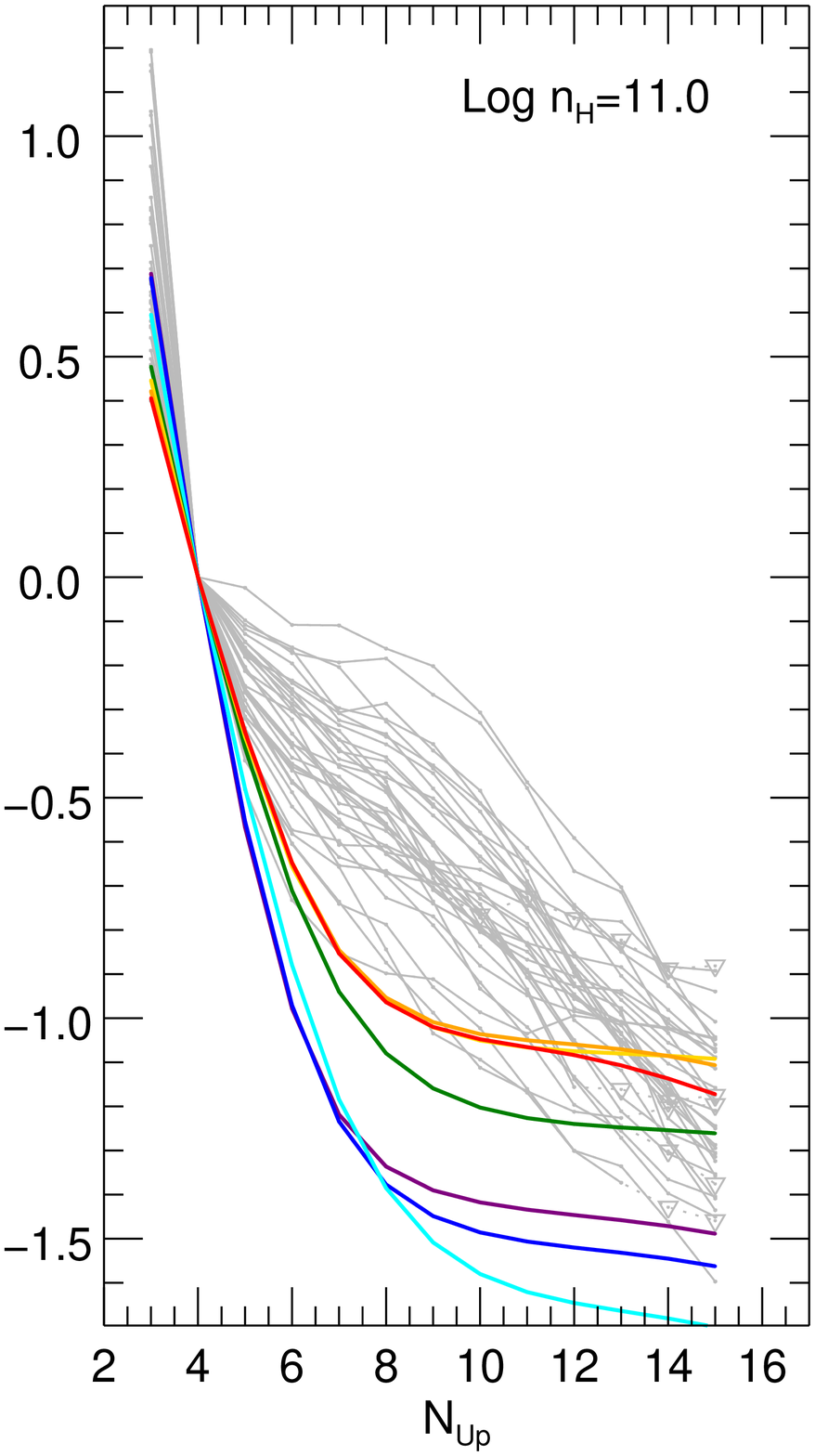}
\includegraphics[width=4.4cm]{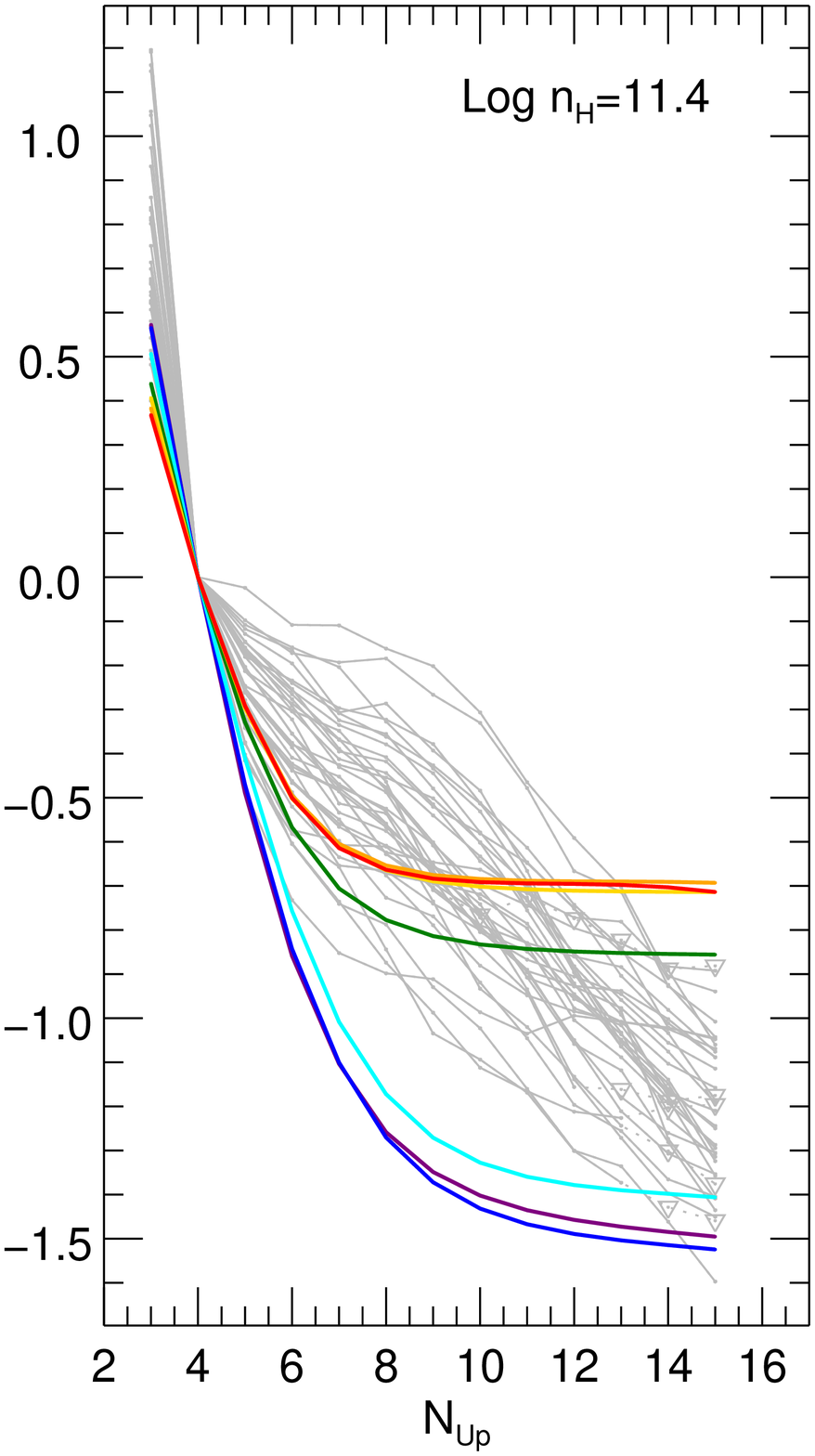}
\includegraphics[width=4.4cm]{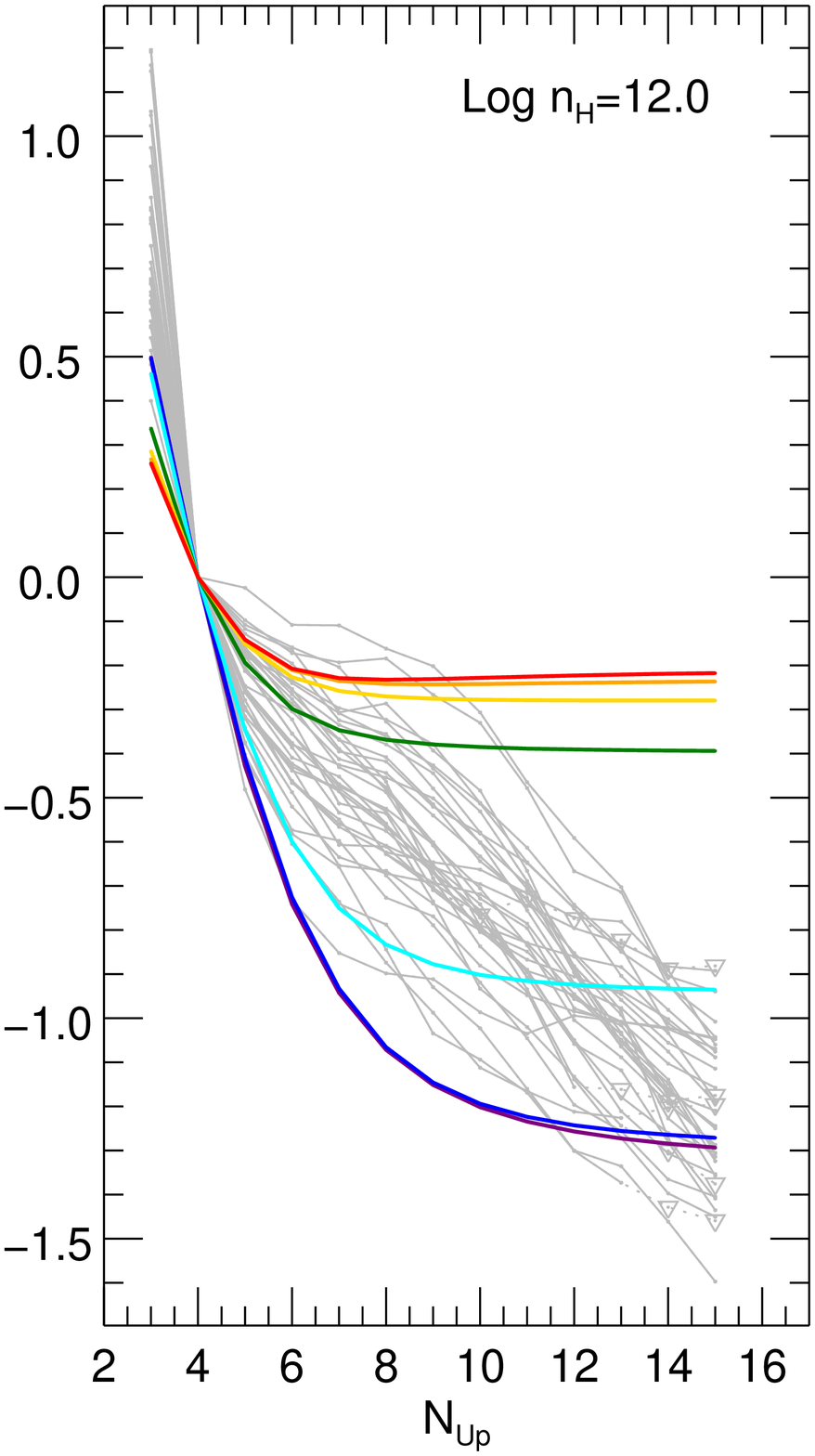}
\caption{\label{fig:kf} Balmer decrements predicted by the local line excitation calculations of Kwan \& Fischer (2011) are superposed on the decrements observed in our sample (grey lines). The panels display results for different values of the hydrogen number density $n_H$ (indicated), while the solid lines of different colours refer to the gas temperatures reported in the legend.} 
\end{figure*}

\begin{figure*}[t]
\centering
\includegraphics[width=4.4cm]{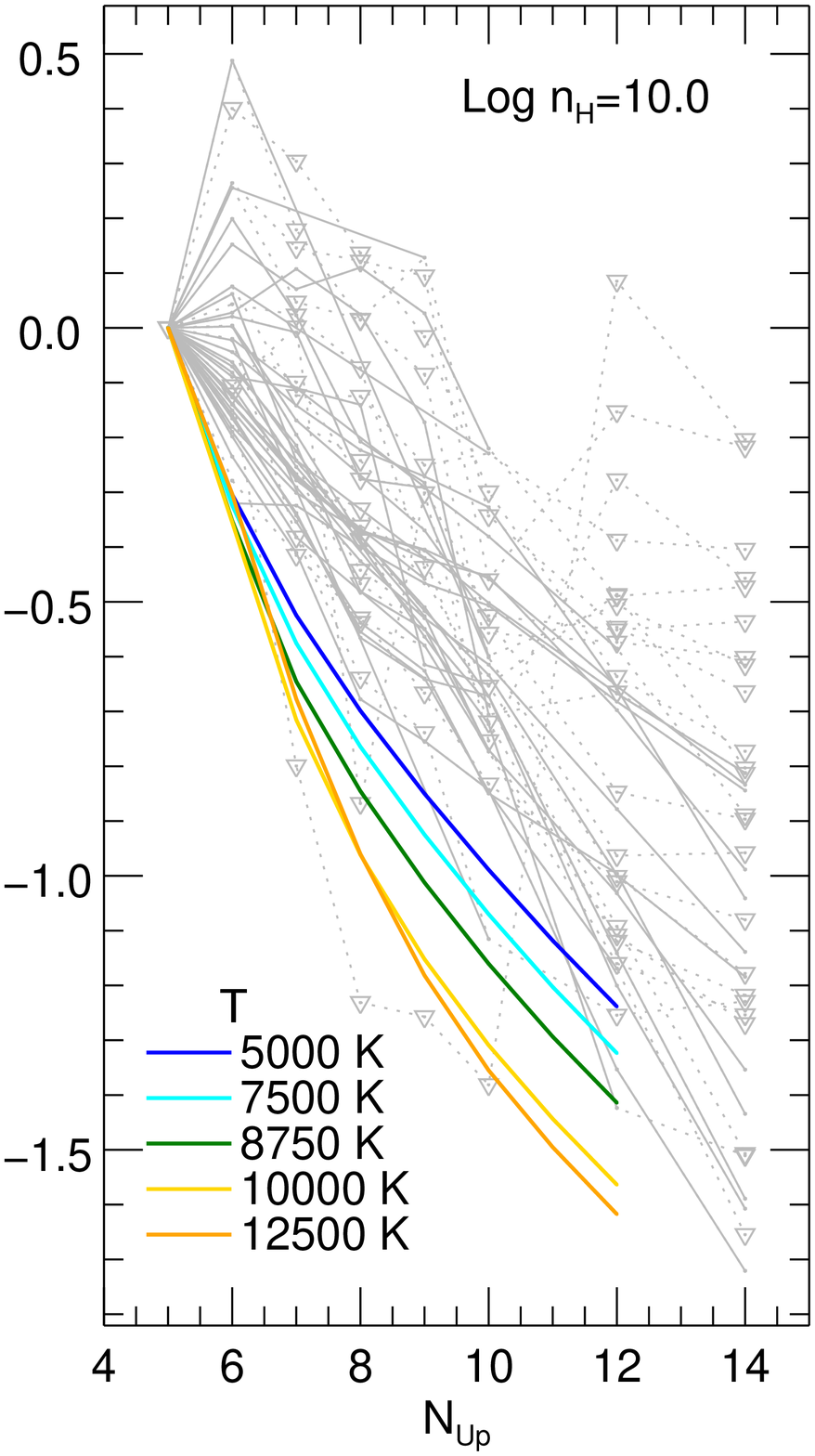}
\includegraphics[width=4.4cm]{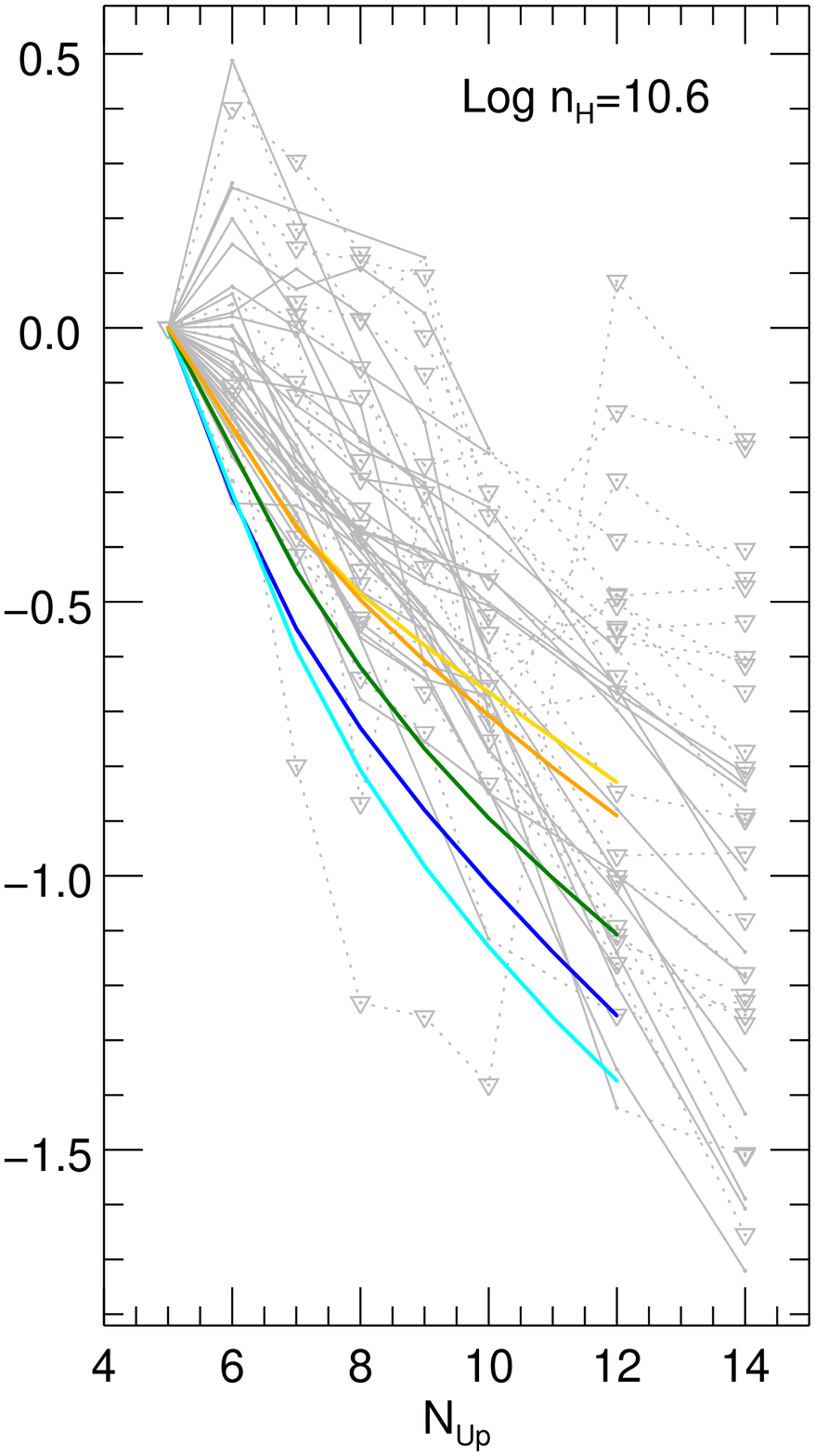}
\includegraphics[width=4.4cm]{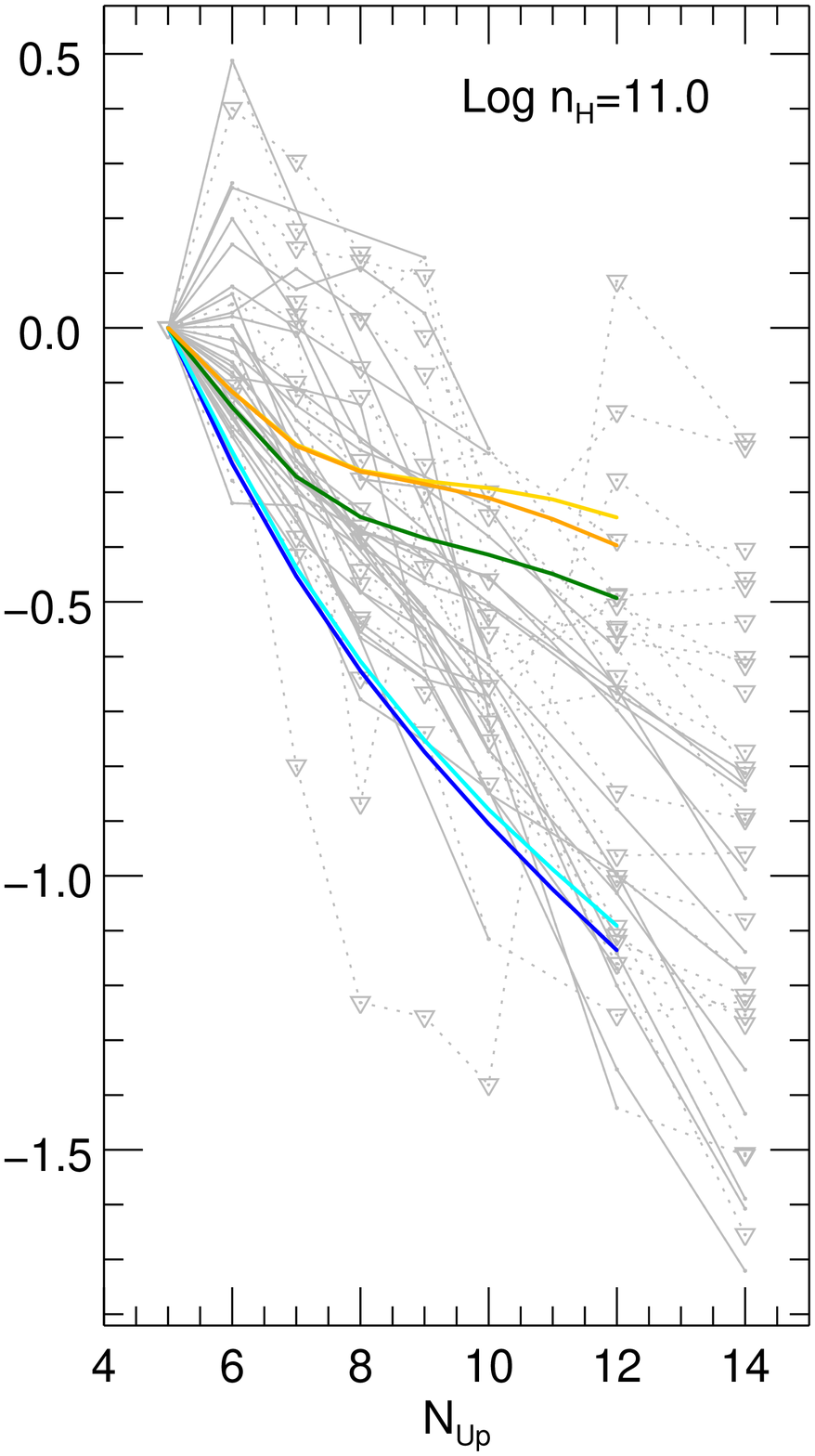}
\includegraphics[width=4.4cm]{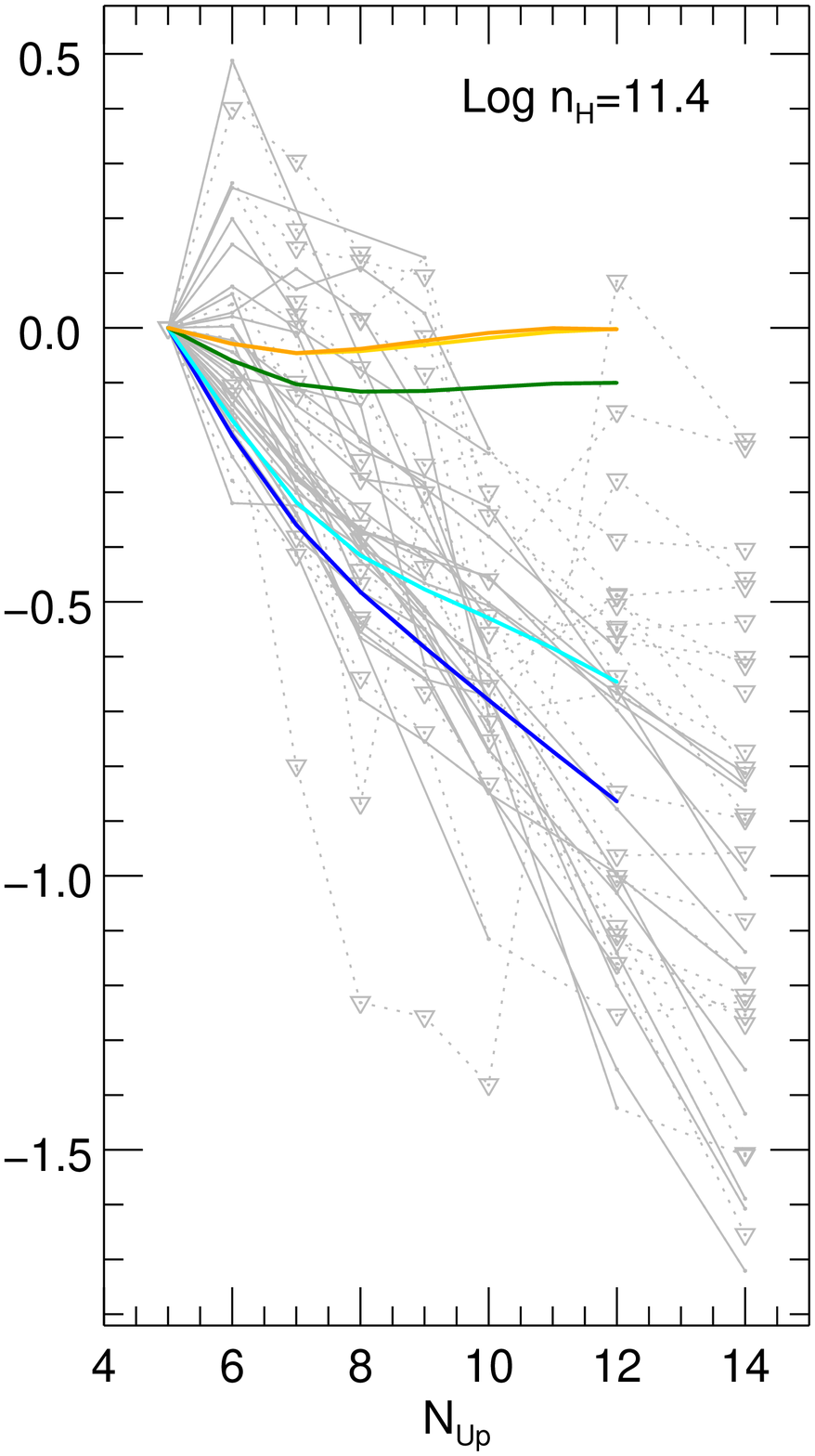}
\caption{\label{fig:kf:pa} Paschen decrements predicted by the local line excitation calculations of Kwan \& Fischer (2011) are superposed to the decrements observed in our sample (grey lines). The panels display results for different values of the hydrogen number density $n_H$ (indicated), while the solid lines of different colours refer to the gas temperatures reported in the legend.} 
\end{figure*}

\subsubsection{Decrements}
\label{sec:models:kf:dec}

The computed model grids for Balmer lines consider a hydrogen total density spanning from log $n_H$=8 to log $n_H$=12 cm$^{-3}$ 
and temperatures in the range $T$=3750-15\;000 K. A direct comparison of KF predictions and the Lupus Balmer decrements is shown in Fig.~\ref{fig:kf}.  
For values of the gas density that vary in the range log $n_H$ 9$\div$10 cm$^{-3}$, the model curves cover the entire locus of the observed decrements. Such a rapid variation in the shape of the decrements is due to the fact that this is the range of densities where the lines of the series become optically thick, starting from the first lines up to high $N_{up}$ lines \citep{edwards13}. Once all lines of the series are optically thick (log $n_H$ $>$11 cm$^{-3}$) the decrement curves assume a characteristic L-shape, reminiscent of that of type 4 objects.

Most of the observed decrement shapes are reproduced in a reasonable way by the KF models. In particular, type 2 decrements are in good agreement with curves expected for densities in the range log $n_H$ = 9.4$\div$9.8 cm$^{-3}$, with temperatures that however cannot be well constrained. 

Similarly to Case B, the type 3 decrements are the most difficult to be reproduced. The region where type 3 curves are located is roughly covered in KF models by decrements with gas at densities log $n_H$ = 9.6-9.8 cm$^{-3}$ and fairly low temperatures ($T < 5000$ K), but shapes that are very similar to type 3 cannot be obtained in any case.

As mentioned above, decrement curves very similar to the type 4 shape are well reproduced in KF models for high hydrogen densities (log $n_H$ $>$11 cm$^{-3}$) and various temperatures (in general $T > 5000$ K, see last three panels of Fig.~\ref{fig:kf}), that is, in a regime where all lines of the series have become optically thick.

Since \ha\ has not been considered in our definition of the decrement shapes, we have checked the predicted flux for this line against the observations in the cases best reproduced by the KF models, namely type 2 and 4 objects. For type 4 sources the observed \ha\ is in good agreement with the KF predictions, while for type 2 curves we find that \ha\ is overestimated in few objects (see Fig.~\ref{fig:decs:h} in the appendix). 
Observed fluxes greater than those predicted may result from additional emission components to the line (e.g. chromospheric and jet emission), while lower fluxes are more difficult to explain and might be related to the strong self-absorptions in the line being not well modelled. 

Paschen decrements computed from KF models are compared with observations in Fig.~\ref{fig:kf:pa}. Curves with total hydrogen densities of order 10$^{10}$--10$^{11}$ cm$^{-3}$ are consistent with the straight types A and B. In particular, the predicted model curves that best match the observed type B decrements are obtained for densities log $n_H$>11 cm$^{-3}$ and temperatures around 7000-10\;000 K, with a shape that varies very rapidly with the temperature in this regime. The high total densities are the same required for type 4 Balmer decrements, which is consistent with the simultaneous occurrence of type 4 and type B decrements (Table~\ref{tab:type_occ}).

It can be noticed that for the same value of $n_H$ the gas temperatures needed to obtain a typical type 4 curve are somewhat higher than those required for the typical type B shape (Figs.~\ref{fig:kf} and \ref{fig:kf:pa}). Additionally, the type A curves frequently associated with the type 2 Balmer decrements are obtained at higher densities than the relative type 2-like curves.
These discrepancies may indicate that the region responsible for the bulk of the Balmer and Paschen line emission is not exactly the same in such cases (temperature and density gradients are indeed expected) or, alternately, might depend on the limitations of the current set of KF computations.
More on the results from the simultaneous analysis of Balmer and Paschen data is given in Sect.~\ref{sec:models:kf:ratios}.

None of the model curves reproduces the type C Paschen decrements with the bump or plateau, although we notice that for high-density high-temperature gas (i.e. in an optically thicker regime) we obtain flat decrements that are in rough agreement at least with what we observe in the first lines of the series. Flux inversion between \pab\ and the following lines is indeed expected in the case of blackbody-like emission at these temperatures \citep[e.g.][]{nisini04}.

The relatively narrow range of gas densities required to reproduce the observed spread of the decrements (both Balmer and Paschen) is in agreement with the results by \citet{edwards13} from the comparison of KF models with the Paschen decrements of a sample of T Tauri stars, although the densities inferred by \citet{edwards13} (2\,10$^{10} < n_H < 2\,10^{11}$ cm$^{-3}$) appear higher than those associated with type 2 decrements and more in line with those derived for type 4-B decrements.
Indeed, most of the objects of the \citet{edwards13} sample present a Paschen decrement that complies with our type B definition (see their Fig.~5) and have on average higher accretion levels than our Lupus objects.

Finally, we point out that the mentioned discrepancies between predictions and data might also indicate that the input parameters of the KF models need to be tuned to take into account our observations, so as to better describe the physical conditions in the line emission region of some of the targets.
For instance, a velocity gradient of 150 km s$^{-1}/2$ \rstar\ appears too high for the sources that display narrow symmetric lines. 
Although with different input parameters we expect to obtain decrements qualitatively similar to those of the current models (i.e. with the same basic shapes), 
some of the features of the decrement curves might change, such as the prominence of the bump in the type 3-like decrements. 
In addition, in the case of a smaller velocity gradient we may expect that the same shapes (i.e. the same line ratios) are found at lower gas densities 
than those of the standard KF models that we are adopting here, because the line opacities scale as $n_{H}/(dv/dl)$ \citep[see][]{kwan11}.
A more in-depth analysis with updated sets of KF models considering different values of the input parameters and possible additional information from lines of other species (the KF models also describe HeI, OI, CaII, and NaI transitions, thus offering the opportunity to study the line emission region conditions from many emission lines simultaneously) is beyond the scope of this article and is therefore postponed to a future paper.

\subsubsection{Line ratios}
\label{sec:models:kf:ratios}

\begin{figure}[t]
\centering
\includegraphics[width=9cm]{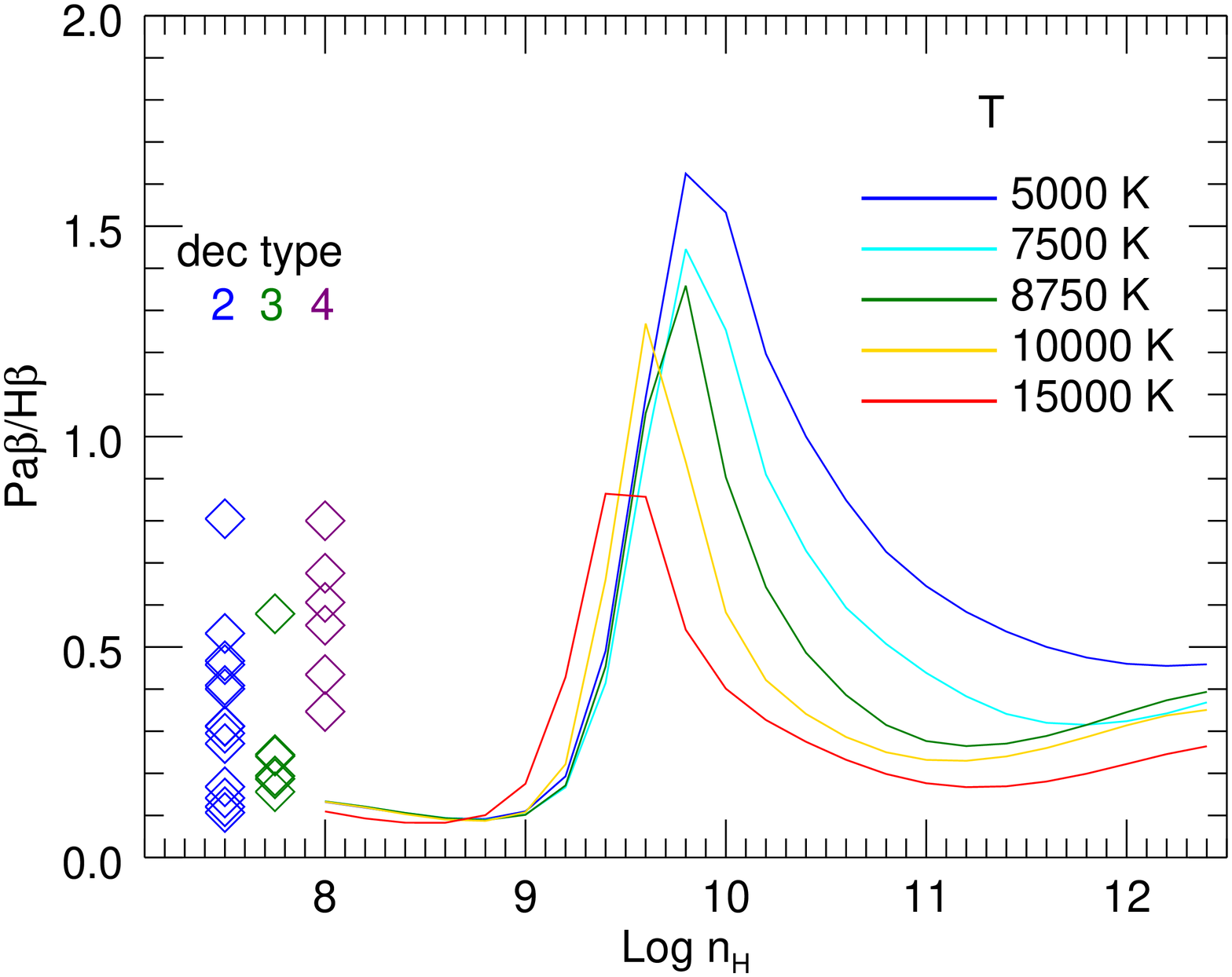}\\
\includegraphics[width=9cm]{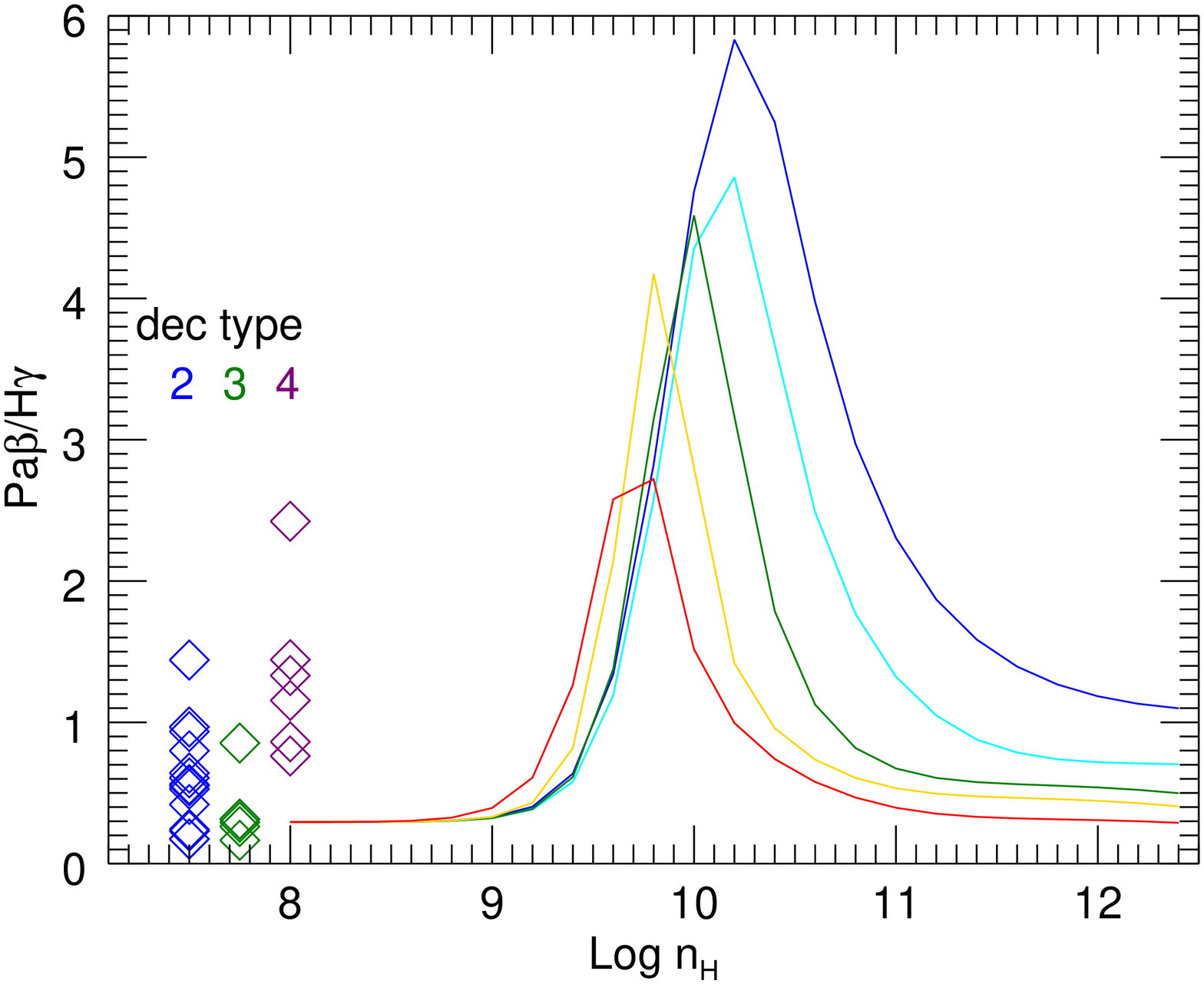}\\
\caption{\label{fig:models:ratios} Measurements of the Pa$\beta$/H$\beta$ (top) and Pa$\beta$/H$\gamma$ (bottom) ratios, which are displayed in the left portion of the plot and divided by Balmer decrement types (we do not include type 1 based on considerations in Sect.\ref{sec:decs:raod}), are compared with KF model curves that are plotted versus the total hydrogen density for different gas temperatures (see legend in the upper panel).}
\end{figure}

To investigate whether the KF models are able to simultaneously account for the observed Balmer and Paschen lines, 
we have checked the KF predictions against line ratios involving both Pa and H lines. In Fig.~\ref{fig:models:ratios} the measured Pa$\beta$/H$\beta$ and Pa$\beta$/H$\gamma$ ratios (divided by Balmer decrement type) are compared with the predicted model curves as a function of the total hydrogen density for different temperatures. The peaked shape of the model curves depends on the different opacity of the two lines. At low densities both lines are optically thin, then the Balmer line becomes optically thick so that the ratio increases until the optical depth of the Paschen line becomes greater than one. Eventually, we enter in a regime were both lines are thick. 

The plots confirm that only a few points are compatible with densities lower than log $n_H$ $\sim$ 9.5 and lines that are optically thin, namely those relative to some type 2 and type 3 sources. Among these, we basically find all NS objects for which it was possible to measure the ratios. It is difficult to constrain the temperature for these objects, although we observe that a temperature as high as 15\,000 K would imply total hydrogen densities lower than Log $n_H$ $\sim$ 9, which does not agree with the higher values indicated in the decrement analysis.
The remaining data points indicate higher total densities, in global agreement with the separate analysis of the Balmer and Paschen decrements, either in the range before the peak of the model curves (this is presumably the case of type 2 objects, see Fig.~\ref{fig:kf}) or after the peak where both lines have become optically thick (type 4 sources).

Ratios measured in type 4 objects, if put in conjunction with the decrement analysis that suggested high densities in these targets (log $n_H \gtrsim$ 11 cm$^{-3}$) to obtain the L-shape curve, provide indication for only moderately high gas temperatures ($T \lesssim$ 9000K), with lower temperatures required as the density increases. For instance, if temperatures as low as 7500 K are considered, the Pa$\beta$/H$\gamma$ indicates that in type 4 objects the densities cannot exceed log $n_H$ $\sim$ 11 cm$^{-3}$, which is a more stringent constraint than the one derived from the Balmer decrement analysis only, where typical type 4 decrements were actually obtained from KF models for total densities in the range log = $n_H$ 11$\div$12 cm$^{-3}$.

Finally, we notice that the value of the expected ratios in the optically thin and thick regimes (before and after the peak) are very similar for gas at high temperature ($T\sim15\,000$ K), especially in the case of the Pa$\beta$/H$\gamma$ ratio. On this basis, the ratios observed in type 3 objects that were in principle compatible with optically thin emission (see also Fig.~\ref{fig:ratios}) might actually be associated with very high-density and high-temperature gas. This would be consistent with the evident opacity effects observed in the first lines of the H series of Sz103 and Sz106, but it is in contrast with the indications, albeit uncertain, from the previous decrement analysis (log $n_H$ $\lesssim$ 10 cm$^{-3}$).

\section{Discussion}
\label{sec:discussion}

Despite the many combinations of Balmer (and Paschen) decrement shapes and line profiles (Tables \ref{tab:type_occ}), we are able to identify at least two groups of objects with common \hi\ observational characteristics, which we can associate with two different \hi\ emission regimes.

The first group contains the narrow symmetric line sources (11/36), which share highly similar \hi\ line profiles with FWHM around 100 \kms and display a type 2 Balmer decrement. They represent a subsample of the sources that show Balmer type 2 decrements and Paschen A decrements, which are largely the most common in our T Tauri sample. Indeed, 23 sources out of 36 have a Balmer 2 decrement, if we also include the three type 1 Balmer decrement sources, which we interpret as "reddened" type 2 curves.
Based on predictions of the KF models, such decrements would indicate gas with a total density in the range Log $n_H$ = 9$\div$10 cm$^{-3}$, with temperatures that, however, remain not well constrained ($T$ = 5\,000-15\,000 K). 
In the NS objects, the line ratios indicate an optically thin emission regime. Additionally, the NS objects are also associated with lower accretion rates (\lacc $< 10^{-9}$\msunyr, Fig.~\ref{fig:macc_width}). We can speculate that these sources are fairly quiescent objects in which the accretion proceeds at moderate levels through geometrically simple circumstellar structures. In this scheme, the narrow symmetric line emission would represent the basic \hi\ emission manifestation in moderately accreting CTTSs.

In the second group we find the sources with type 4-B (Balmer-Paschen) decrements (6/36). Based on the good agreement with predictions of KF models, emission in such objects can be well described assuming higher hydrogen densities ($n_H > 10^{11}$ cm$^{-3}$). At these densities the lines are expected to be optically thick, in agreement with the observed ratios of Paschen and Balmer lines (Sects.~\ref{sec:decs:raod} and \ref{sec:models:kf:ratios}). These ratios suggest also that gas temperatures cannot be too high ($T \lesssim$ 9\,000K). The very wide line profiles displayed by most of these objects signal the presence of gas moving at high velocities. 
Four type 4-B targets out of six are associated with the highest accretion rates in our sample, suggesting that there is a direct connection between gas at higher densities (more massive accretion flows) and higher \macc\ values. 
At variance with narrow symmetric line objects, Type 4-B sources therefore appear as the most active objects of the sample, with clear evidence of high-density circumstellar structures.

It is more difficult to delineate common recurrent characteristics for the remaining objects of the sample, which show a great variety of line profiles. 
Type 3 Balmer decrements cannot be well reproduced with neither the KF models or Case B, thus it is not clear if the \hi\ gas in sources that display type 3 Balmer decrements have properties in between those of the previous two groups. 
On the one hand, decrements that show a bump partially similar to that of type 3 curves are obtained in both Case B and KF models for temperatures T $<$ 5000K. These fairly low gas temperatures, however, do not agree with predictions of standard accretion models \citep[e.g.][]{muzerolle01}, so that in this scenario the \hi\ lines might originate in a different region.
On the other hand, our analysis suggests in some cases (e.g. Pa/H ratios, Paschen C decrements) optically thick emission from gas at high temperature and density. 
These contrasting interpretations and the lack of a good agreement between observations and the predicted decrement shapes  may simply indicate that the considered models are not adequate to describe the \hi\ emission in type 3 sources.
The tentative scheme described above is summarised in Table~\ref{tab:recap}, where the main combinations of decrements and profiles are connected with the gas properties and \hi\ emission characteristics.  

The Lupus low-mass star sample investigated in this paper, albeit incomplete, is well representative of the Class II population of the Lupus clouds in terms of both stellar, accretion, and emission line properties (Alcal\'a et al. 2017, submitted).
As we find evidence of a connection between circumstellar accretion activity and decrement types (types 2-A versus 4-B), it is plausible that the fraction of stars showing these types of decrement varies depending on average accretion rate of a given star-forming sample. 
This is for example in agreement with the mentioned fact that the T Tauri stars investigated by \citet{edwards13}, which are on average characterised by higher accretion rates than our Lupus sources, present typical Paschen decrements of type B (Sect.~\ref{sec:models:kf:dec}).
  
In addition, because the accretion rate decreases in time \citep[e.g.][]{antoniucci14d}, we might expect the relative fracion of decrement types is correlated to the age of the investigated sample. In this case, the decrement distribution we observe should be representative of young populations with a mean age similar to that of the Lupus cloud \citep[e.g. $\sim$3-4 Myr][]{mortier11}. 
Investigations similar to the one performed in this paper are encouraged to study the decrements in samples of different age.     

We have no knowledge about the possible variability of the observed decrements and its time-scales. This kind of investigation would certainly provide additional information on the properties of the \hi-emitting gas and help to shed light on those cases that are currently difficult to interpret.

The \hi\ emission lines are commonly associated with gas in the magnetospheric accretion columns \citep[e.g.][]{muzerolle98b,muzerolle00}. 
The range of gas densities and temperatures derived from the decrement analysis ($n_H > 10^{9}$ cm$^{-3}$, $T$ = 5\,000-15\,000 K) is indeed in good agreement with accretion models \citep[e.g.][]{muzerolle01}.
Many recent works have considered, however, that an important contribution to the line may come also from winds that originate in the inner regions of the sources \citep[e.g.][]{lima10,kurosawa12}. The presence of additional (wind) emission components is indeed suggested by some of the features evidenced in the profile analysis (Sect.~\ref{sec:lines}). 
These models do a reasonable job in reproducing the observed line shapes and put in evidence that the great variety of line profiles depends not only on the emitting gas properties but also on geometrical factors such as the inclination and the extent of the magnetospheric/wind region \citep[e.g.][]{kurosawa11}. 
We can indirectly evaluate a possible contribution from winds based on the gas properties derived in our decrement analysis, in particular for the NS sources, which show the simplest line profiles.
In our previous work, \citet{natta14} studied the so-called slow winds in our Lupus sample. Based on the analysis of the slightly blue-shifted low velocity component of various optical forbidden lines (e.g. \oi, \sii), which have typical FWHM  in the range 50-100 \kms and signal the presence of outflowing matter at low velocities, \citet{natta14} derived $n_H > 10^{8}$ cm$^{-3}$ and $T$ = 5000-7000 K for the gas in the slow winds. These conditions are compatible with those derived for the type 2 decrements of the NS sources and suggest that the same gas can be responsible for at least part of the \hi\ emission.
On the contrary, we can rule out a significant contribution to \hi\ emission from photo-evaporative winds, because the line widths expected for such winds (up to $\sim 20$\kms, e.g. Ercolano \& Owen 2010 \nocite{ercolano10}) are much smaller than those observed and the profiles show no evidence of a narrow component at the line core.

\begin{table*}[t]
\begin{small}
\begin{center}
\caption{\label{tab:recap} Physical conditions of the \hi-emitting gas for the main decrement-profile combinations, as suggested by comparison with models.}
\begin{tabular}{l | c | c | c | c | l}
\hline
\hline
Decrement & Profile & $T$ (K) & log $n_{HI}$(cm$^{-3}$) & \# of objects & Comments\\
\hline
1     & W,MP  & ...   &  ...   & 3 & only in sub-luminous sources; probably reddened type 2 decrements\\
2-A   & NS    & $5\,000-15\,000$   &  9.4-9.8   & 11 & optically thin, weakest accretors\\
2-A   & W,MP  & $5\,000-15\,000$   &  9.4-9.8   & 9 & ... \\
3     & W,MP  & ? &  ?  & 7 & decrements not reproduced by models\\
4-B   & W,MP  & $\lesssim$9000     &  $\gtrsim$ 11 & 6 & optically thick, very wide profiles, strongest accretors\\
\hline
\end{tabular}
\end{center}

\end{small}
\end{table*}

\section{Summary}
\label{sec:outro}

We presented a study of the \hi\ recombination lines observed in the X-Shooter spectra of a sample of 36 T Tauri objects in the Lupus star-forming region, 
for which we had already derived in a previous work the stellar and accretion properties in a self-consistent and uniform fashion. 
In our analysis we focus attention on the \hi\ series decrements and their connection with both line profiles and source properties. The aim is to obtain information on the physical conditions of the \hi\ gas so as to draw a consistent picture of the \hi\ emission in a sample representative of T Tauri objects. 
The main conclusions of our work can be summarised as follows.

The Balmer lines present a great variety of profiles, indicative of various kinematic conditions in the circumstellar environment.
About one third of the objects display profiles that are fairly narrow (FWHM $\sim$ 100 km/s) and symmetric. We refer to these objects as narrow symmetric line sources. All remaining objects present more complex \hi\ profiles that we divide into wide symmetric and asymmetric or multi-peaked. 
We observe a roughly equal distribution of the red and blue asymmetries.
Paschen lines are observed with a good S/N only in sources with wide or asymmetric Balmer lines. They appear in general more symmetric, although they often retain signatures of the same absoprtion features seen in the Balmer lines. 
Several features observed in many \hi\ line profiles (e.g. very extended wings, blue/red-shifted absorptions, inverse P-Cygni profiles in \pab)  
are indicative of both infalling and outflowing matter in the circumstellar region of the targets. 

We have empirically classified the observed Balmer decrements into four types (1 to 4) that present different shapes. 
Type 2 is the most common and all narrow symmetric sources present this kind of shape, supporting the idea that these objects share the same \hi\ emission modality.
We found no clear relationship between Balmer decrement type and source properties, except for a tentative connection with the mass accretion rate. Indeed, stronger accretors of our sample display similar Balmer decrement shapes that tend to type 4 and display very wide line profiles.
Type 1 Balmer decrements are found only in three sub-luminous sources viewed edge-on and this suggests that the observed shape might be the result of neglecting a (residual) amount of extinction in the line emission region. If this scenario is correct, then the actual decrement in these objects is probably of type 2.
Despite the lower quality of the near-IR data, we can identify three different shapes (named A, B, C) also for Paschen decrements. All the sources with a type 4 Balmer decrement show a type B Paschen decrement, indicating that these decrements are strongly related. A strong connection between type 2 Balmer decrements and type A Paschen decrements is also revealed.

The comparison of the measured decrements with the predictions of both the Case B emission model \citep{baker38} and the local line excitation calculations of Kwan \& Fischer (2011) yields the following results. 
Case B can formally reproduce some of the observed (type 2) decrements, suggesting fairly large electron densities ($n_e \sim 10^{10}$ cm$^{-3}$) and low temperatures ($T\lesssim$ 2000~K), in line with results of previous works \citep[e.g.][]{bary08}. However, these conditions require a strong photo-ionisation regime that is hardly found in the circumstellar region of CTTSs, so that strong doubts remain about the applicability of Case B in this case.
Using the KF models we find a good agreement with type 2 and 4 Balmer decrements and type A and B Paschen decrements, suggesting log $n_H$ = 9.4-9.8 cm$^{-3}$ with  temperatures more difficult to constrain ($T$ = 5\,000-15\,000 K) for type 2-A, and log $n_H$ $\gtrsim$ 11 cm$^{-3}$ and $T \lesssim 9\,000$ K for type 4-B.
Type 3 and type C shapes cannot be well reproduced by current models and remain the most difficult to interpret. 

Even though it is difficult to derive a consistent and comprehensive picture that can explain all the different \hi\ decrement shapes and the rich variety of line profiles, we draw a tentative scheme that connects the observed decrements to the emitting gas properties and to the general degree of accretion of the sources. 
In this scheme the narrow symmetric line sources (11 out of 36 in our sample) appear as the least accretion-active objects, generally associated with lower mass accretion rates.
They present lines that are compatible with an optically thin emission regime and type 2-A decrements (the most common shapes), indicative of gas densities of order 10$^9$ cm$^{-3}$). A plausible origin for these narrow lines are the accretion flows \citep[e.g.][]{muzerolle01} and/or slow winds from the disk \citep[e.g.][]{kurosawa11, natta14}. This type of emission would represent the basic \hi\ line manifestation in CTTSs.
On the other end, we find objects with decrements of type 4-B, which are indicative of log $n_H$ $\gtrsim$ 11 cm$^{-3}$. They present very wide line profiles and are on average associated with the strongest accretors of the sample, which suggests that these targets are the most accretion-active objects with high-density and high-velocity gas in their circumstellar structures.
The gas conditions indicated by the type 2-A decrements are the most common in our sample (23 objects out of 36, if we include also the sub-luminous objects).
We cannot derive clear indications for gas conditions in sources with a type 3 Balmer decrement, for which decrement and line ratio analyses provide incongruous results. 

The discrepancies between models and observations might indicate various emission components from regions with different conditions, so that the description provided by the current KF models may not be accurate in such cases.
New sets of KF models with input parameters tuned on the line properties actually observed in our sample might provide refined predictions that are able to better match the observations.

\subsection*{Acknowledgements}
\begin{footnotesize}
We are very grateful to the anonymous referee for all suggestions and comments, which helped us to improve the quality of the paper.

\noindent
The authors are very grateful to John Kwan for providing the predictions of his local line excitation models and for fruitful discussions.

\noindent
SA acknowledges the funding support from the PRIN INAF 2012 ``Disks, jets, and the dawn of planets" and from the T-REX-project, the INAF (Istituto Nazionale di Astrofisica) national project aimed at maximizing the participation of astrophysicists and Italian industries
to the realization of the E-ELT (European Extremely Large Telescope). The T-REX project has been approved and funded by the Italian Ministry for Research and University (MIUR) in the framework of “Progetti Premiali 2011” and “Progetti Premiali 2012”.

\noindent
AN aknowledges funding from Science Foundation Ireland (Grant 13/ERC/12907)

\end{footnotesize}

\bibliographystyle{aa} 
\bibliography{refs} 

\Online

\section*{Appendix A: Atlas of line profiles}
\setcounter{figure}{0} \renewcommand{\thefigure}{A.\arabic{figure}} 
\setcounter{table}{0} \renewcommand{\thetable}{A.\arabic{table}}

\begin{figure*}[t]
\centering
\includegraphics[width=6.cm]{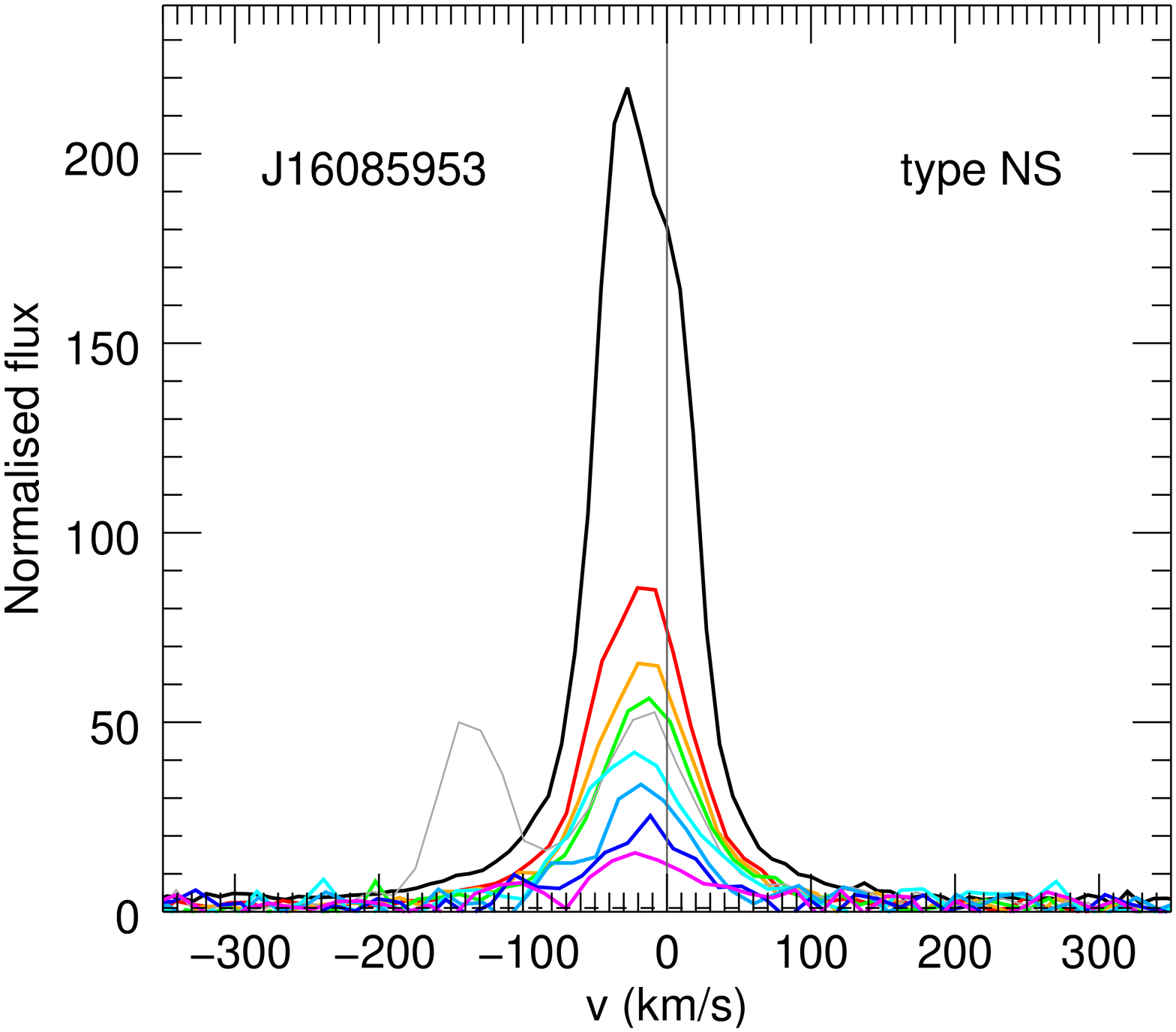}
\includegraphics[width=6.cm]{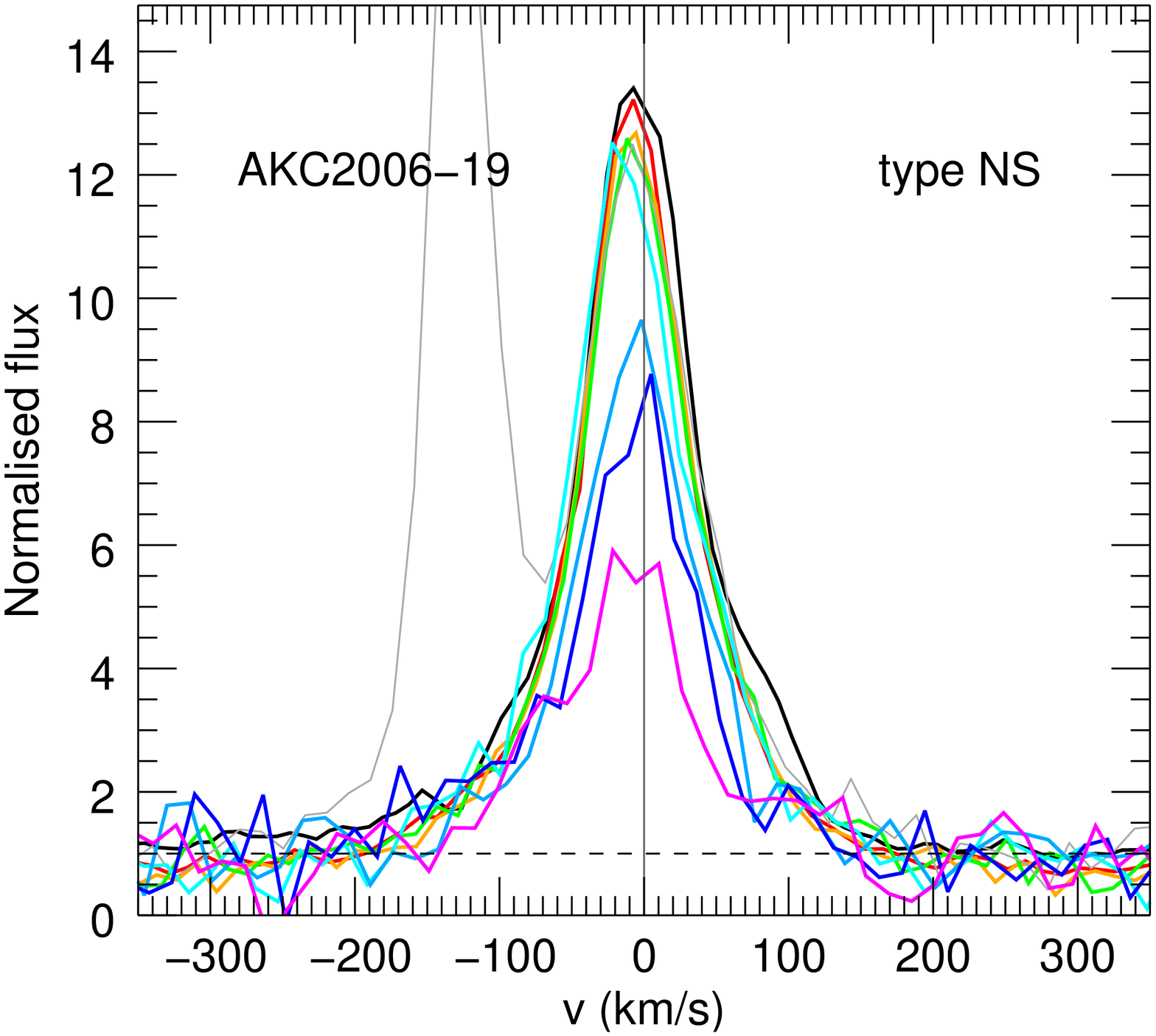}
\includegraphics[width=6.cm]{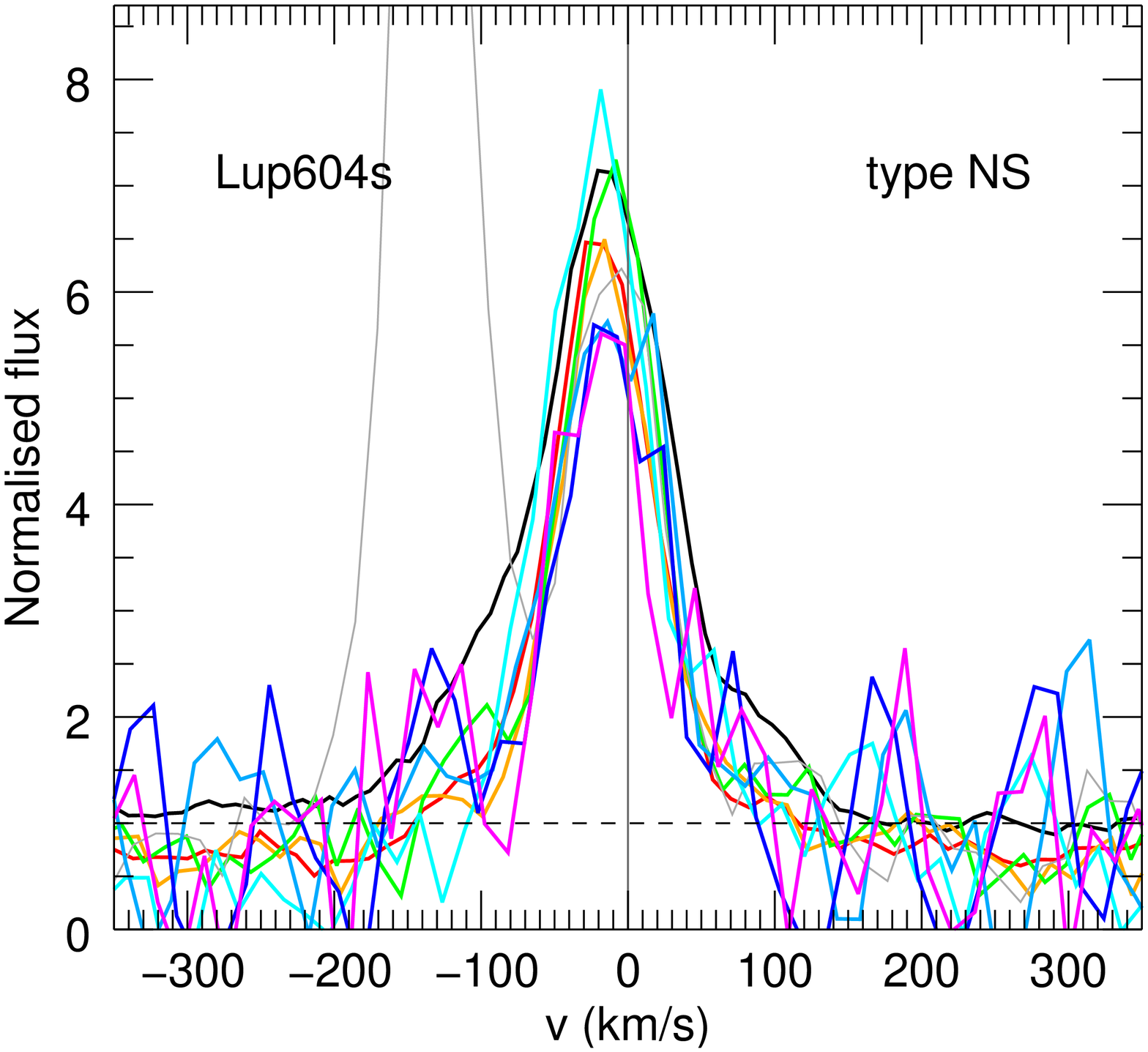}\\
\includegraphics[width=6.cm]{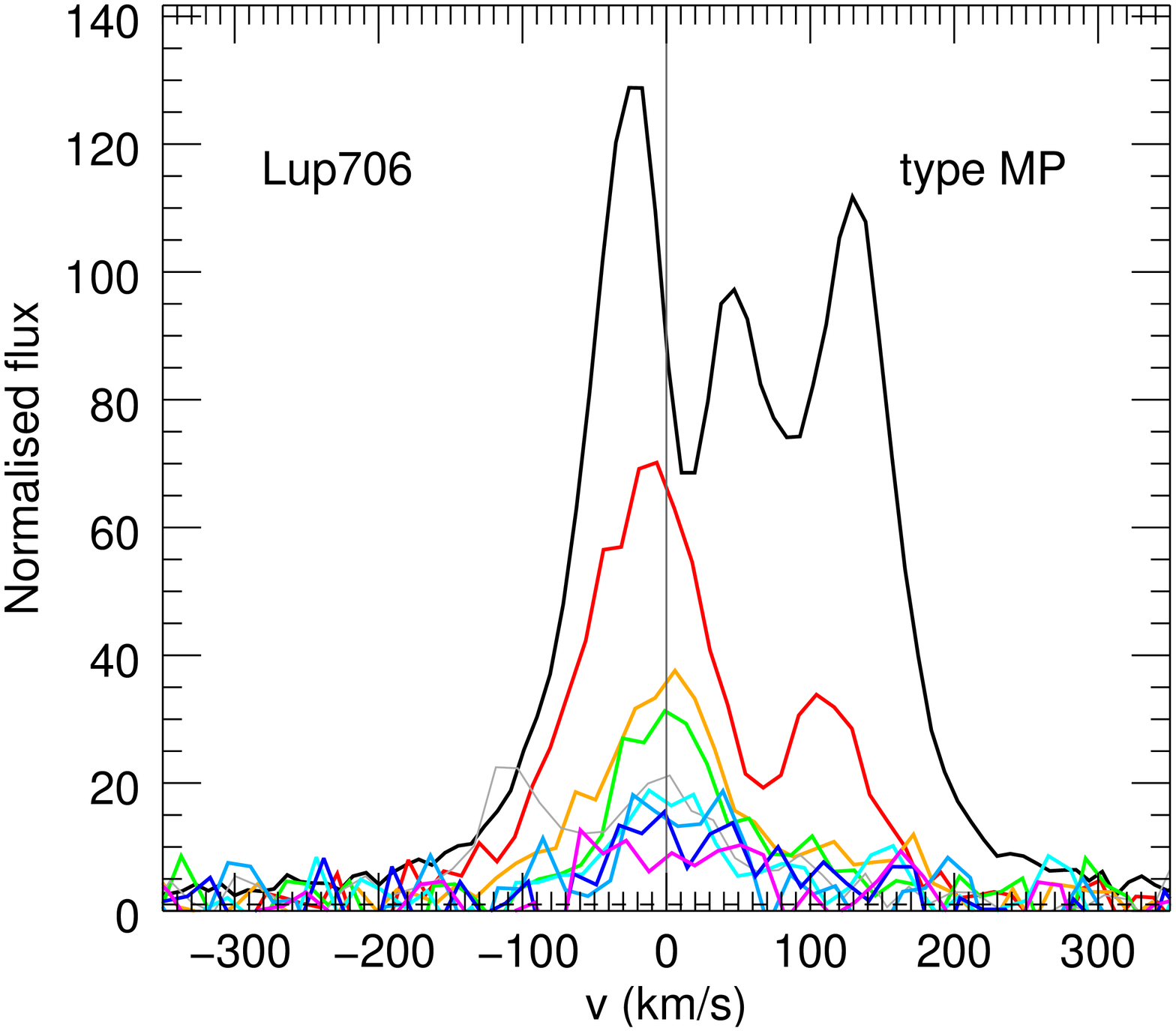}
\includegraphics[width=6.cm]{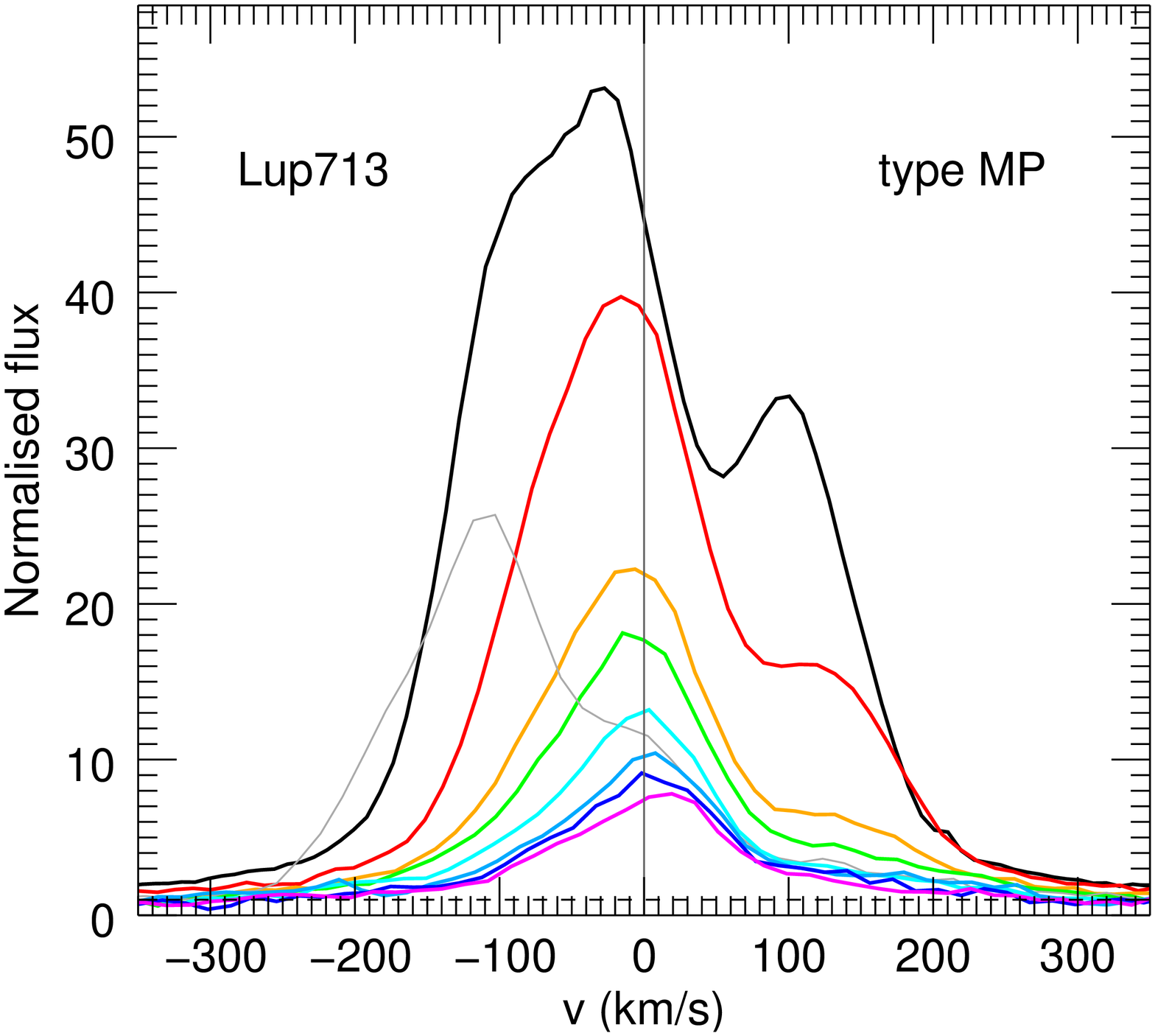}
\includegraphics[width=6.cm]{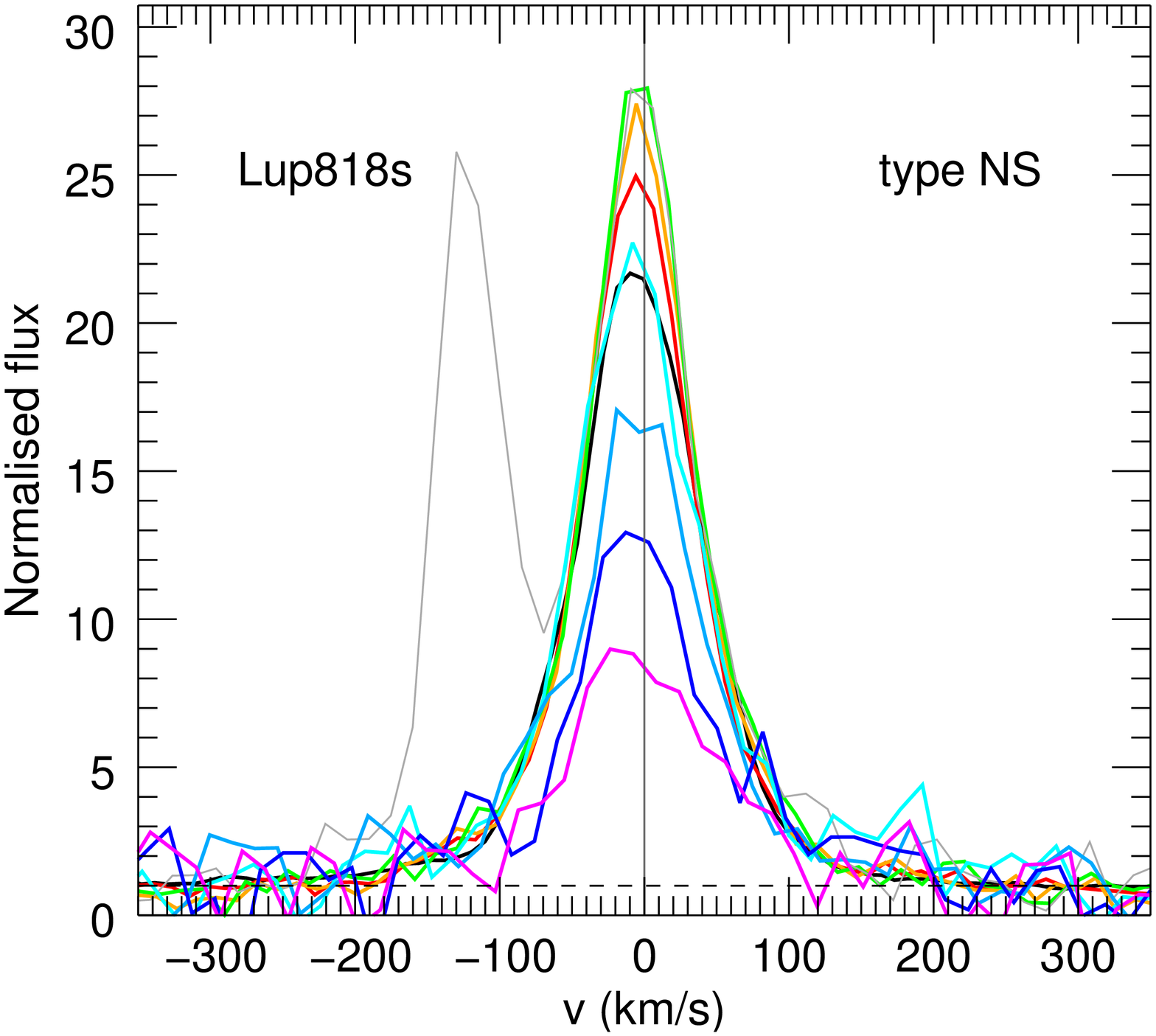}\\
\includegraphics[width=6.cm]{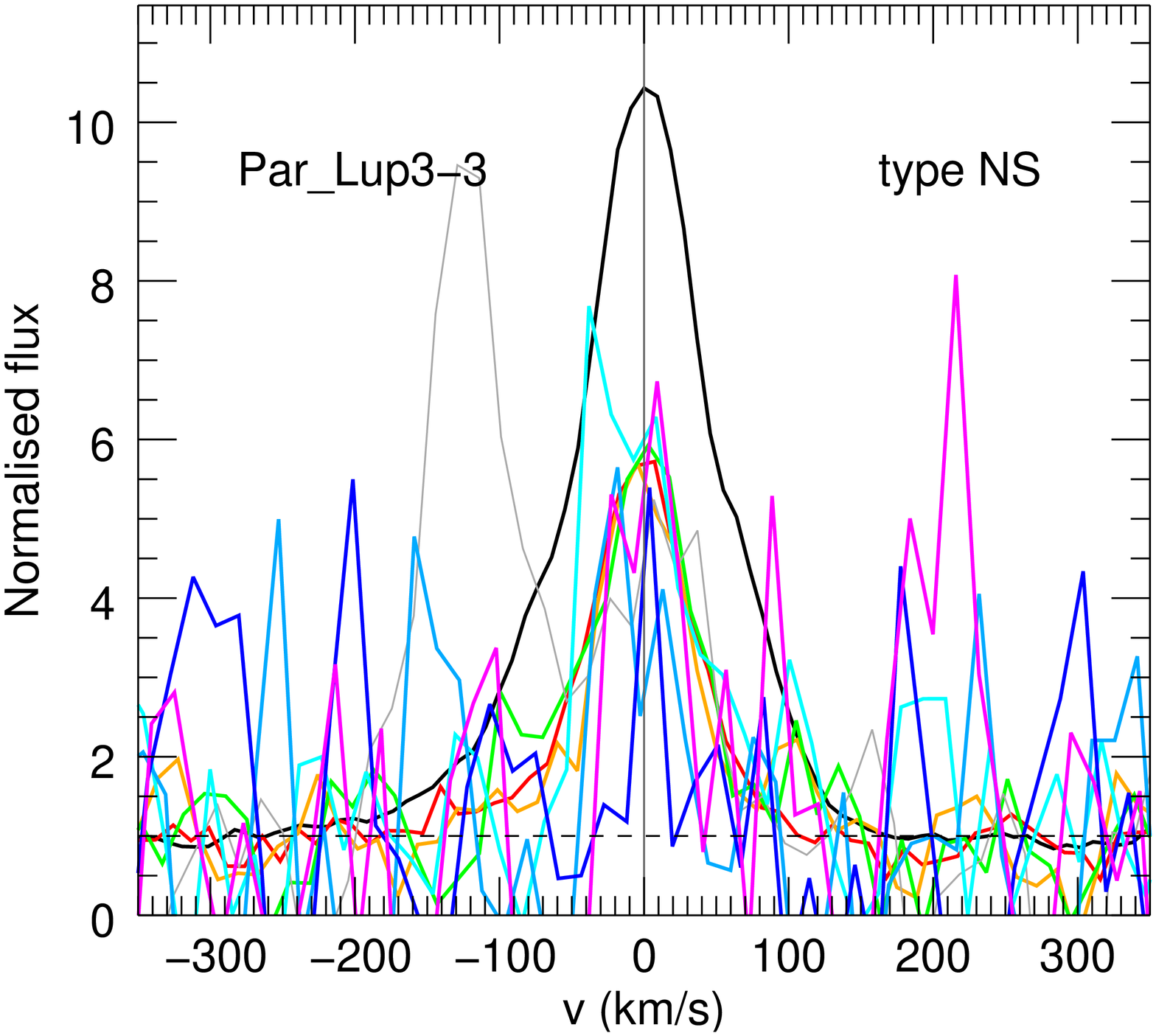}
\includegraphics[width=6.cm]{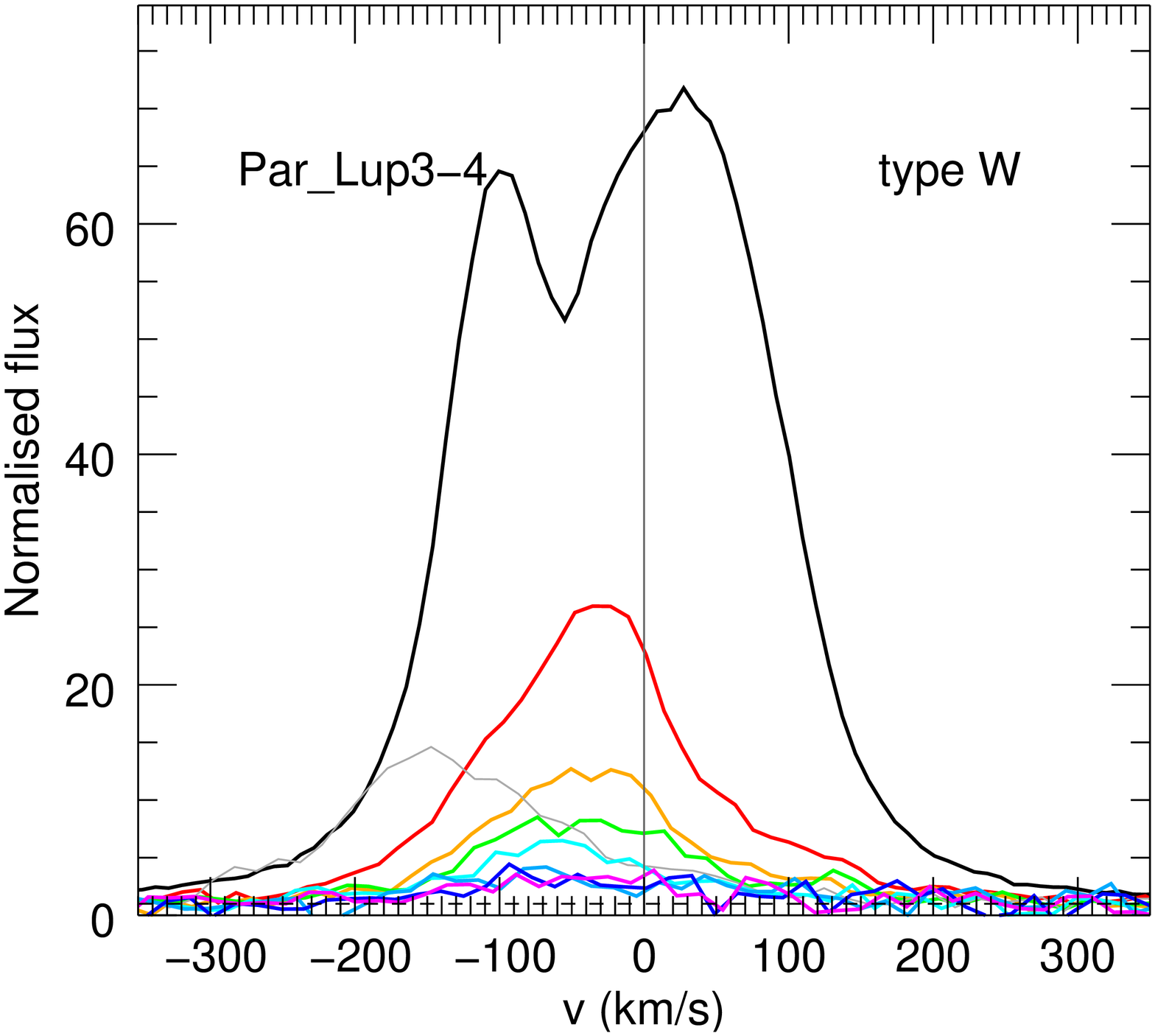}
\includegraphics[width=6.cm]{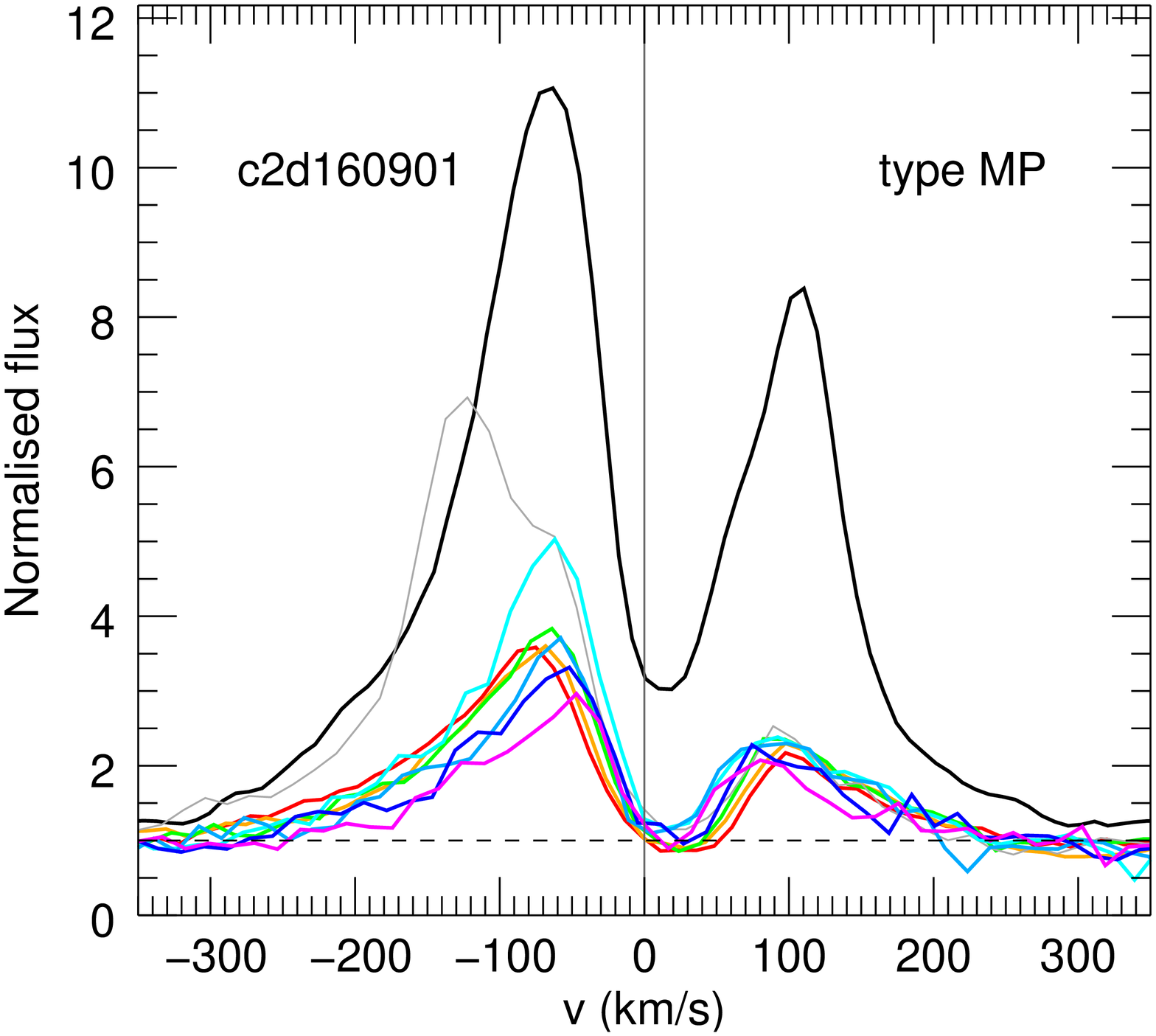}\\
\includegraphics[width=6.cm]{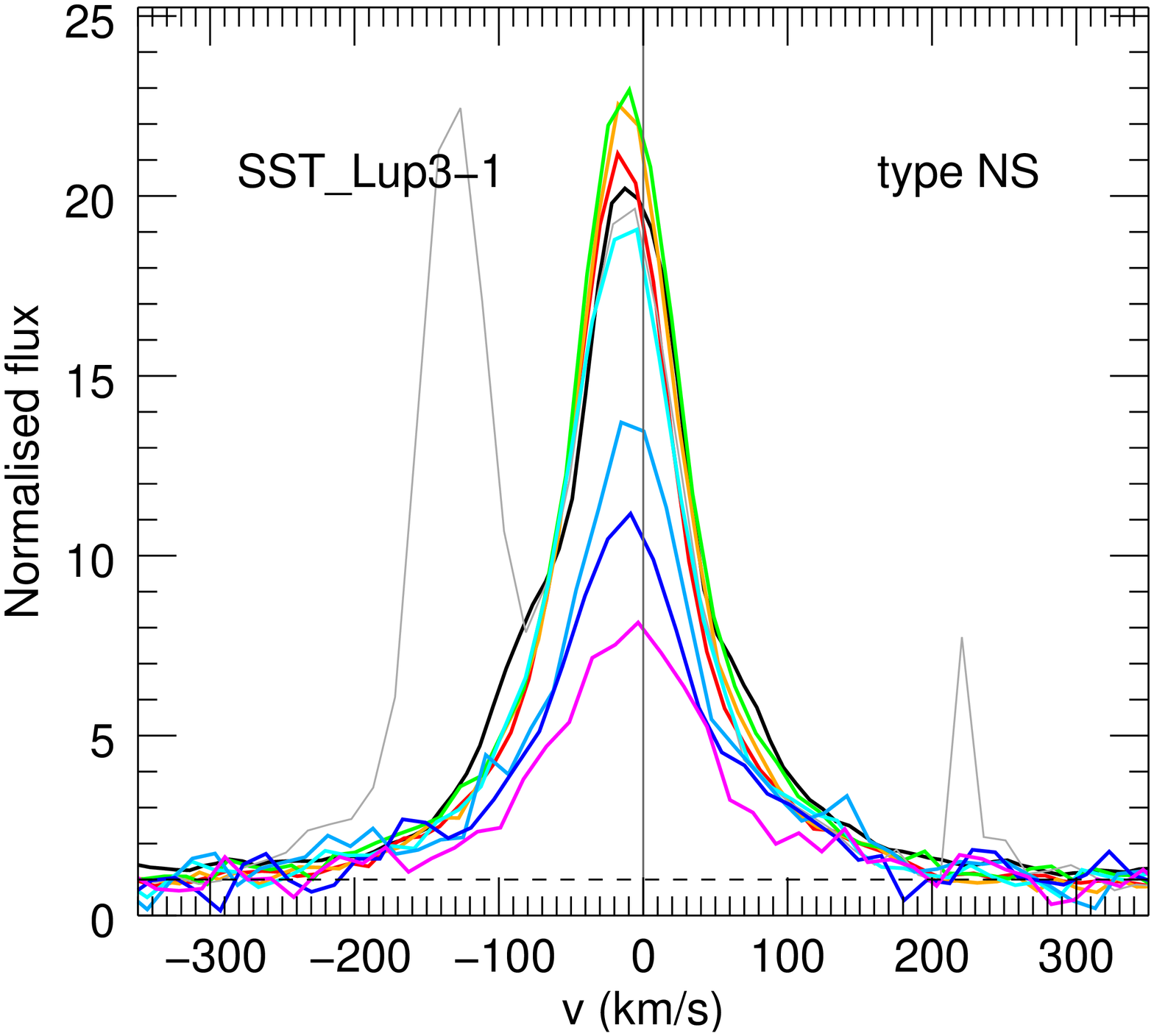}
\includegraphics[width=6.cm]{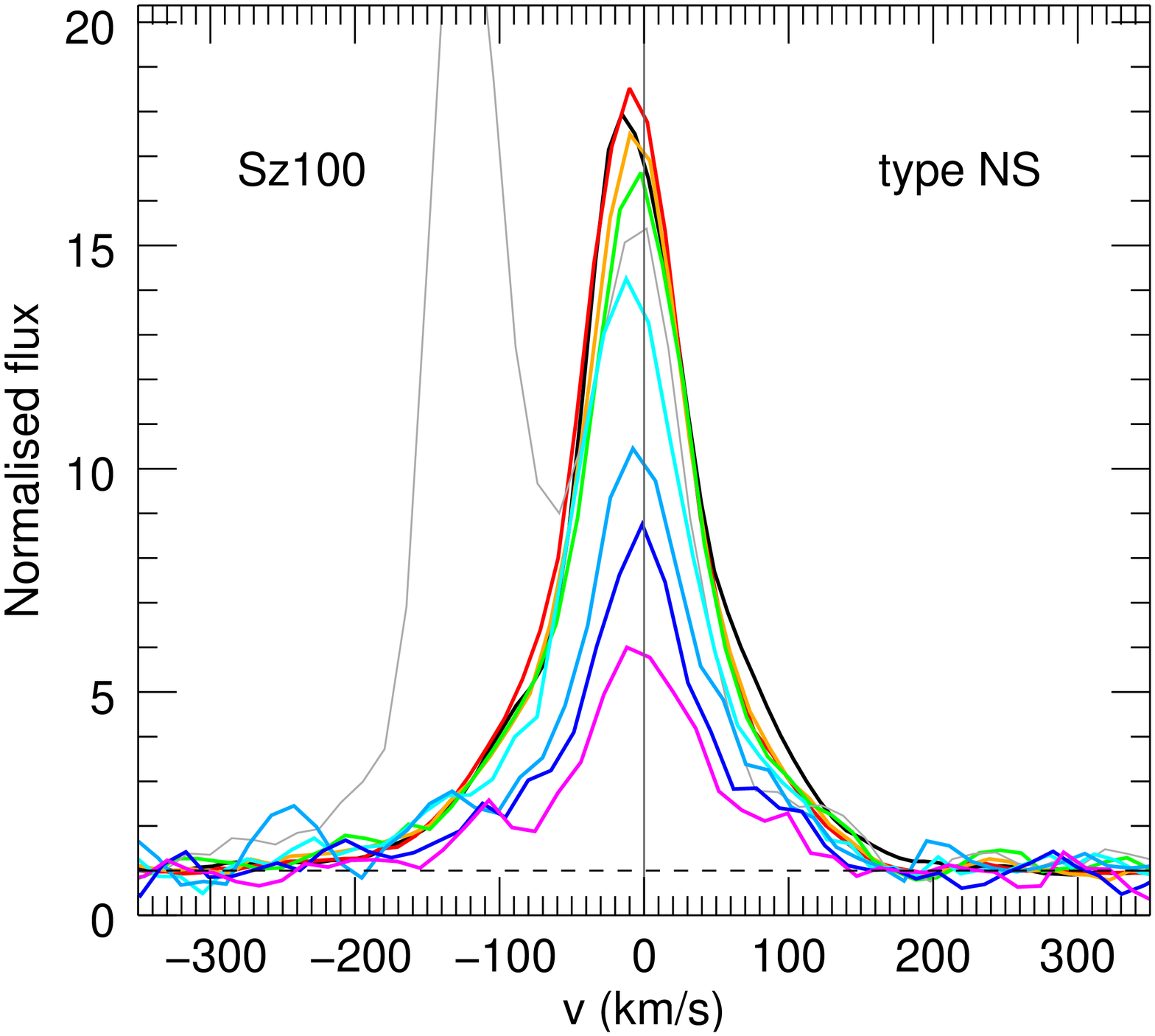}
\includegraphics[width=6.cm]{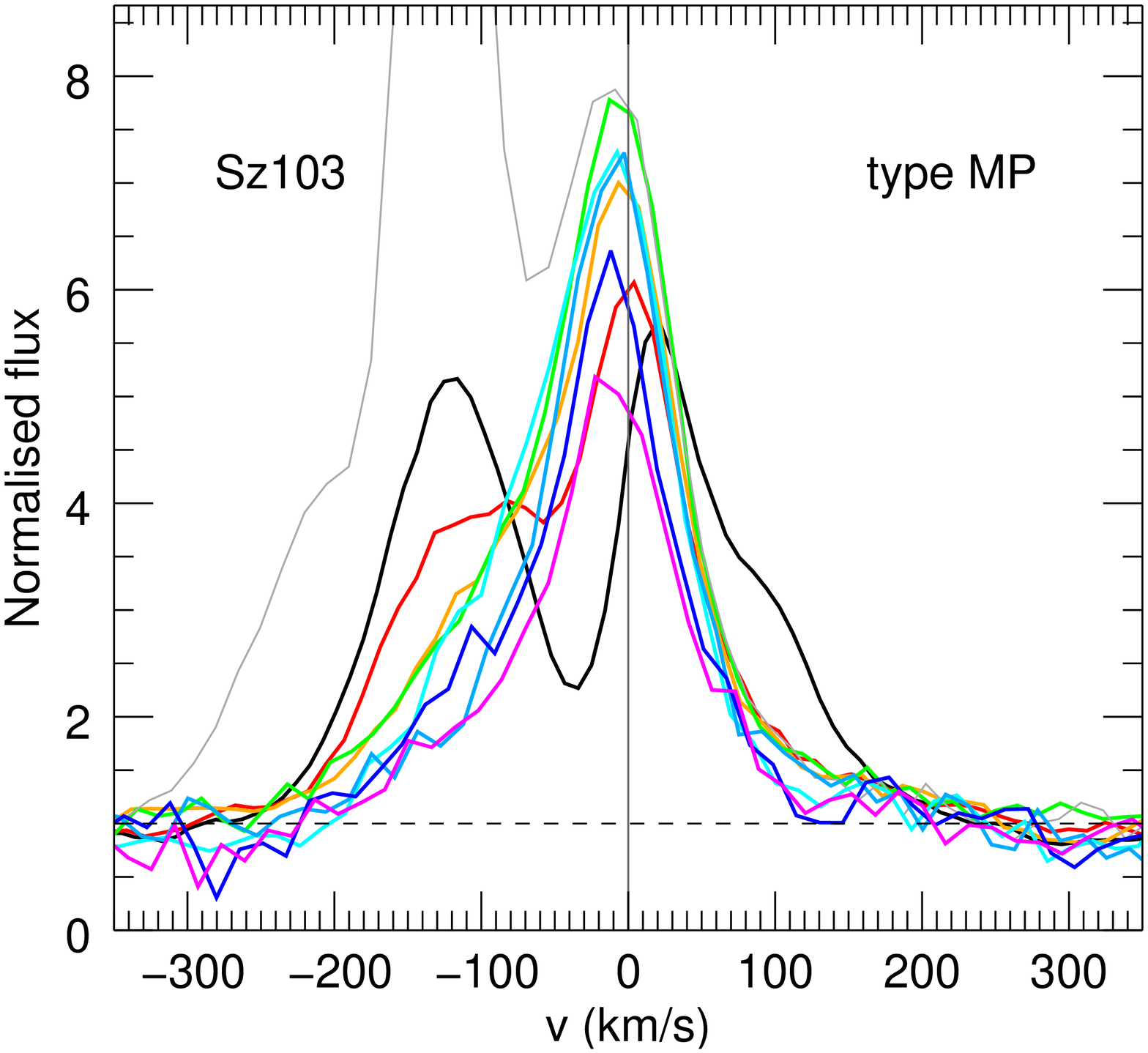}

\caption{\label{fig:lines:h} Continuum-normalised Balmer lines. \ha\ (black), \hb\ (red), H$\gamma$ (orange), H$\delta$ (green), H8 (cyan), H9 (turquoise), H10 (blue), and H11 (magenta) are plotted. The H7 line, which is blended with the CaII H line, is marked in grey.} 
\end{figure*}

\setcounter{figure}{0}
\begin{figure*}[t]
\centering
\includegraphics[width=6.cm]{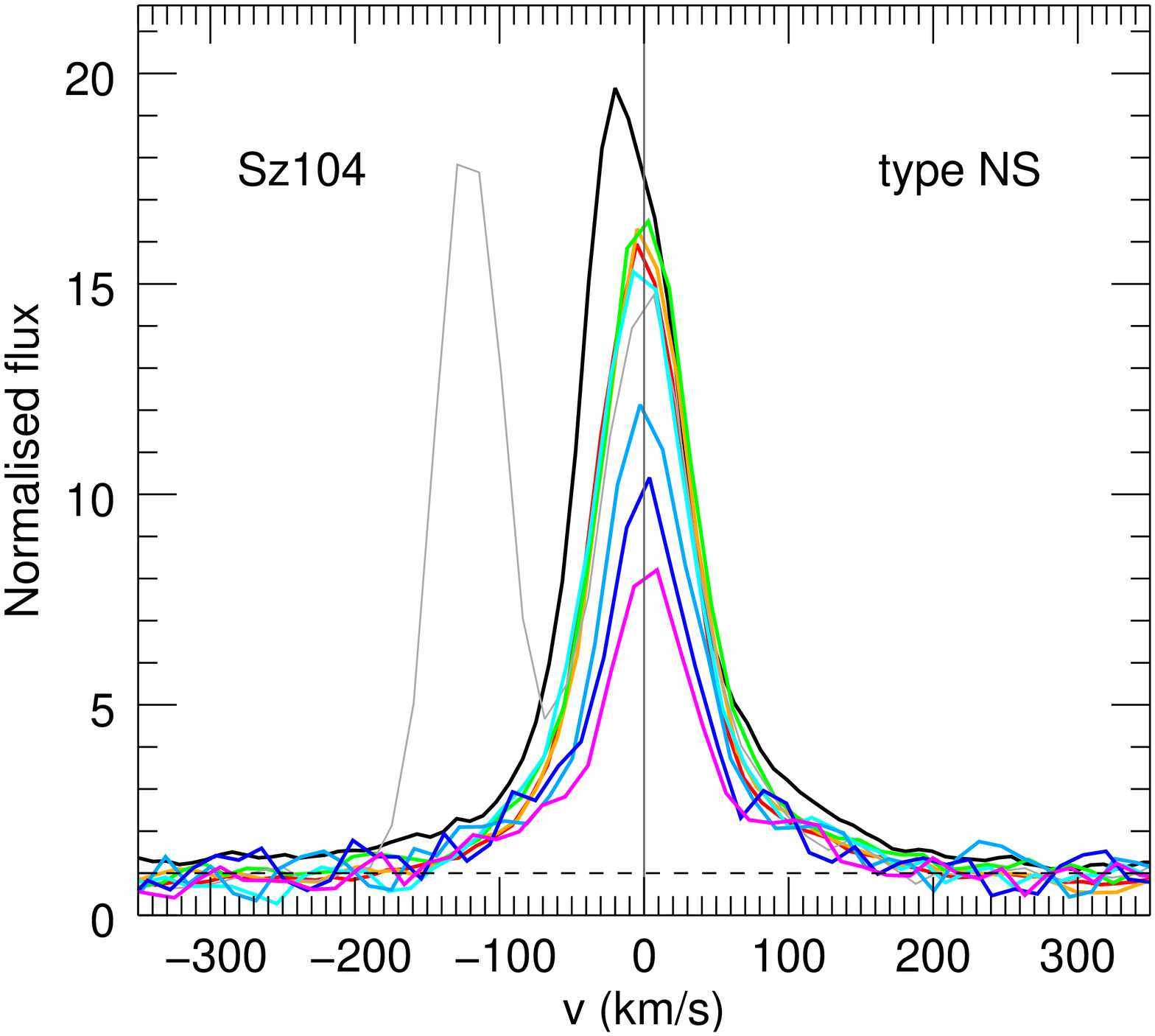}
\includegraphics[width=6.cm]{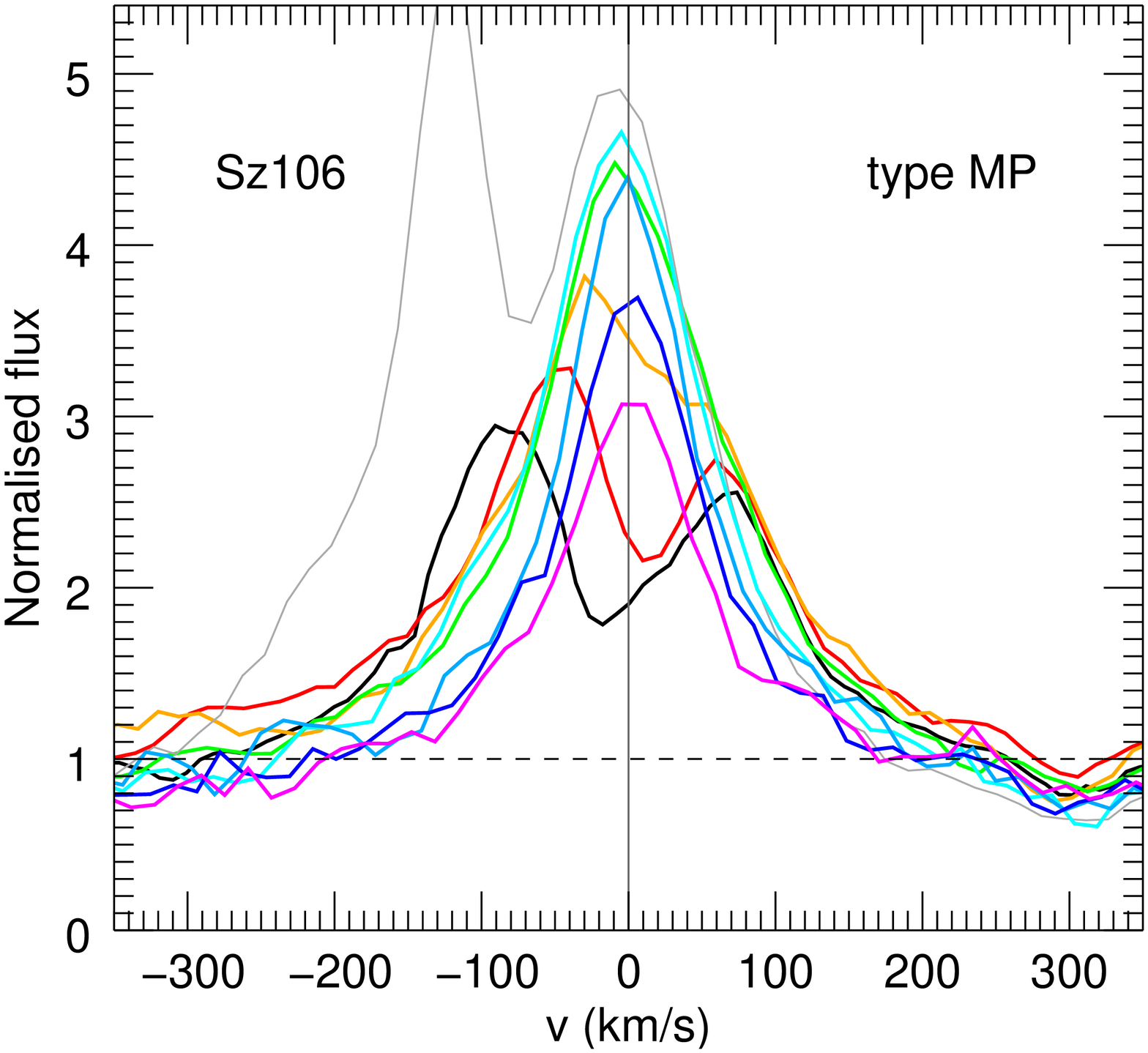}
\includegraphics[width=6.cm]{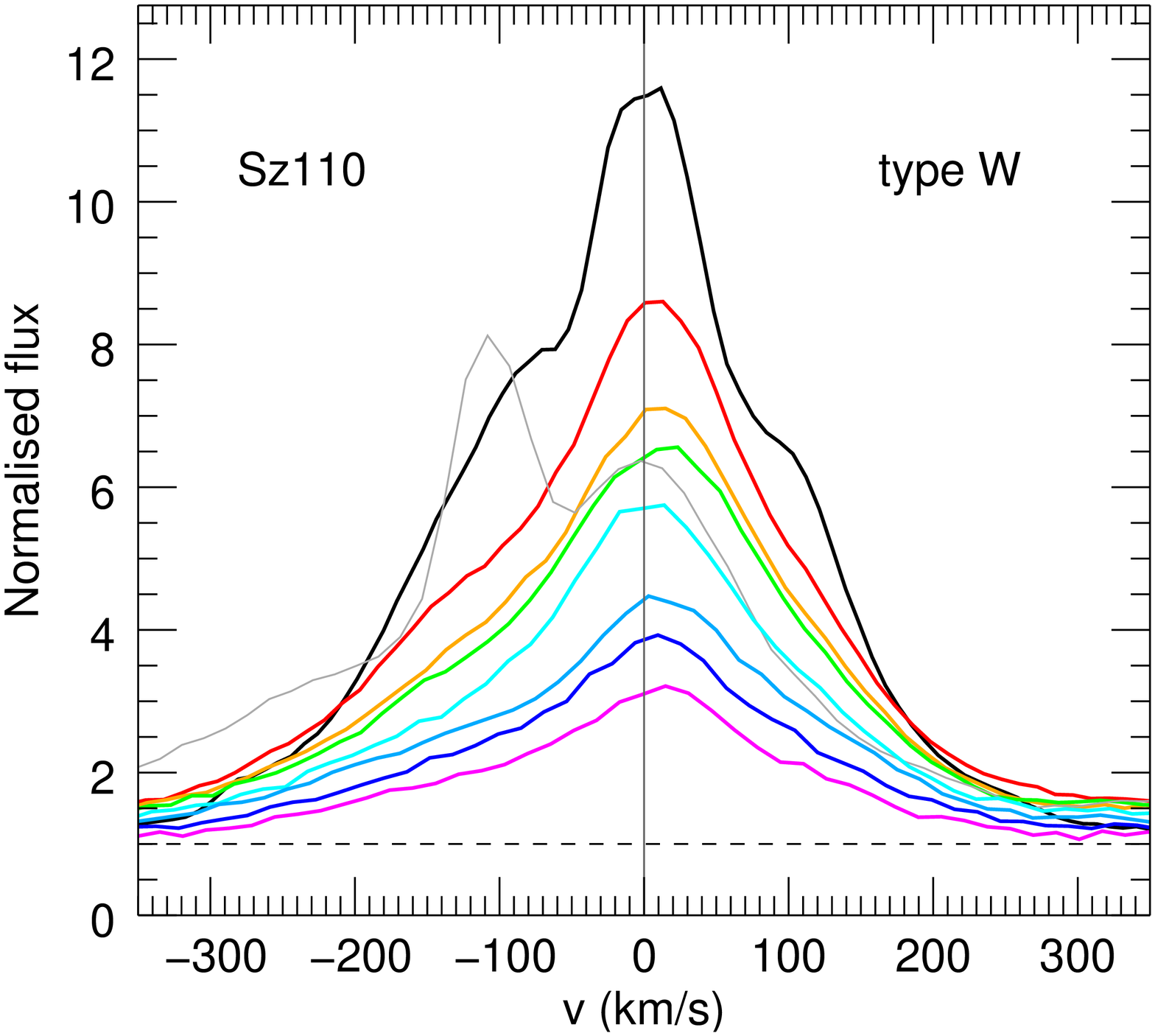}\\
\includegraphics[width=6.cm]{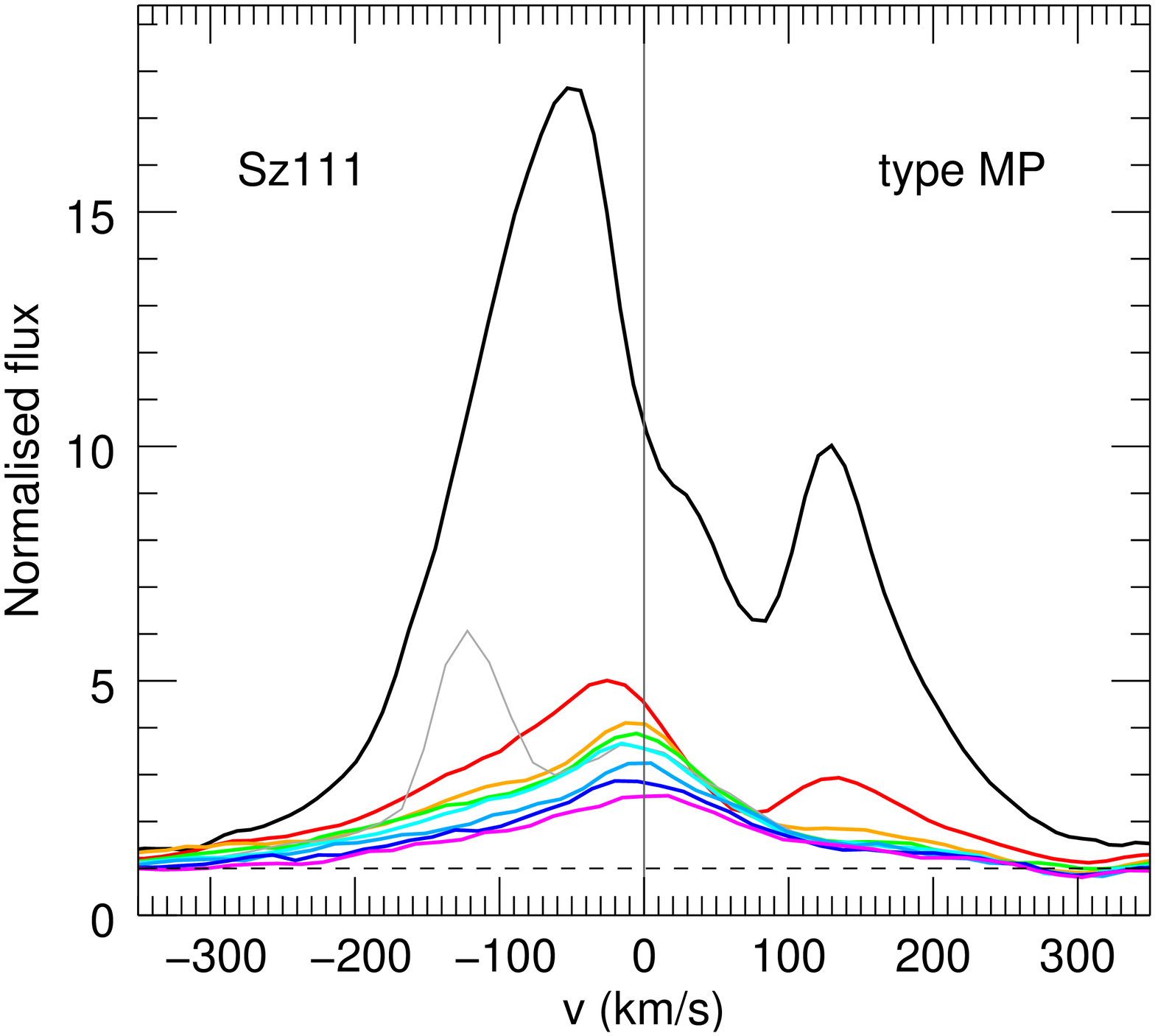}
\includegraphics[width=6.cm]{Sz112_Hprof.eps}
\includegraphics[width=6.cm]{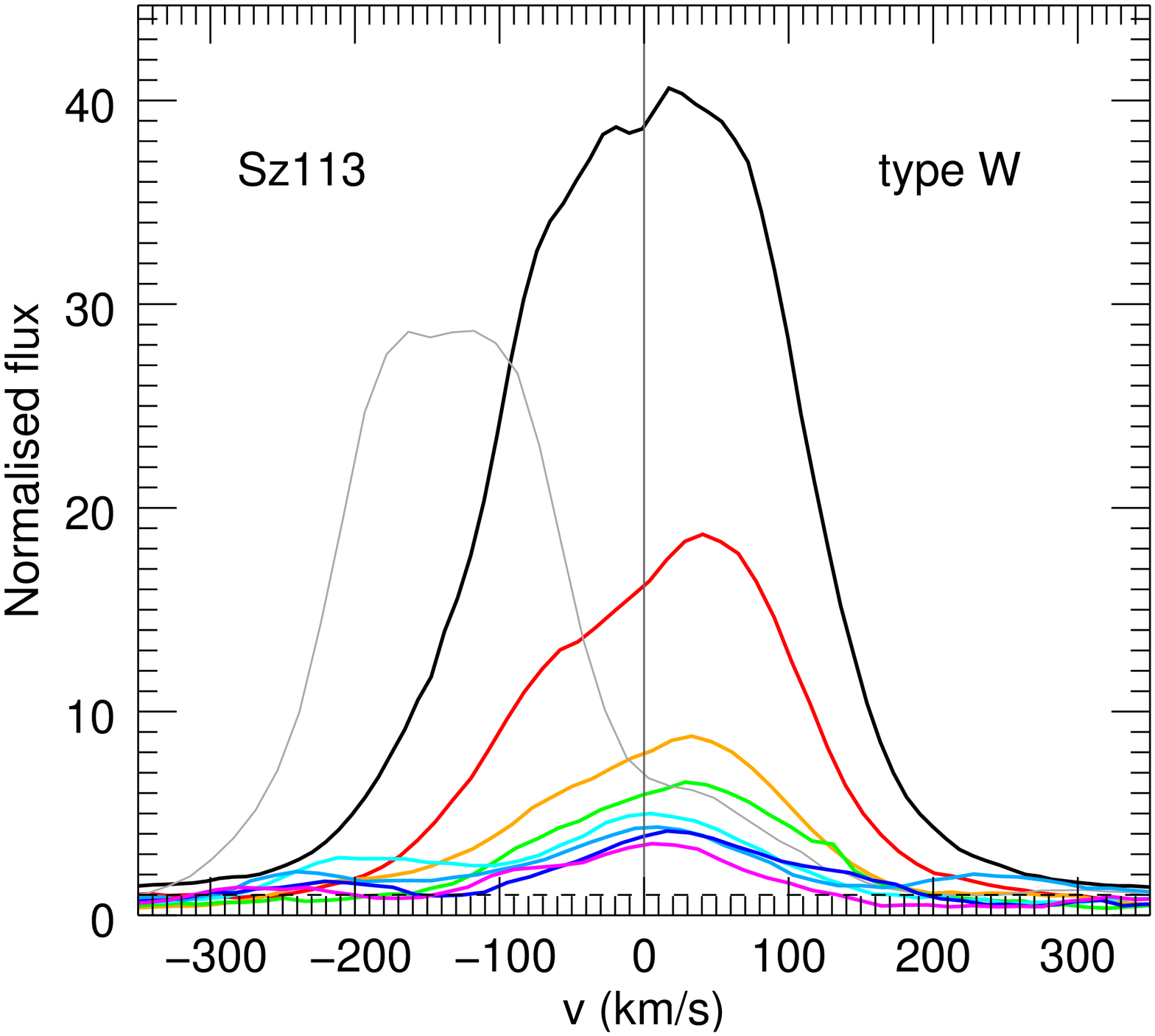}\\
\includegraphics[width=6.cm]{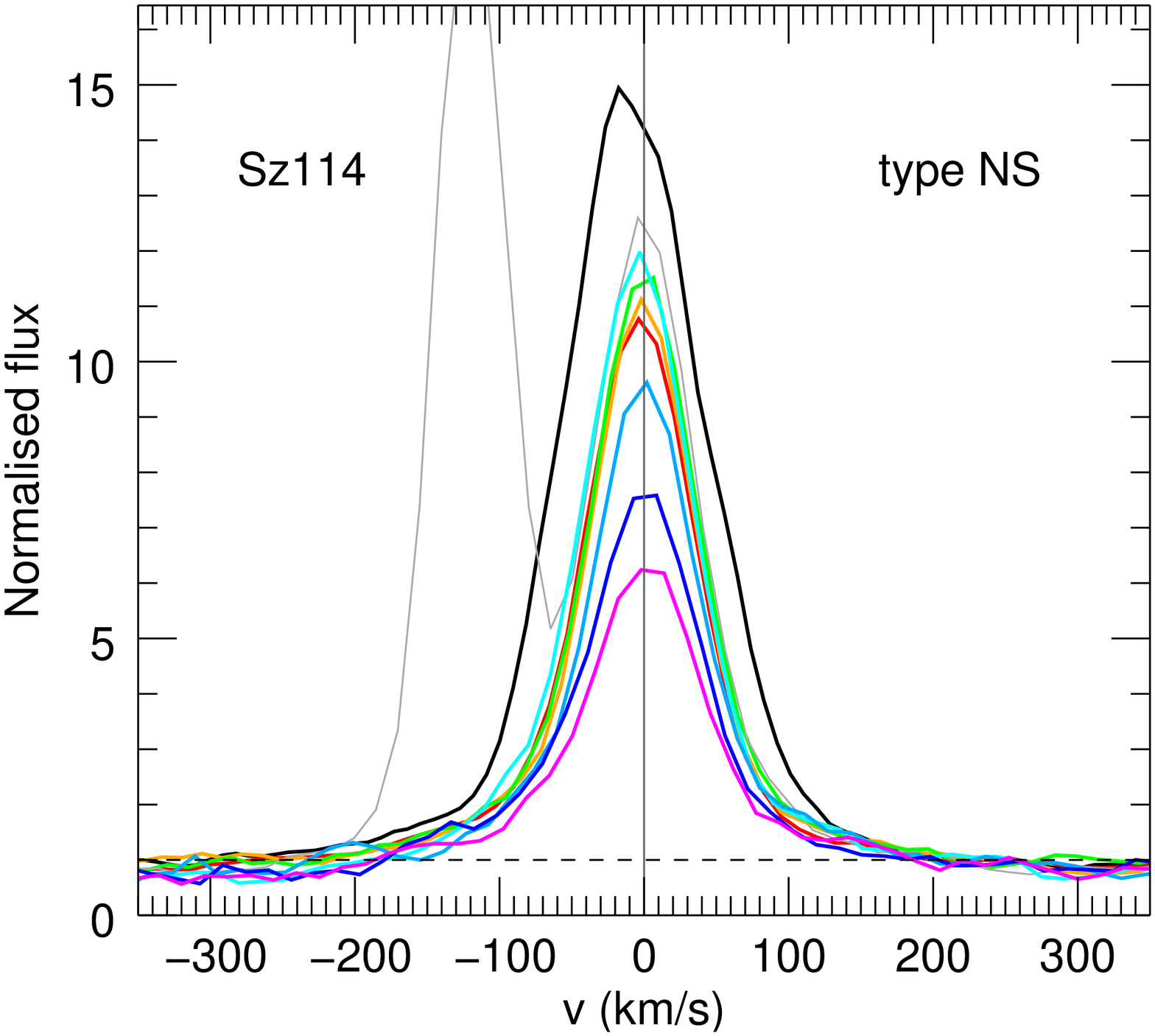}
\includegraphics[width=6.cm]{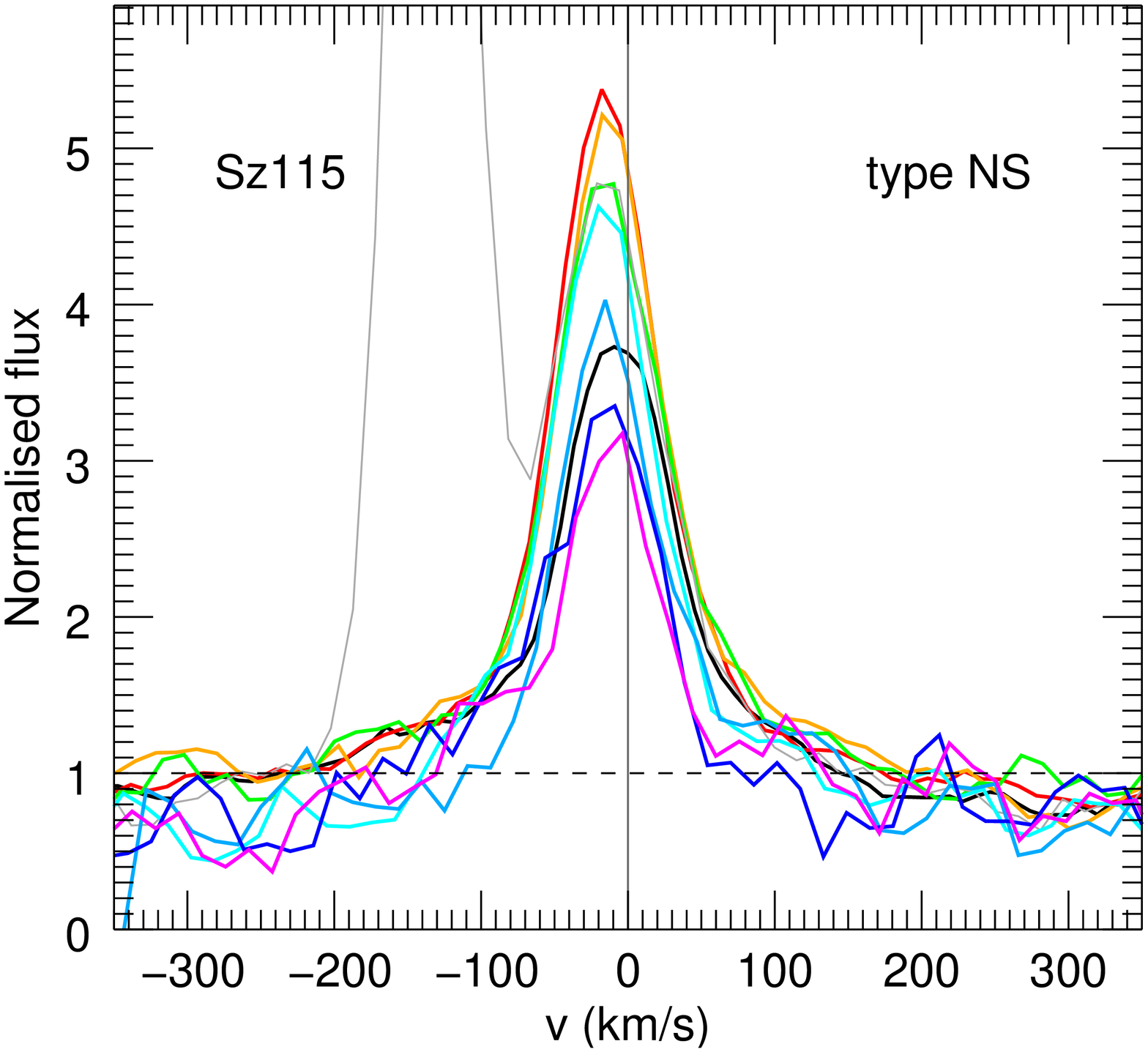}
\includegraphics[width=6.cm]{Sz123A_Hprof.eps}\\
\includegraphics[width=6.cm]{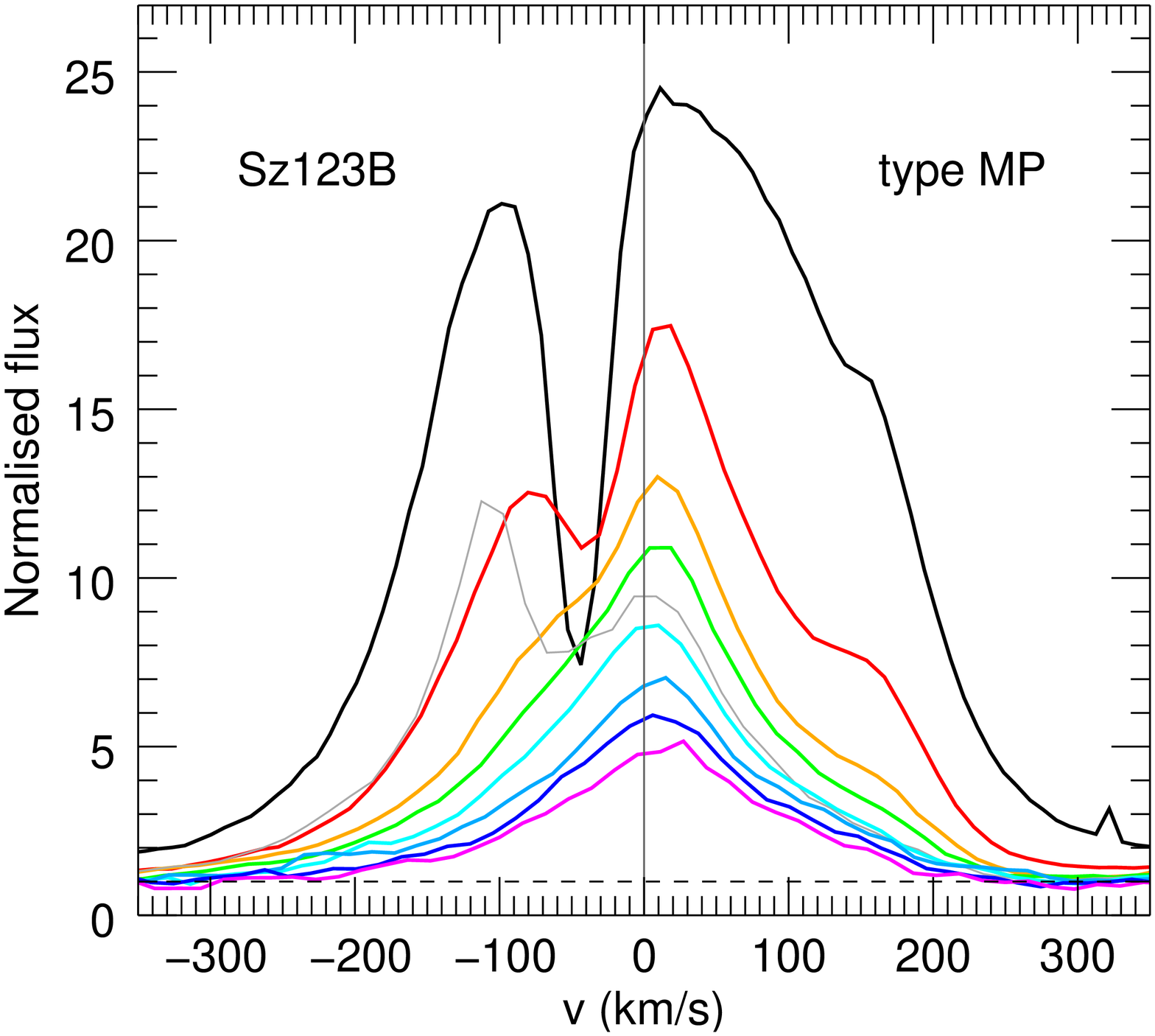}
\includegraphics[width=6.cm]{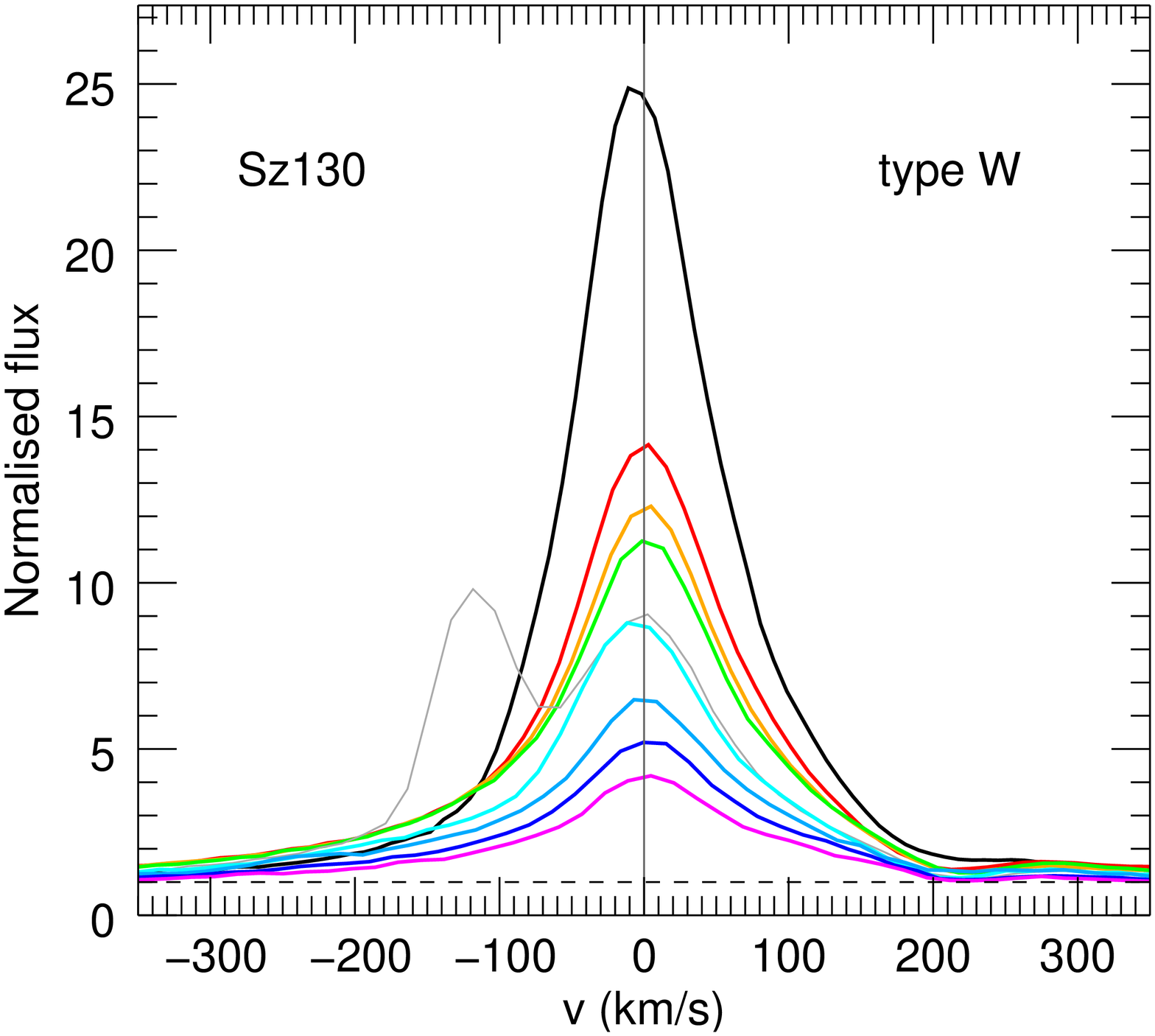}
\includegraphics[width=6.cm]{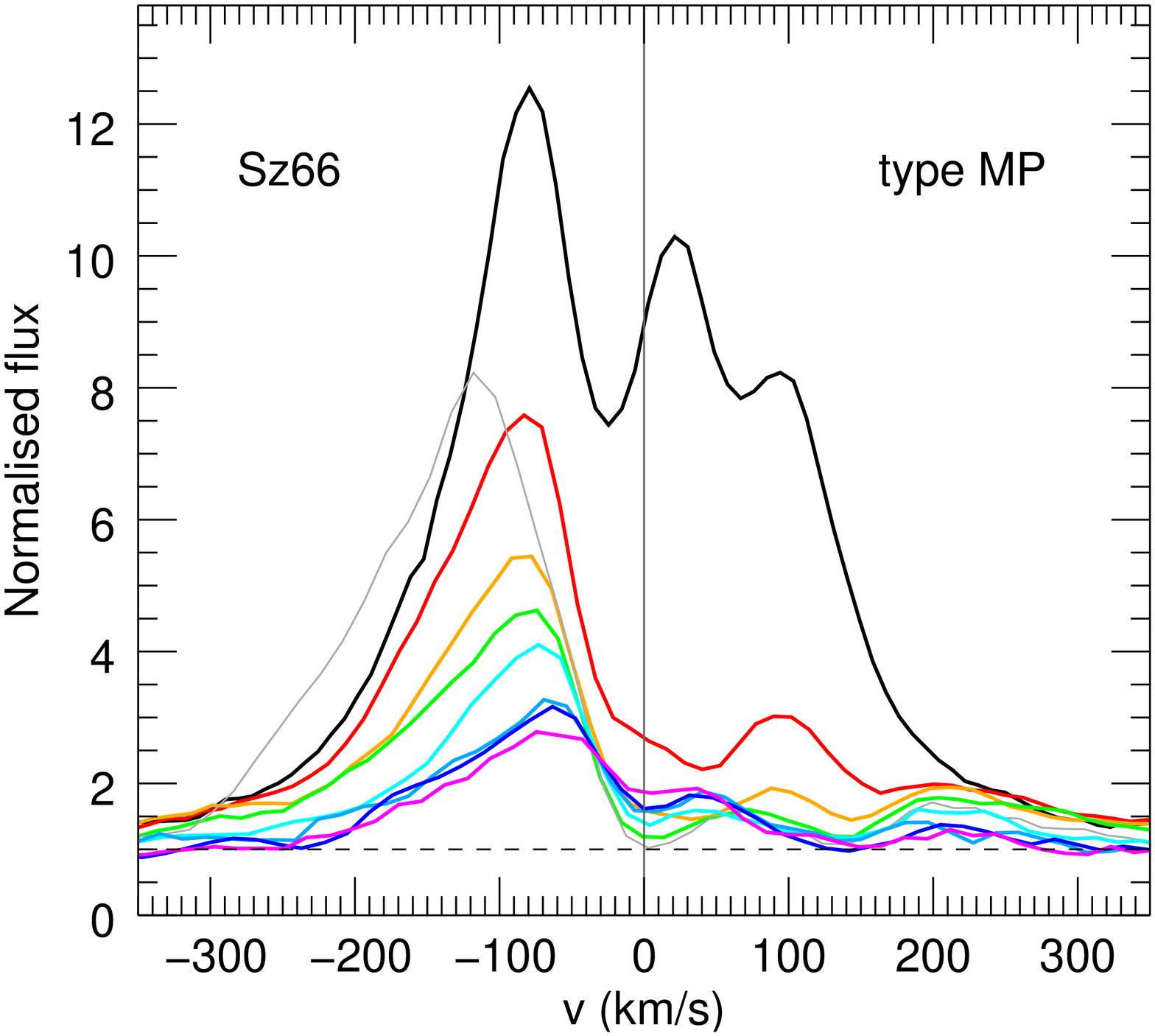}
\caption{Continued.} 
\end{figure*}

\setcounter{figure}{0}
\begin{figure*}[t]
\centering
\includegraphics[width=6.cm]{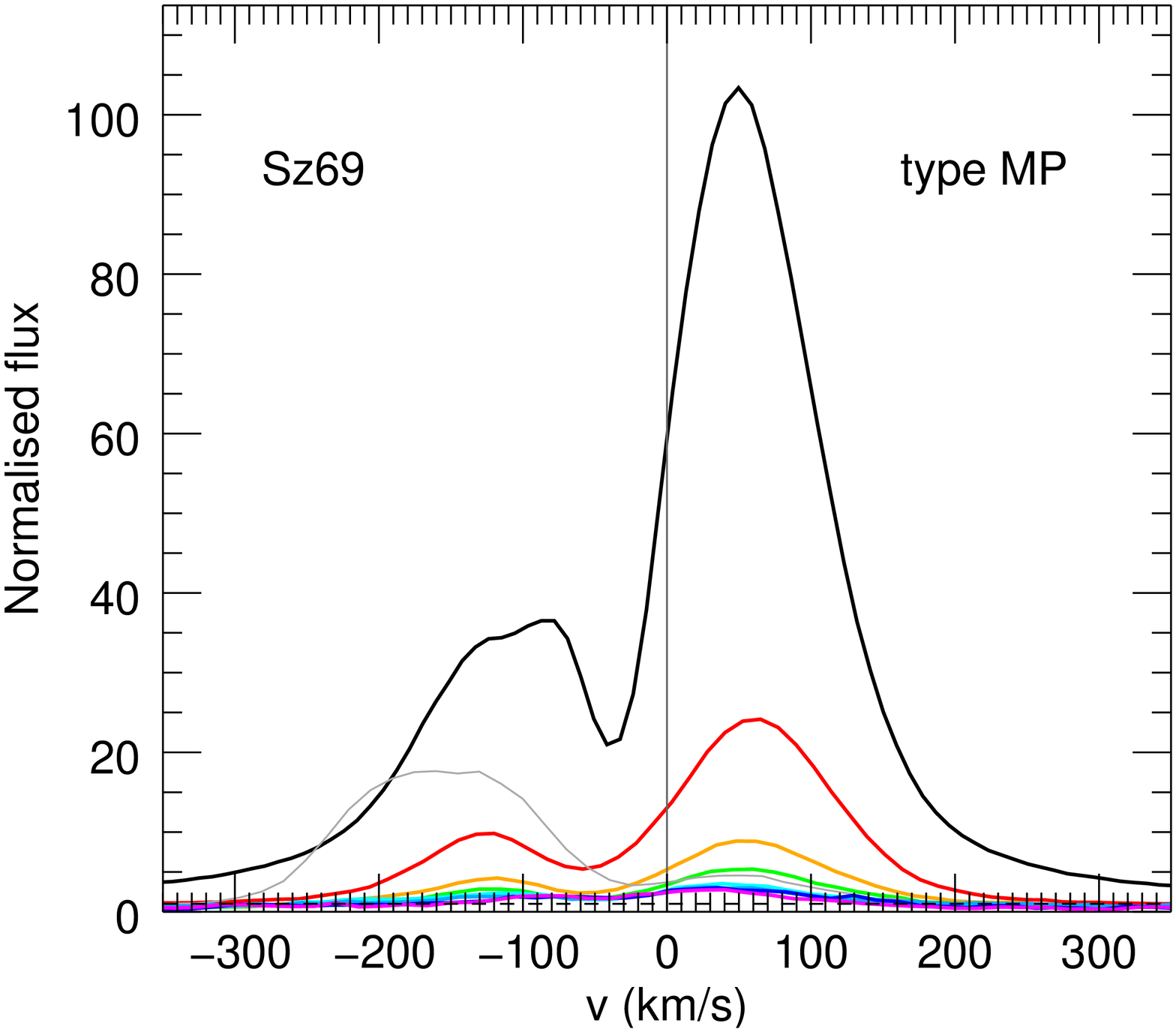}
\includegraphics[width=6.cm]{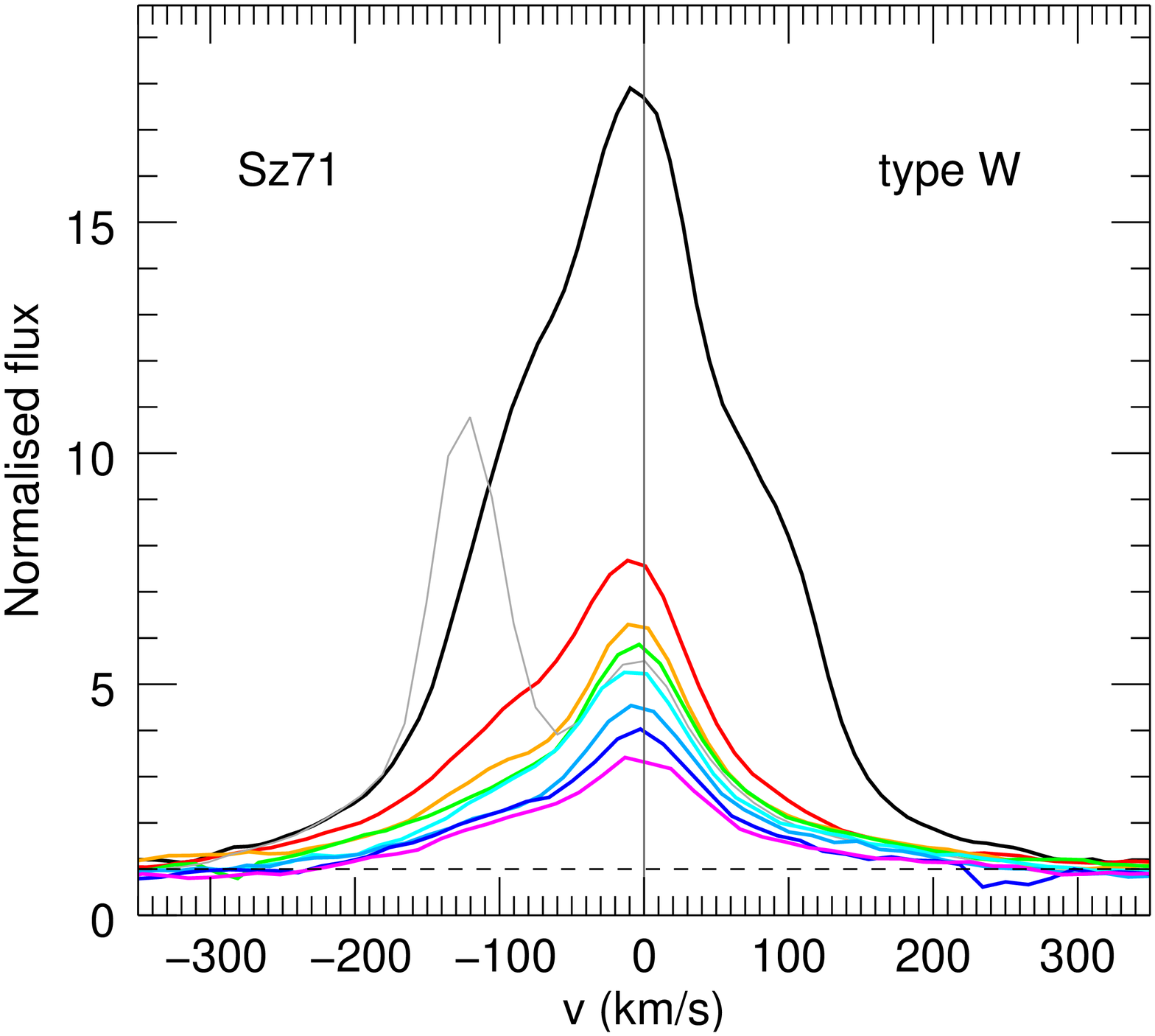}
\includegraphics[width=6.cm]{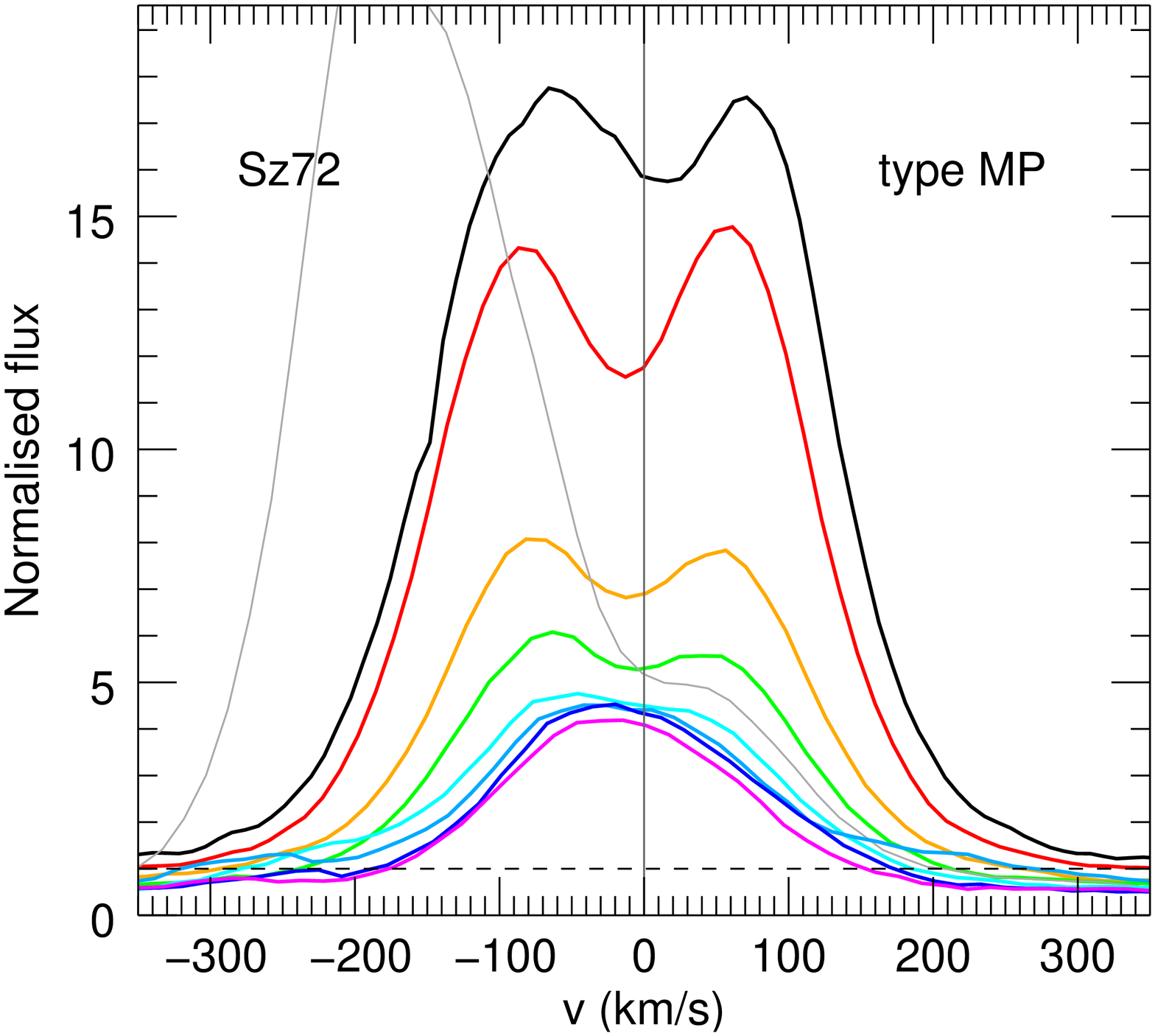}\\
\includegraphics[width=6.cm]{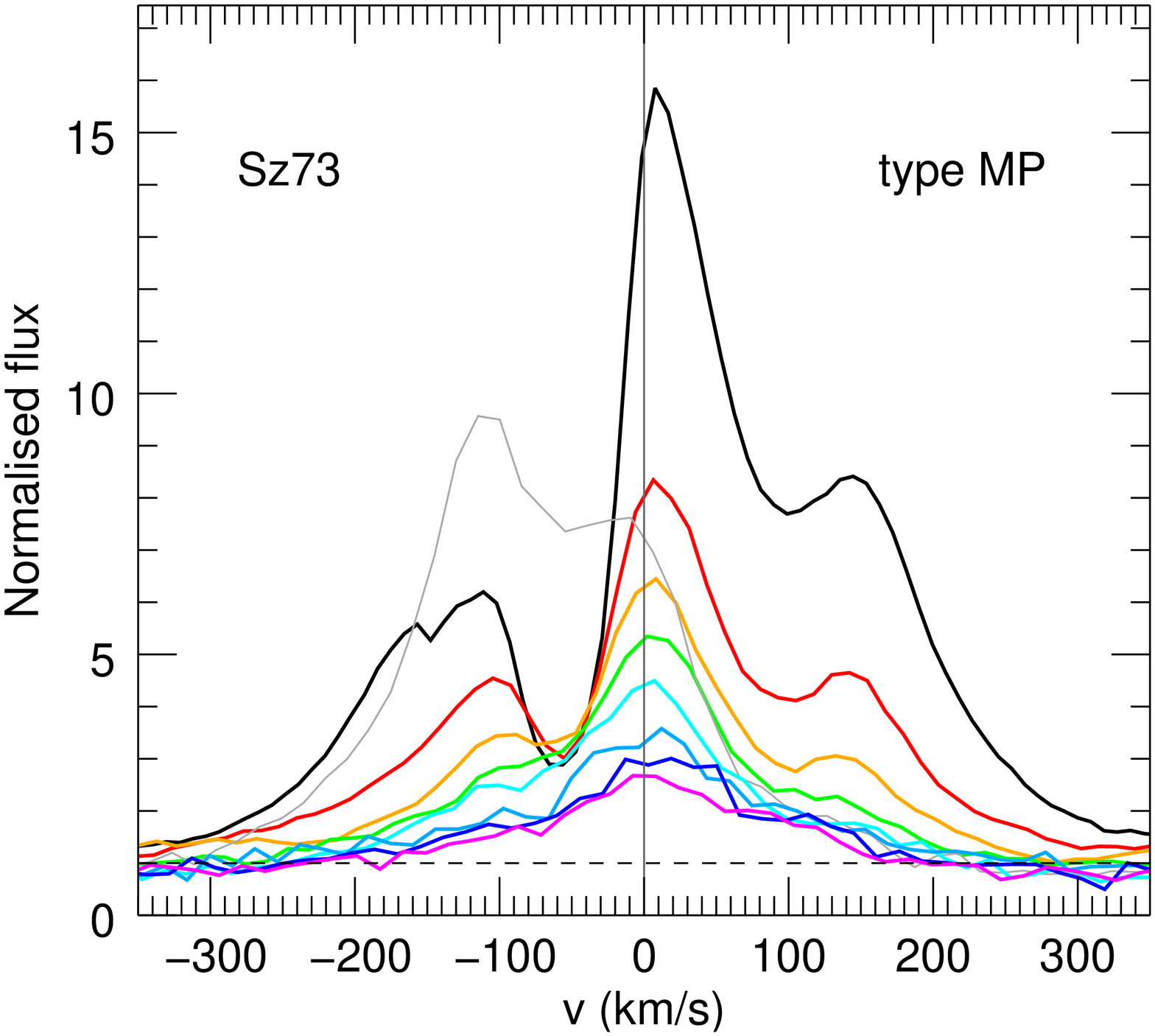}
\includegraphics[width=6.cm]{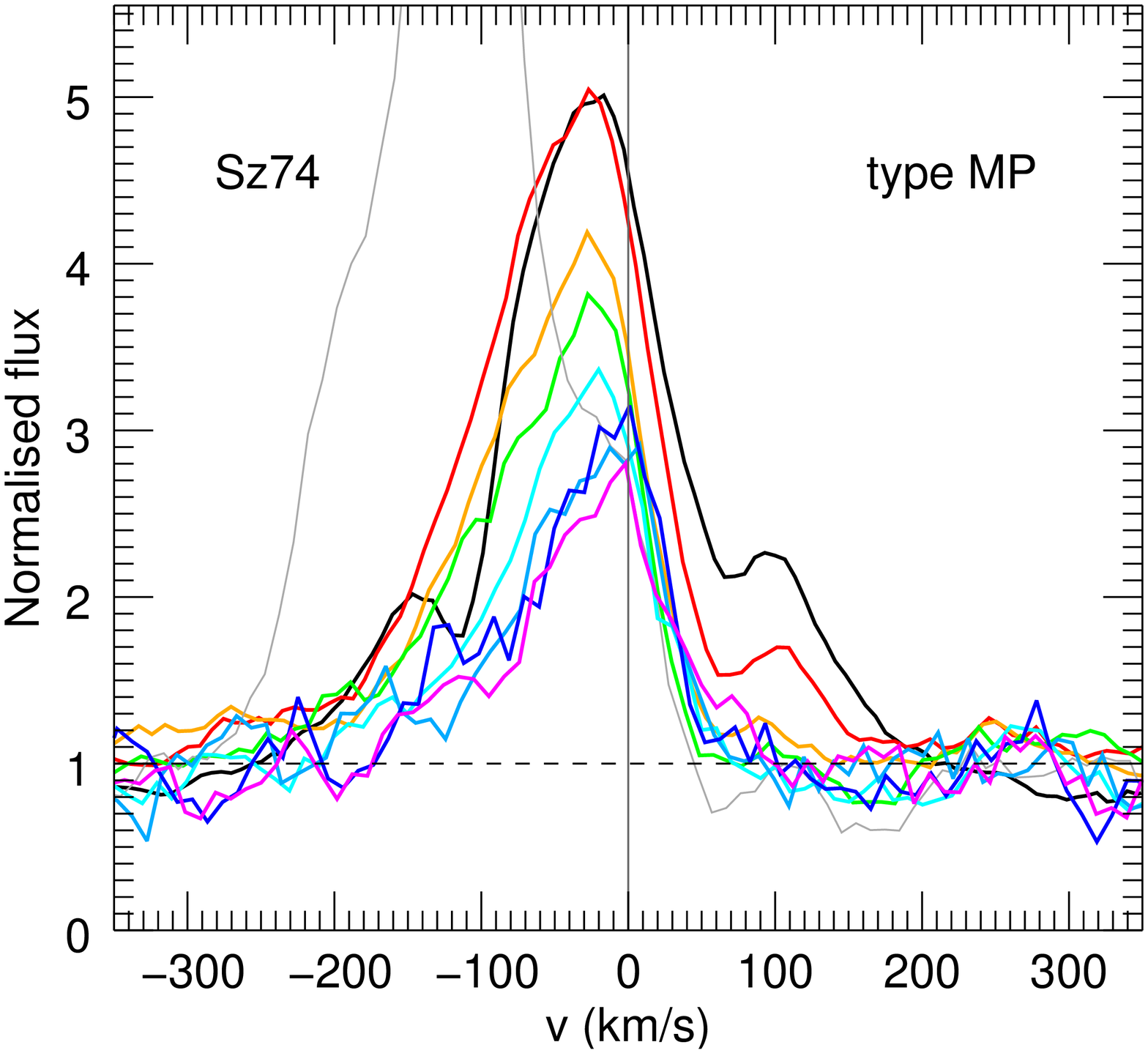}
\includegraphics[width=6.cm]{Sz83_Hprof.eps}\\
\includegraphics[width=6.cm]{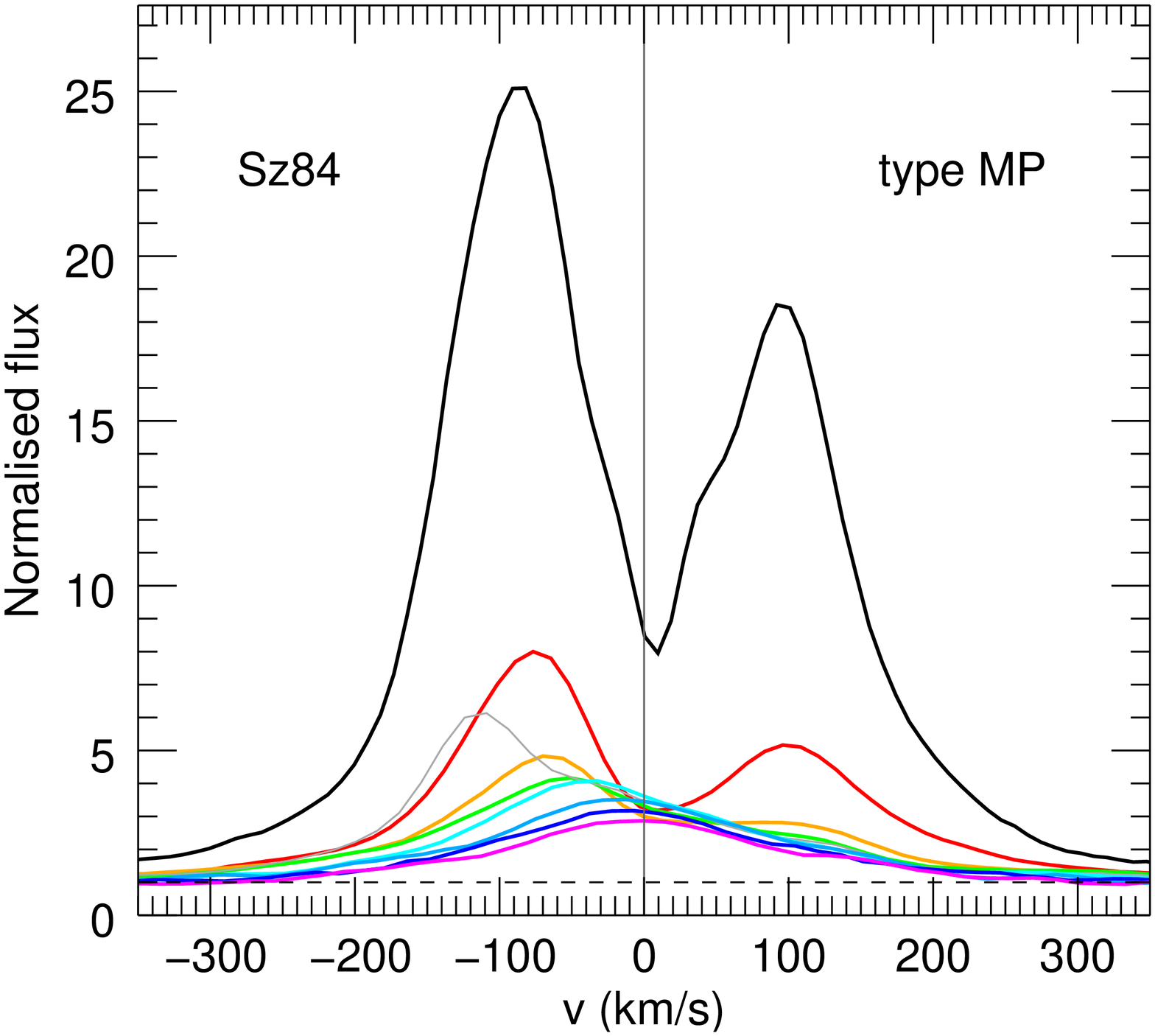}
\includegraphics[width=6.cm]{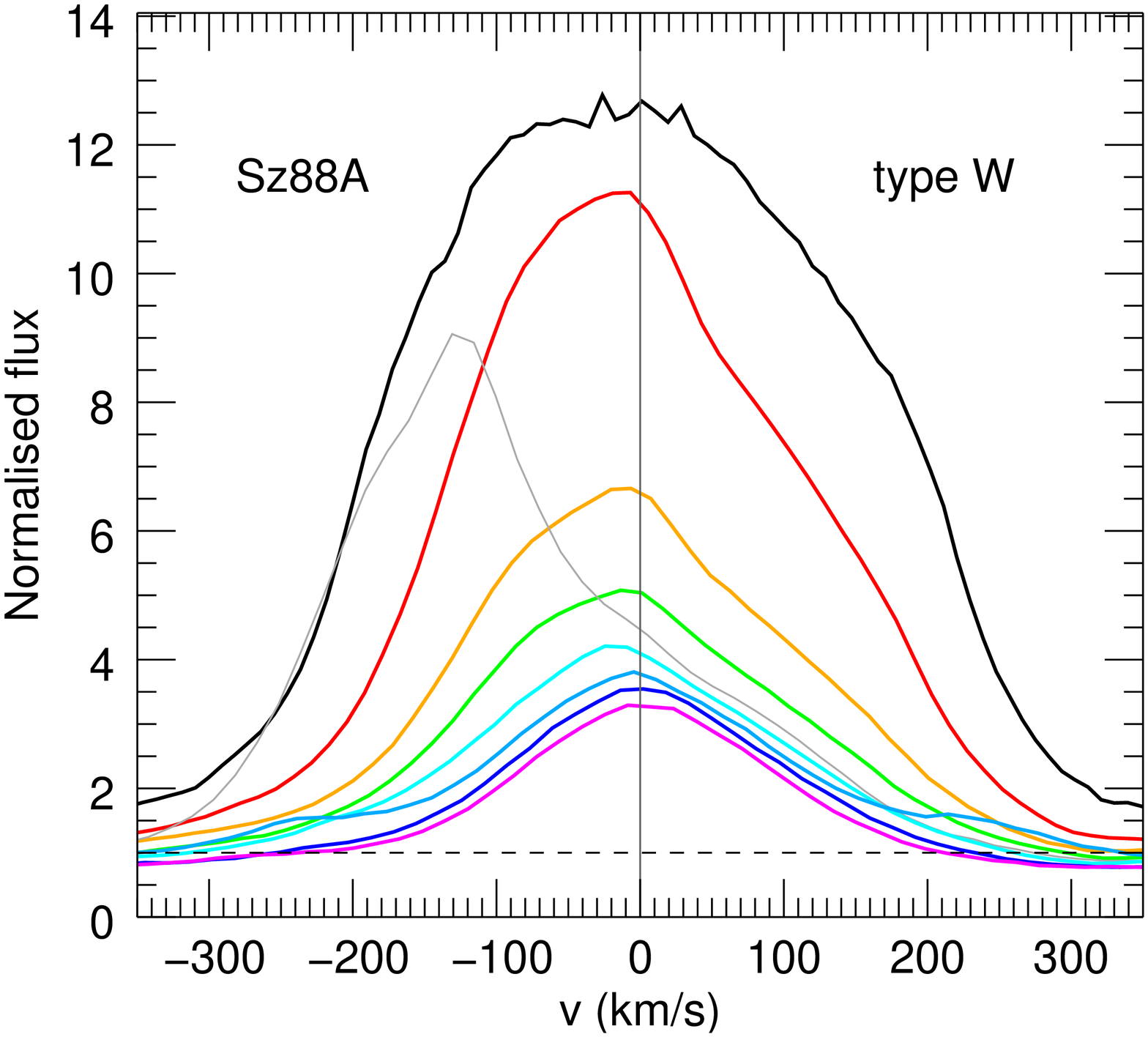}
\includegraphics[width=6.cm]{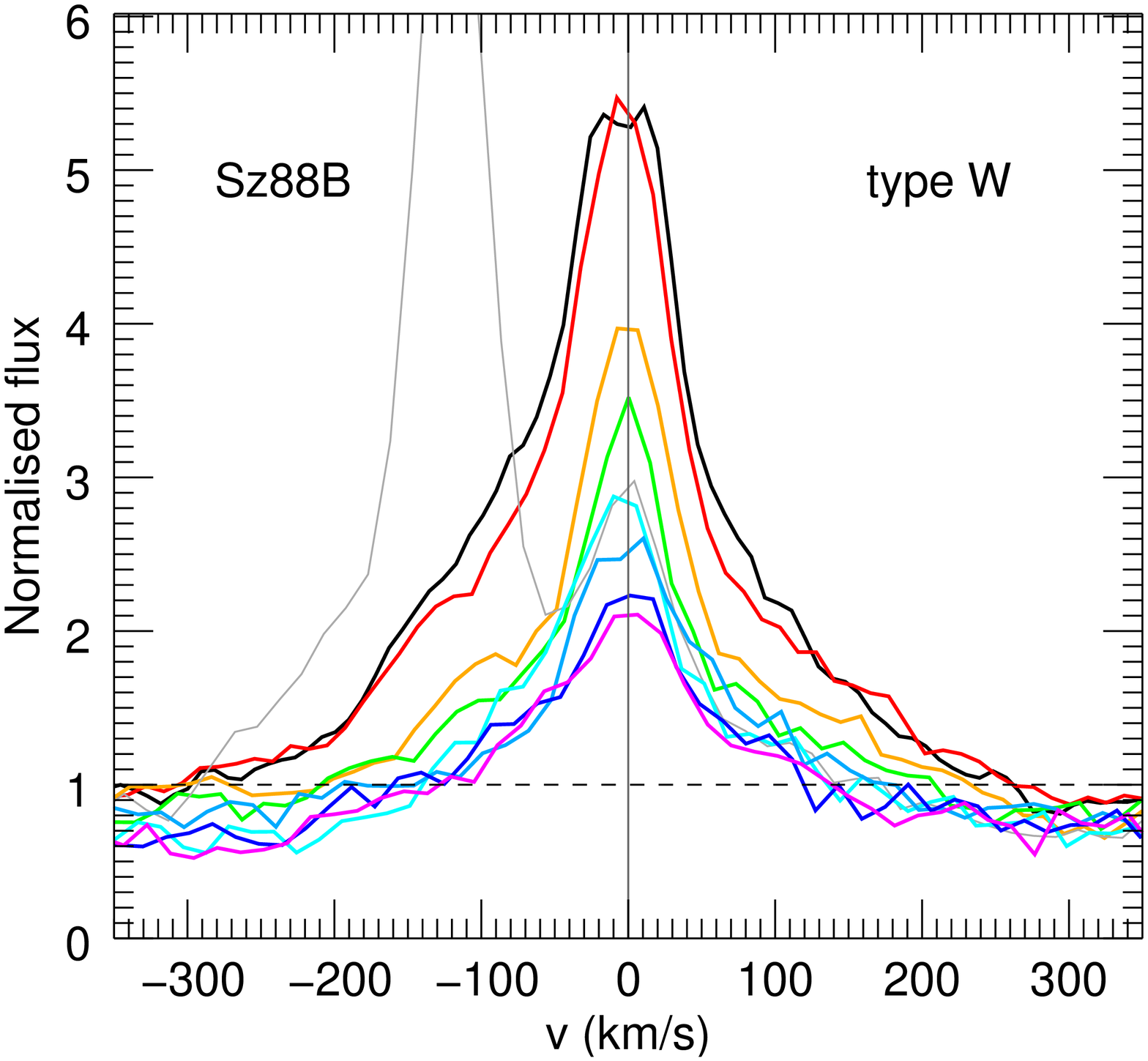}\\
\includegraphics[width=6.cm]{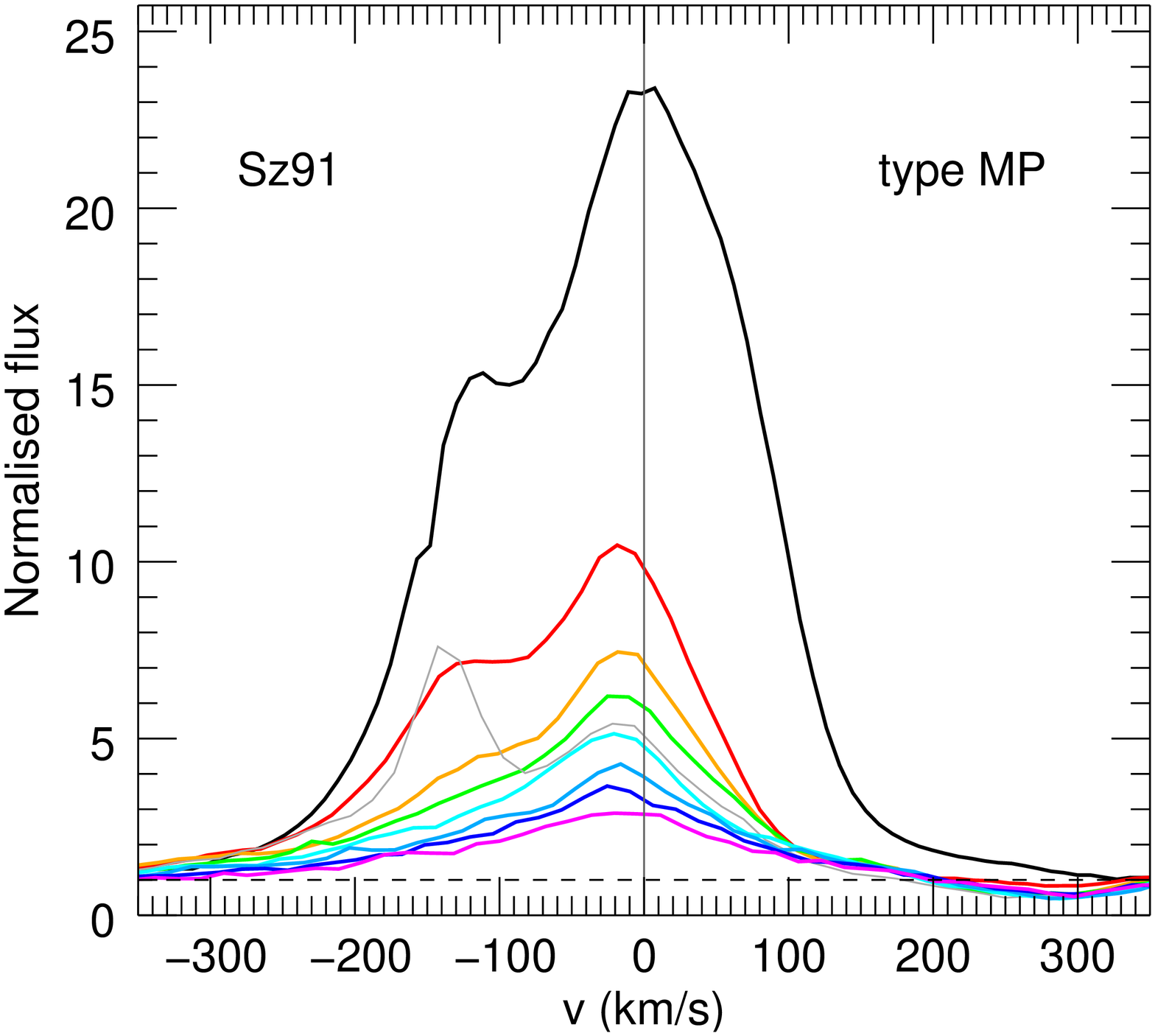}
\includegraphics[width=6.cm]{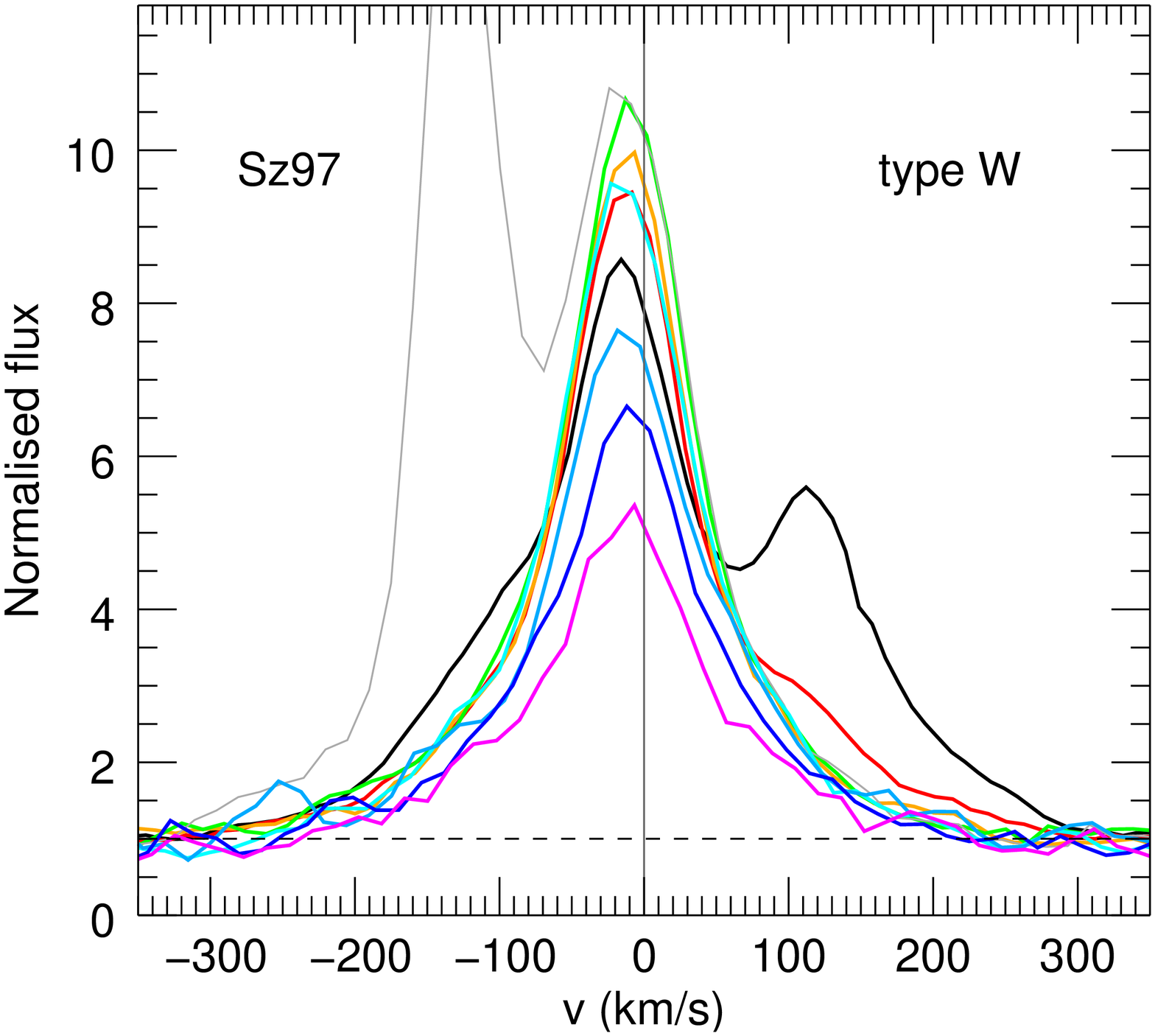}
\includegraphics[width=6.cm]{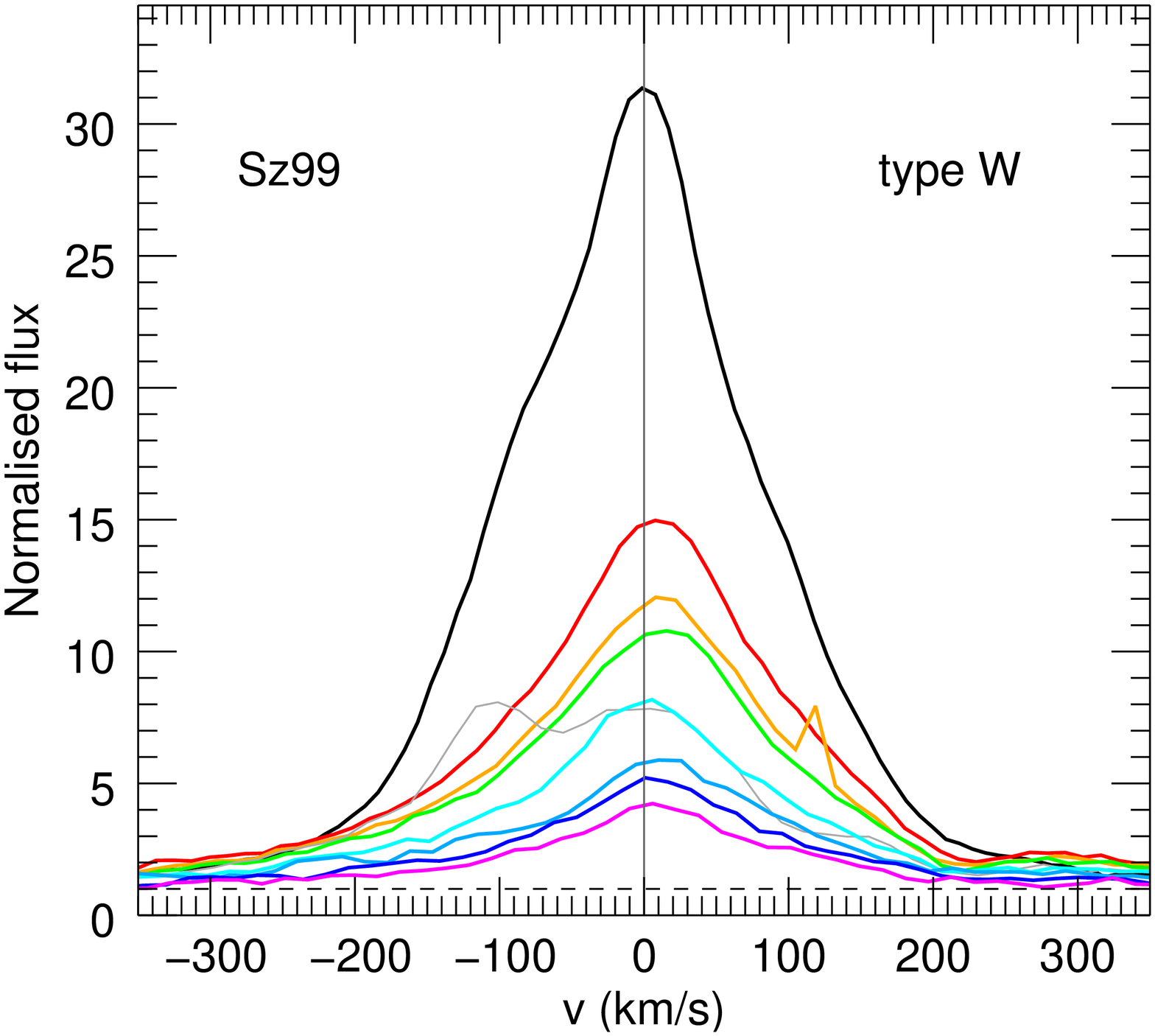}
\caption{Continued.} 
\end{figure*}

\begin{figure*}[t]
\centering
\includegraphics[width=6.cm]{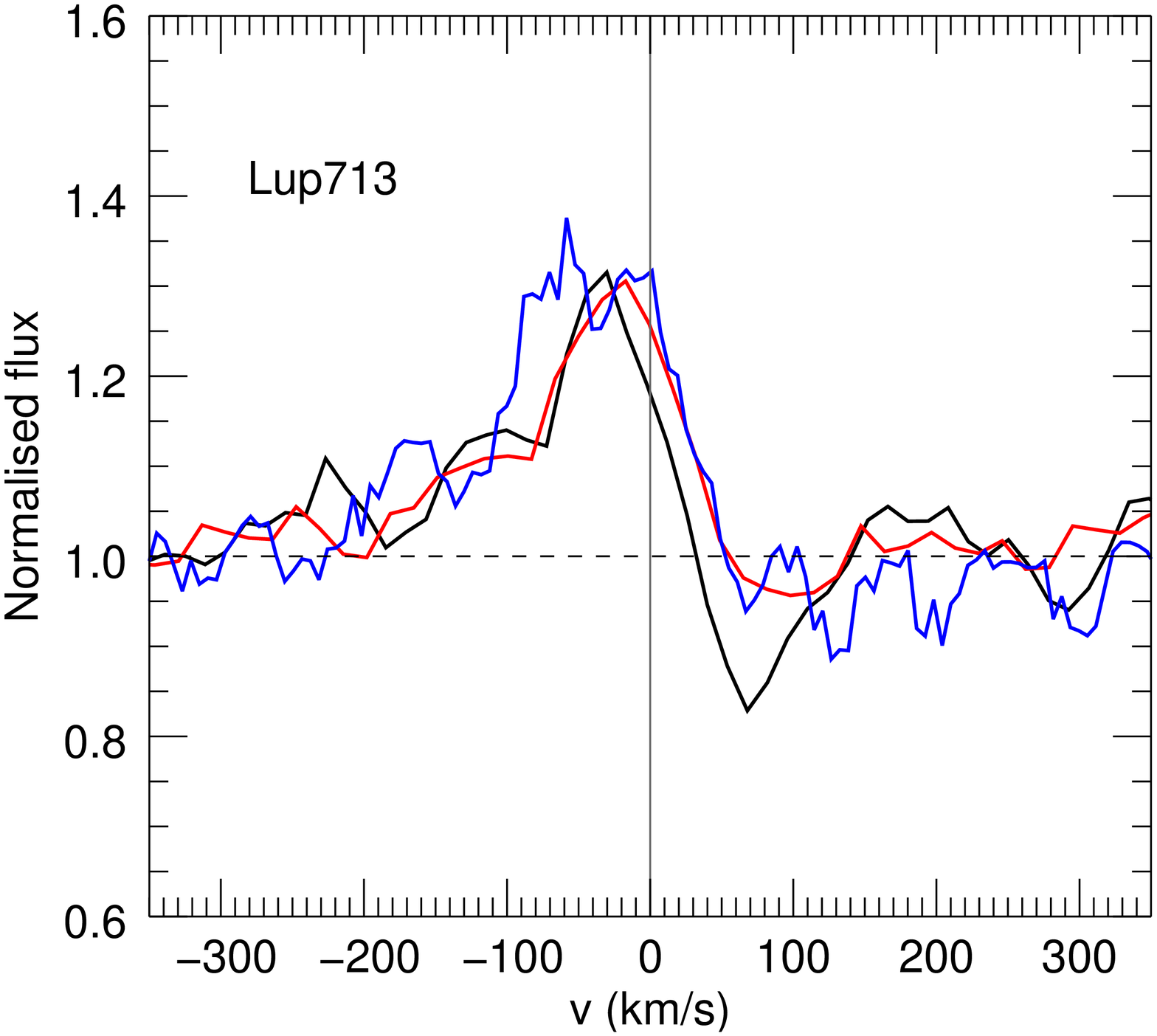}
\includegraphics[width=6.cm]{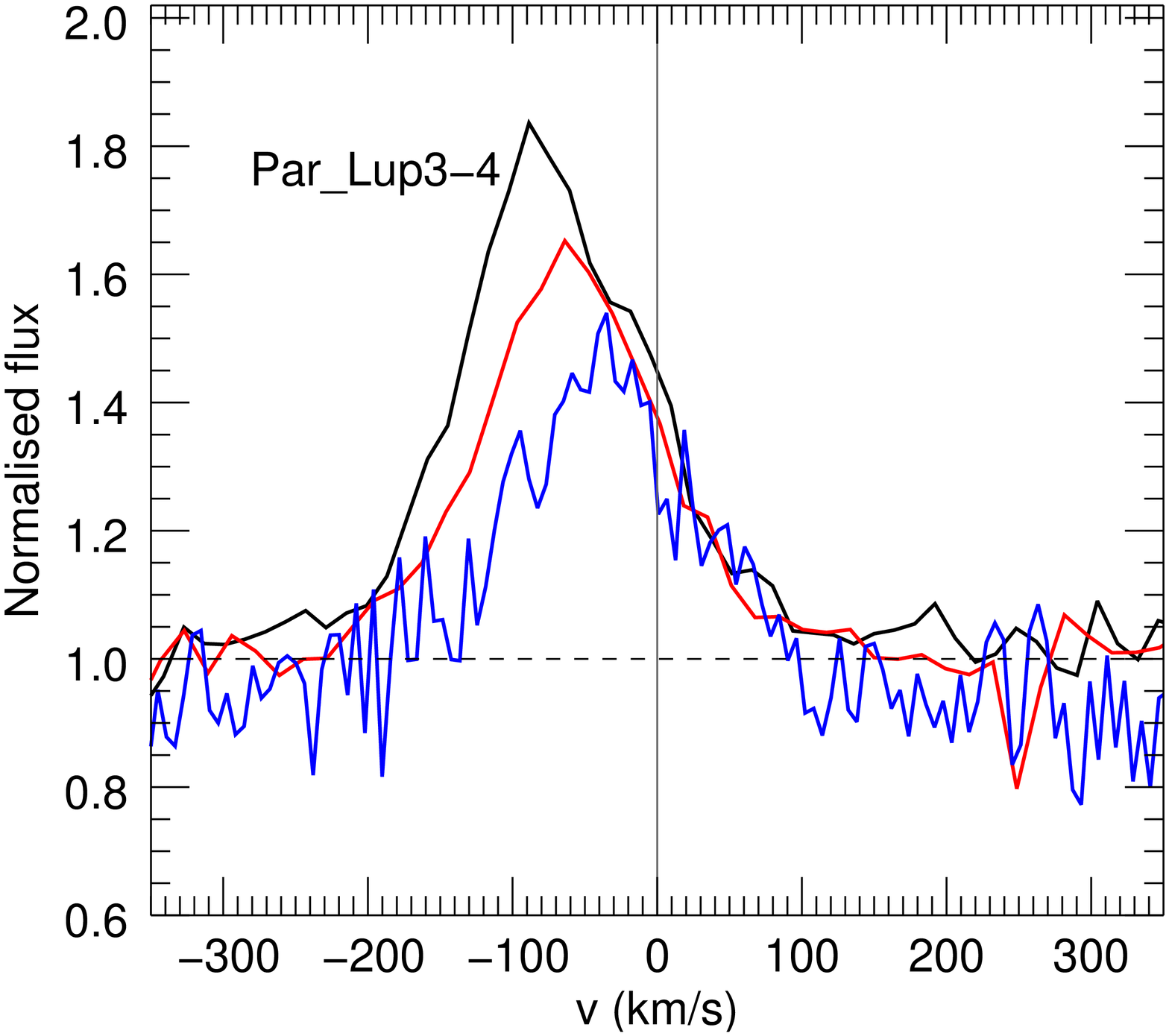}
\includegraphics[width=6.cm]{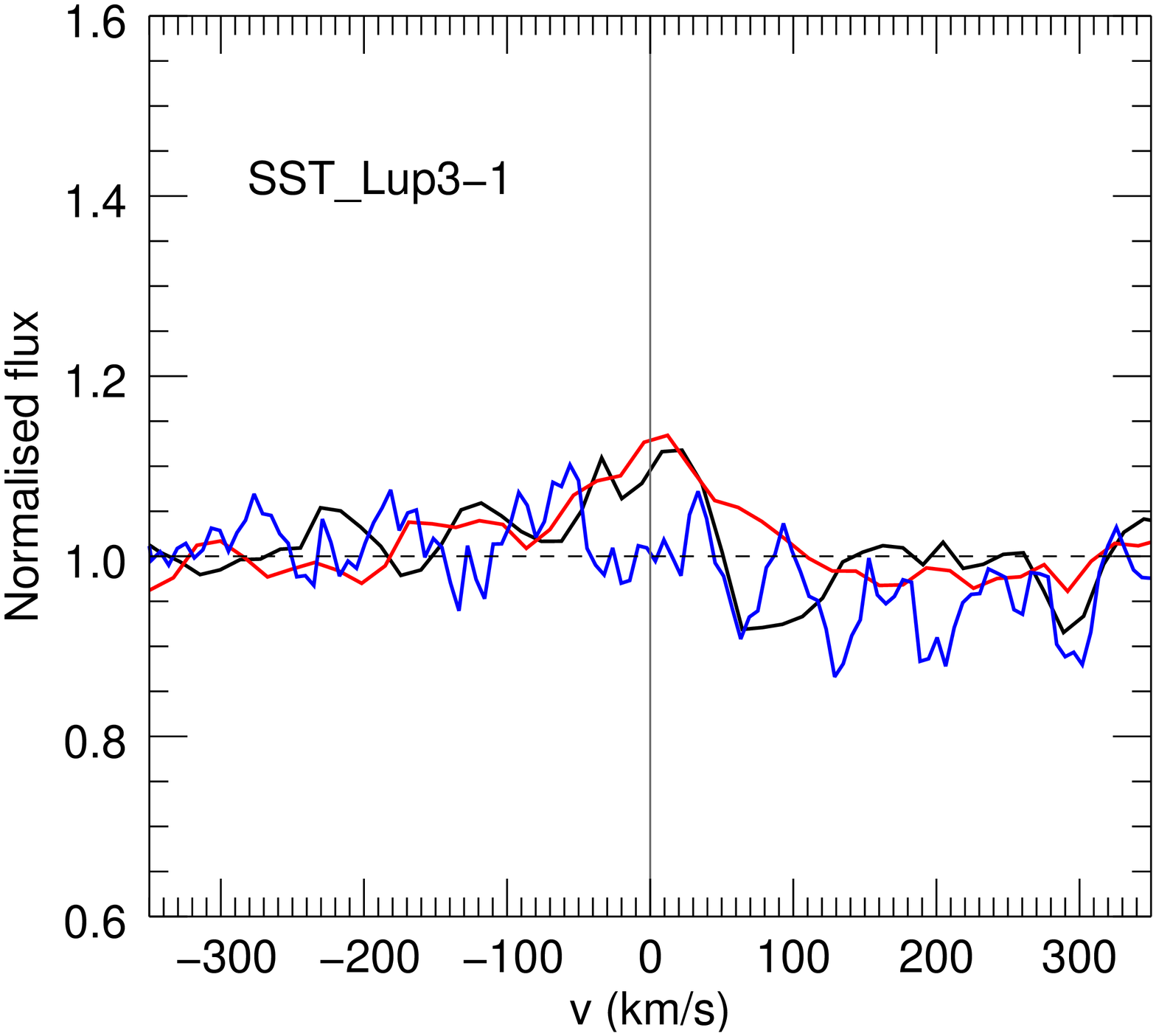}\\
\includegraphics[width=6.cm]{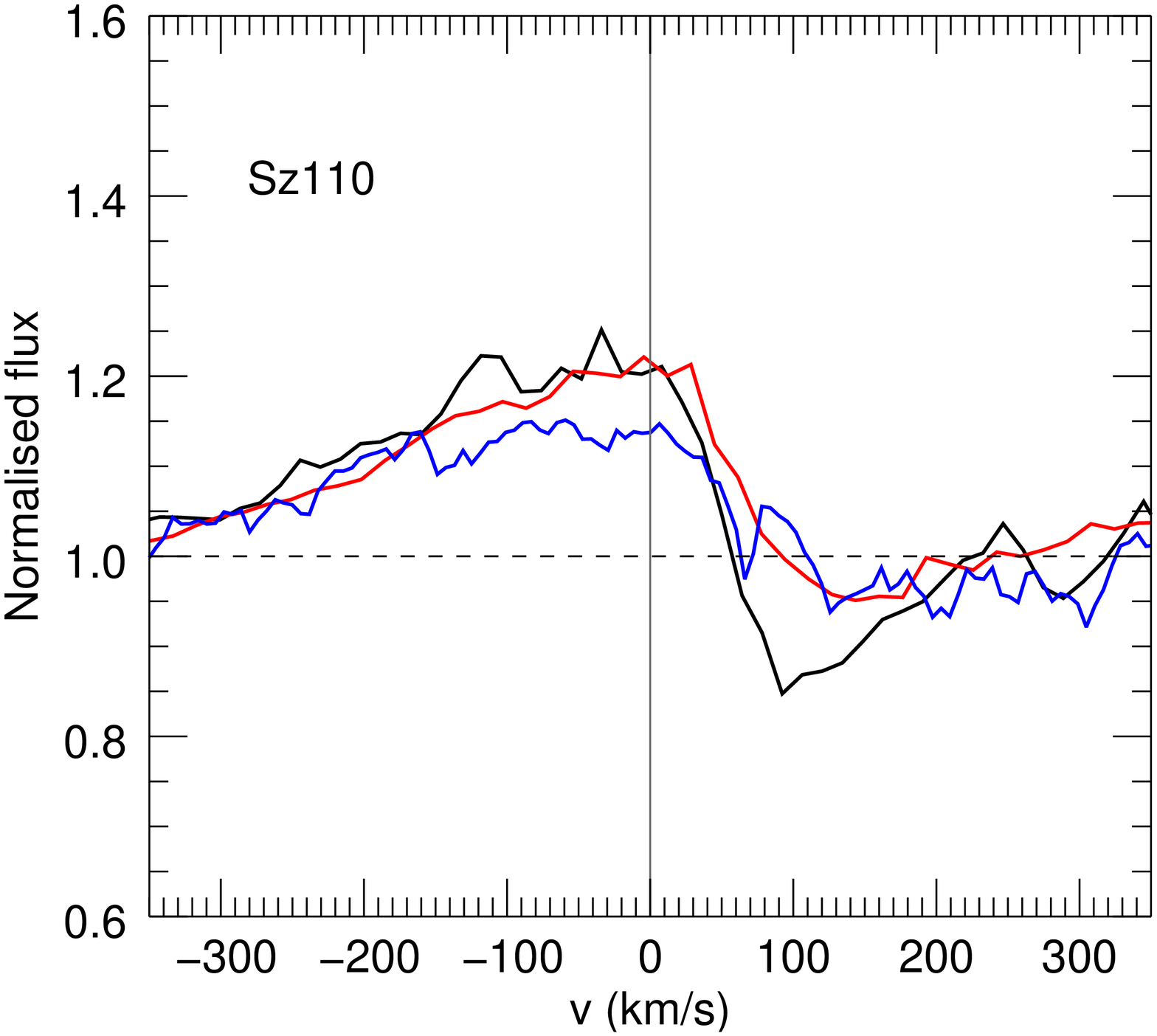}
\includegraphics[width=6.cm]{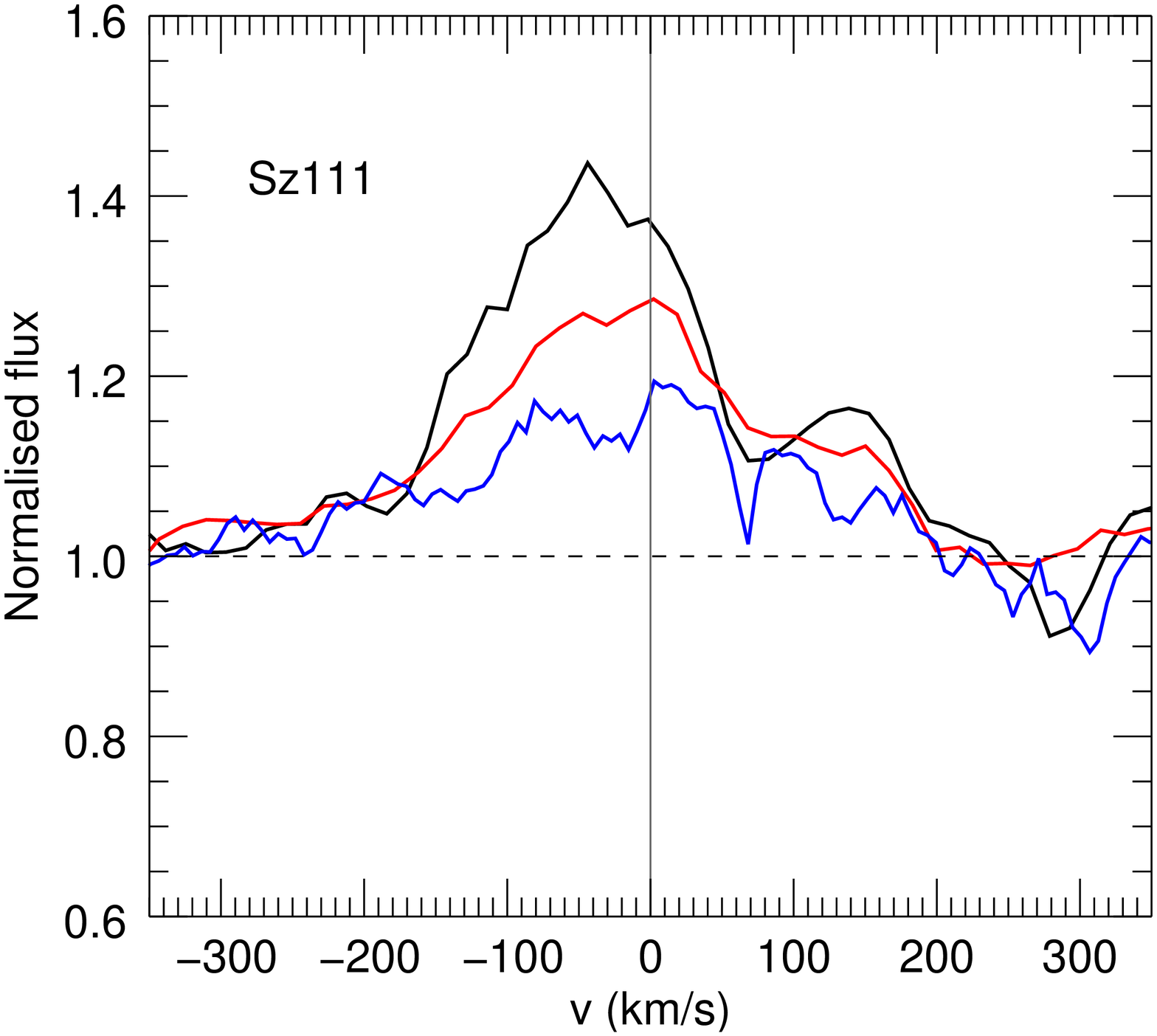}
\includegraphics[width=6.cm]{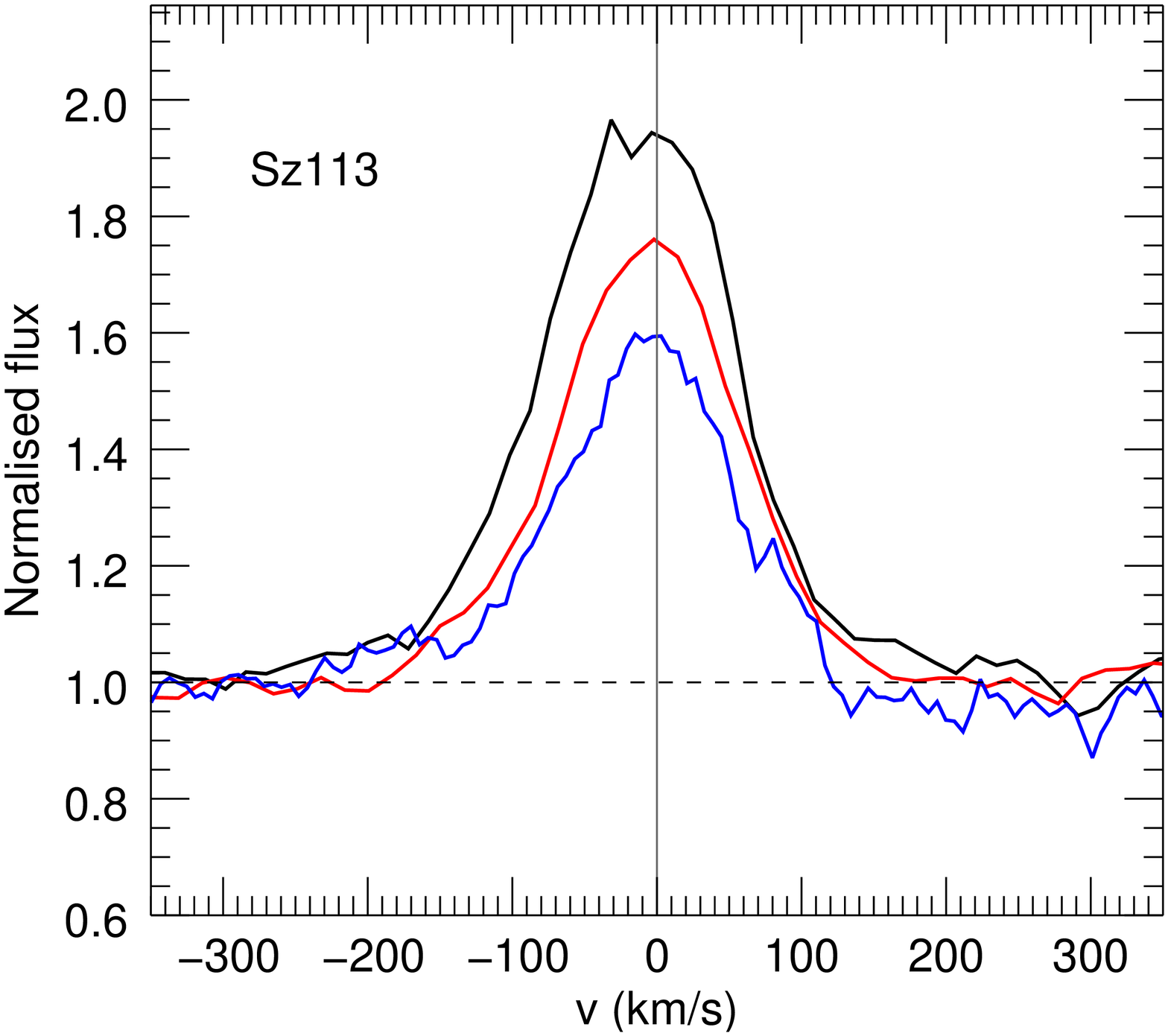}\\
\includegraphics[width=6.cm]{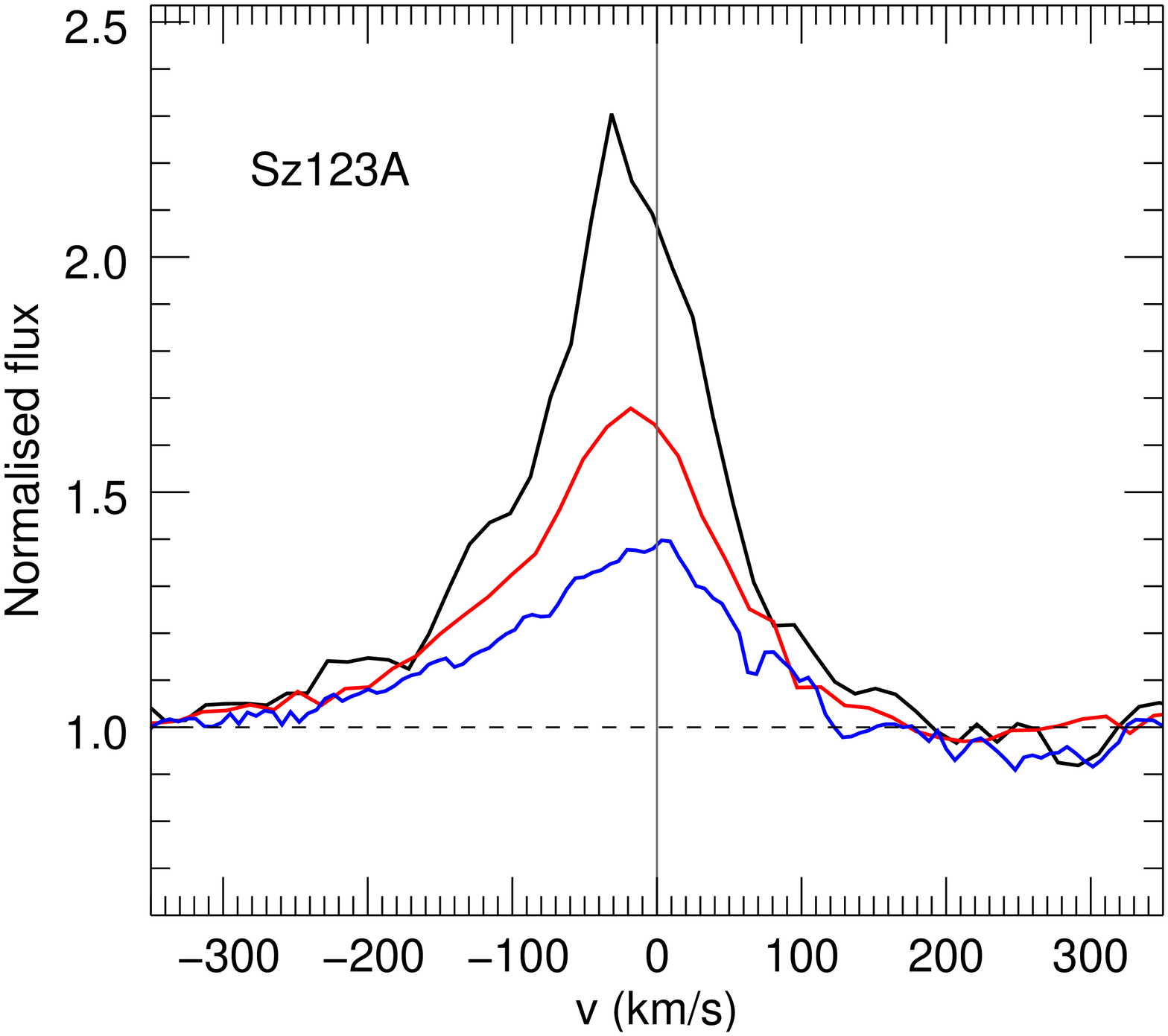}
\includegraphics[width=6.cm]{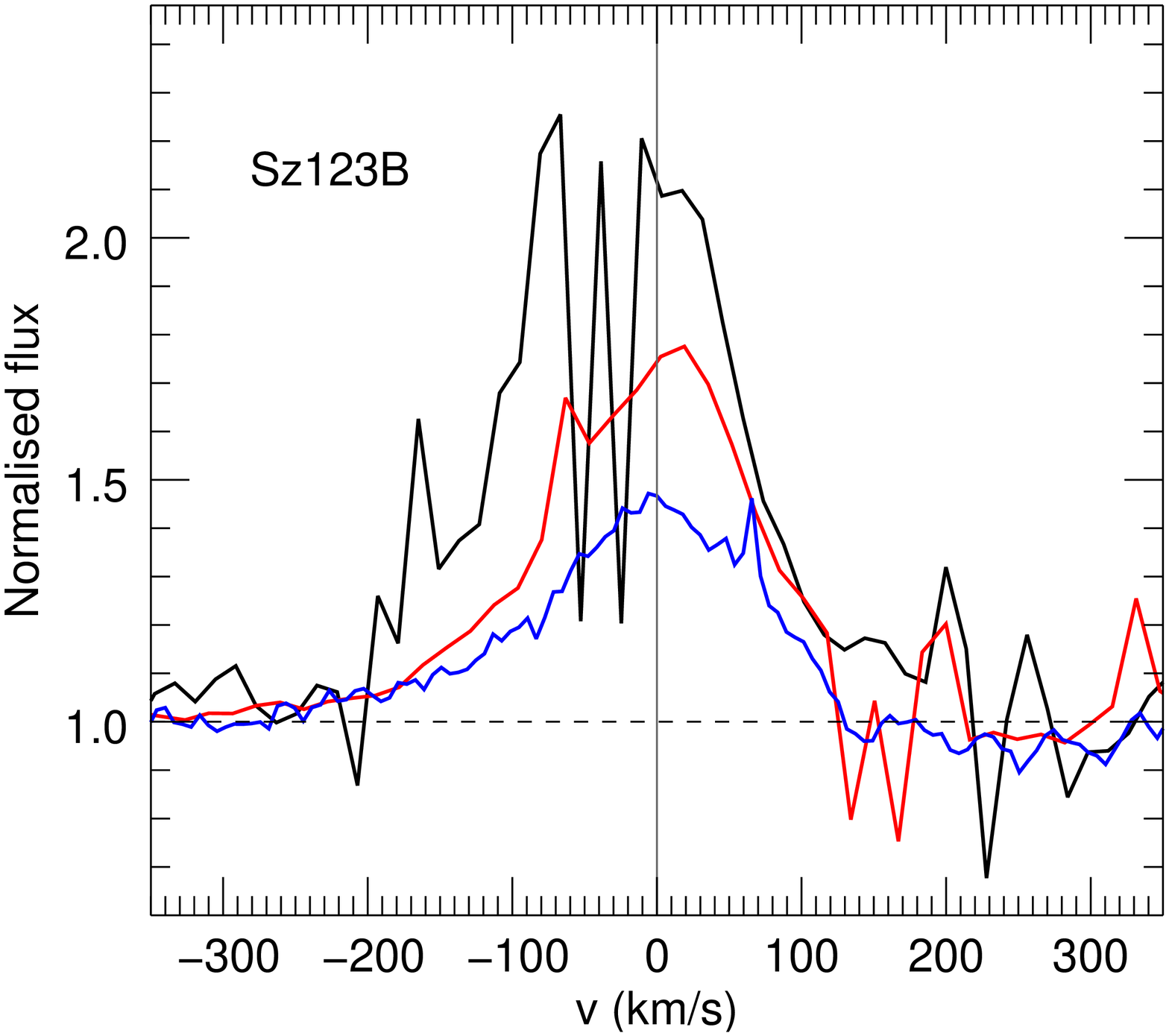}
\includegraphics[width=6.cm]{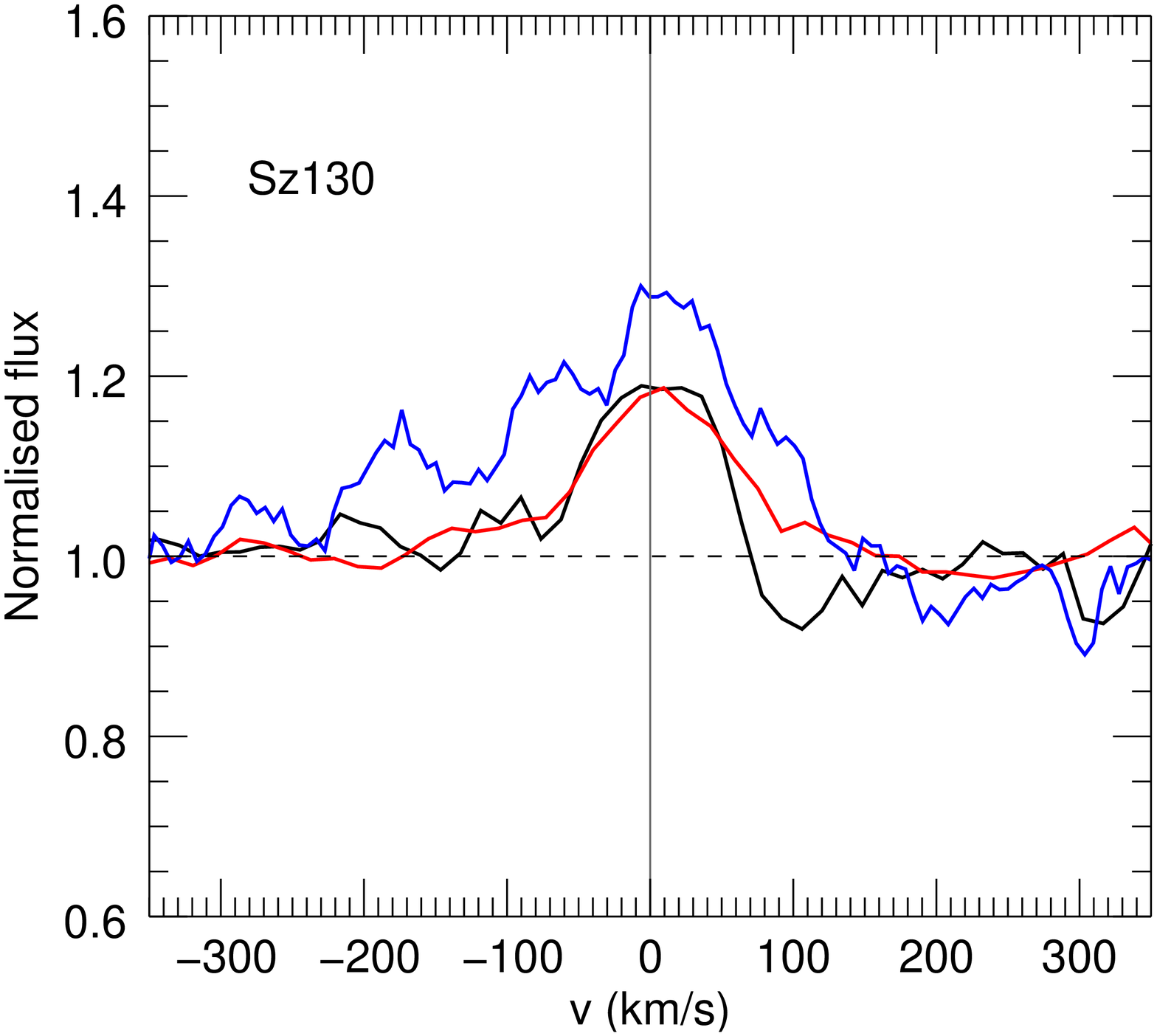}\\
\includegraphics[width=6.cm]{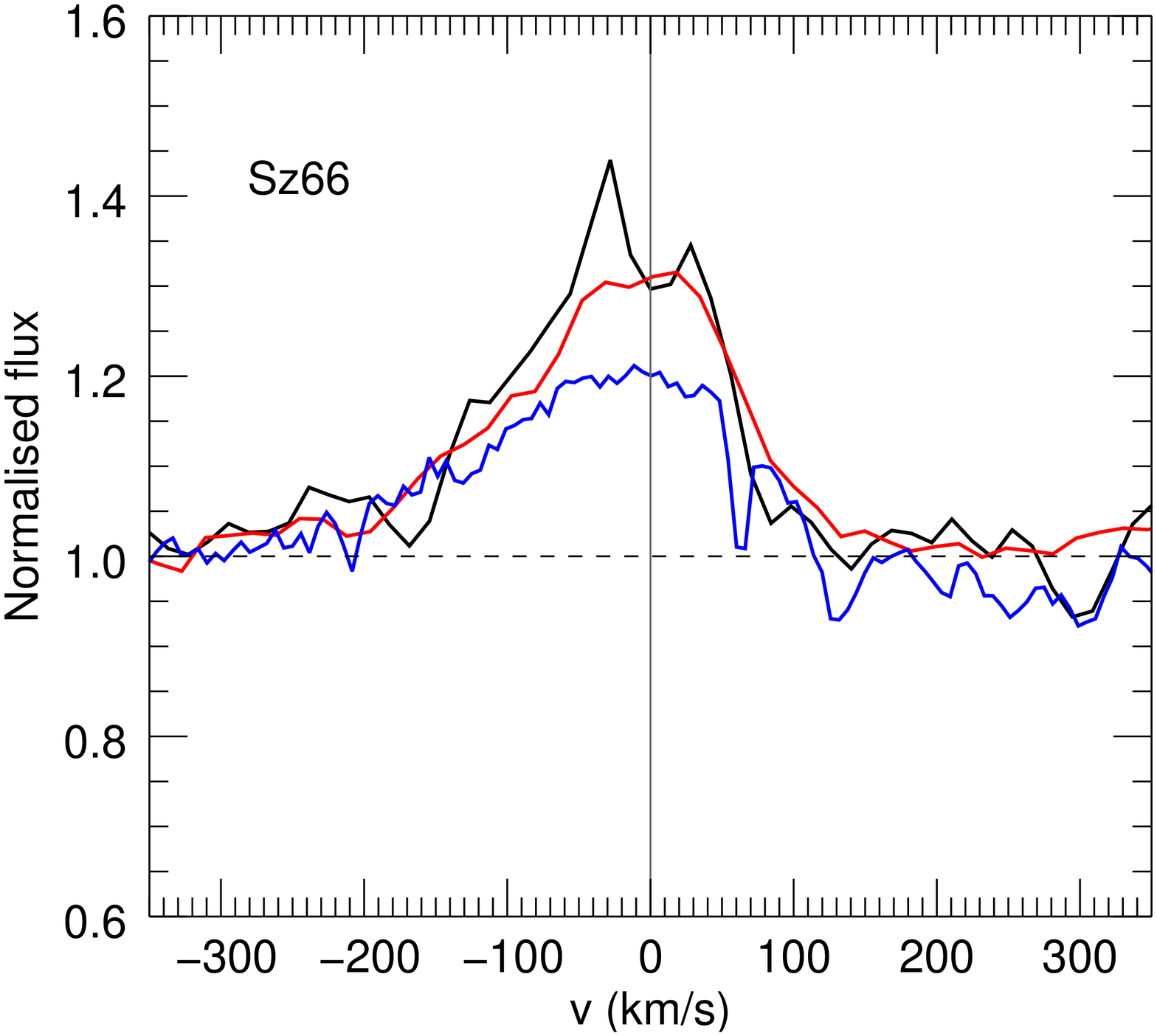}
\includegraphics[width=6.cm]{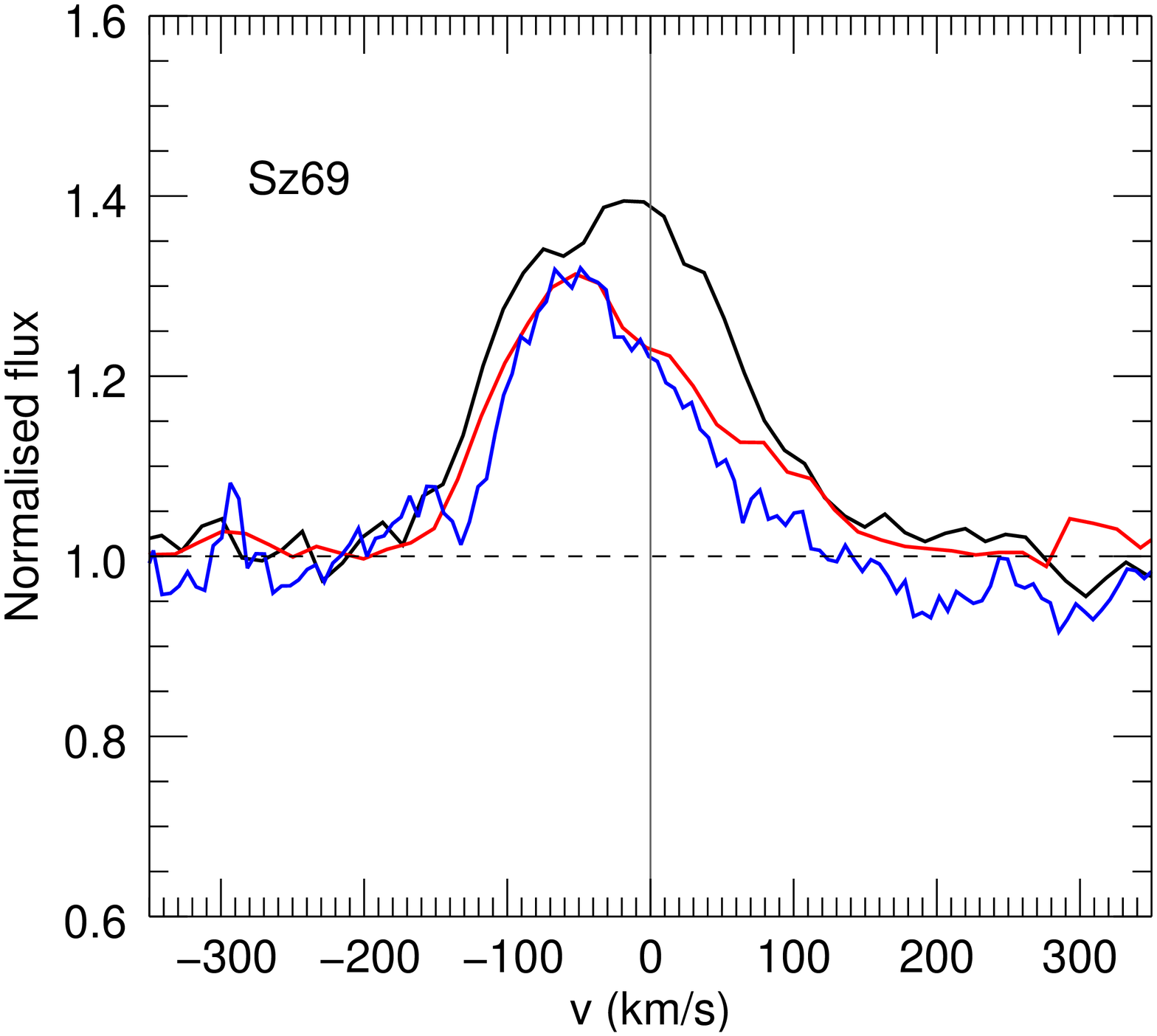}
\includegraphics[width=6.cm]{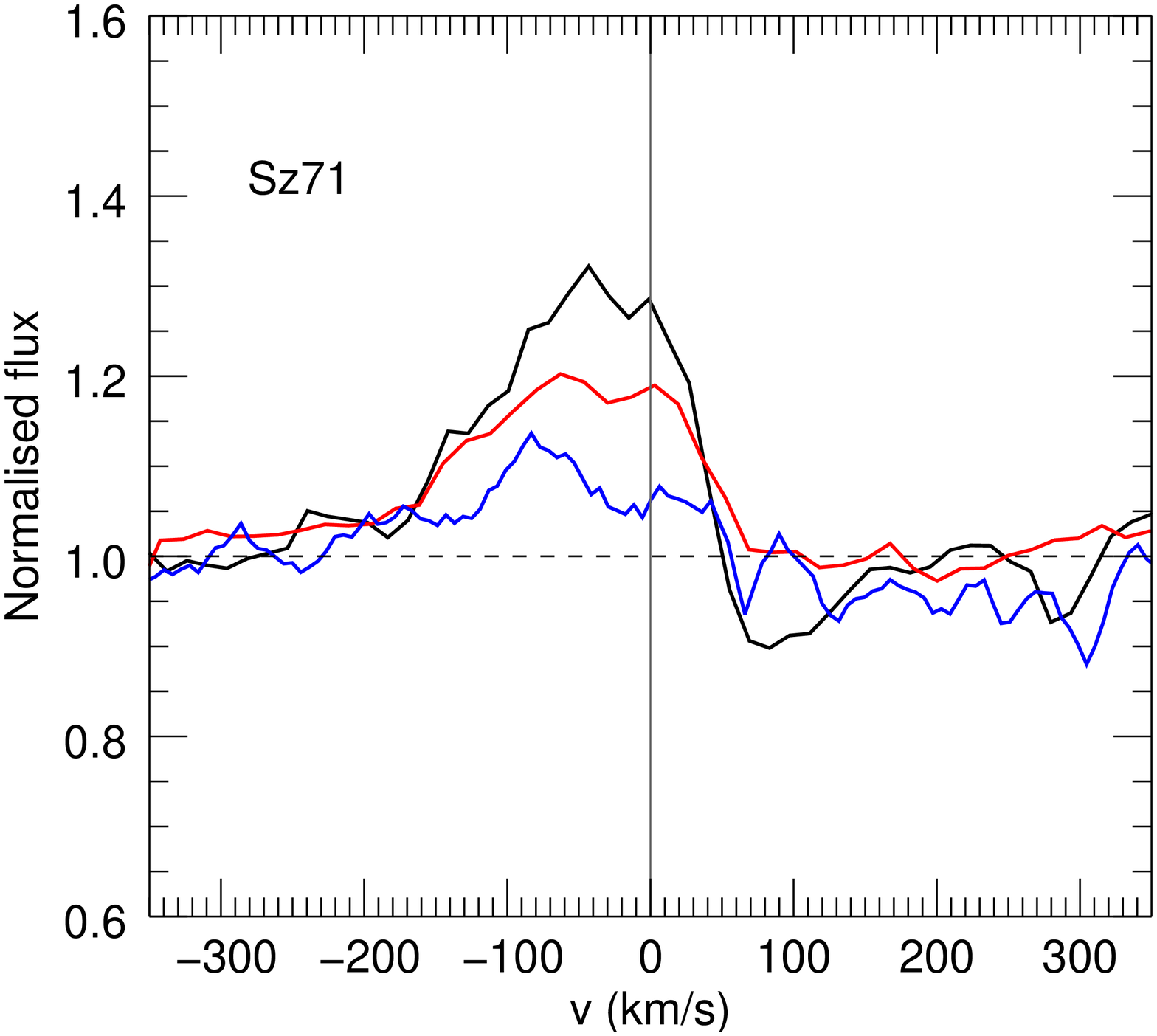}
\caption{\label{fig:lines:pa} Continuum-normalised Paschen lines in sources where \pab\ is detected with a S/N $>$ 5: Pa$\beta$ (black), Pa$\gamma$ (red), and Pa$\delta$ (blue) are shown.} 
\end{figure*}

\setcounter{figure}{1}
\begin{figure*}[t]
\centering
\includegraphics[width=6.cm]{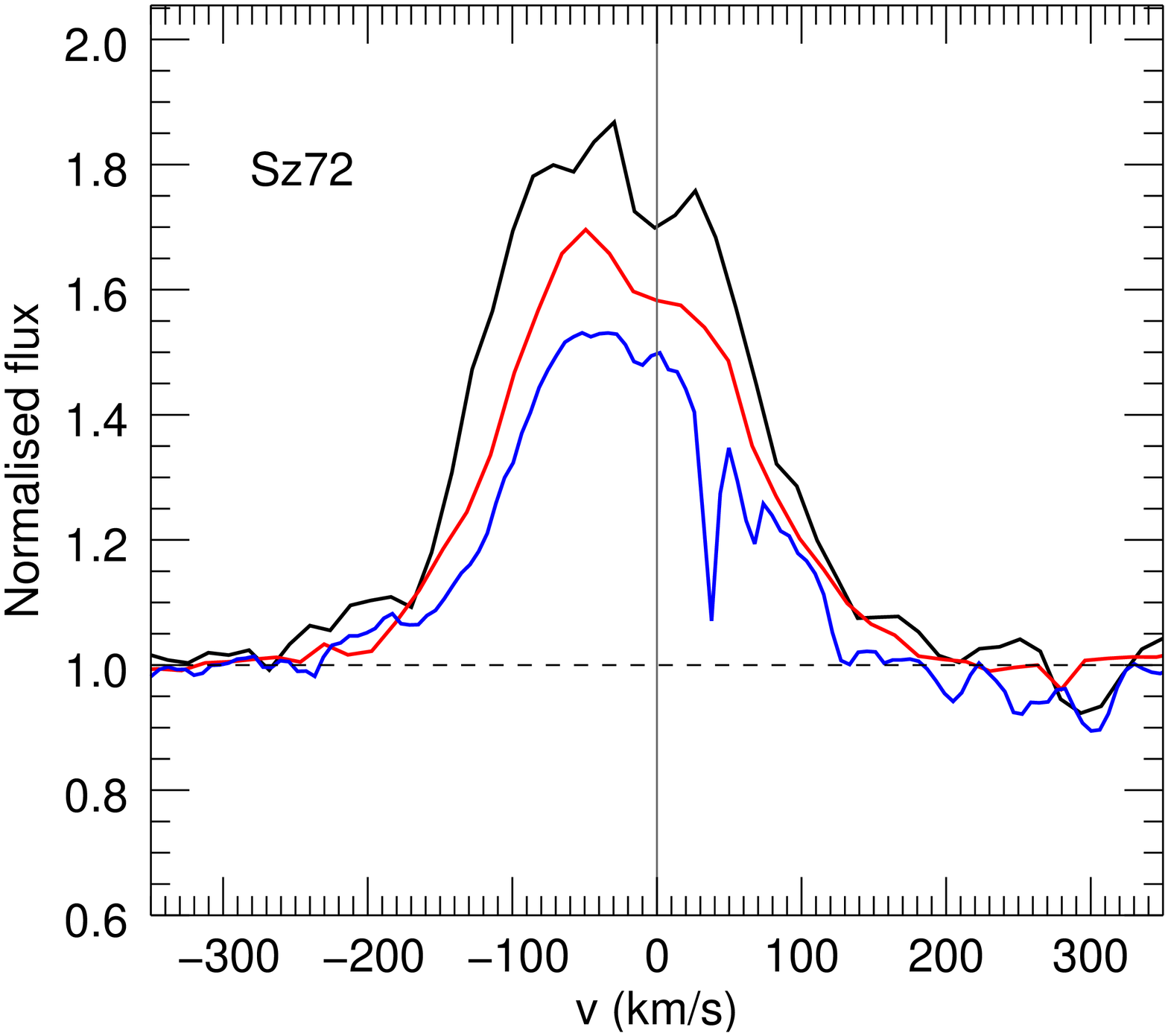}
\includegraphics[width=6.cm]{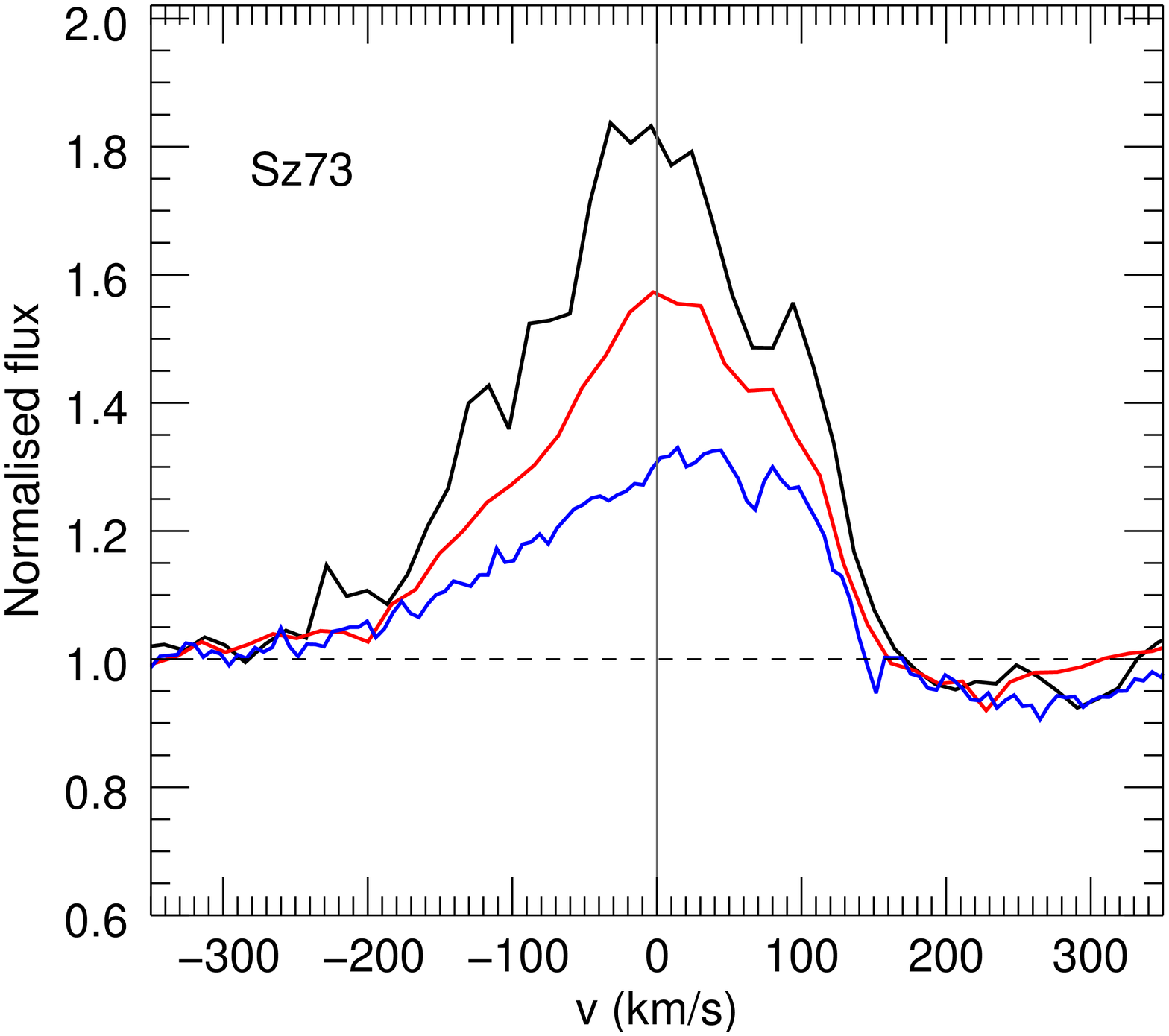}
\includegraphics[width=6.cm]{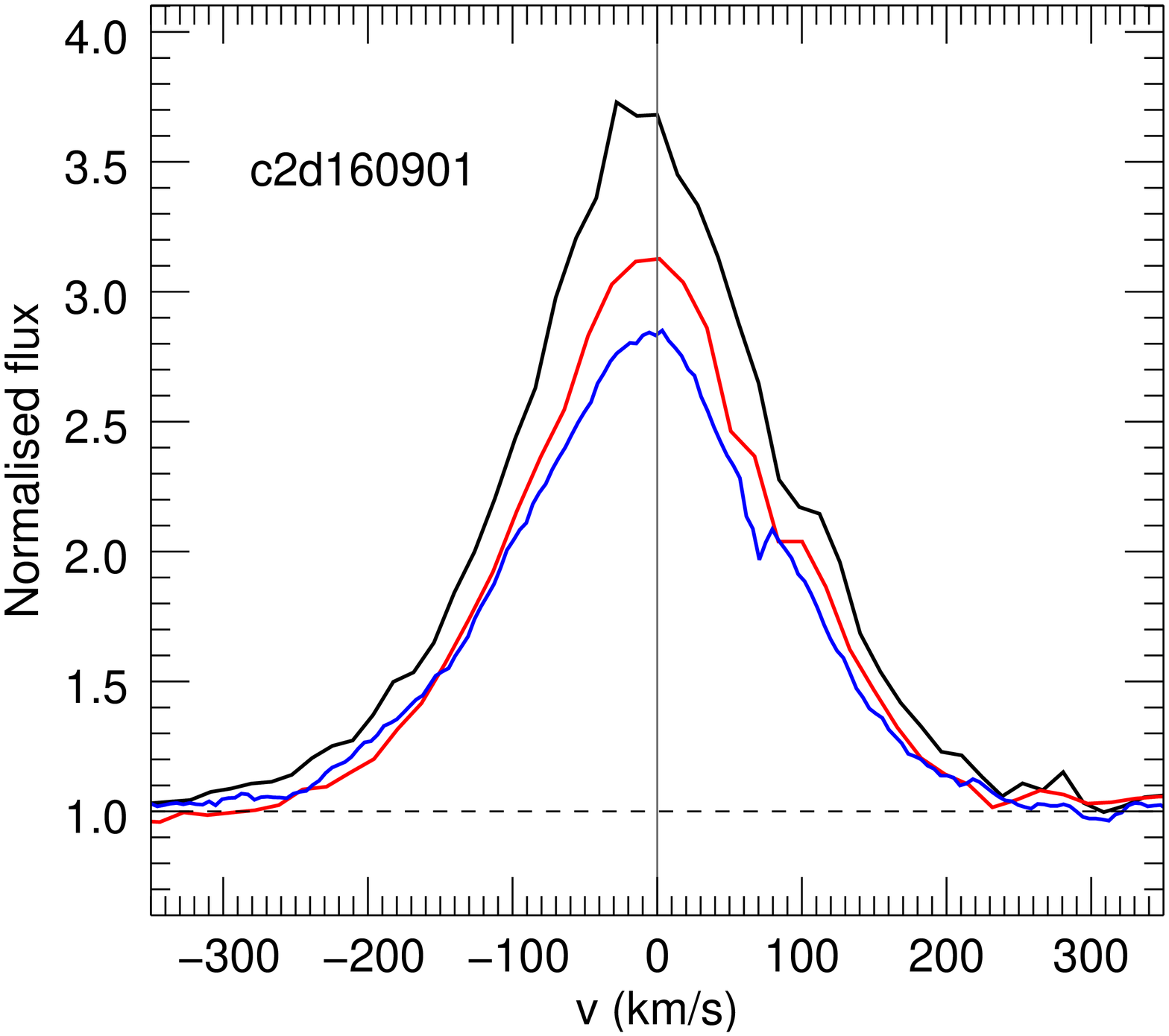}\\
\includegraphics[width=6.cm]{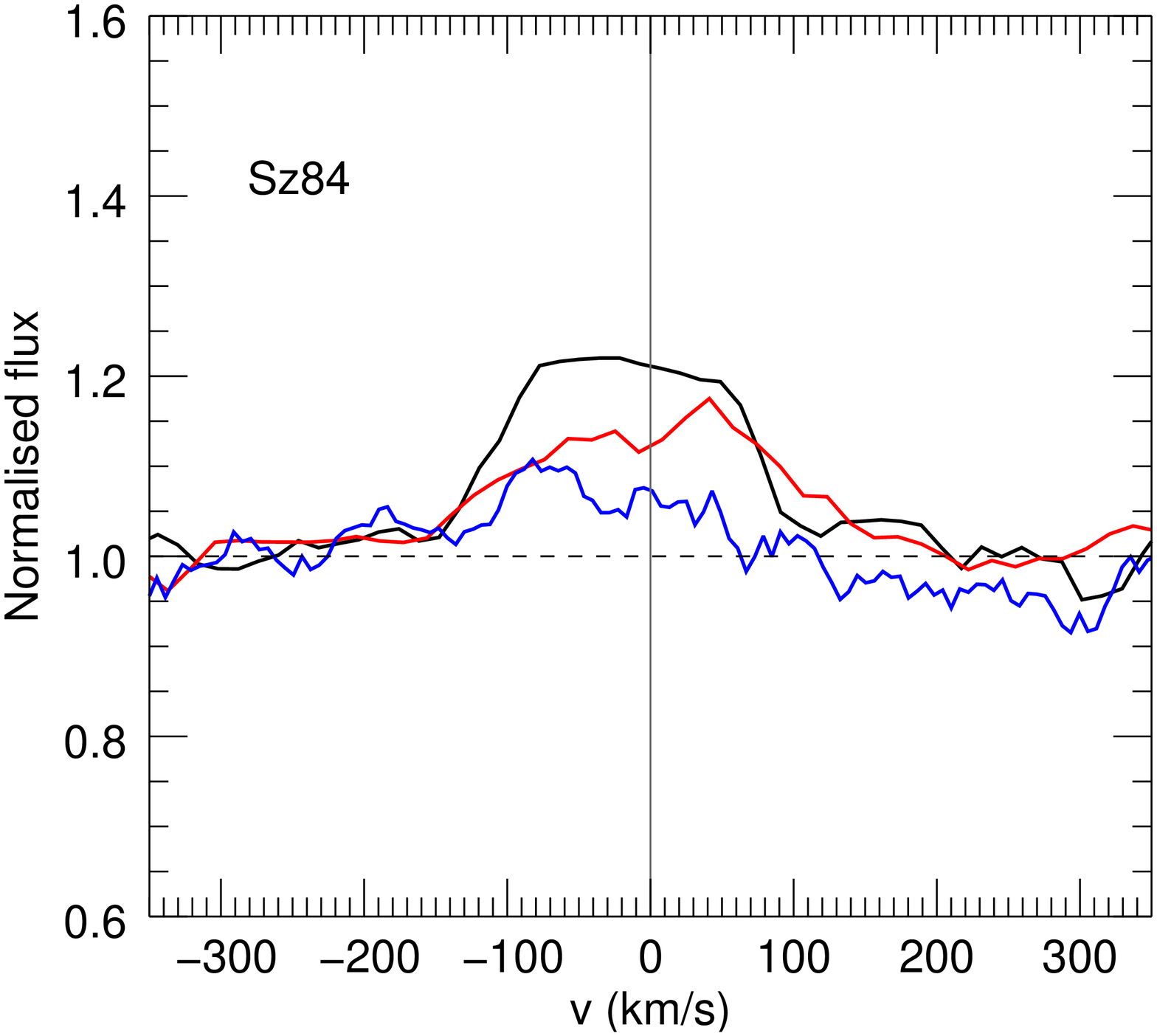}
\includegraphics[width=6.cm]{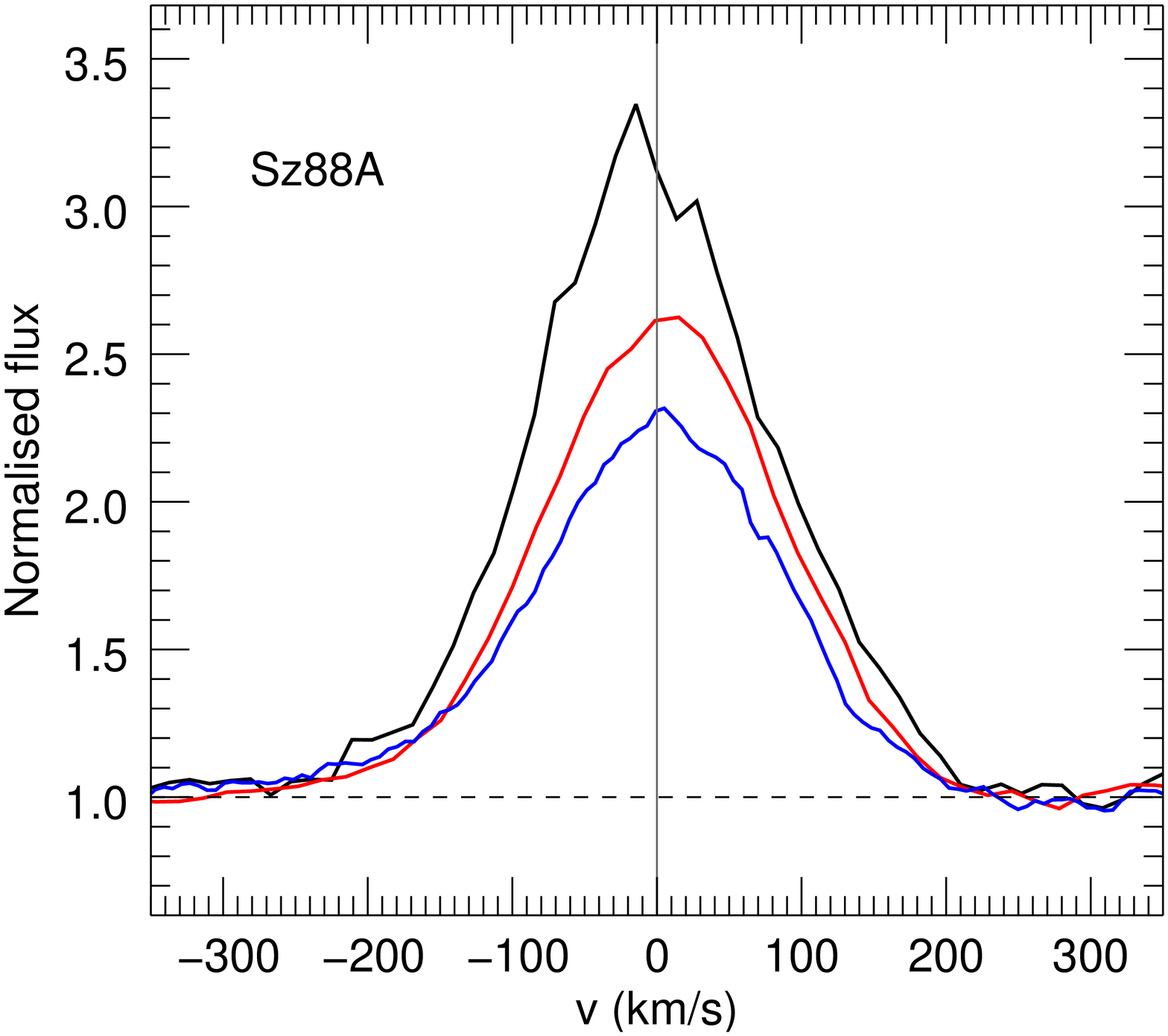}
\includegraphics[width=6.cm]{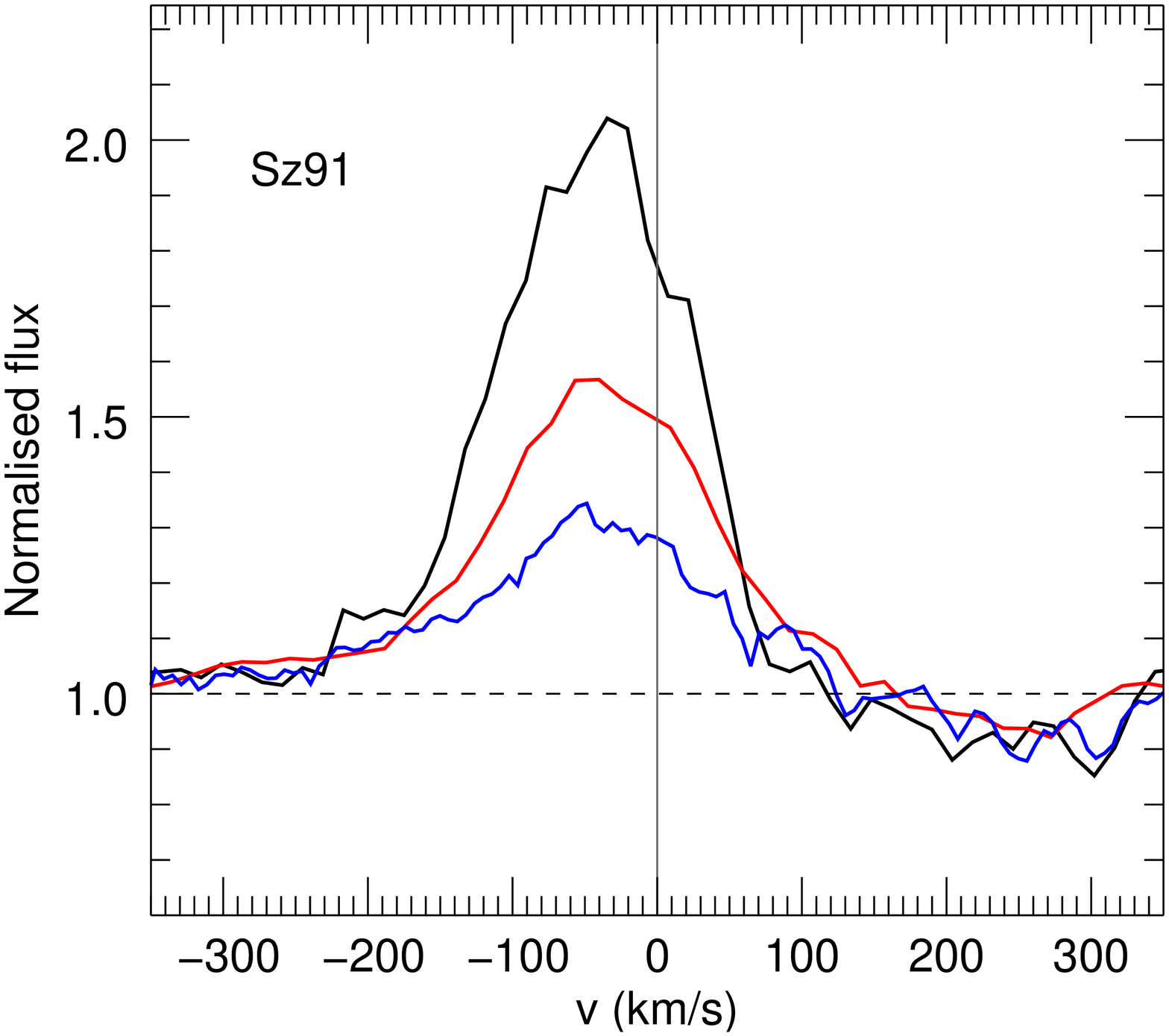}\\
\flushleft \includegraphics[width=6.cm]{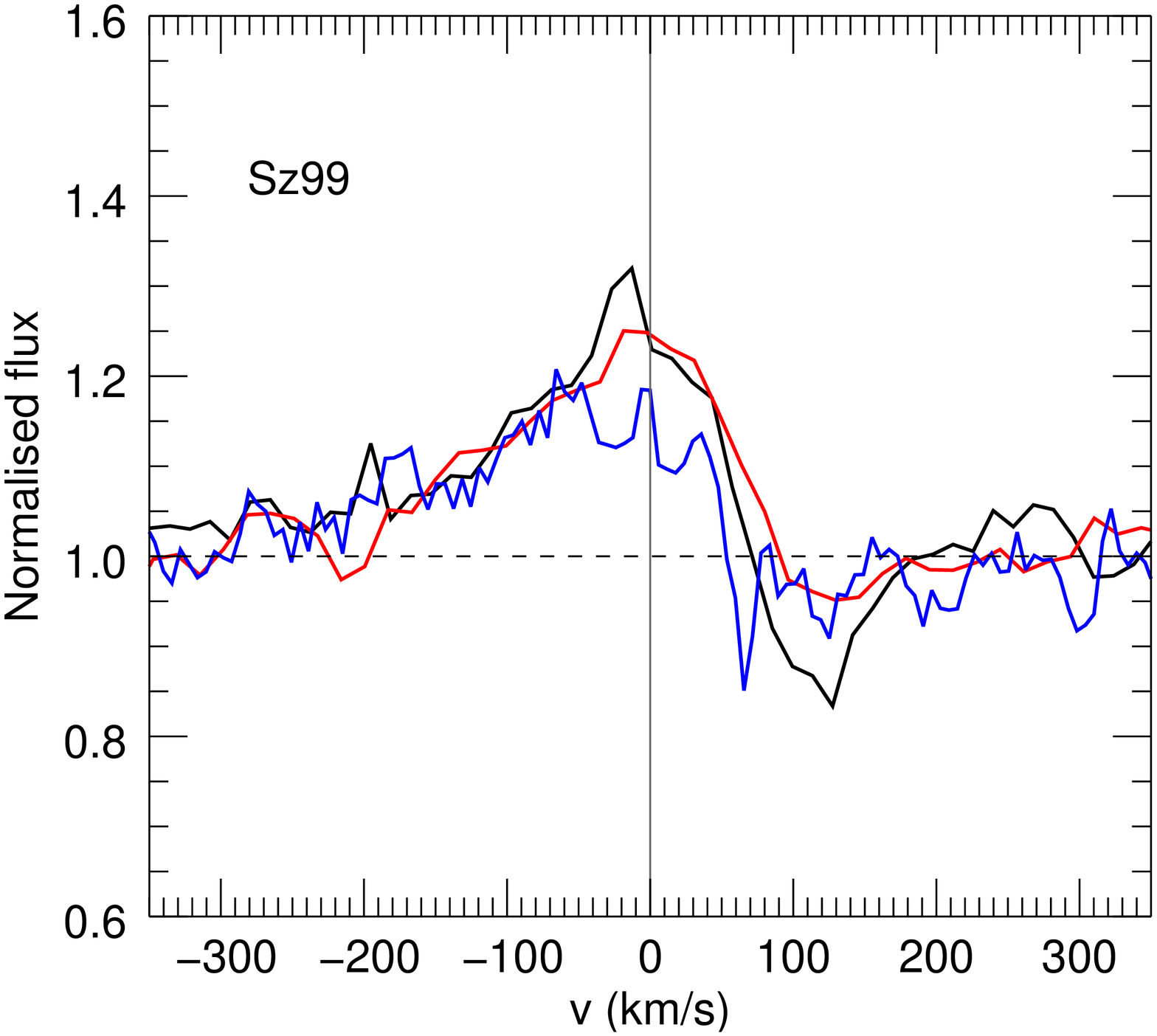}
\caption{Continued.} 
\end{figure*}

\begin{figure*}[t]
\centering

\includegraphics[width=6.cm]{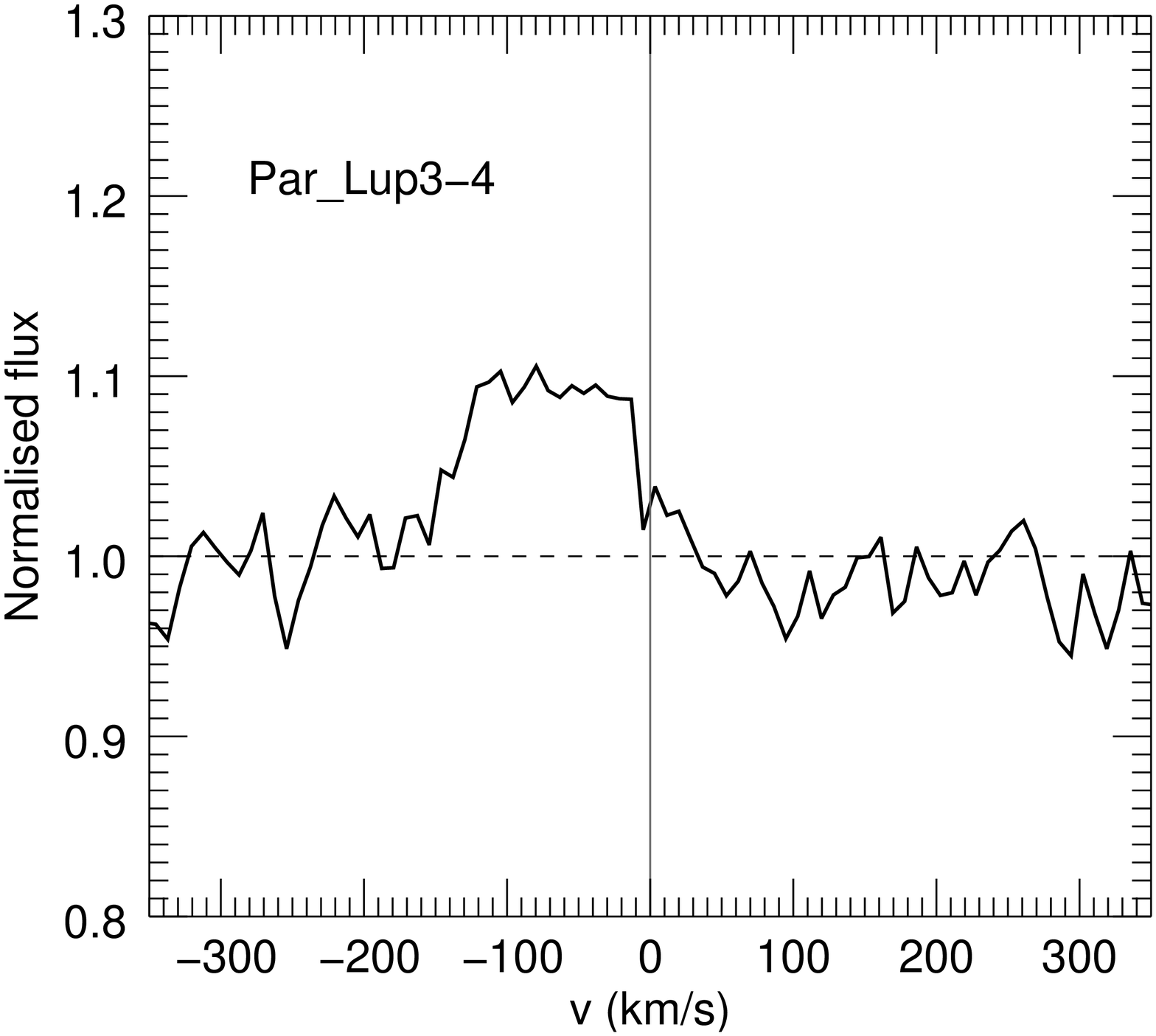}
\includegraphics[width=6.cm]{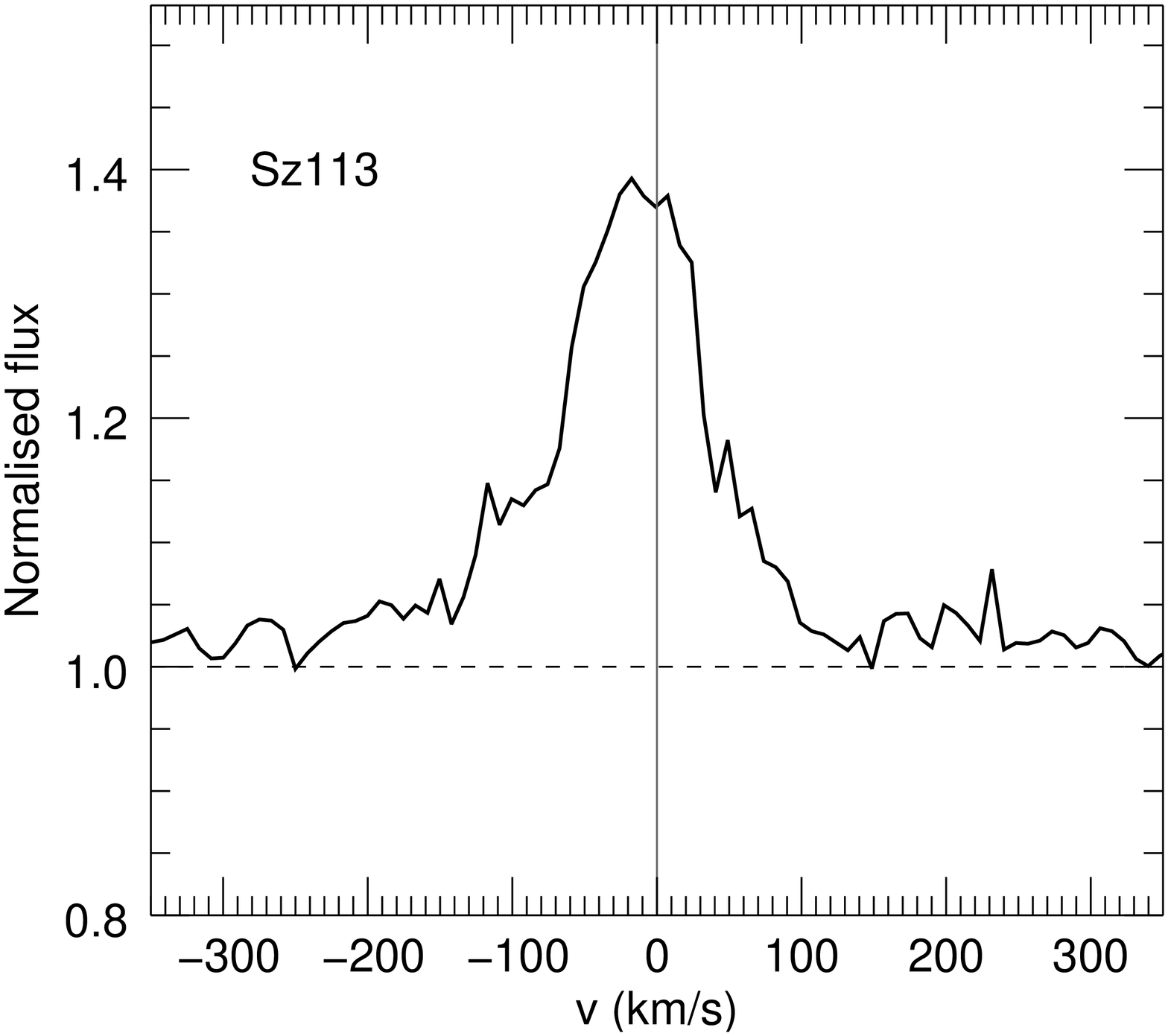}
\includegraphics[width=6.cm]{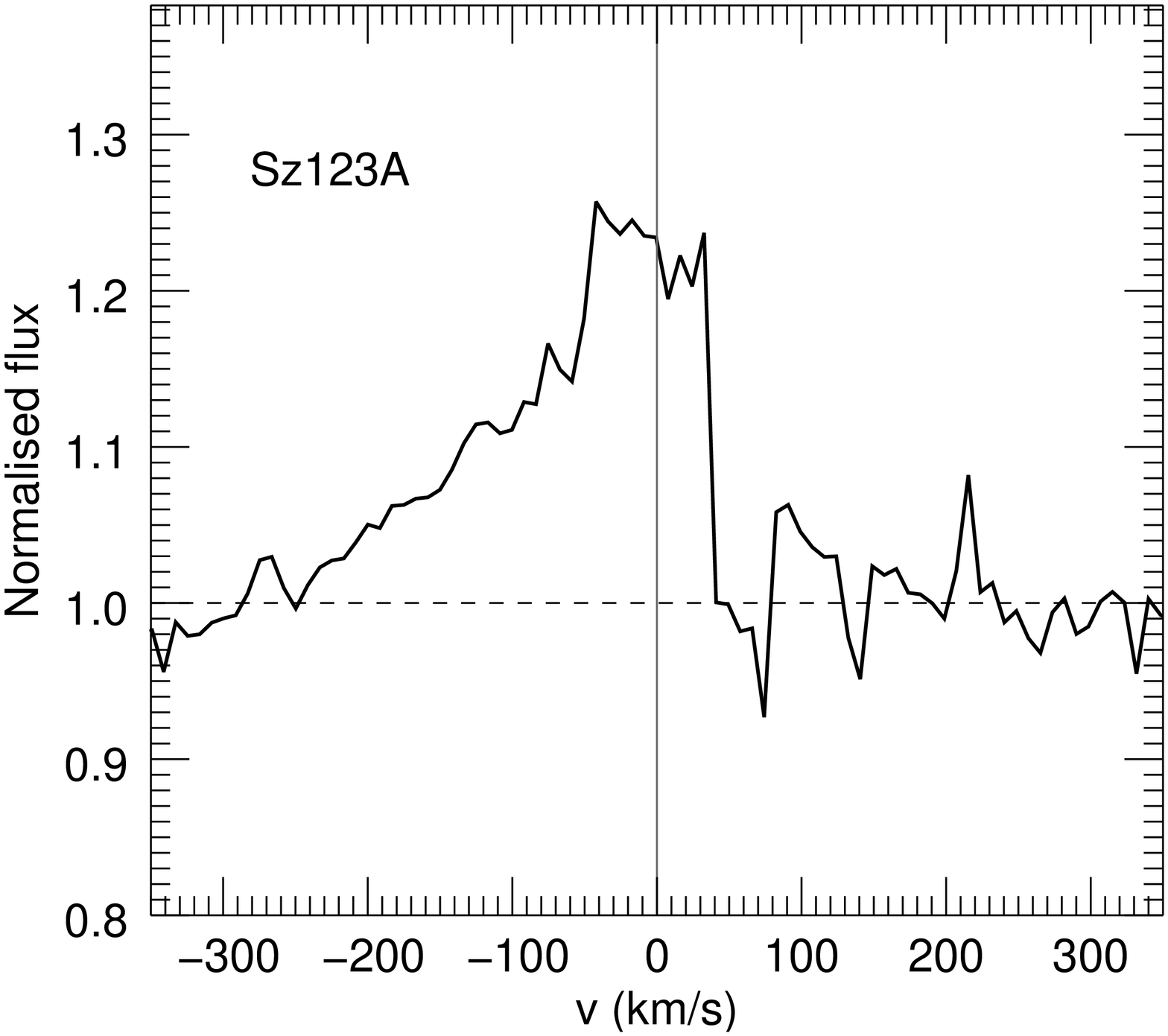}\\
\includegraphics[width=6.cm]{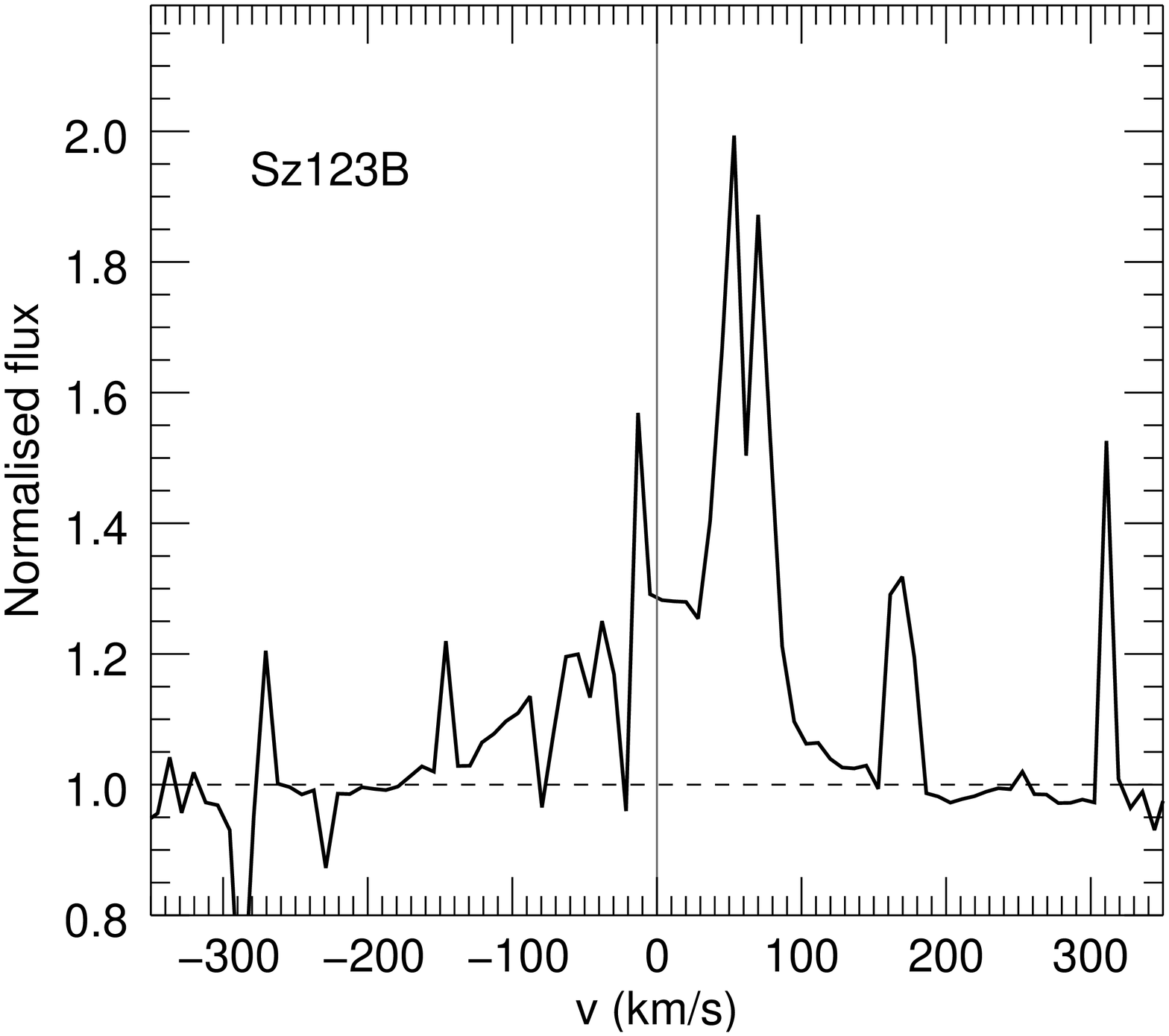}
\includegraphics[width=6.cm]{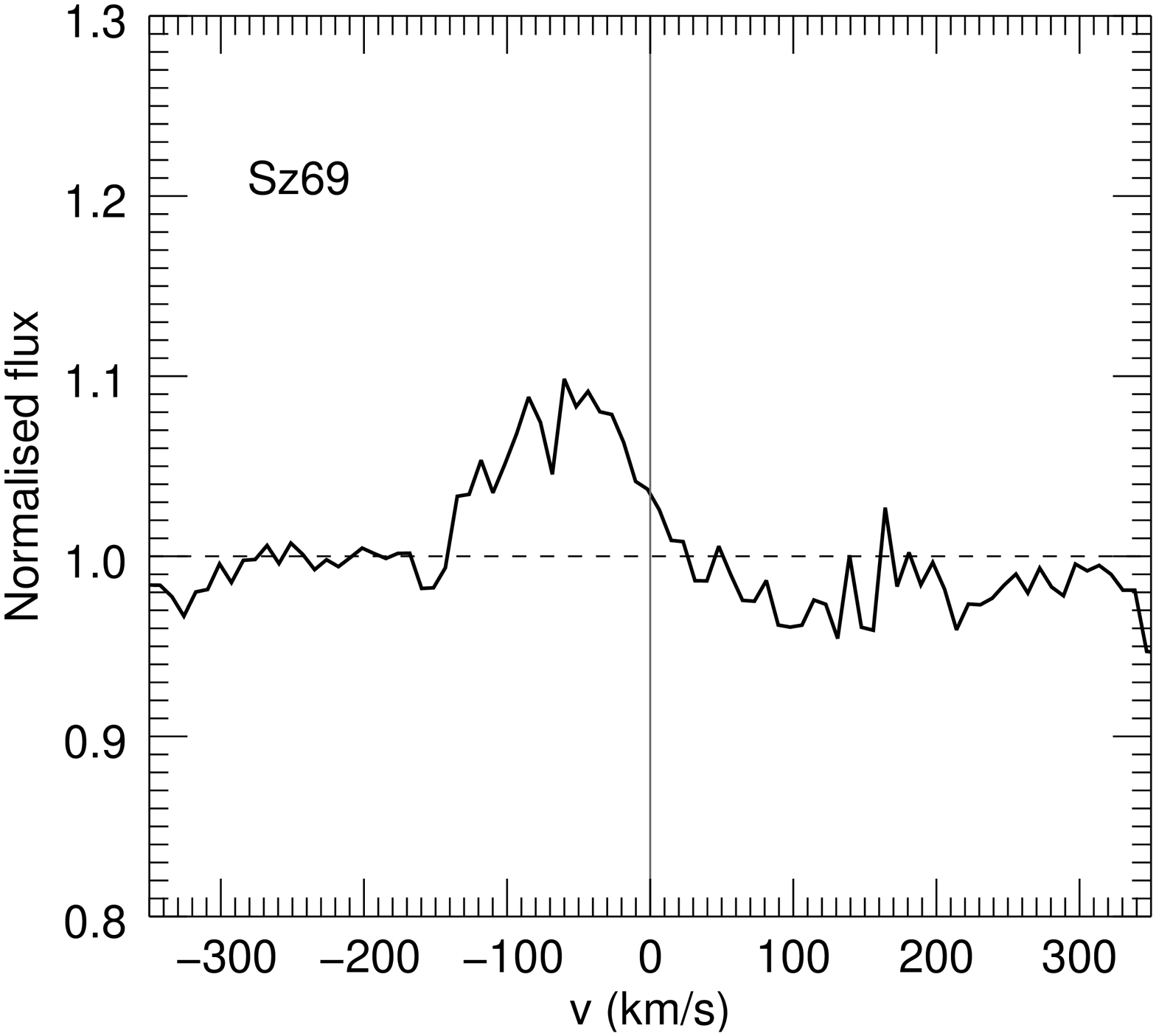}
\includegraphics[width=6.cm]{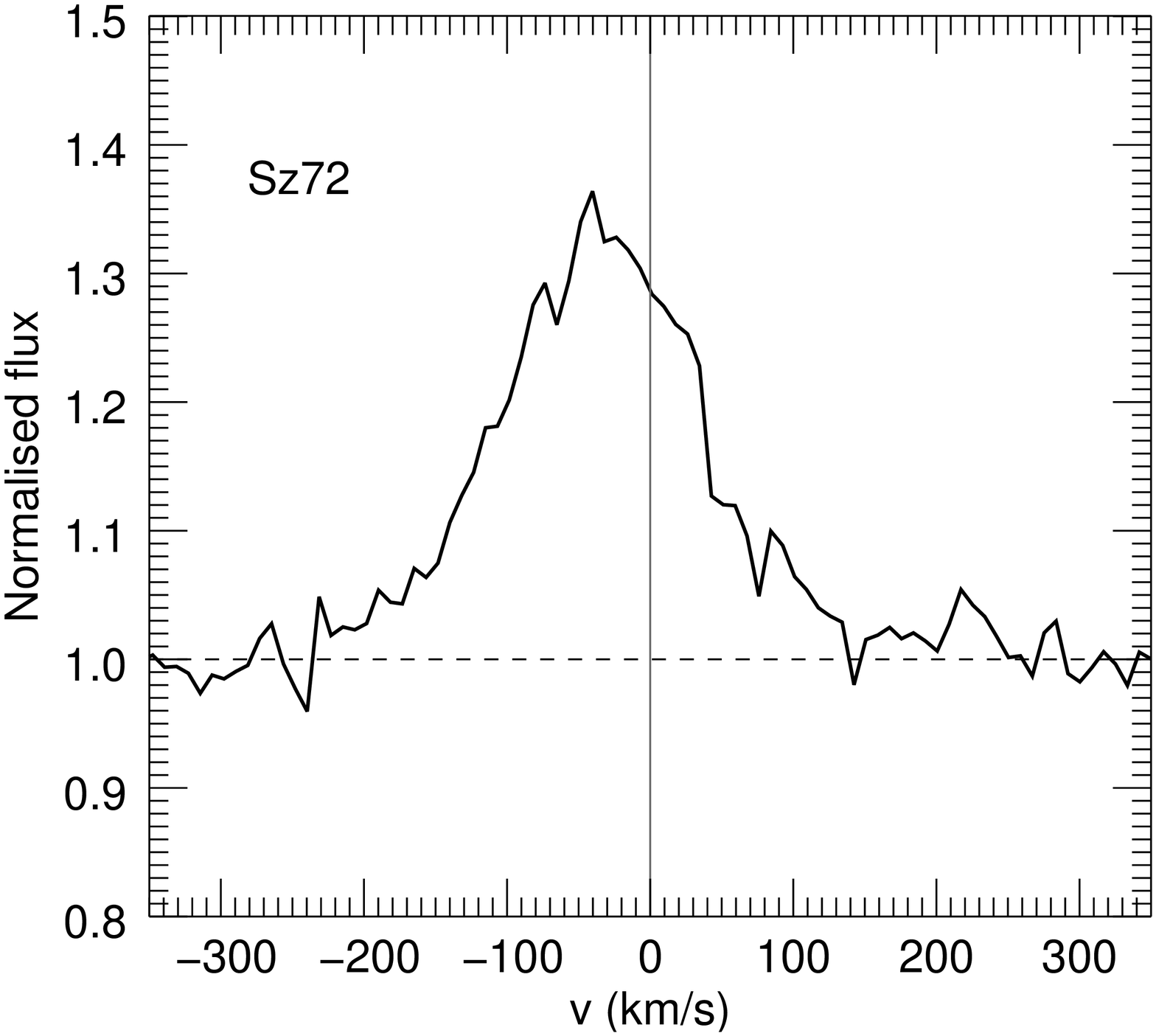}\\
\includegraphics[width=6.cm]{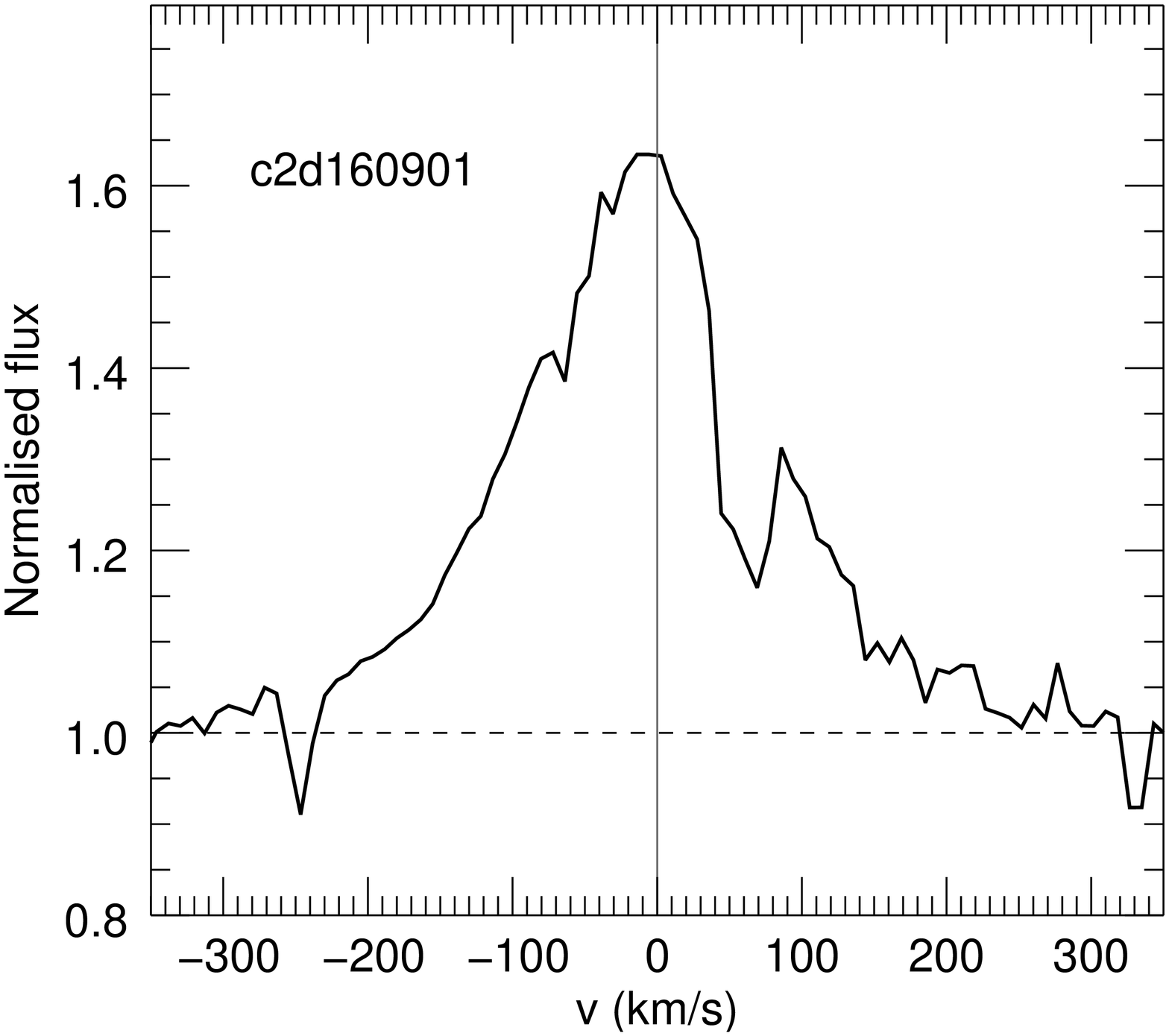}
\includegraphics[width=6.cm]{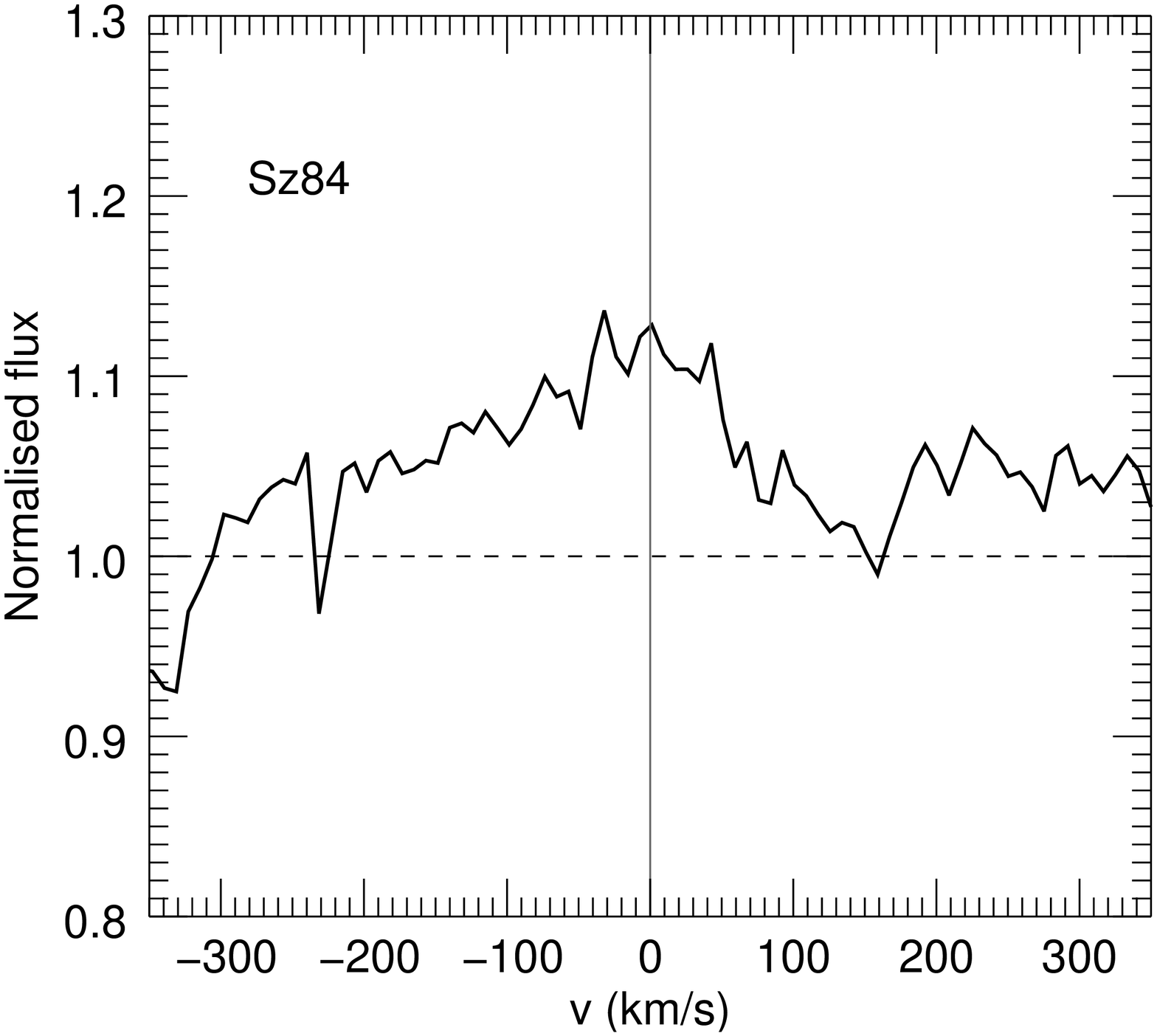}
\includegraphics[width=6.cm]{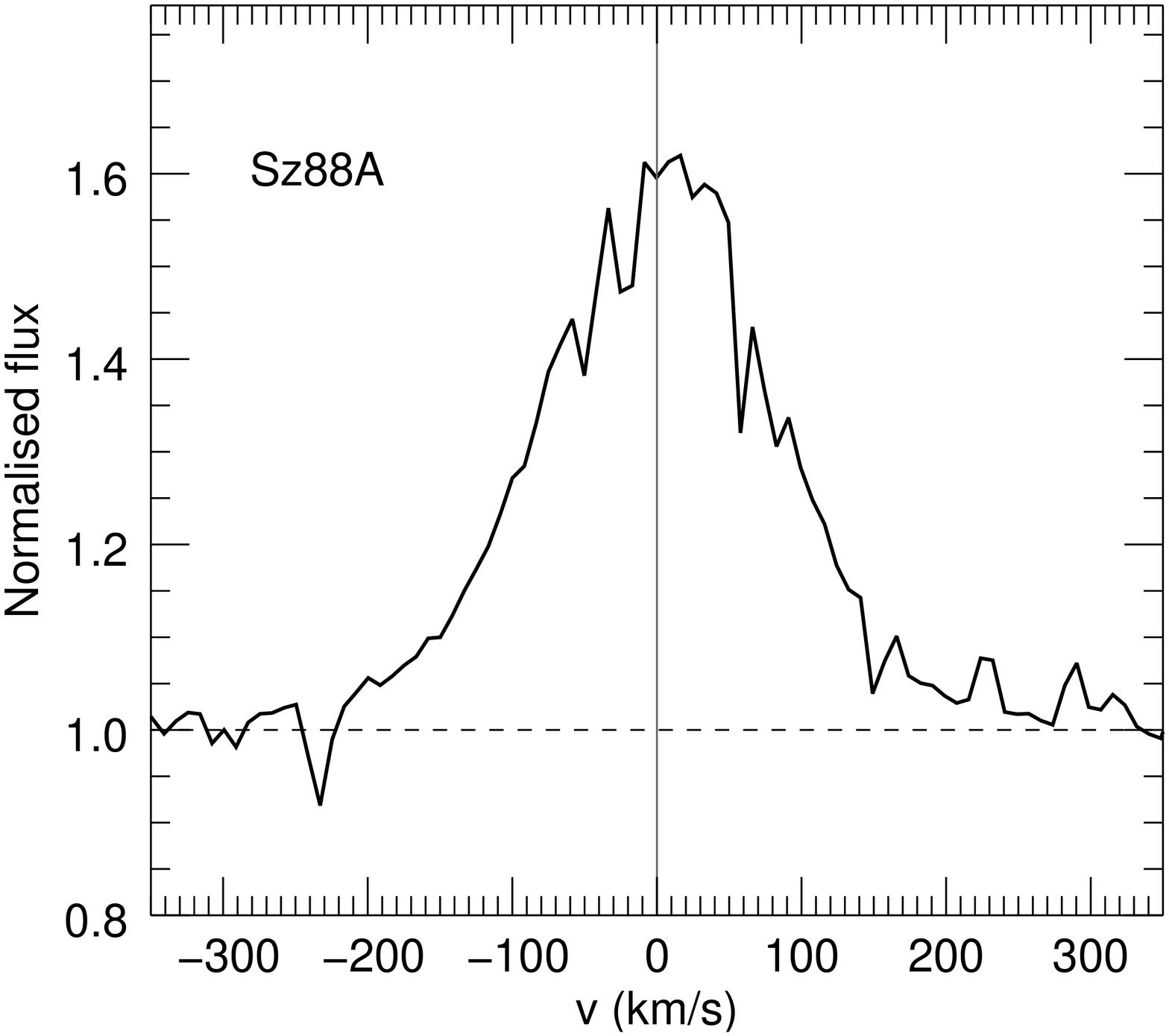}\\
\flushleft \includegraphics[width=6.cm]{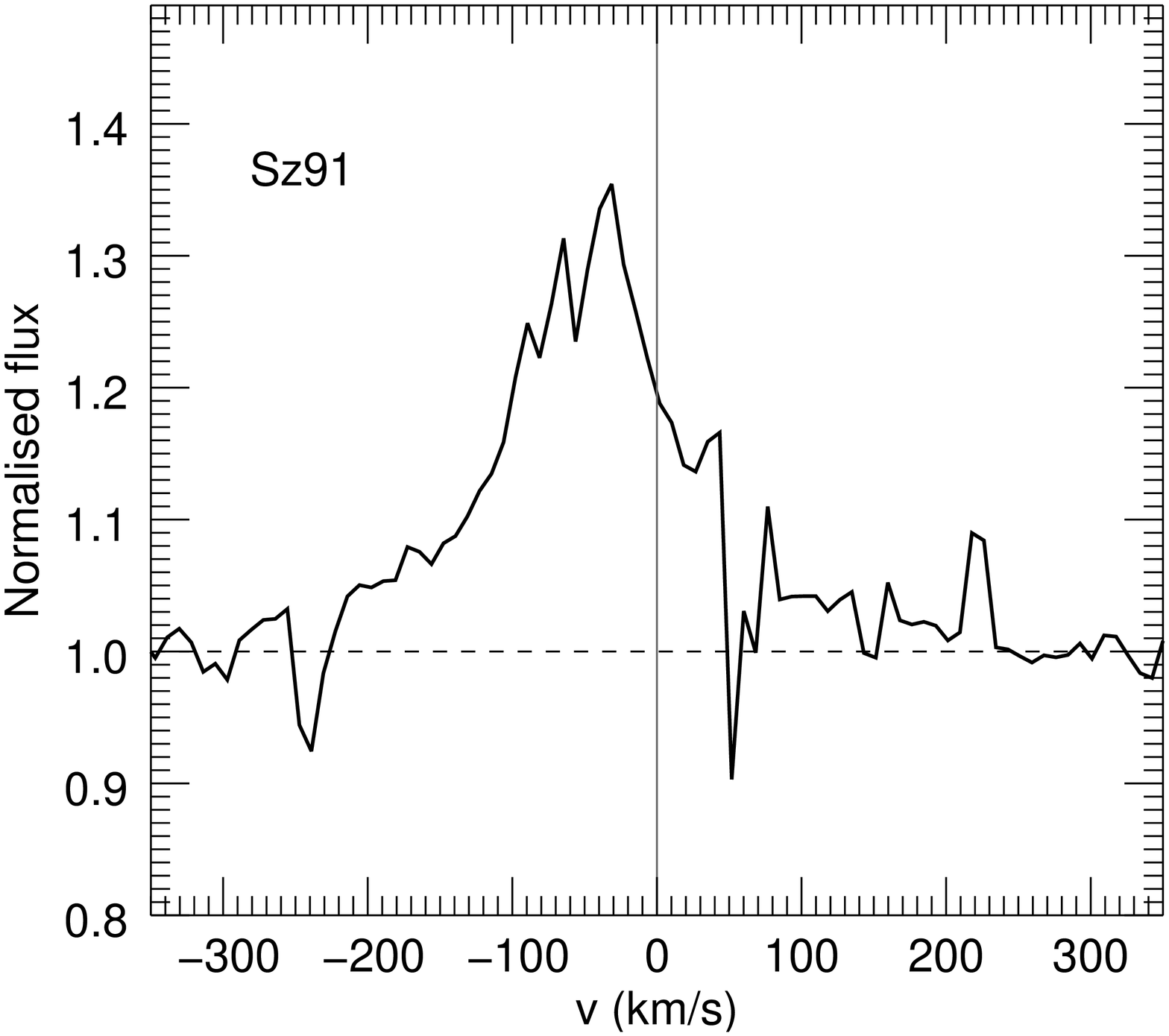}
\caption{\label{fig:lines:br} Continuum-normalised Br$\gamma$ of sources where the line is detected with a S/N $>$ 5.} 
\end{figure*}

\section*{Appendix B: Atlas of Balmer decrements}
\setcounter{figure}{0} \renewcommand{\thefigure}{B.\arabic{figure}} 
\setcounter{table}{0} \renewcommand{\thetable}{B.\arabic{table}}

\begin{figure*}[t]
\centering
\includegraphics[width=4.4cm]{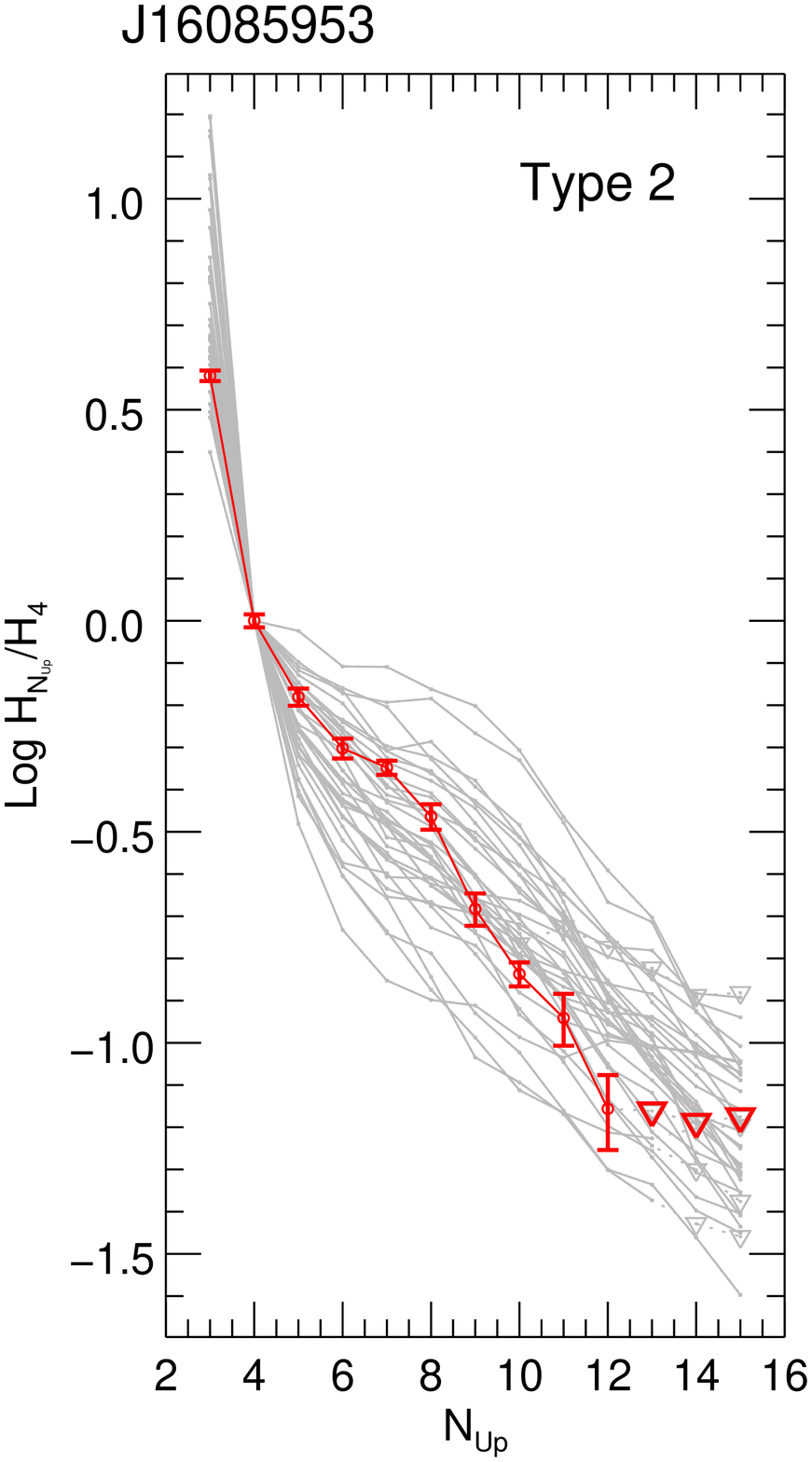}
\includegraphics[width=4.4cm]{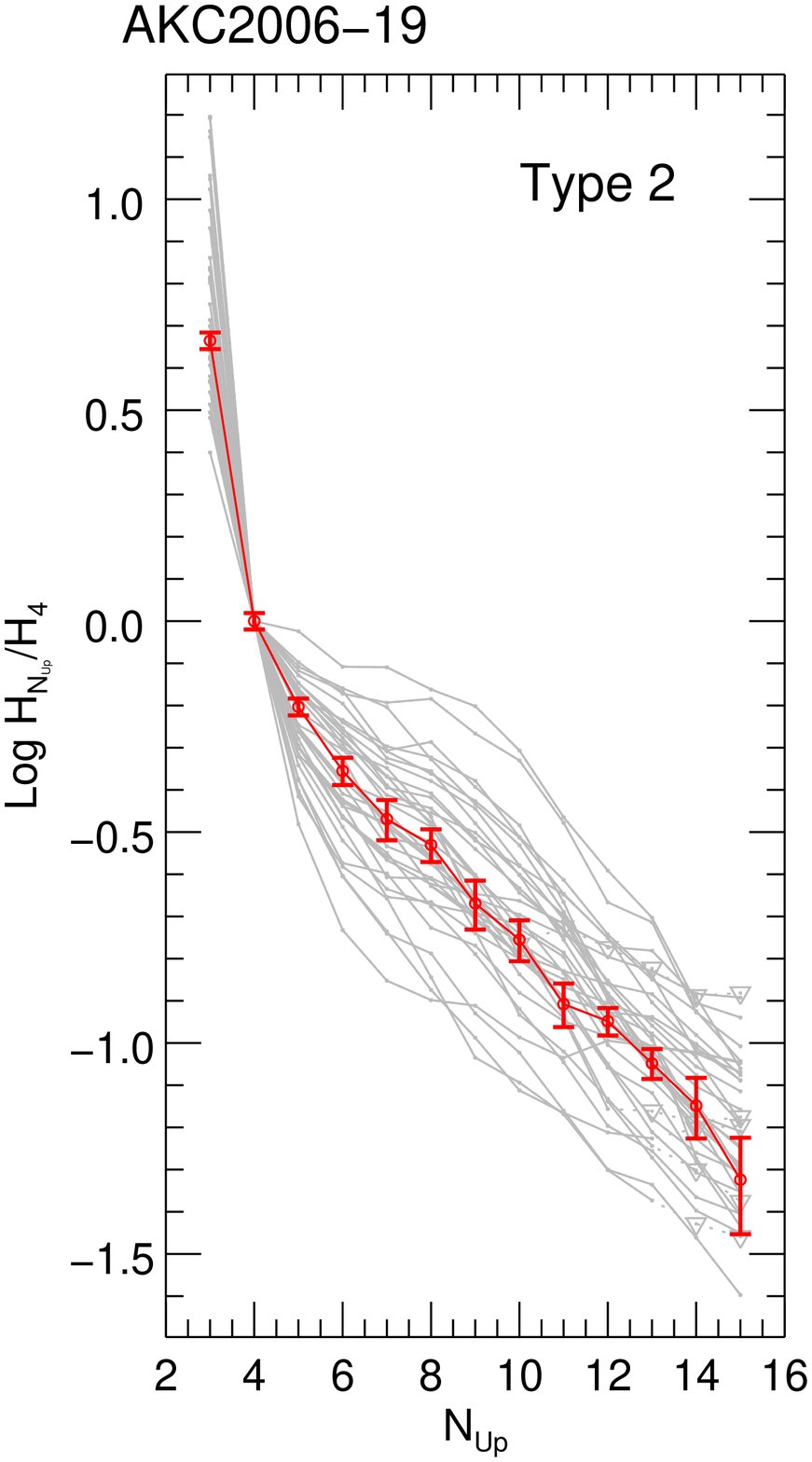}
\includegraphics[width=4.4cm]{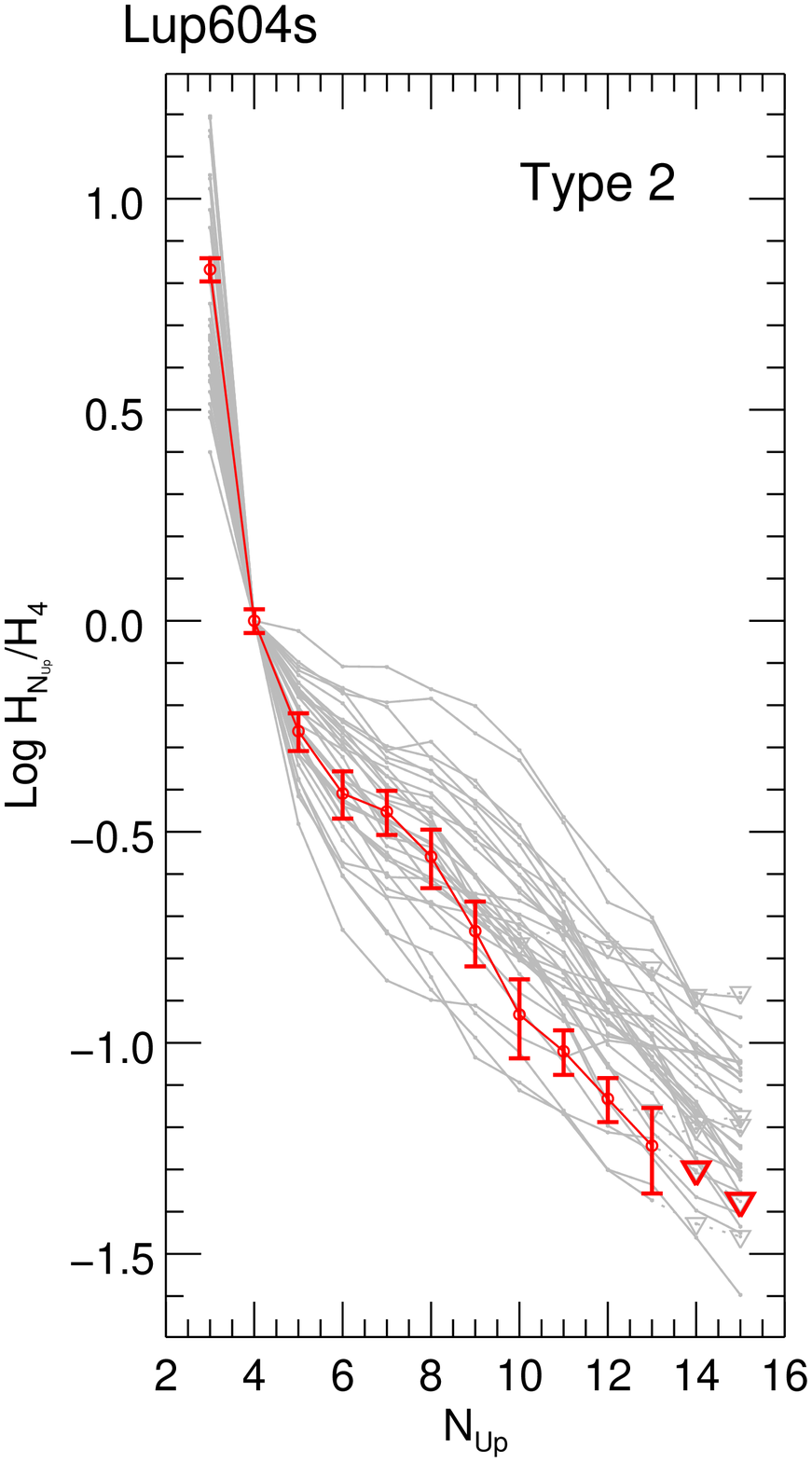}
\includegraphics[width=4.4cm]{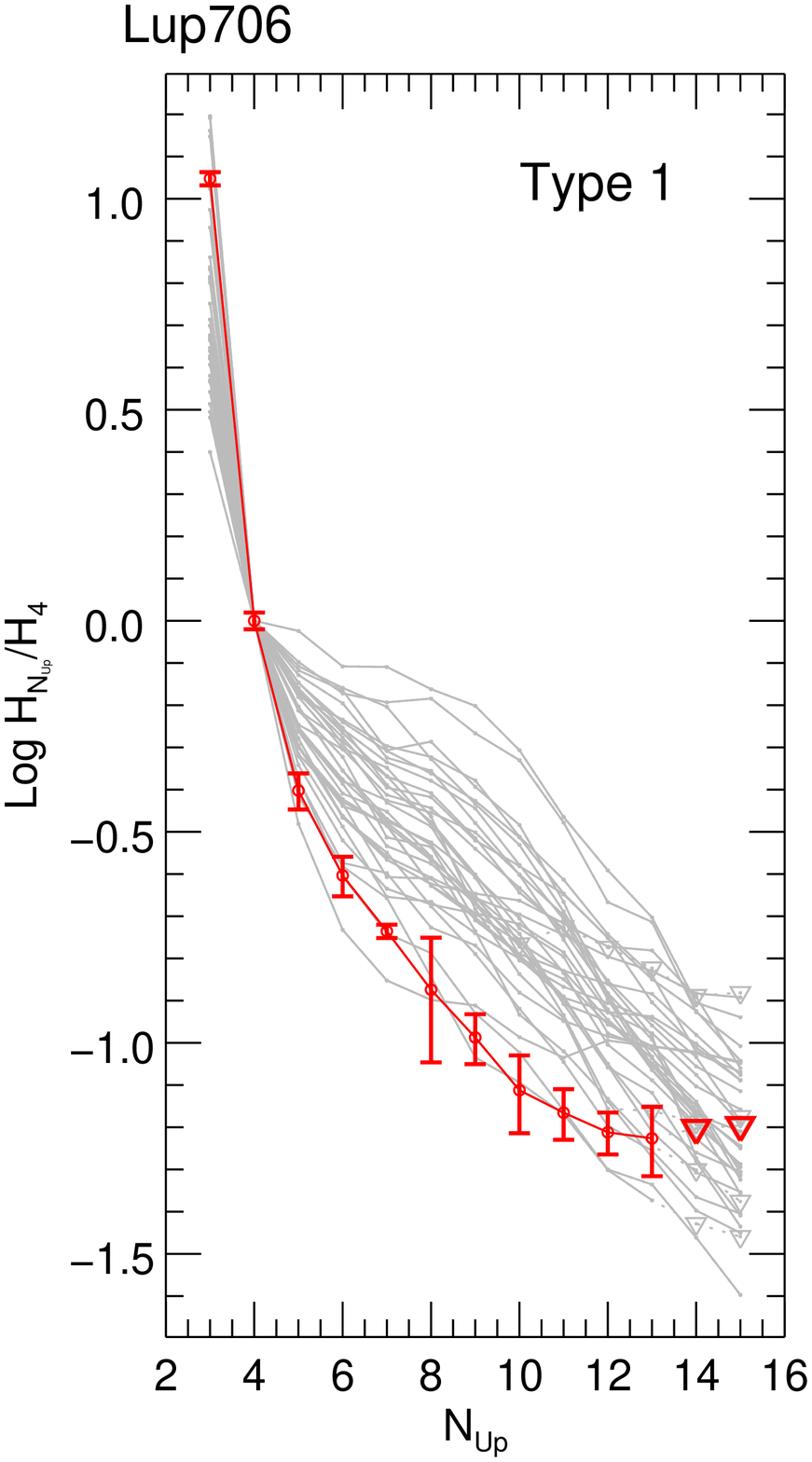}\\
\includegraphics[width=4.4cm]{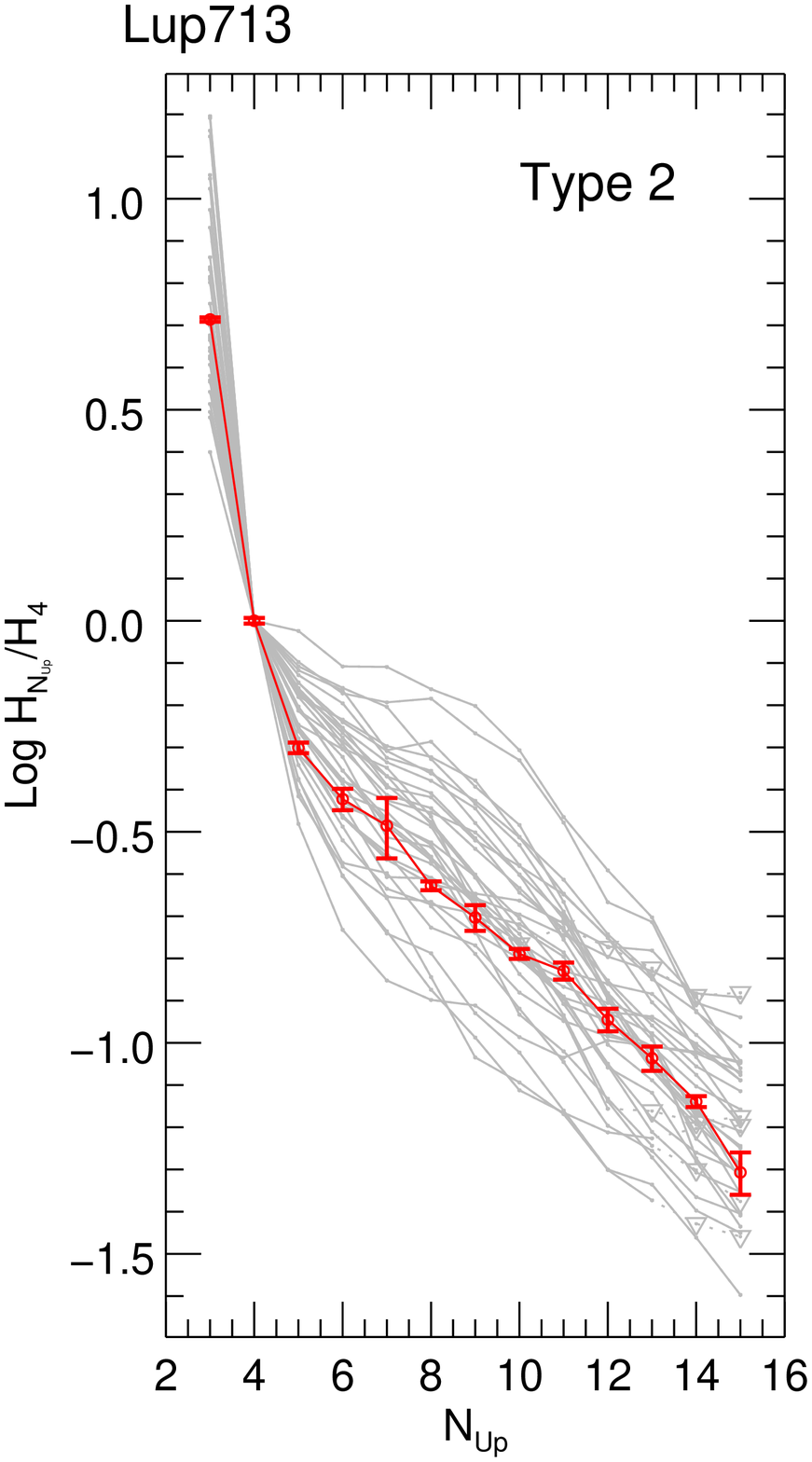}
\includegraphics[width=4.4cm]{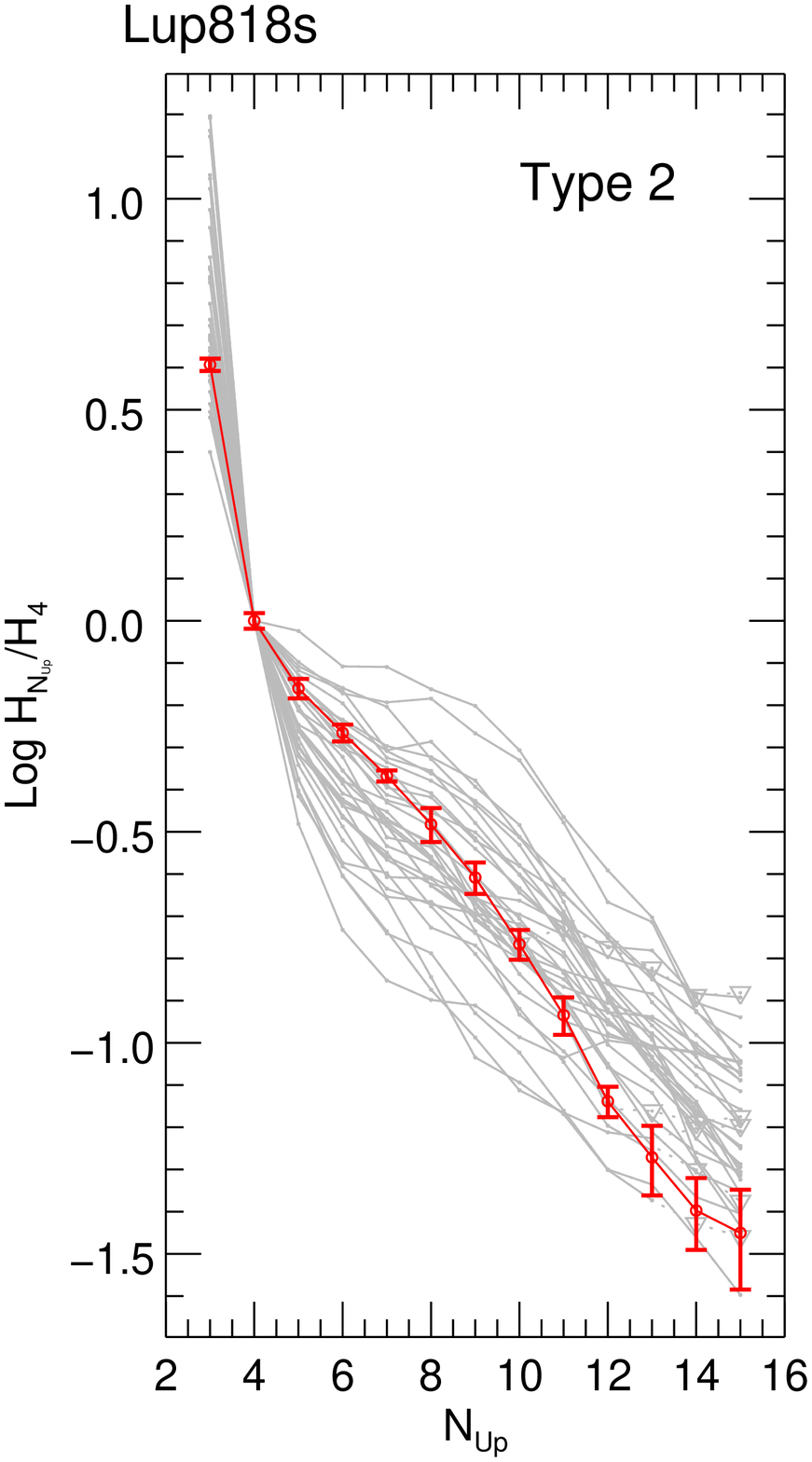}
\includegraphics[width=4.4cm]{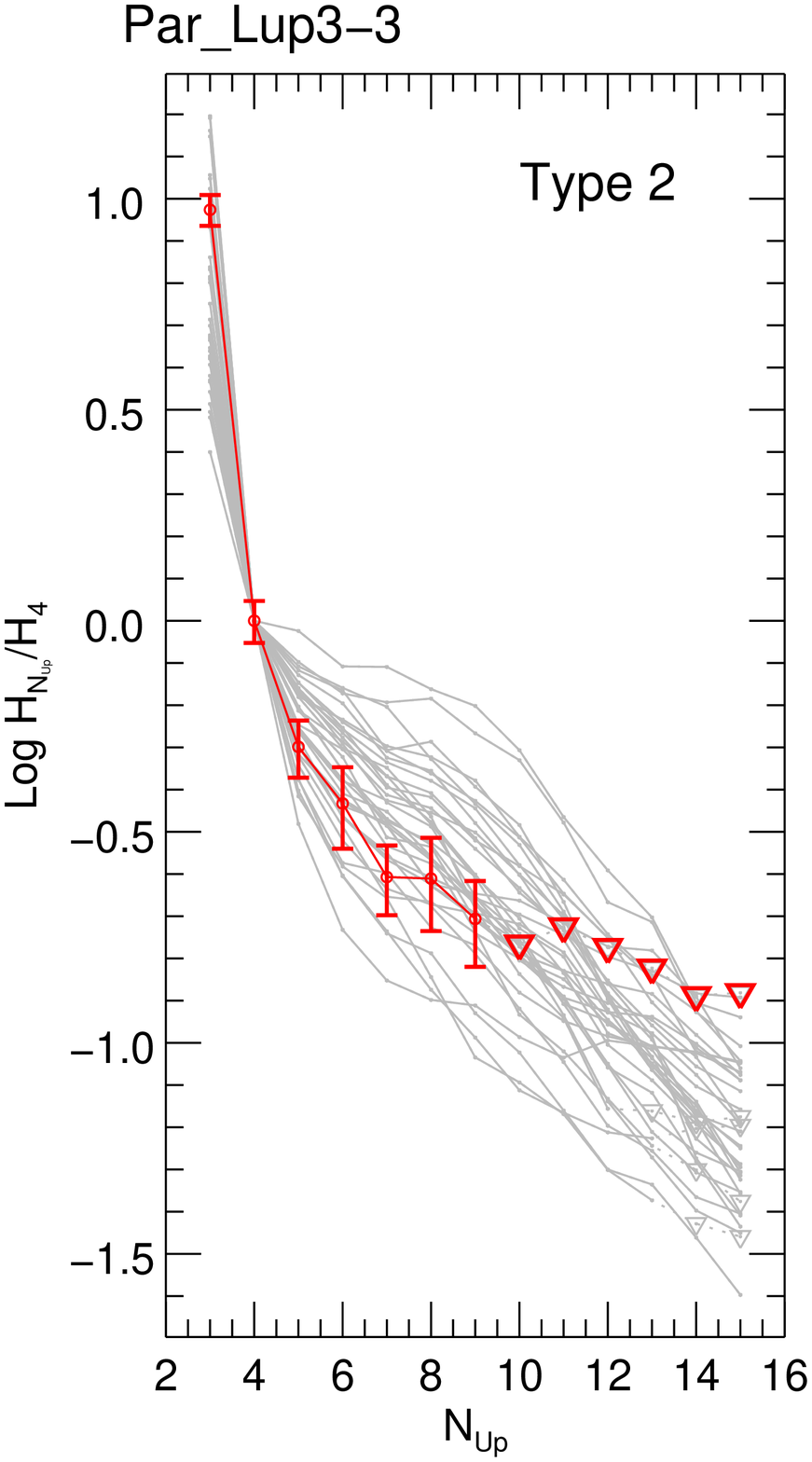}
\includegraphics[width=4.4cm]{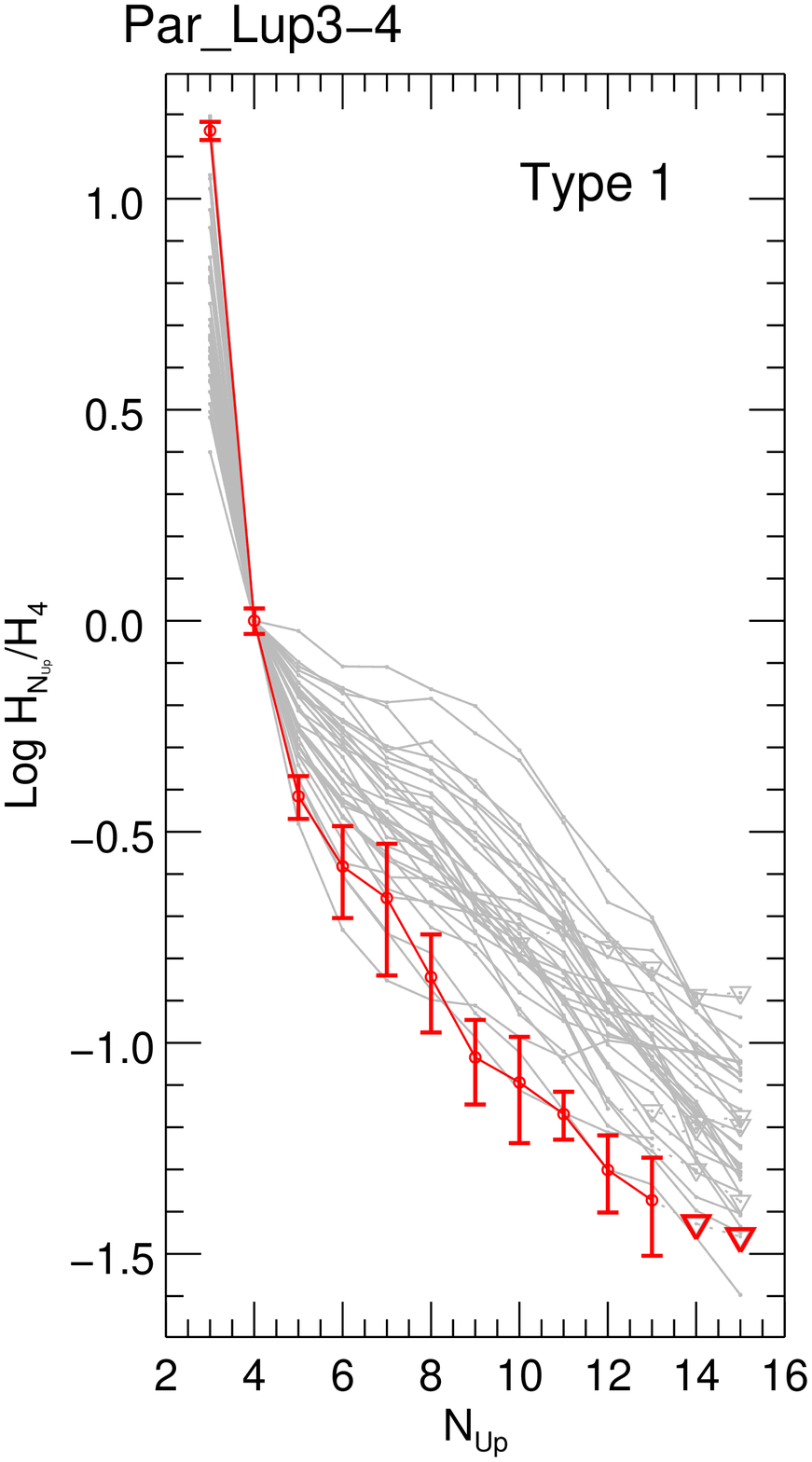}\\
\includegraphics[width=4.4cm]{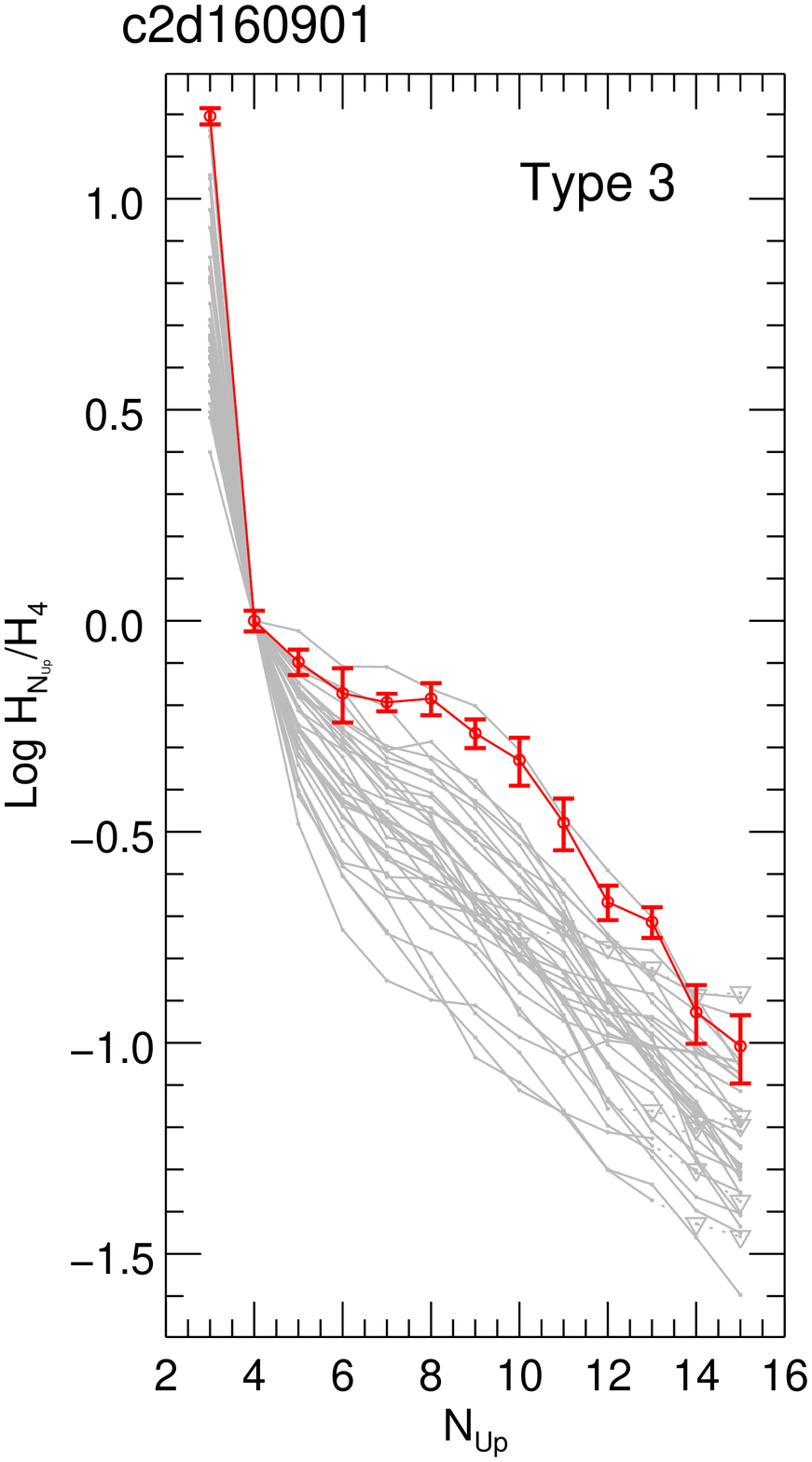}
\includegraphics[width=4.4cm]{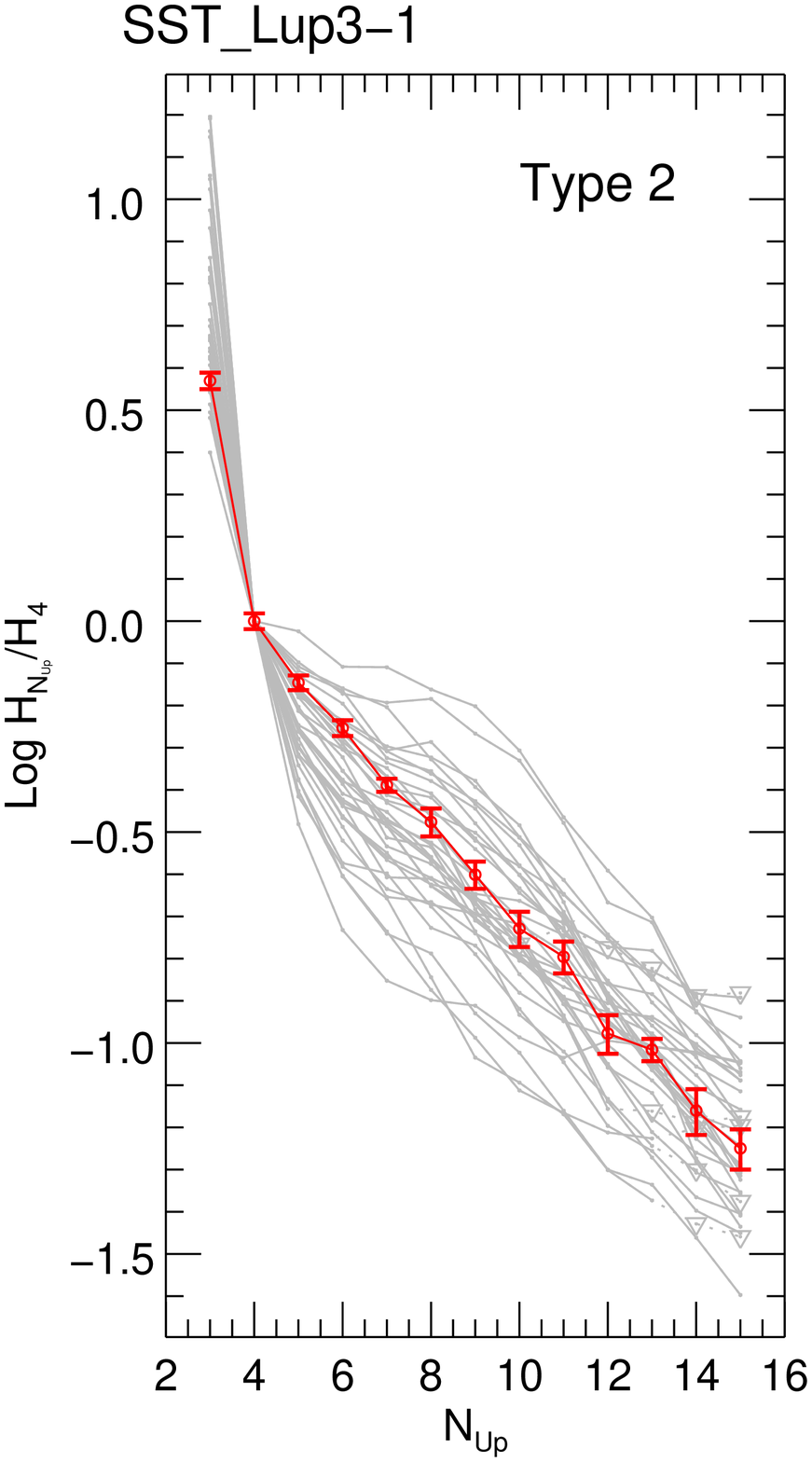}
\includegraphics[width=4.4cm]{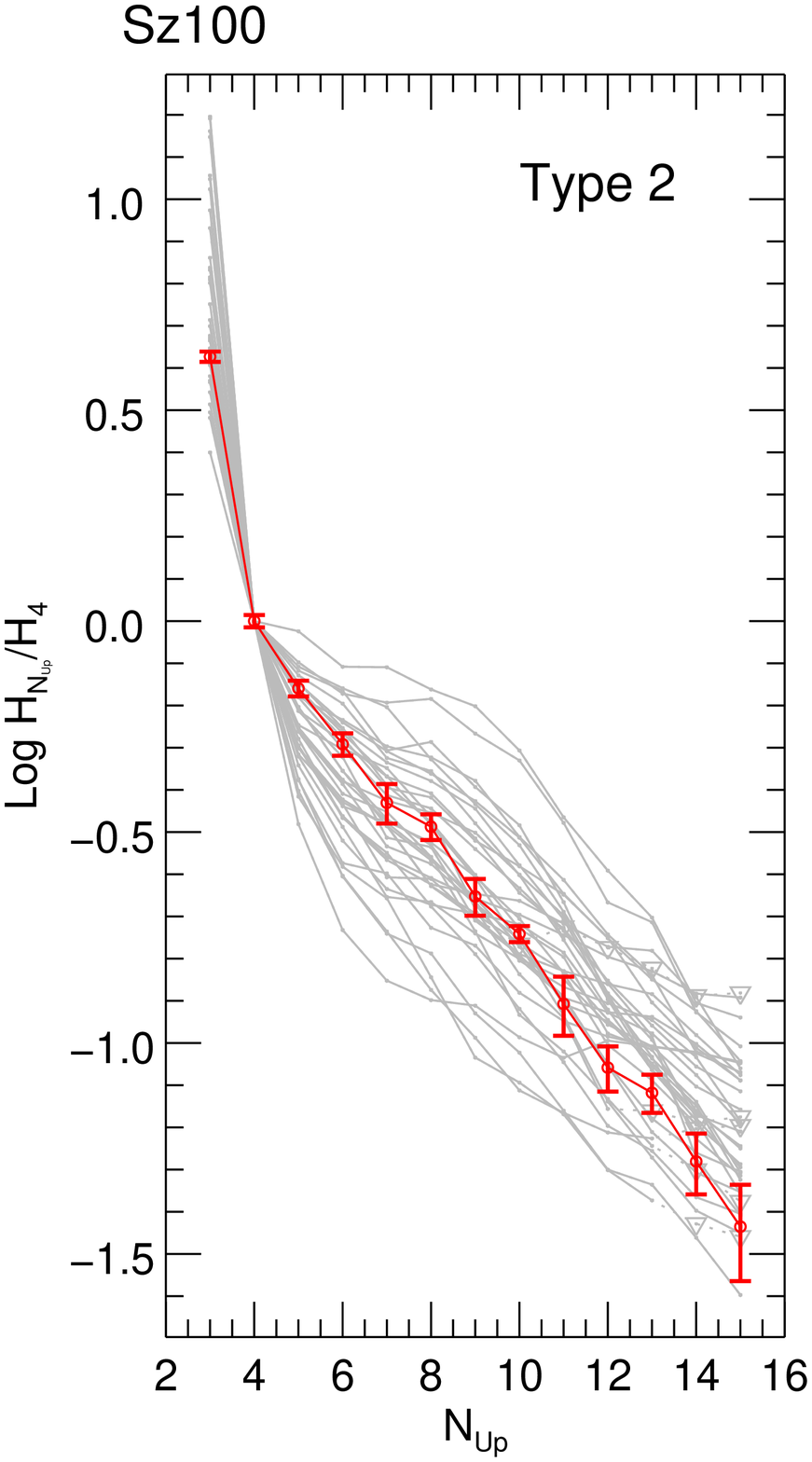}
\includegraphics[width=4.4cm]{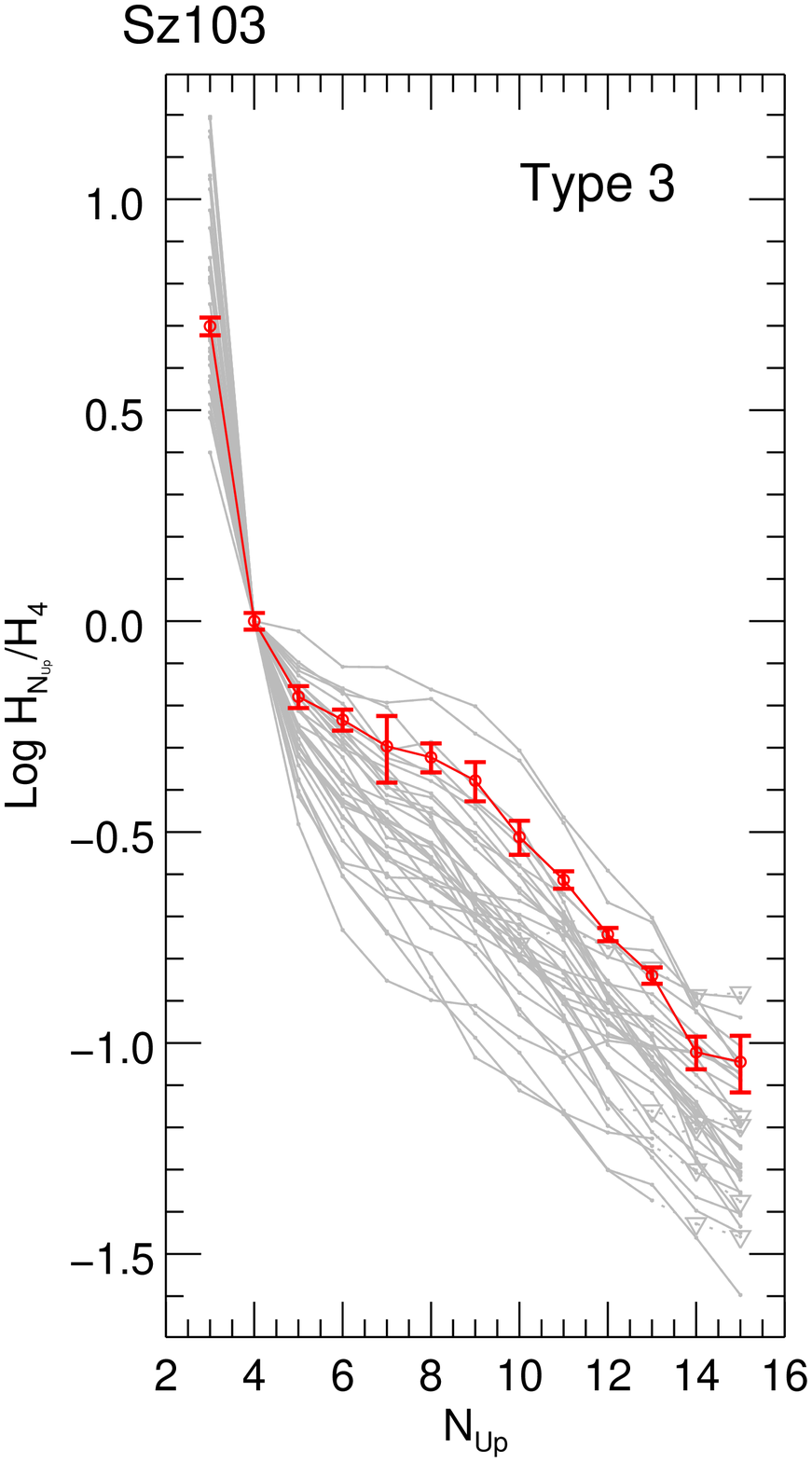}\\
\caption{\label{fig:decs:h} Balmer decrements computed with respect to H$\beta$. For each object, the decrement shape is highlighted in red against all observed decrements, which are plotted as grey curves. Triangles indicate upper limits.} 
\end{figure*}

\setcounter{figure}{0}
\begin{figure*}[t]
\centering
\includegraphics[width=4.4cm]{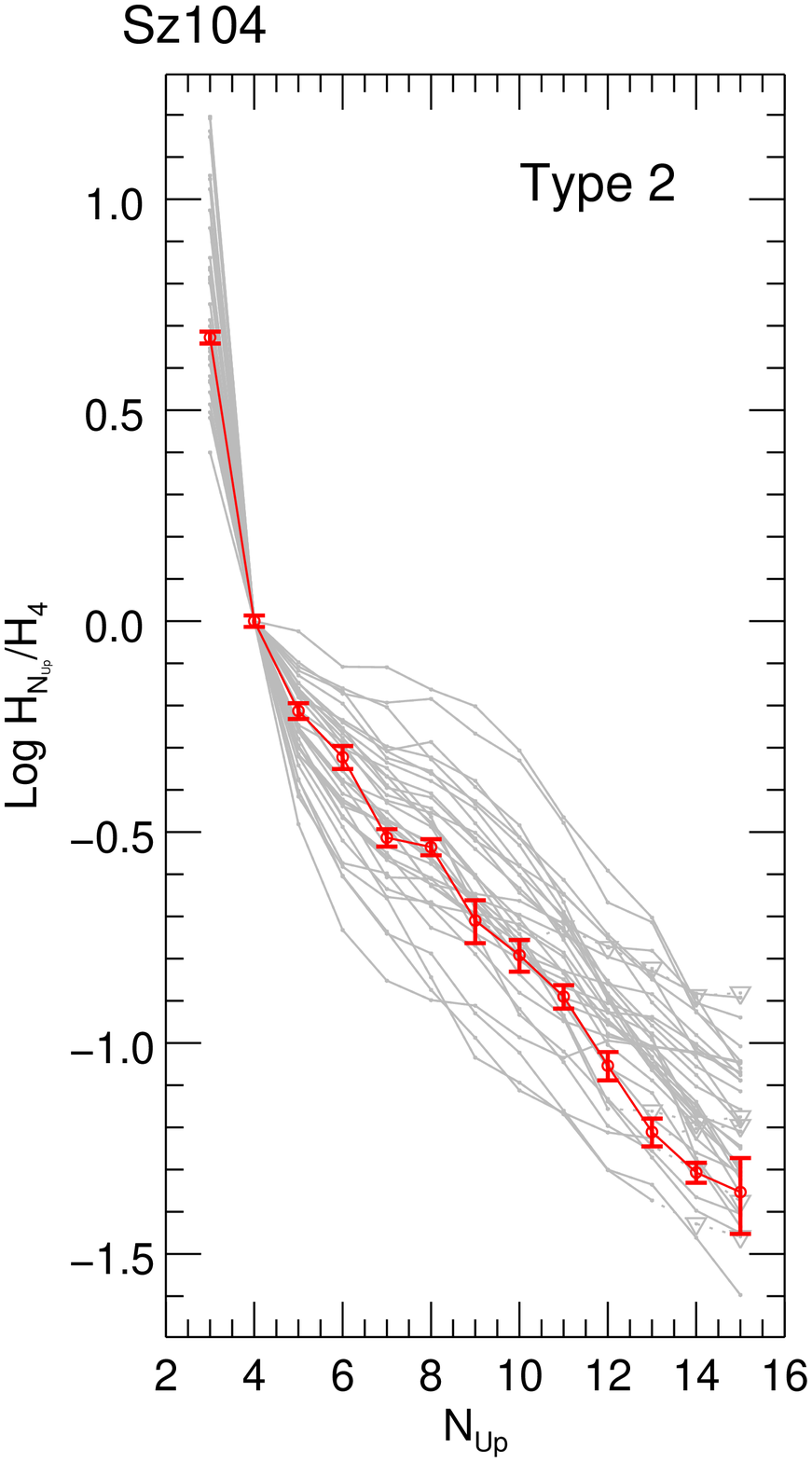}
\includegraphics[width=4.4cm]{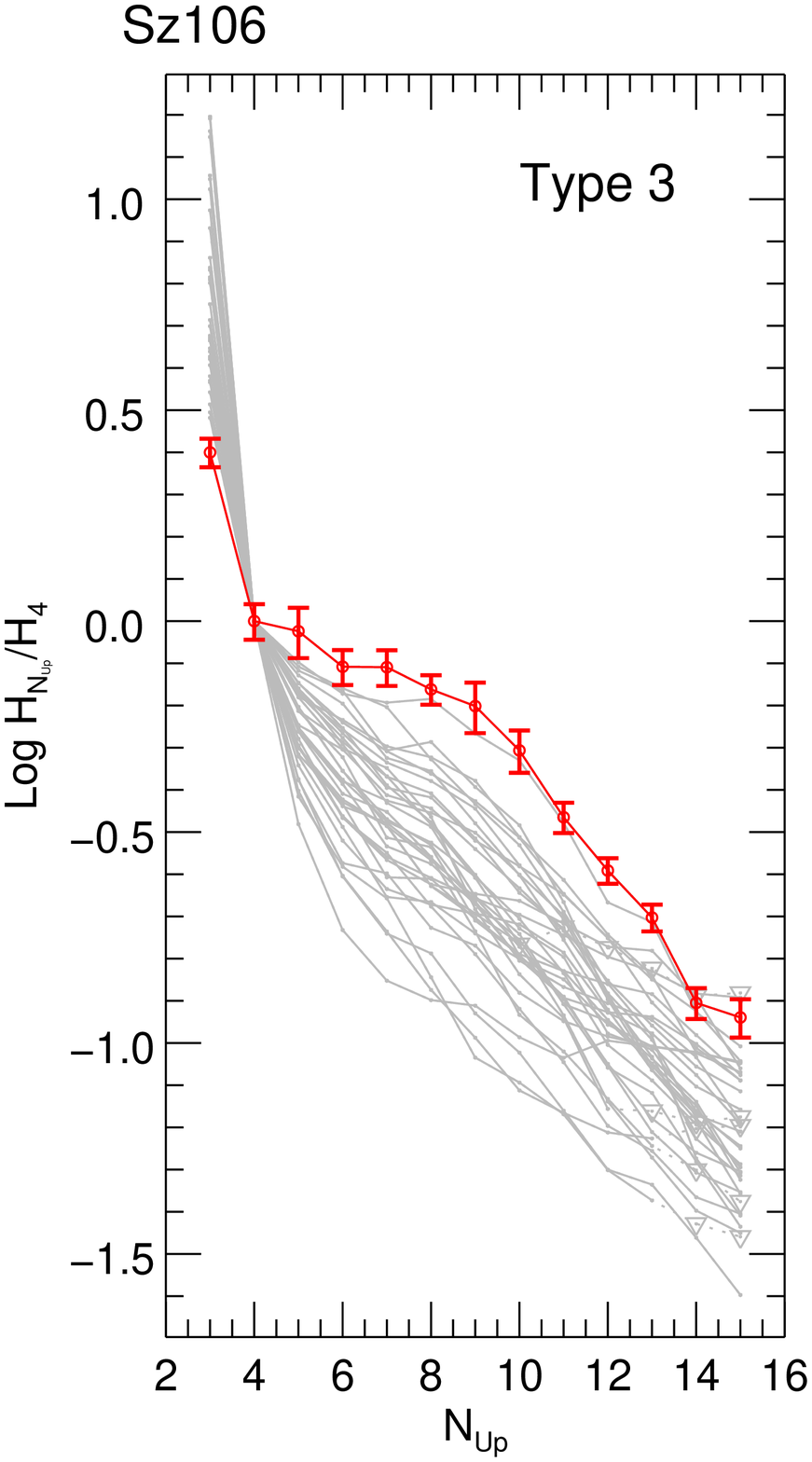}
\includegraphics[width=4.4cm]{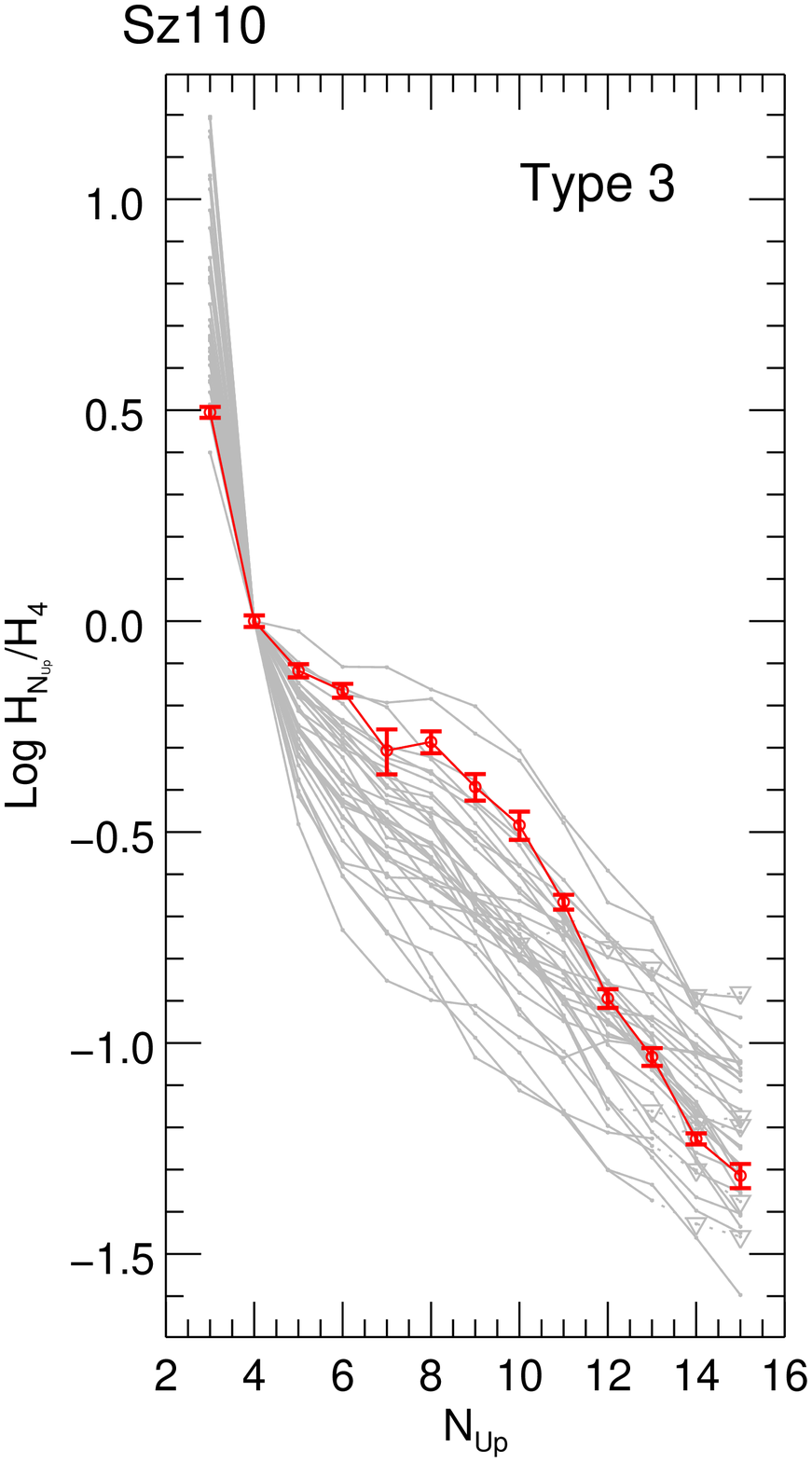}
\includegraphics[width=4.4cm]{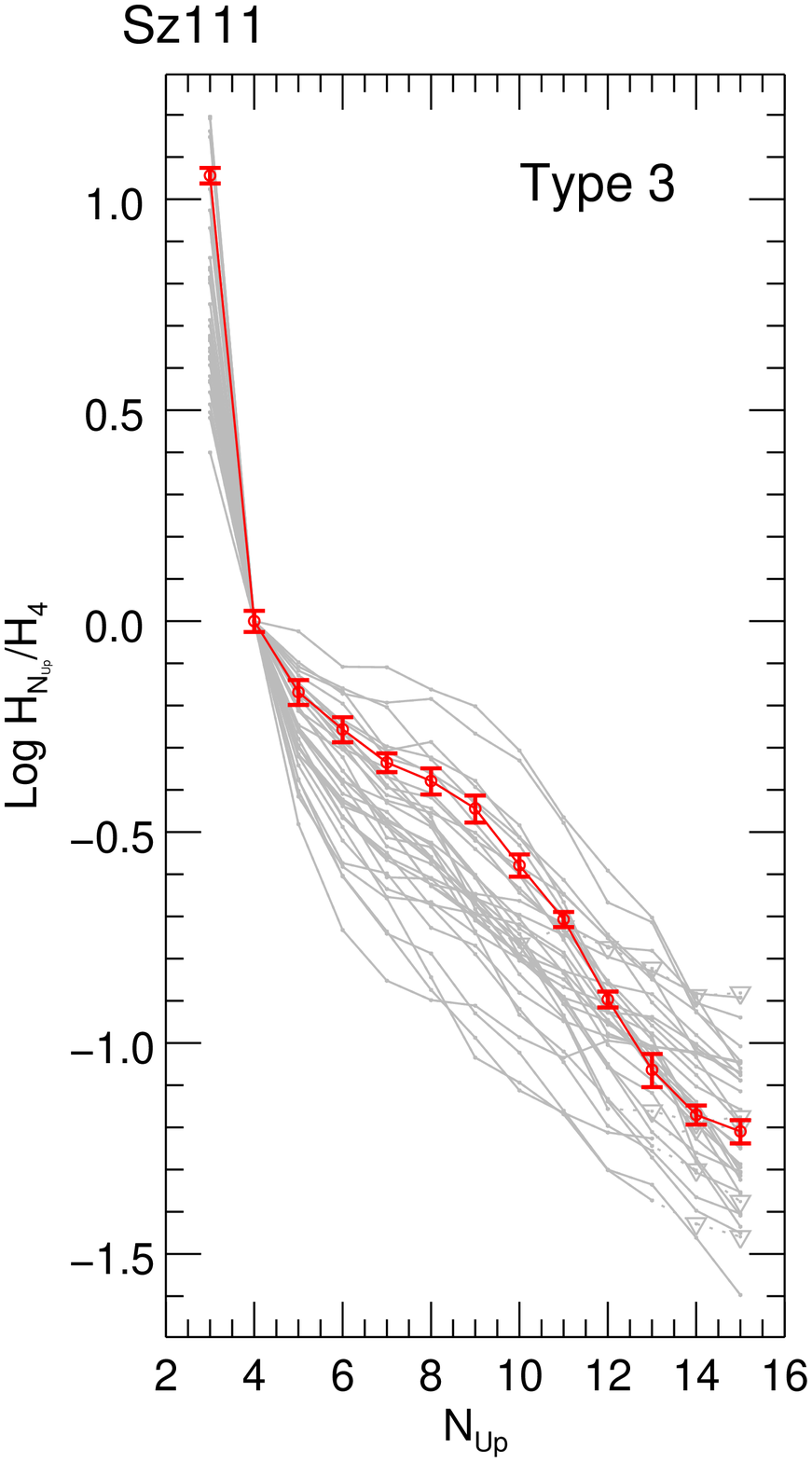}\\
\includegraphics[width=4.4cm]{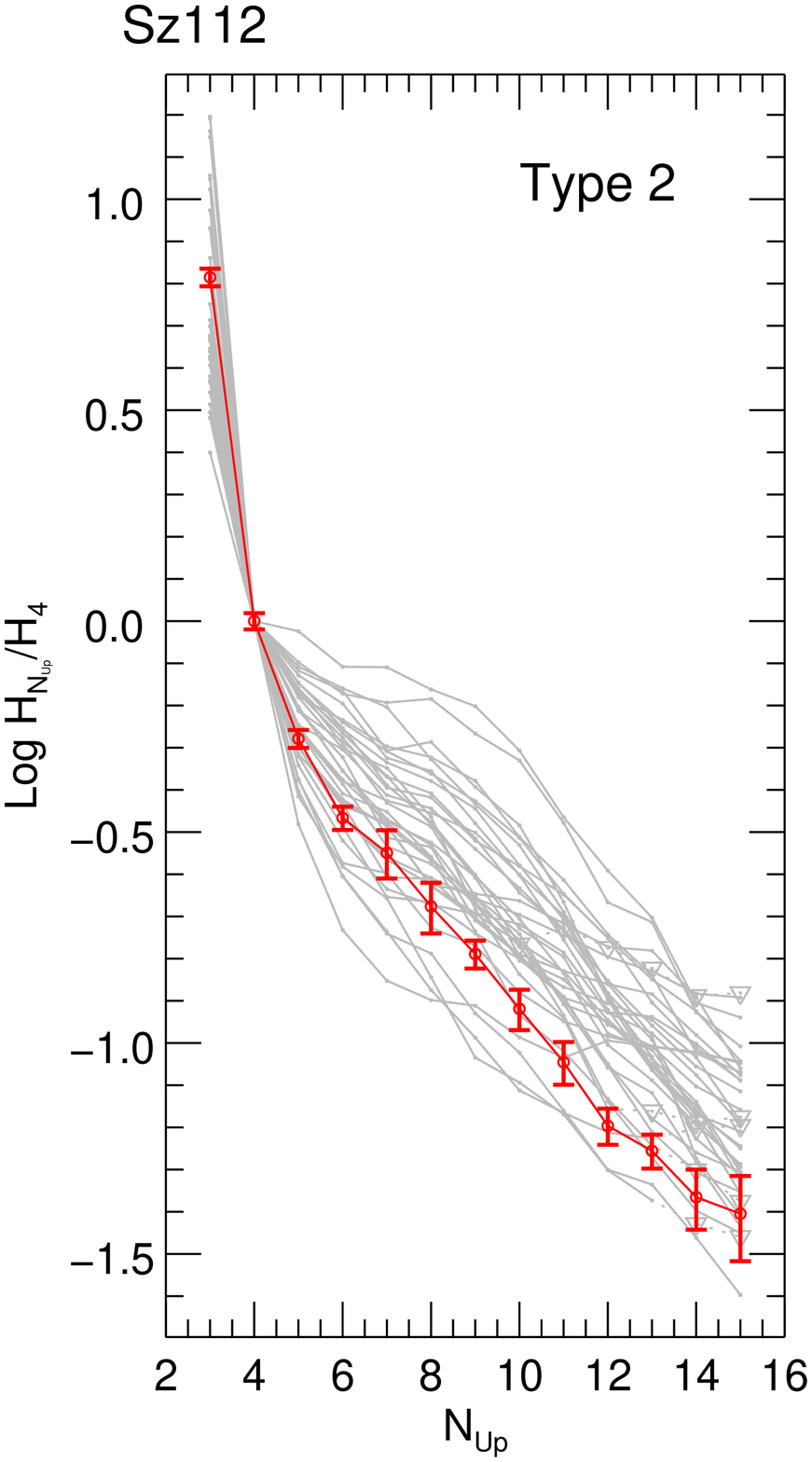}
\includegraphics[width=4.4cm]{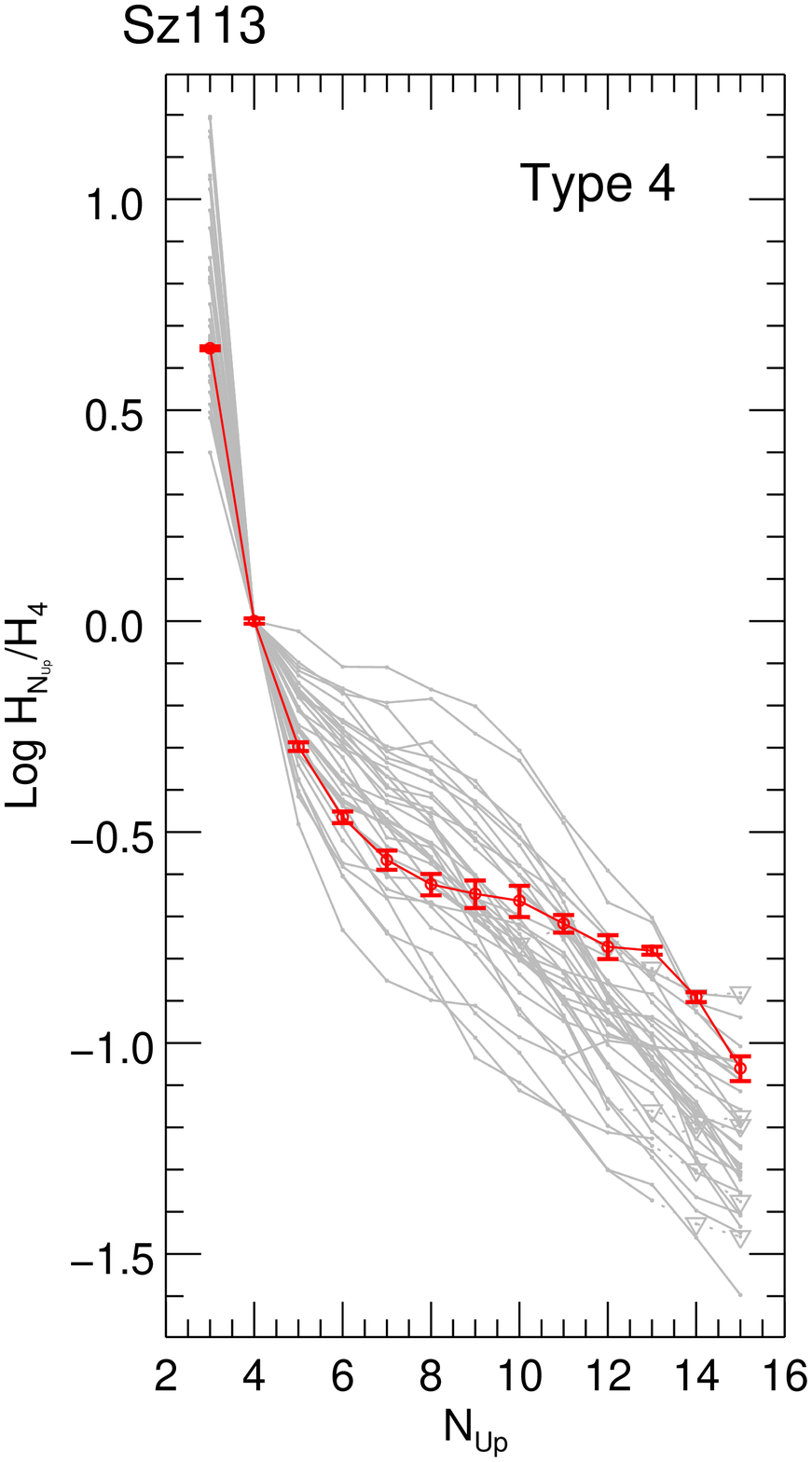}
\includegraphics[width=4.4cm]{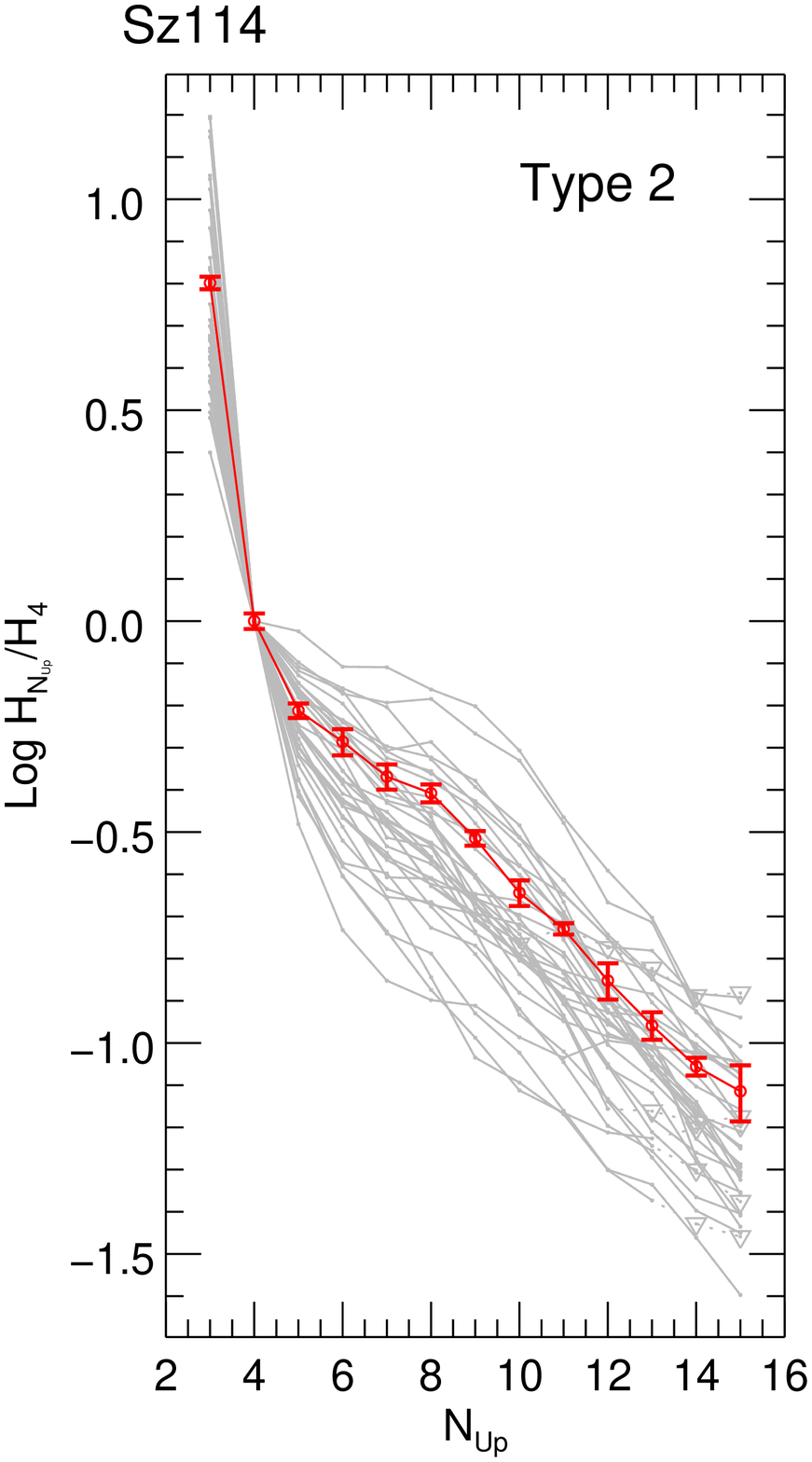}
\includegraphics[width=4.4cm]{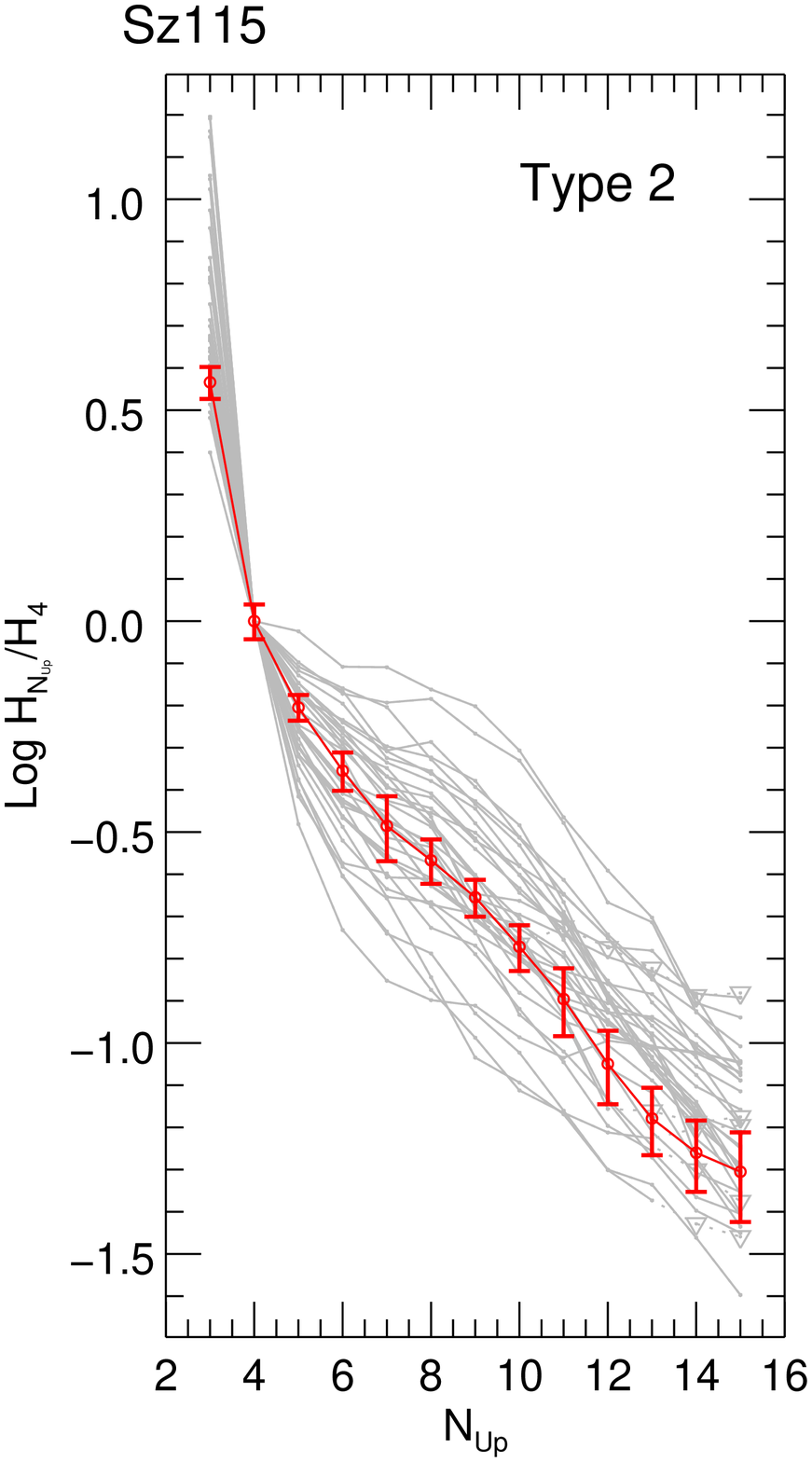}\\
\includegraphics[width=4.4cm]{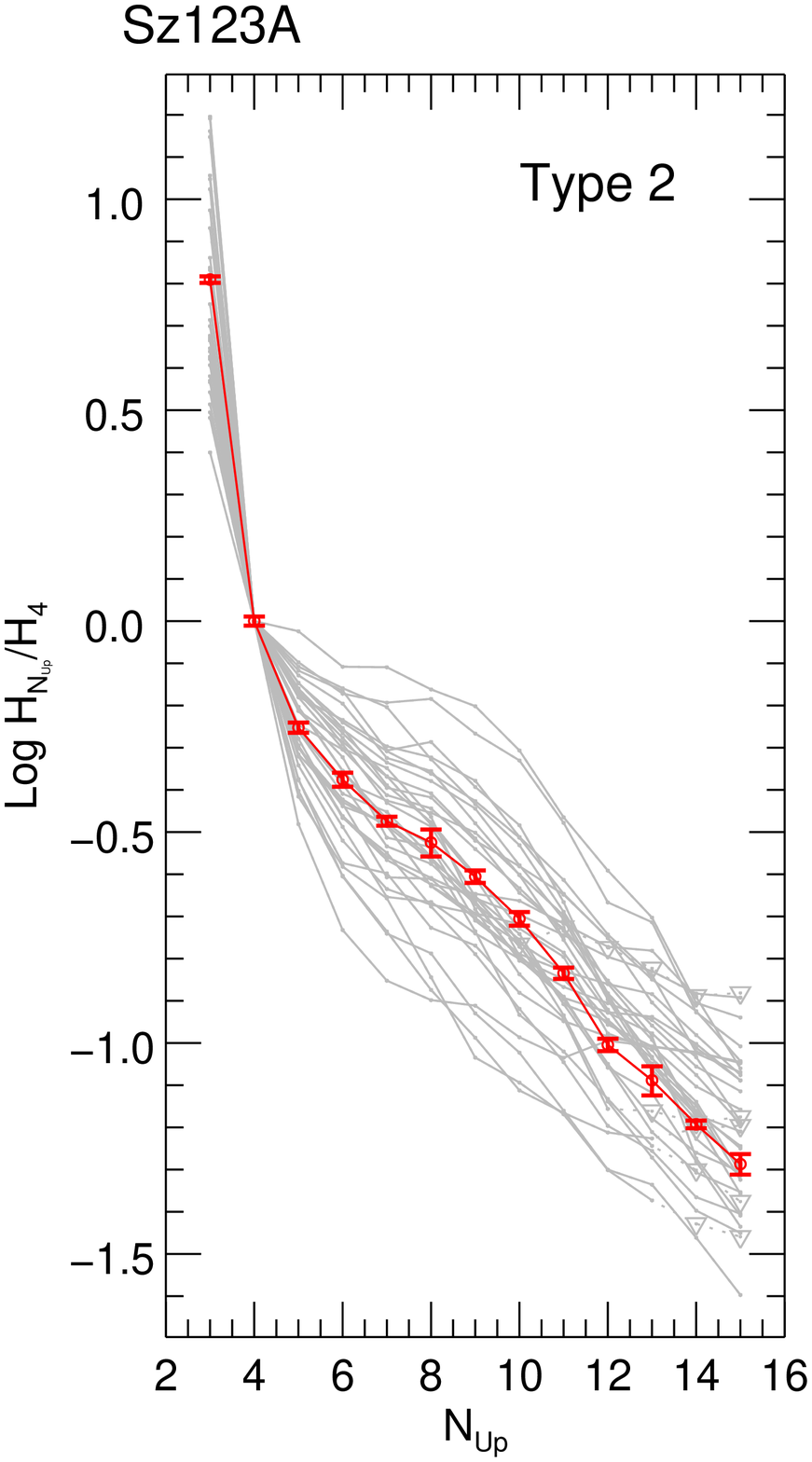}
\includegraphics[width=4.4cm]{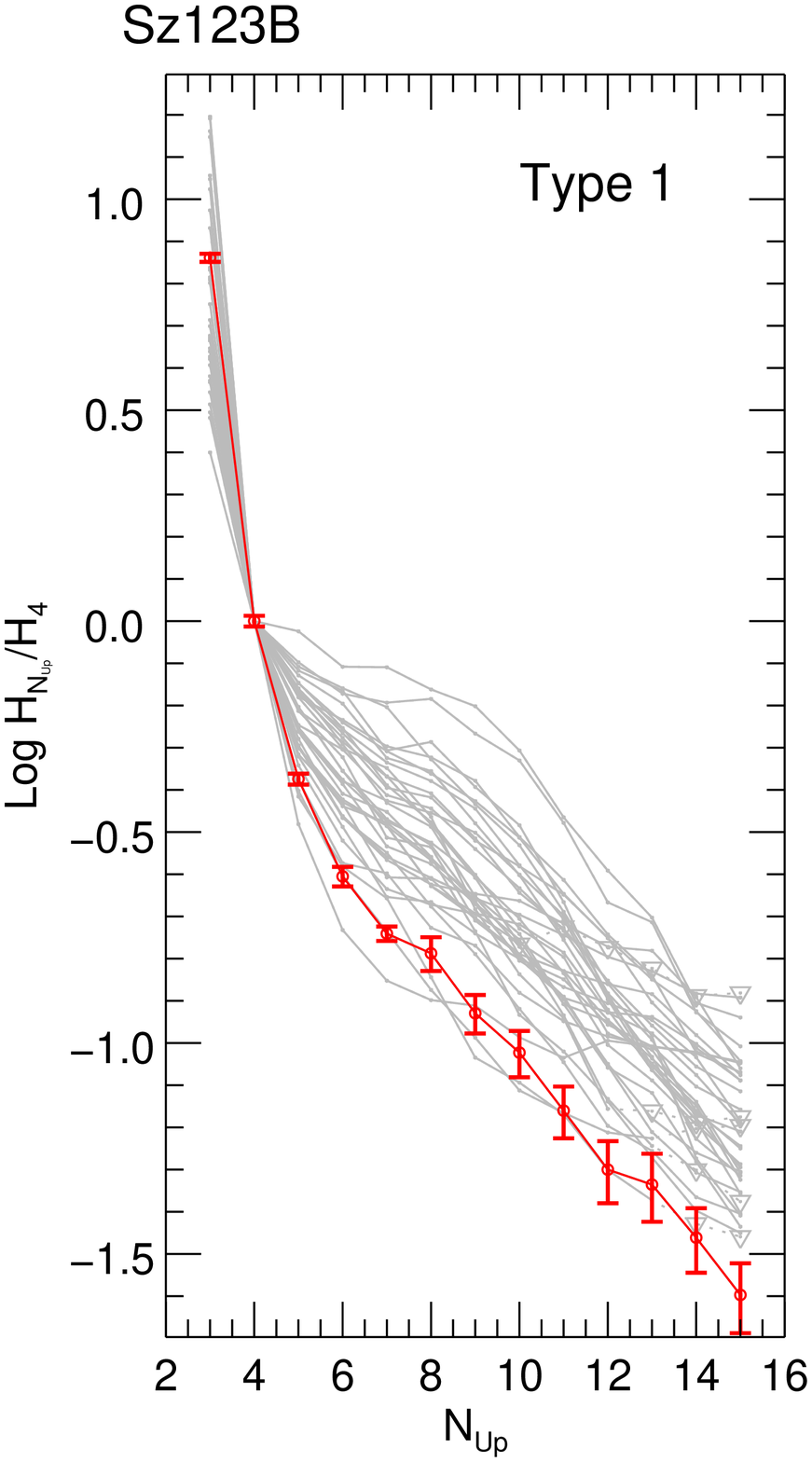}
\includegraphics[width=4.4cm]{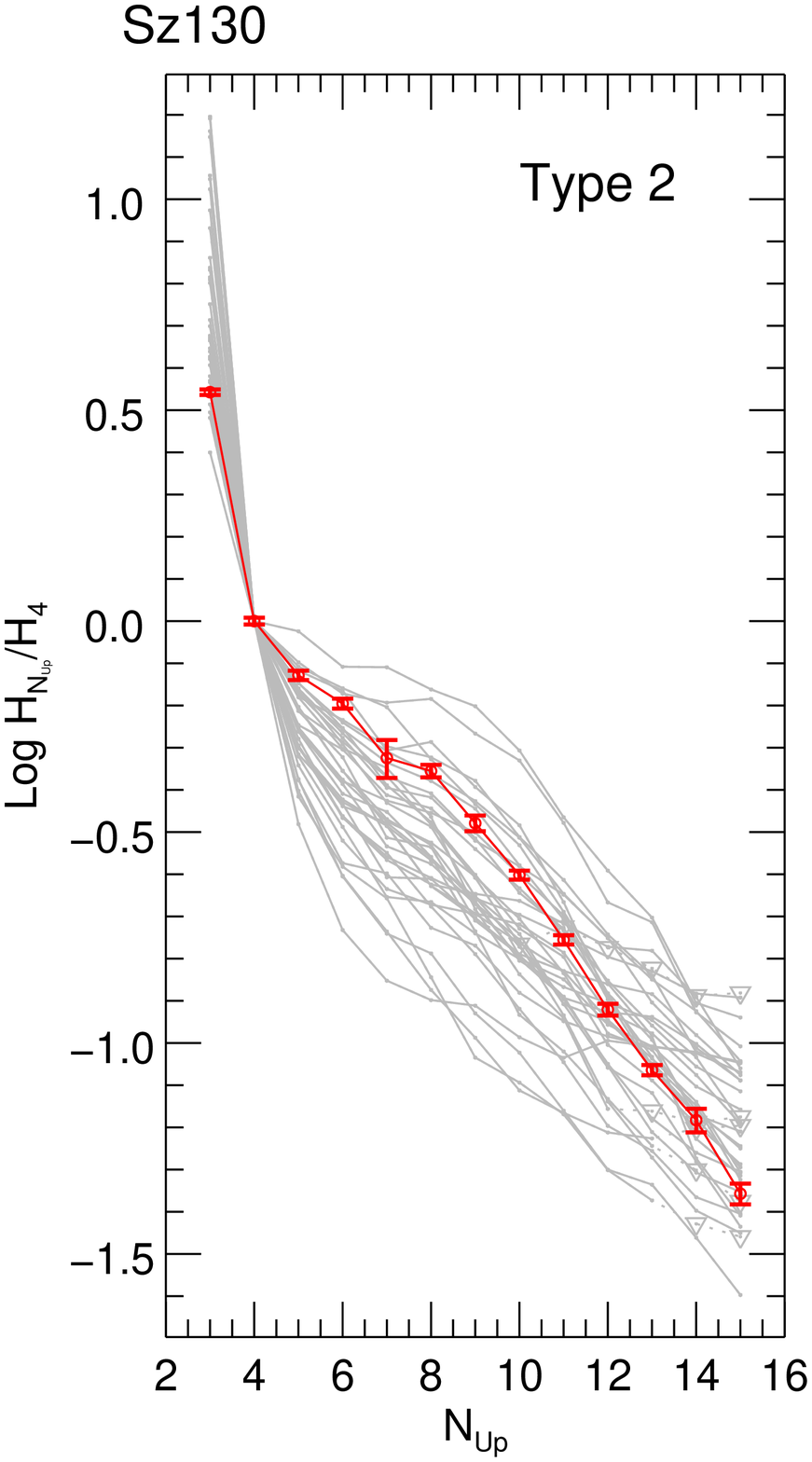}
\includegraphics[width=4.4cm]{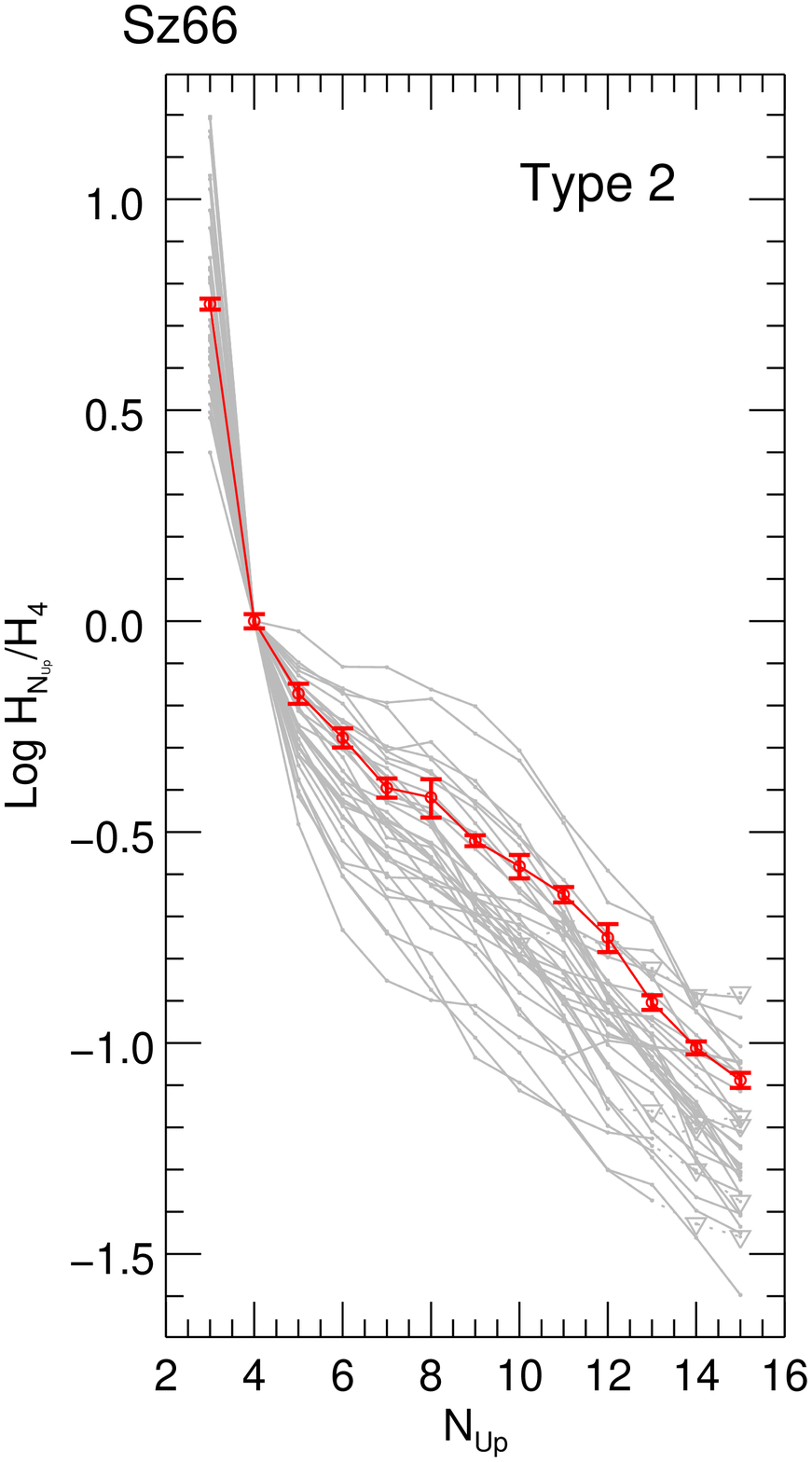}\\
\caption{Continued.} 
\end{figure*}

\setcounter{figure}{0}
\begin{figure*}[t]
\centering
\includegraphics[width=4.4cm]{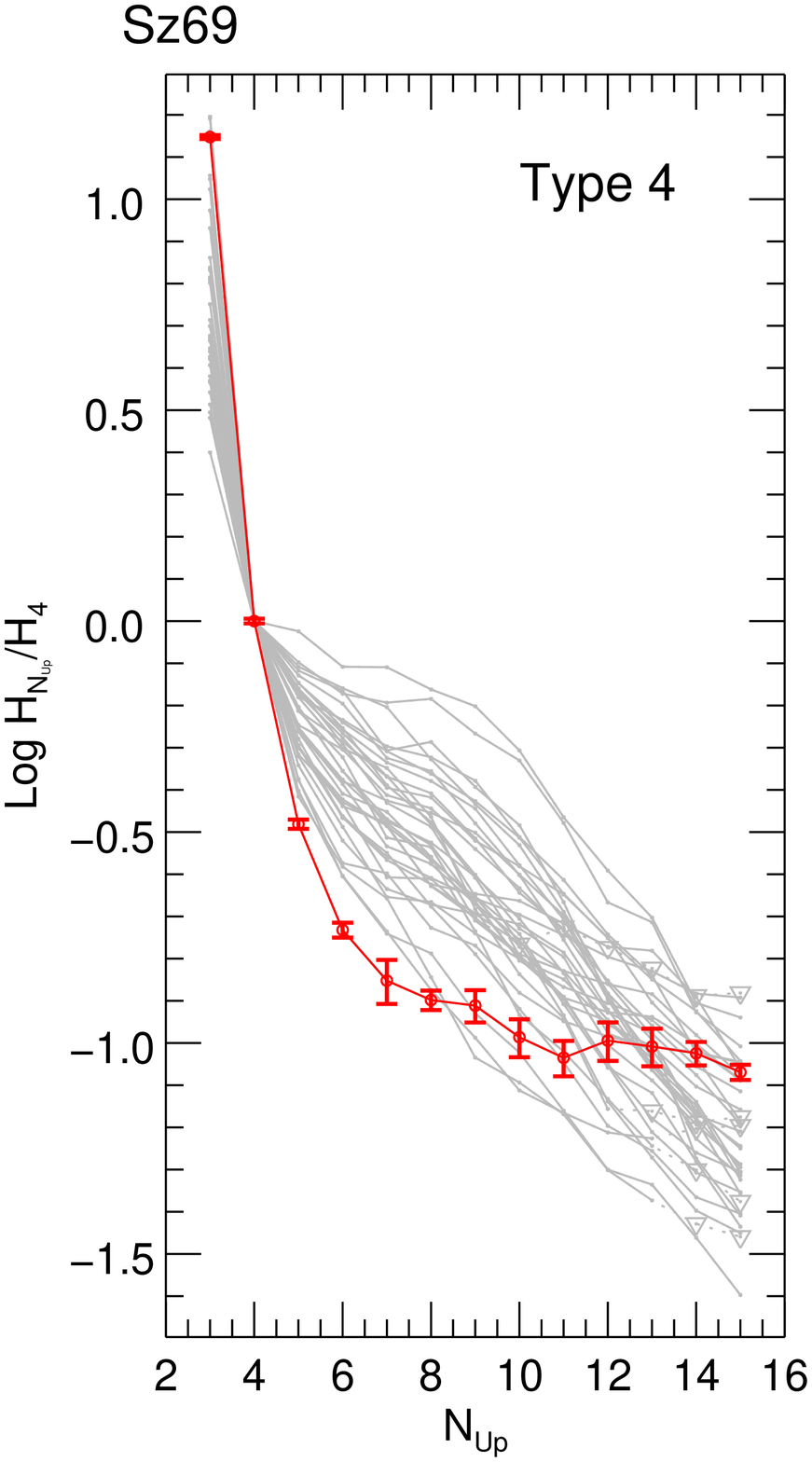}
\includegraphics[width=4.4cm]{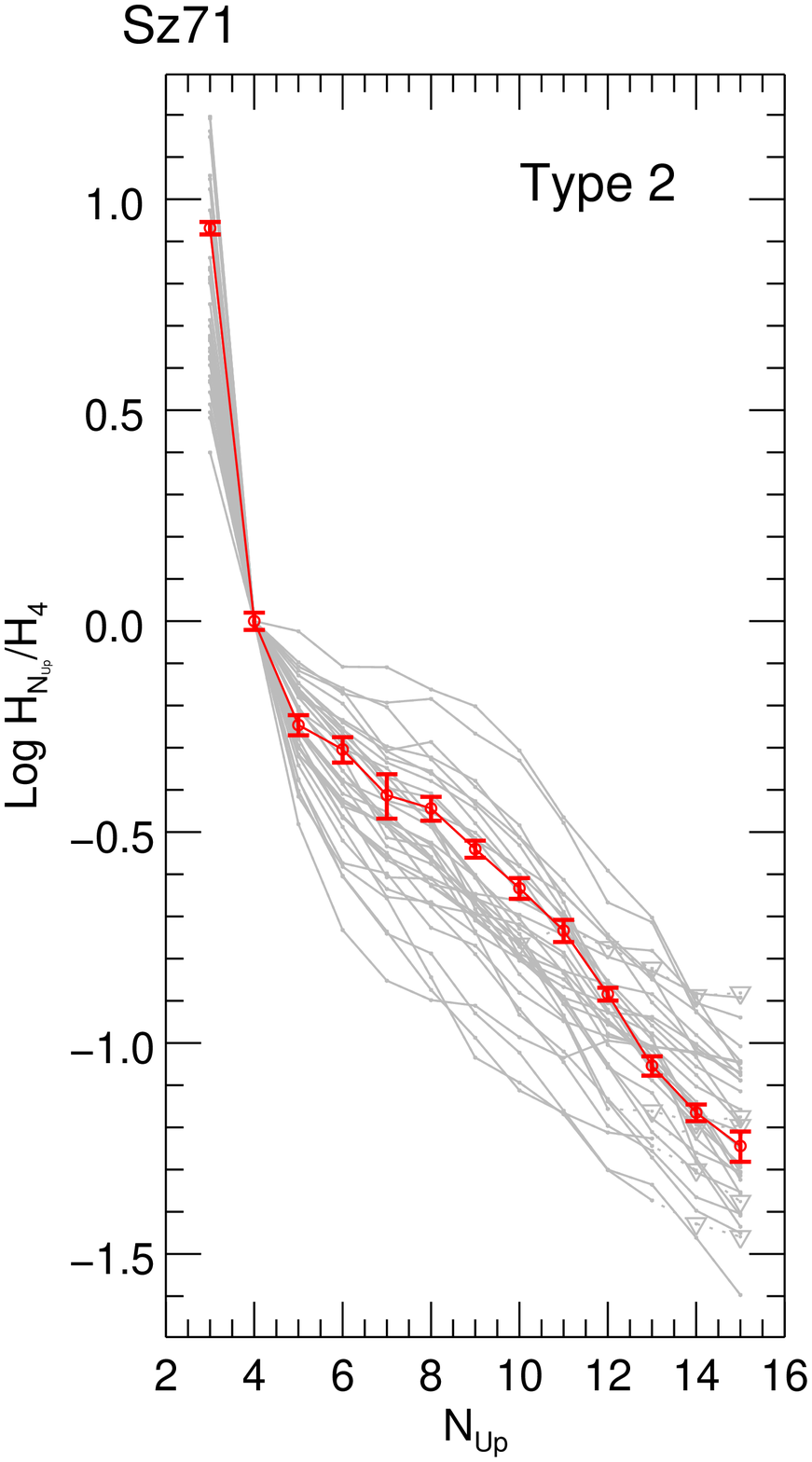}
\includegraphics[width=4.4cm]{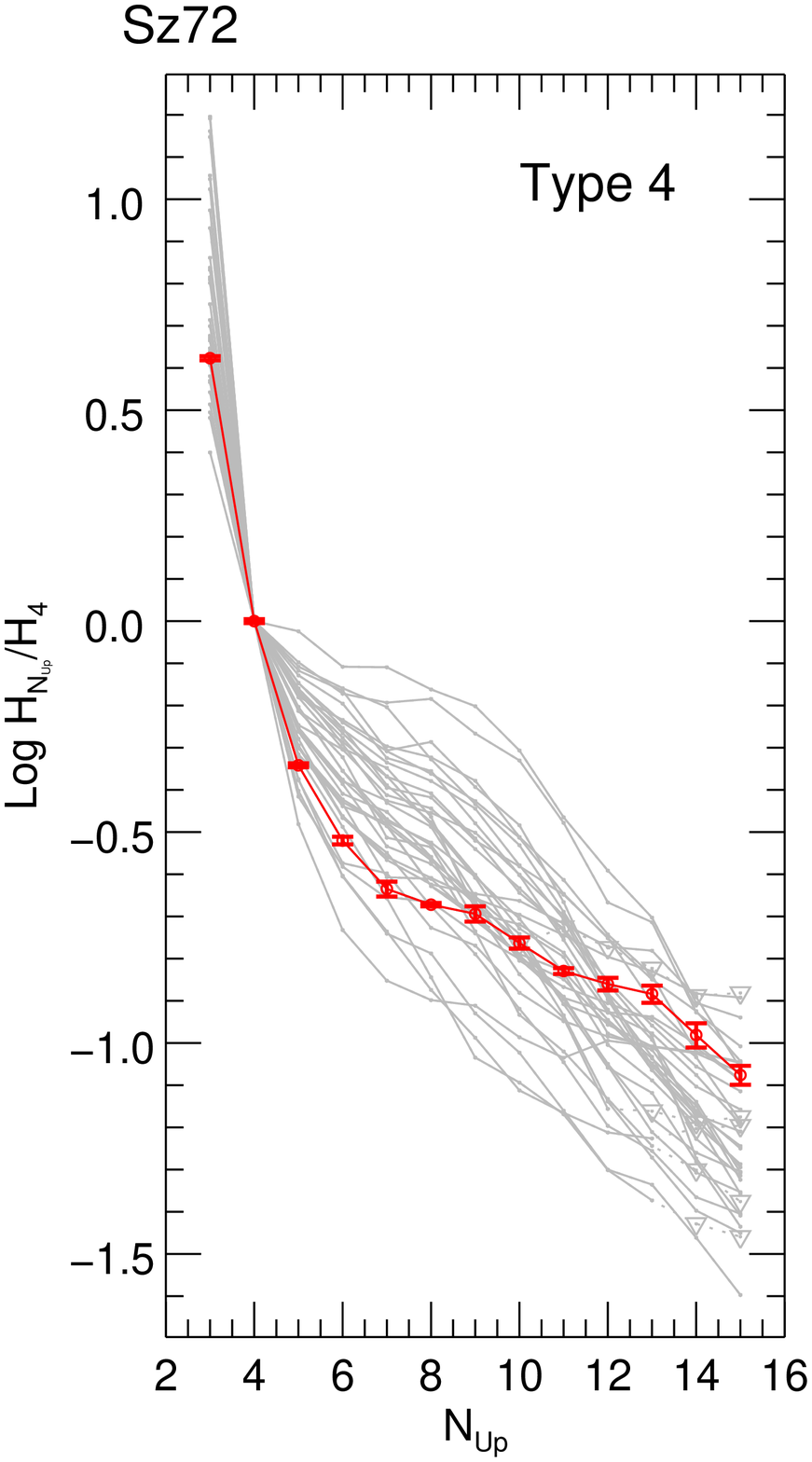}
\includegraphics[width=4.4cm]{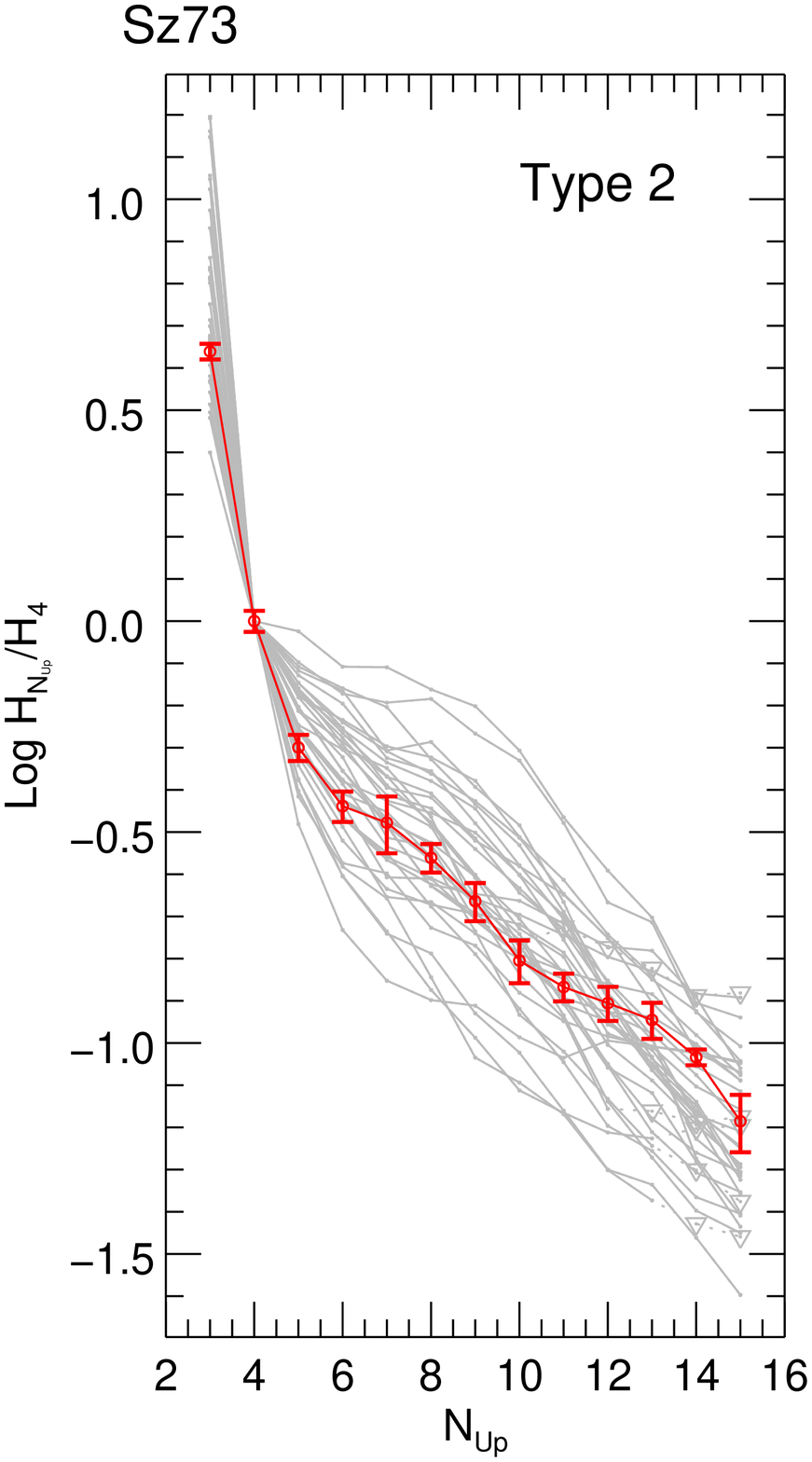}\\
\includegraphics[width=4.4cm]{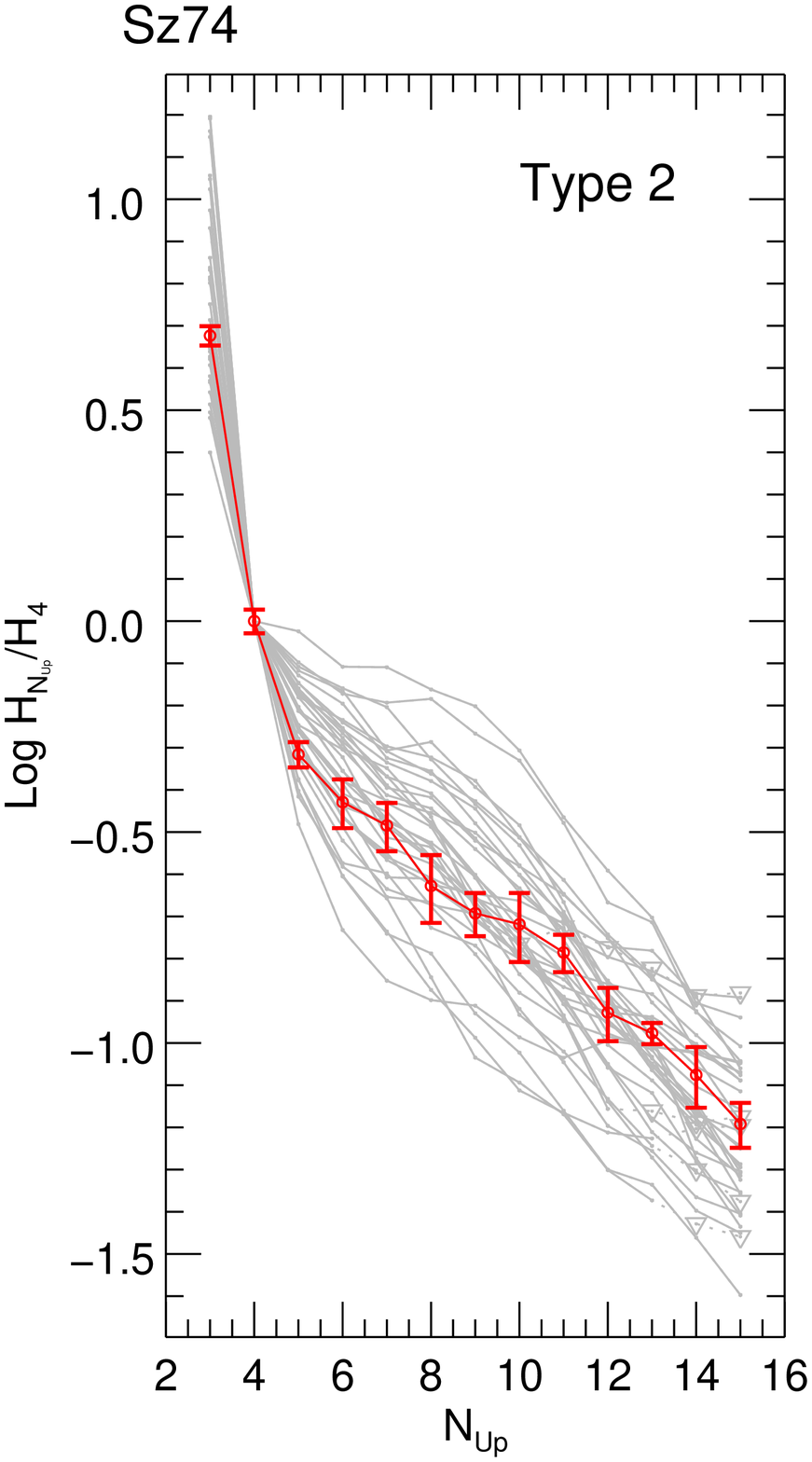}
\includegraphics[width=4.4cm]{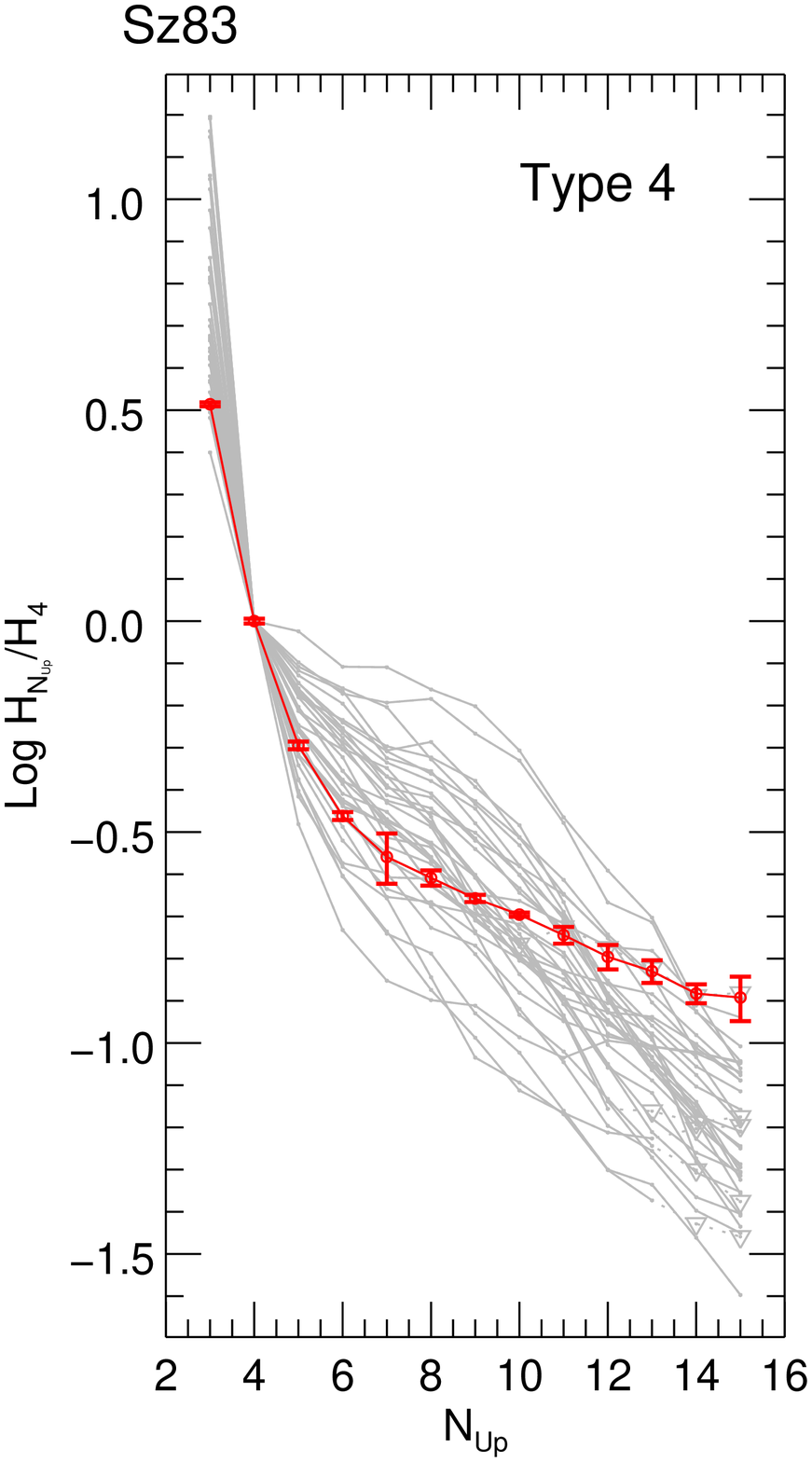}
\includegraphics[width=4.4cm]{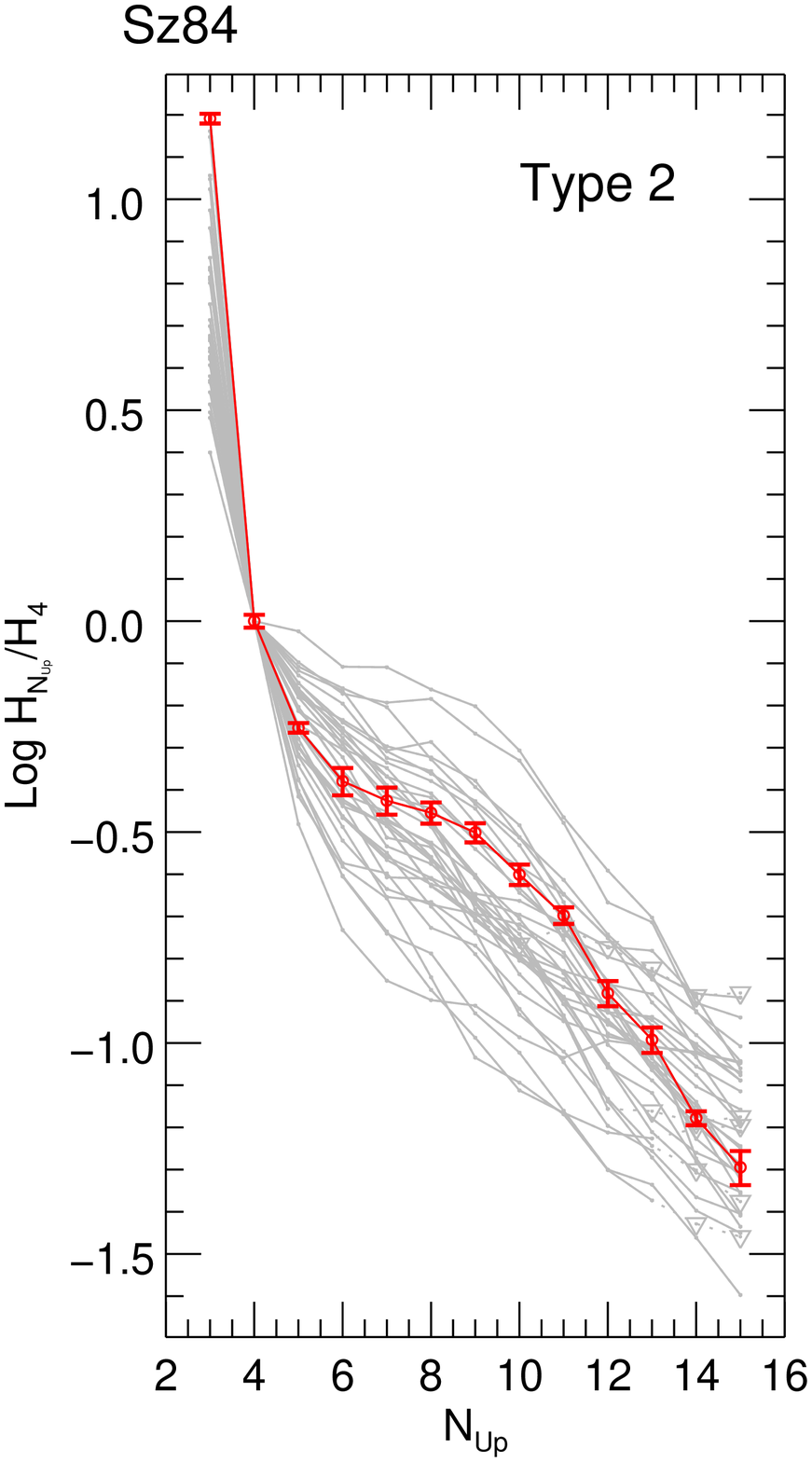}
\includegraphics[width=4.4cm]{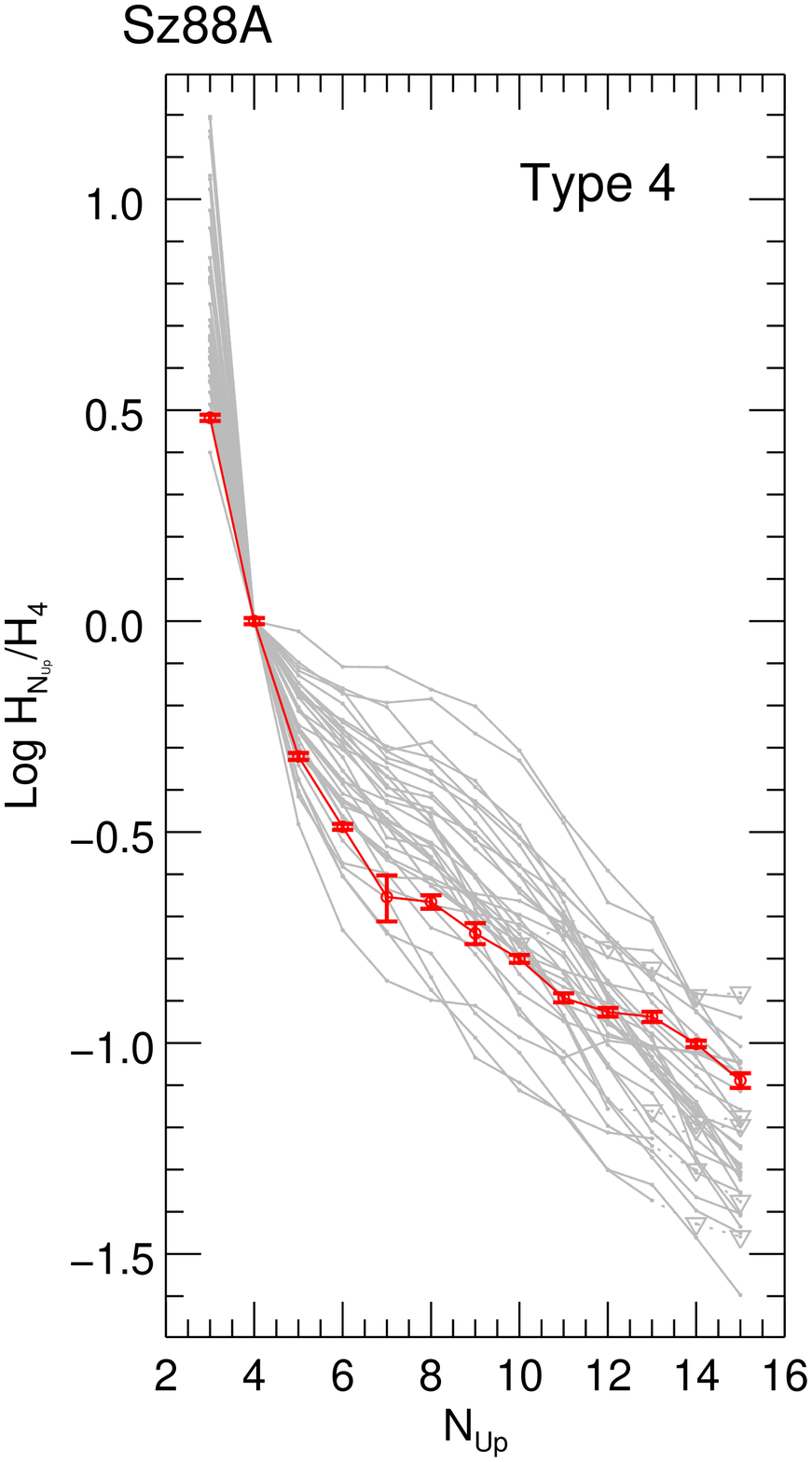}\\
\includegraphics[width=4.4cm]{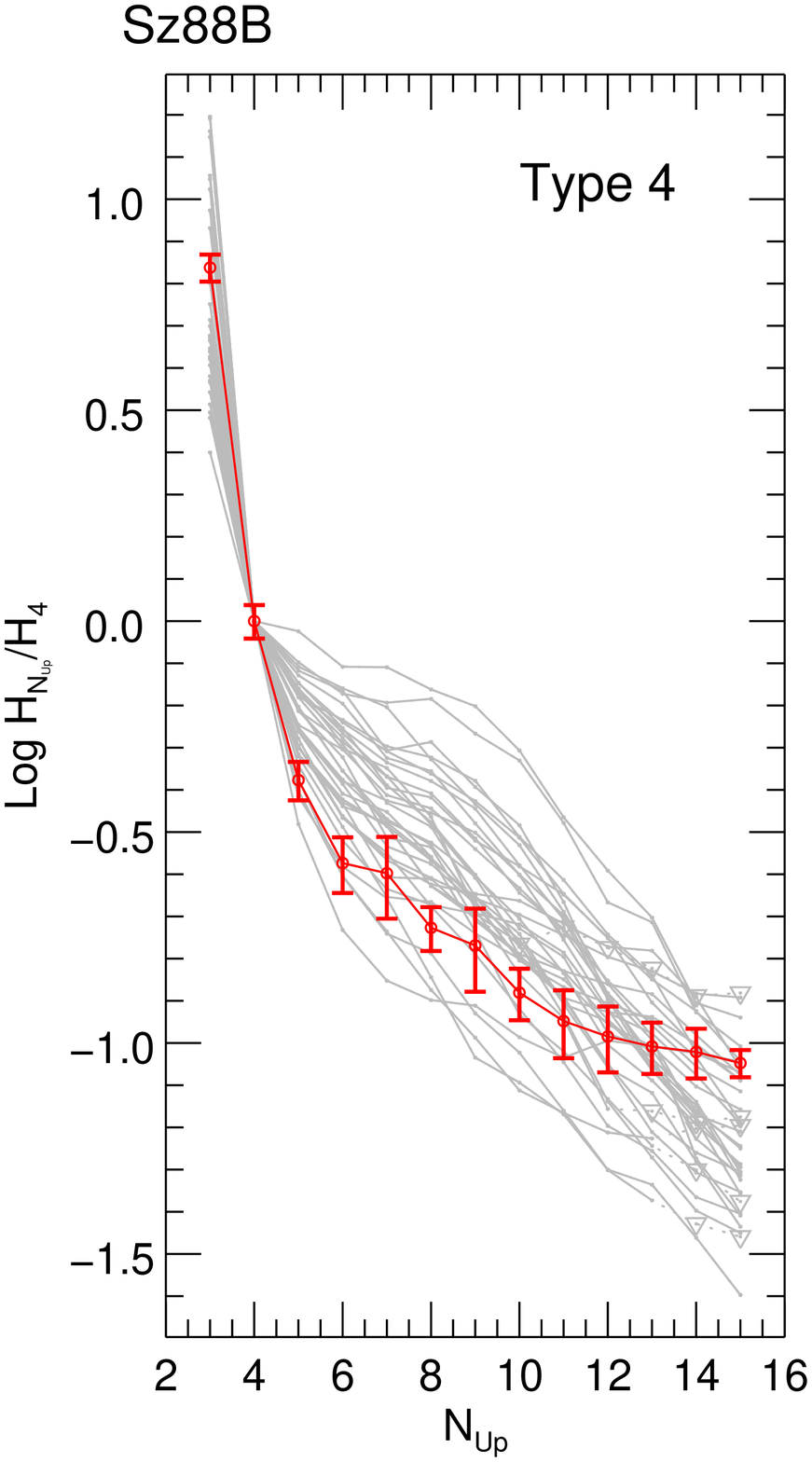}
\includegraphics[width=4.4cm]{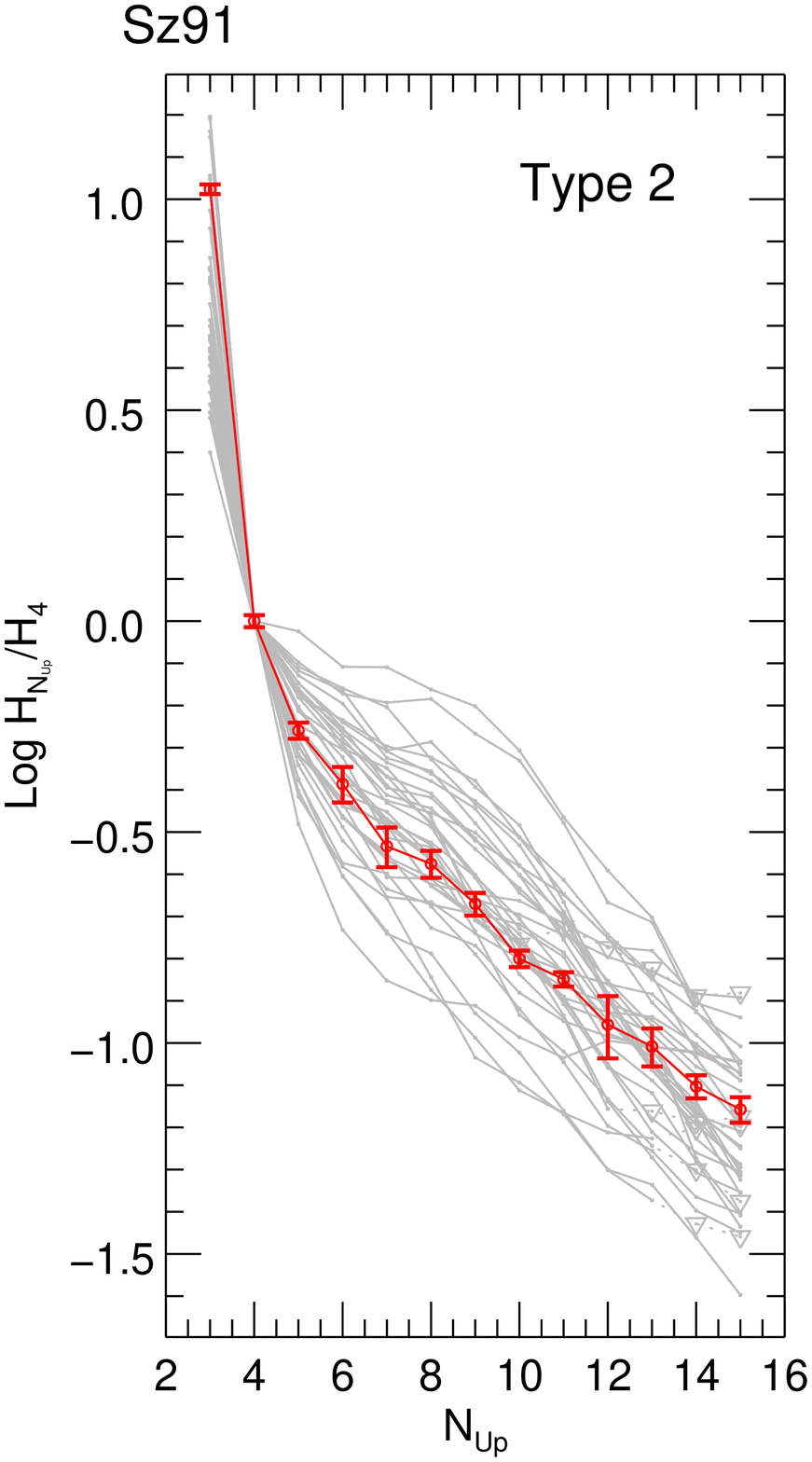}
\includegraphics[width=4.4cm]{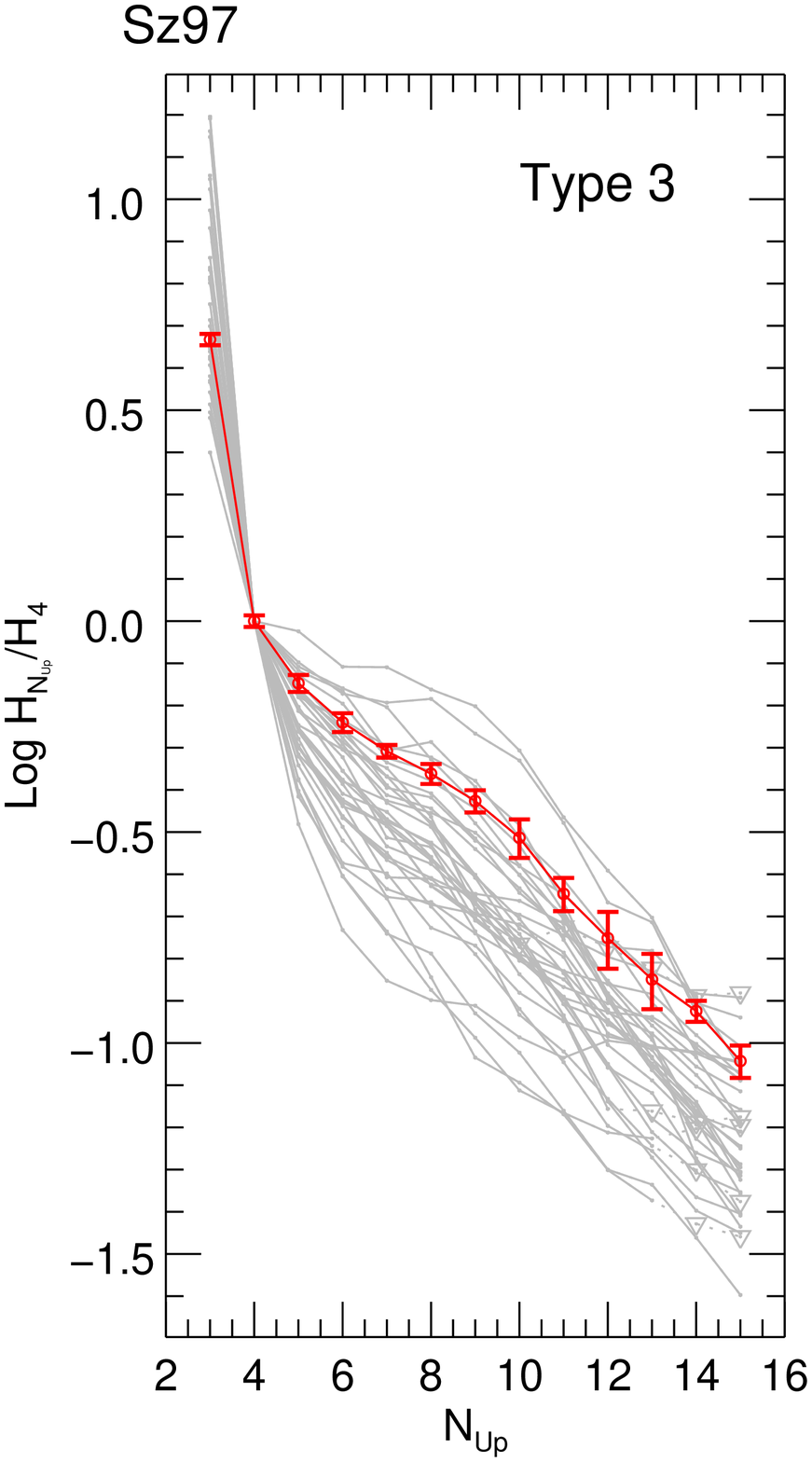}
\includegraphics[width=4.4cm]{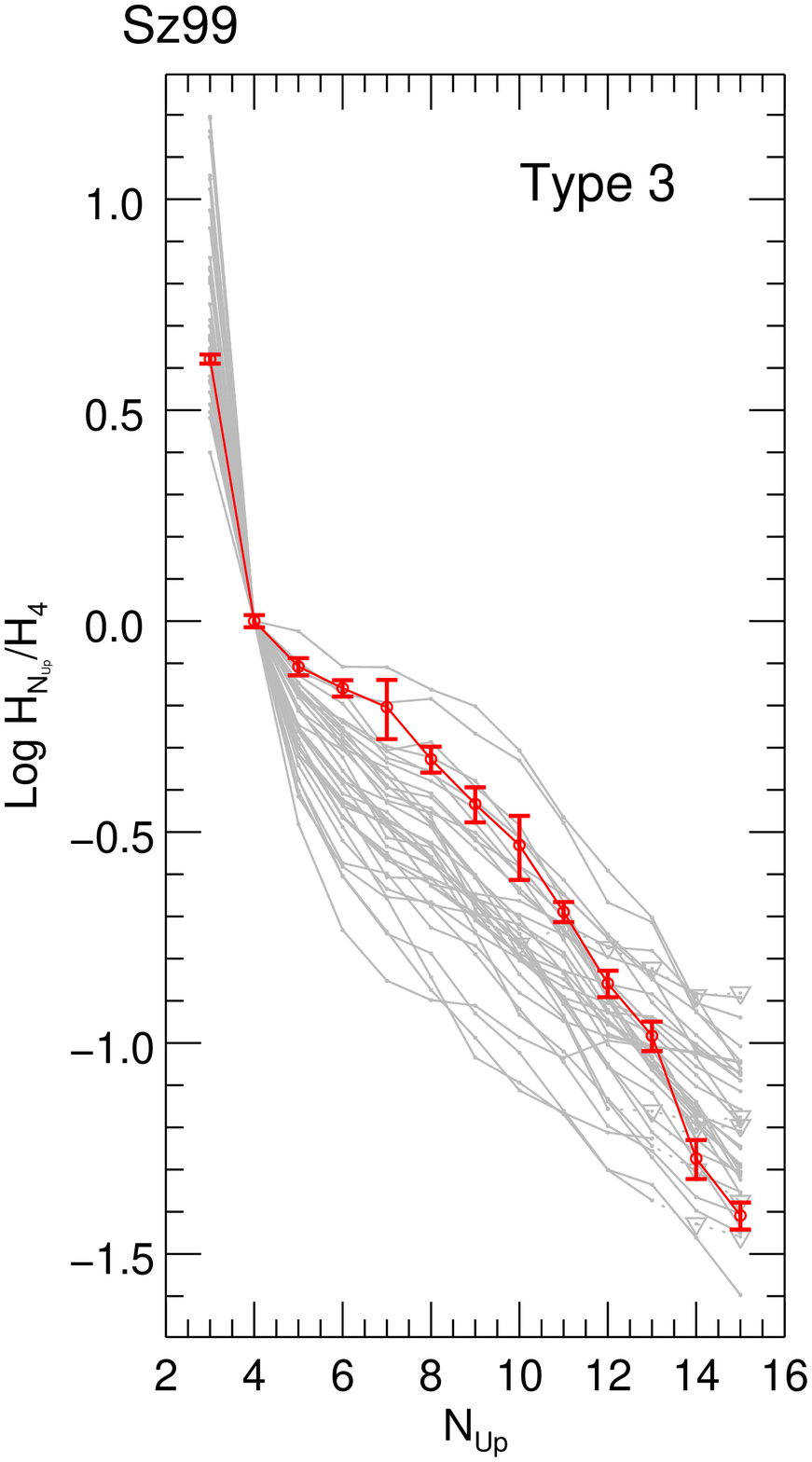}
\caption{Continued.} 
\end{figure*}

\section*{Appendix C: Atlas of Paschen decrements}
\setcounter{figure}{0} \renewcommand{\thefigure}{C.\arabic{figure}} 
\setcounter{table}{0} \renewcommand{\thetable}{C.\arabic{table}}

\begin{figure*}[t]
\centering
\includegraphics[width=4.4cm]{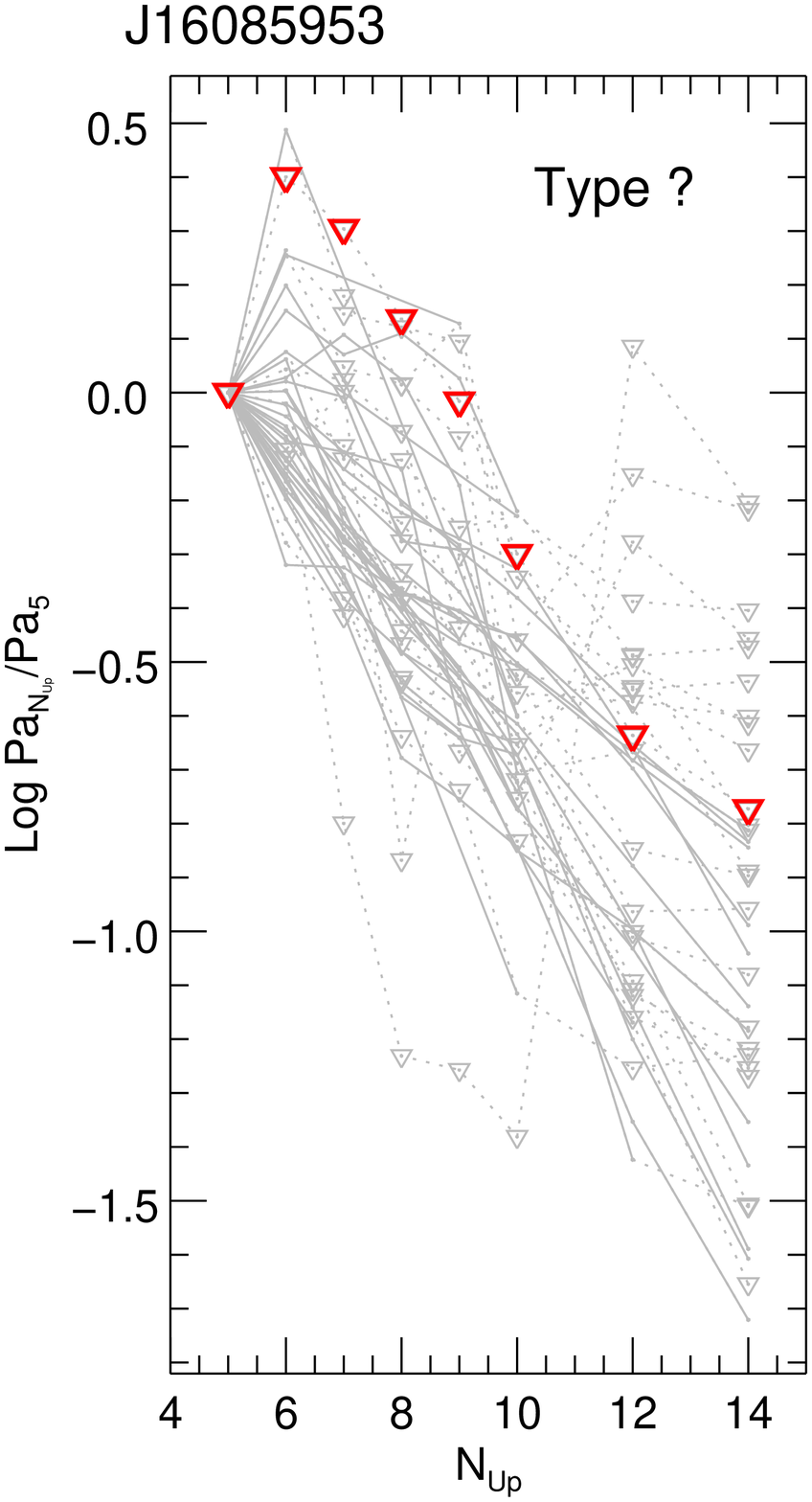}
\includegraphics[width=4.4cm]{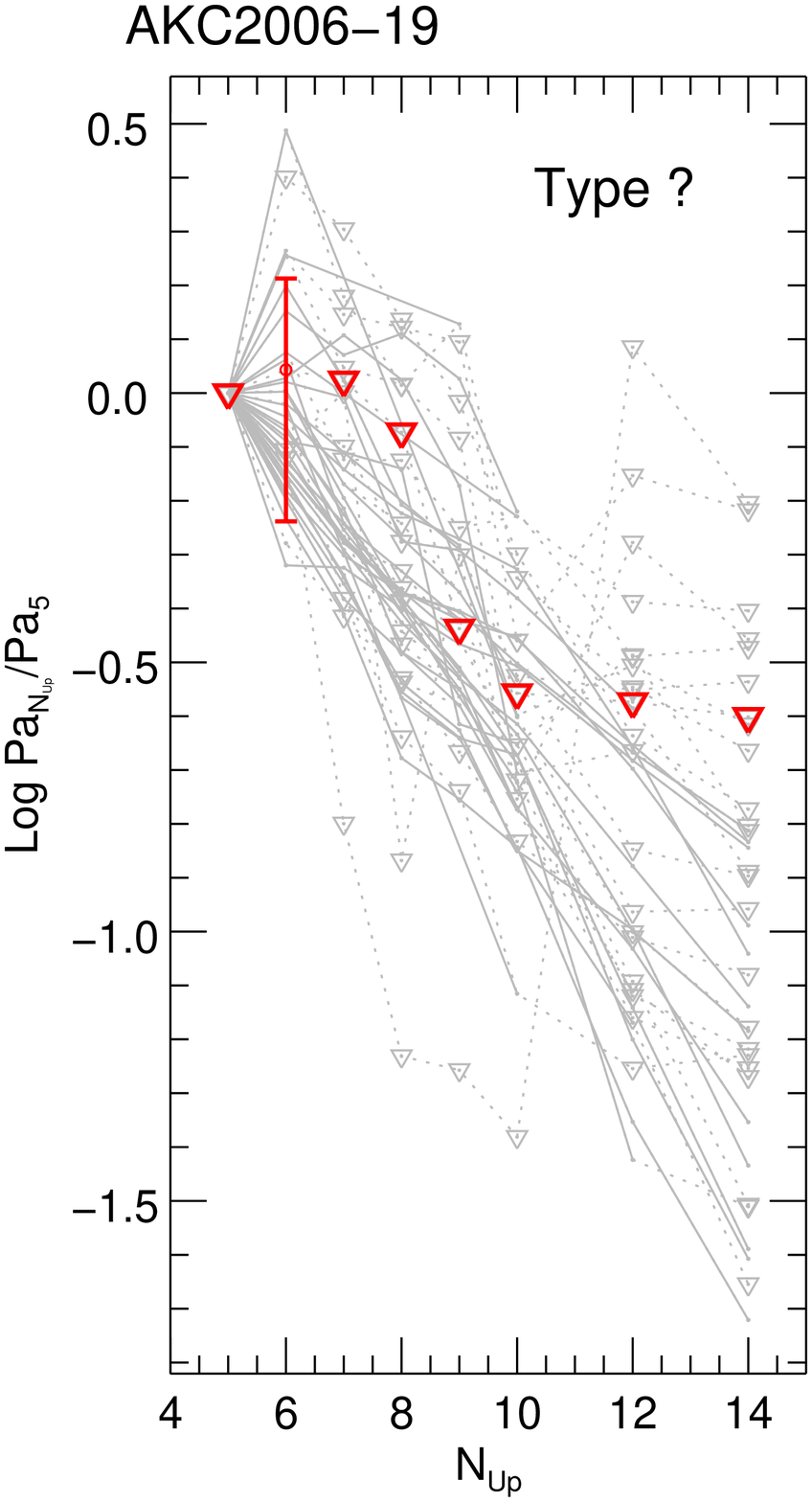}
\includegraphics[width=4.4cm]{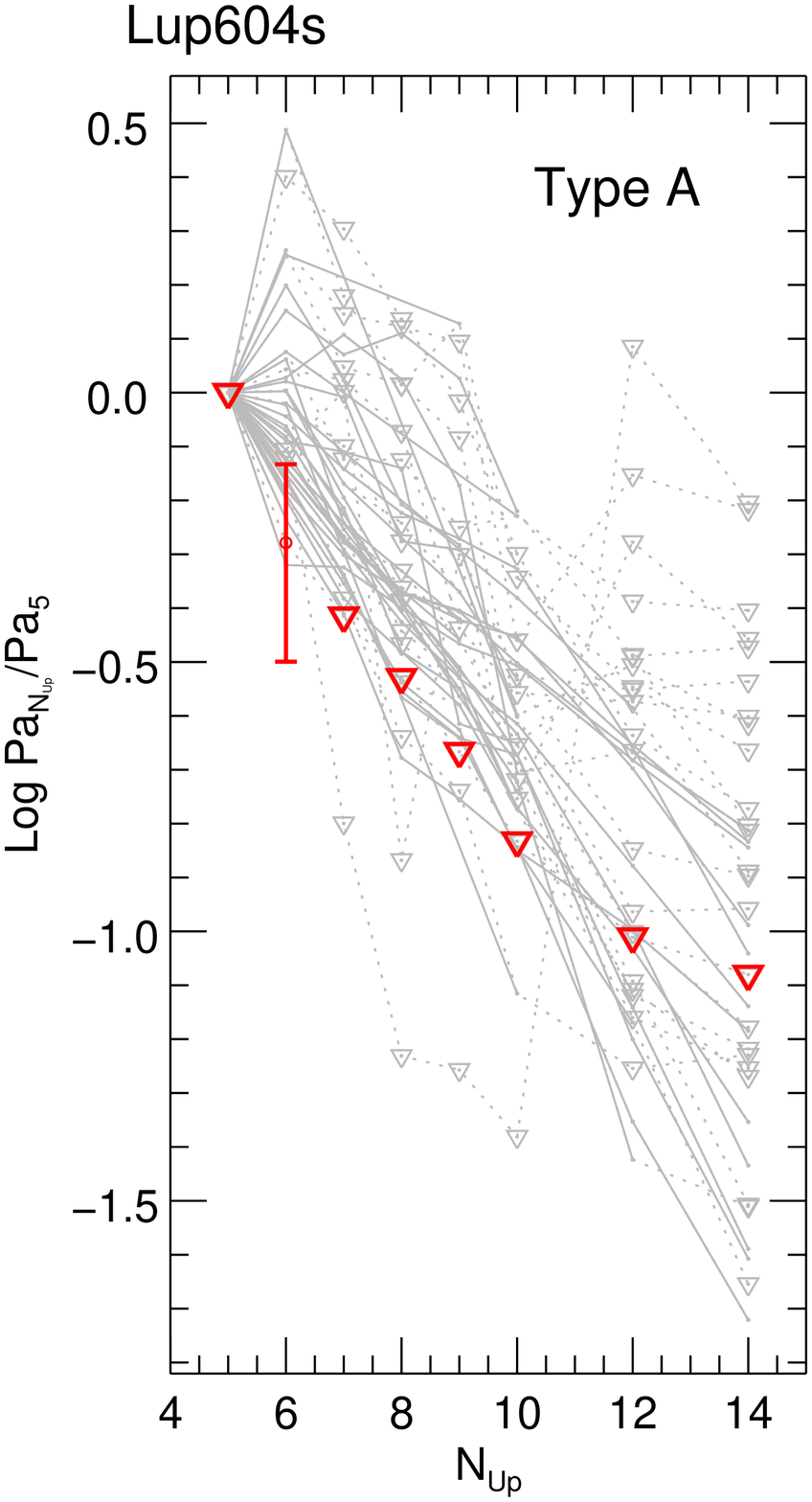}
\includegraphics[width=4.4cm]{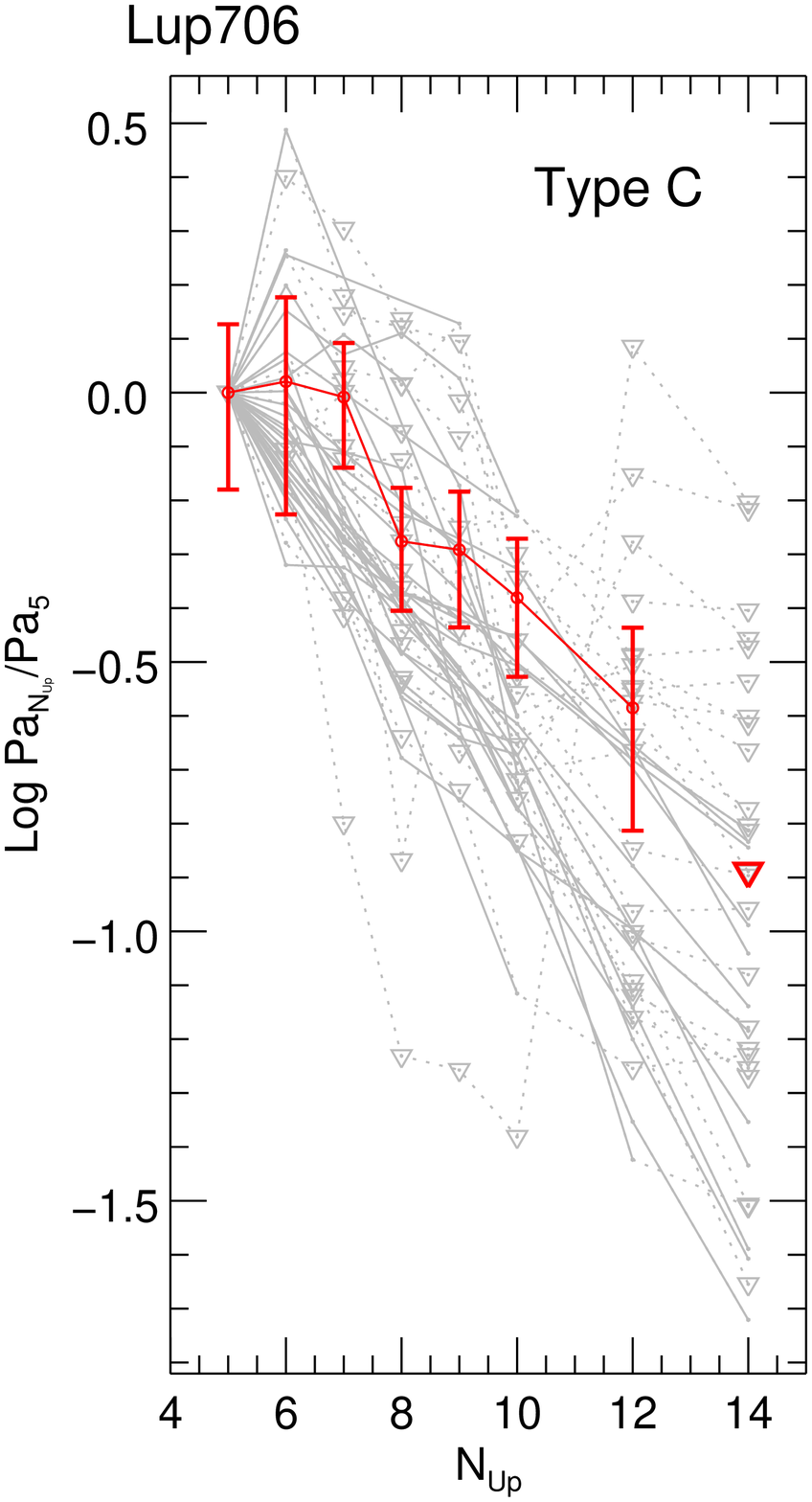}\\
\includegraphics[width=4.4cm]{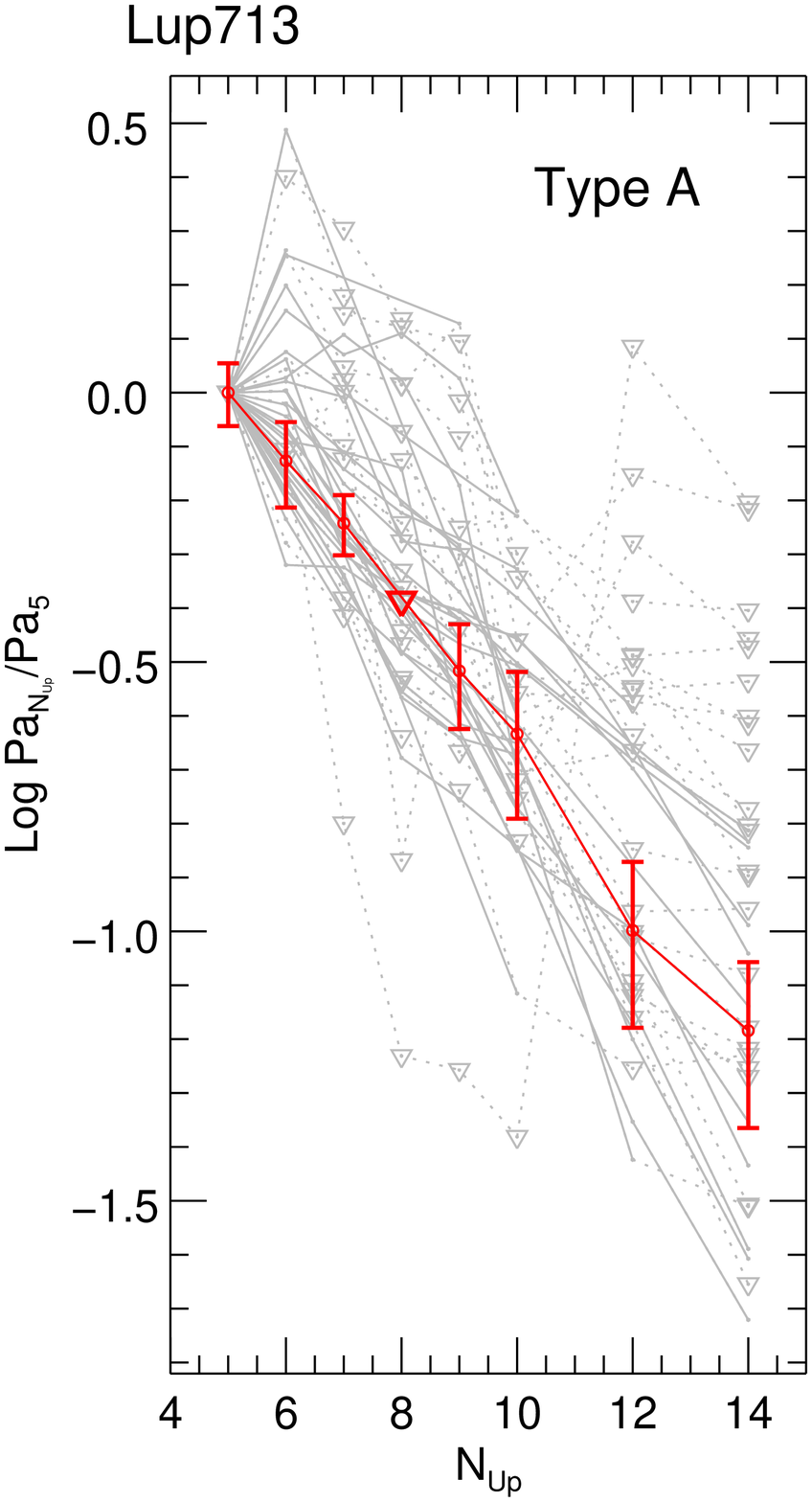}
\includegraphics[width=4.4cm]{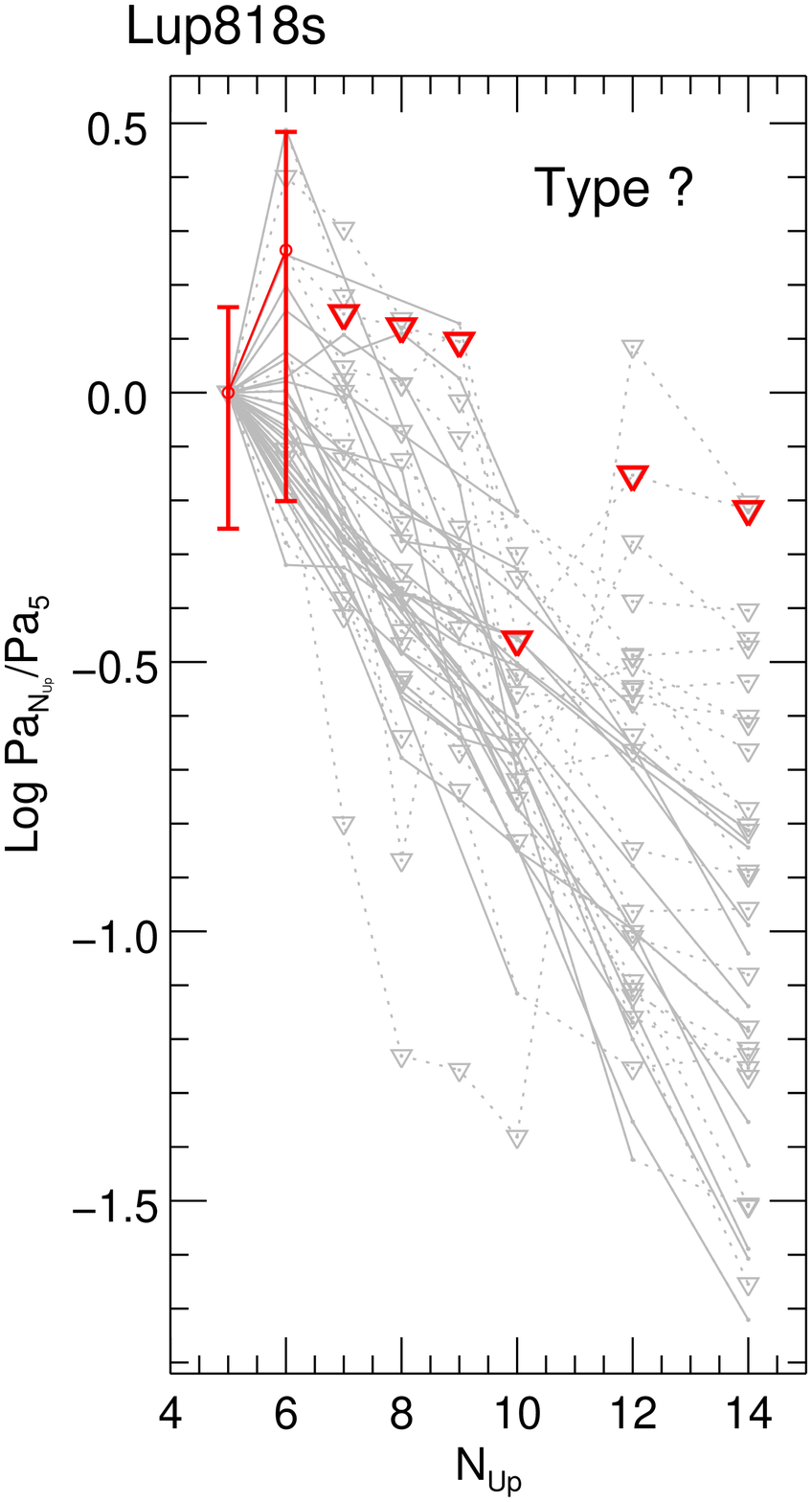}
\includegraphics[width=4.4cm]{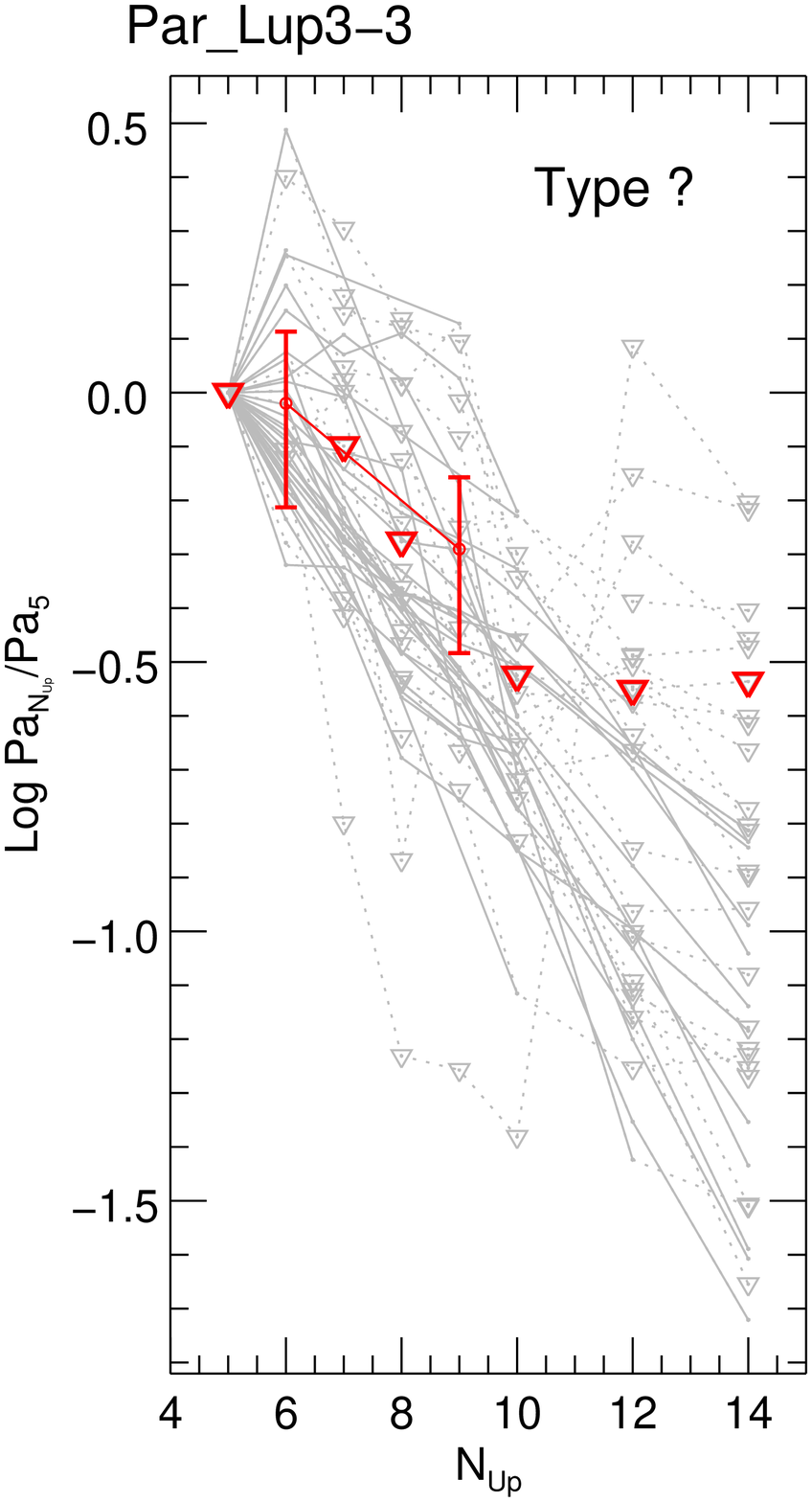}
\includegraphics[width=4.4cm]{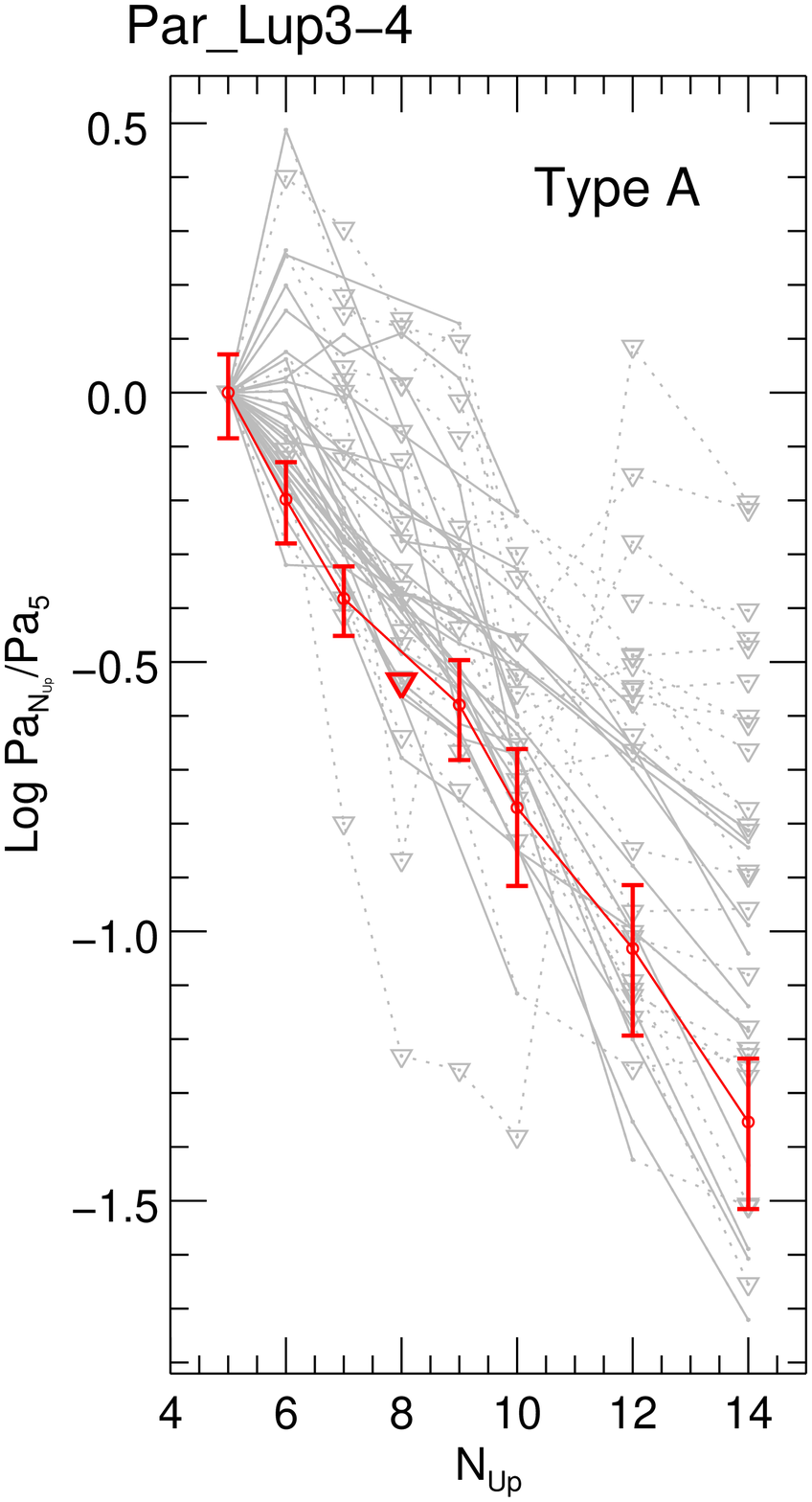}\\
\includegraphics[width=4.4cm]{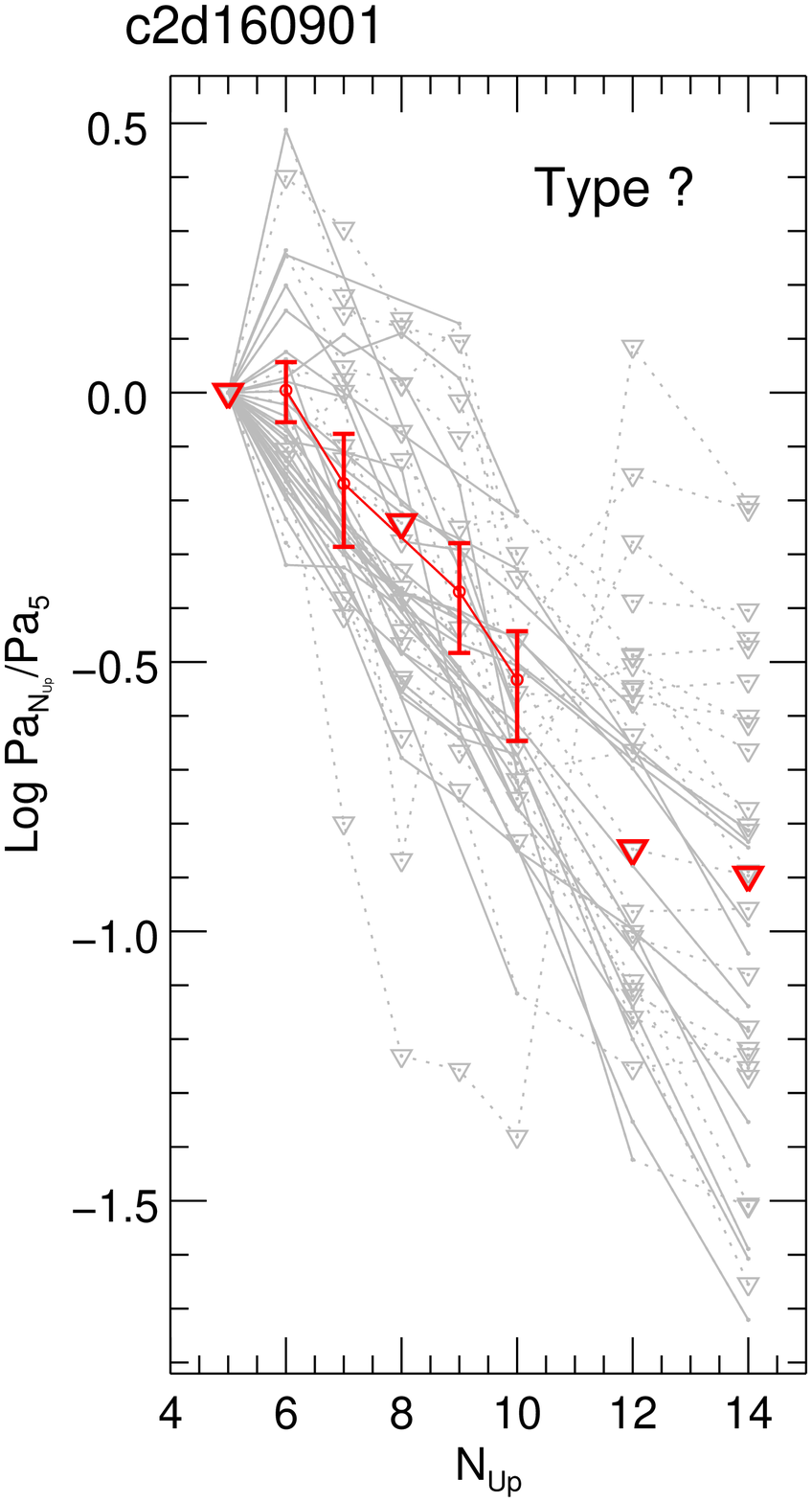}
\includegraphics[width=4.4cm]{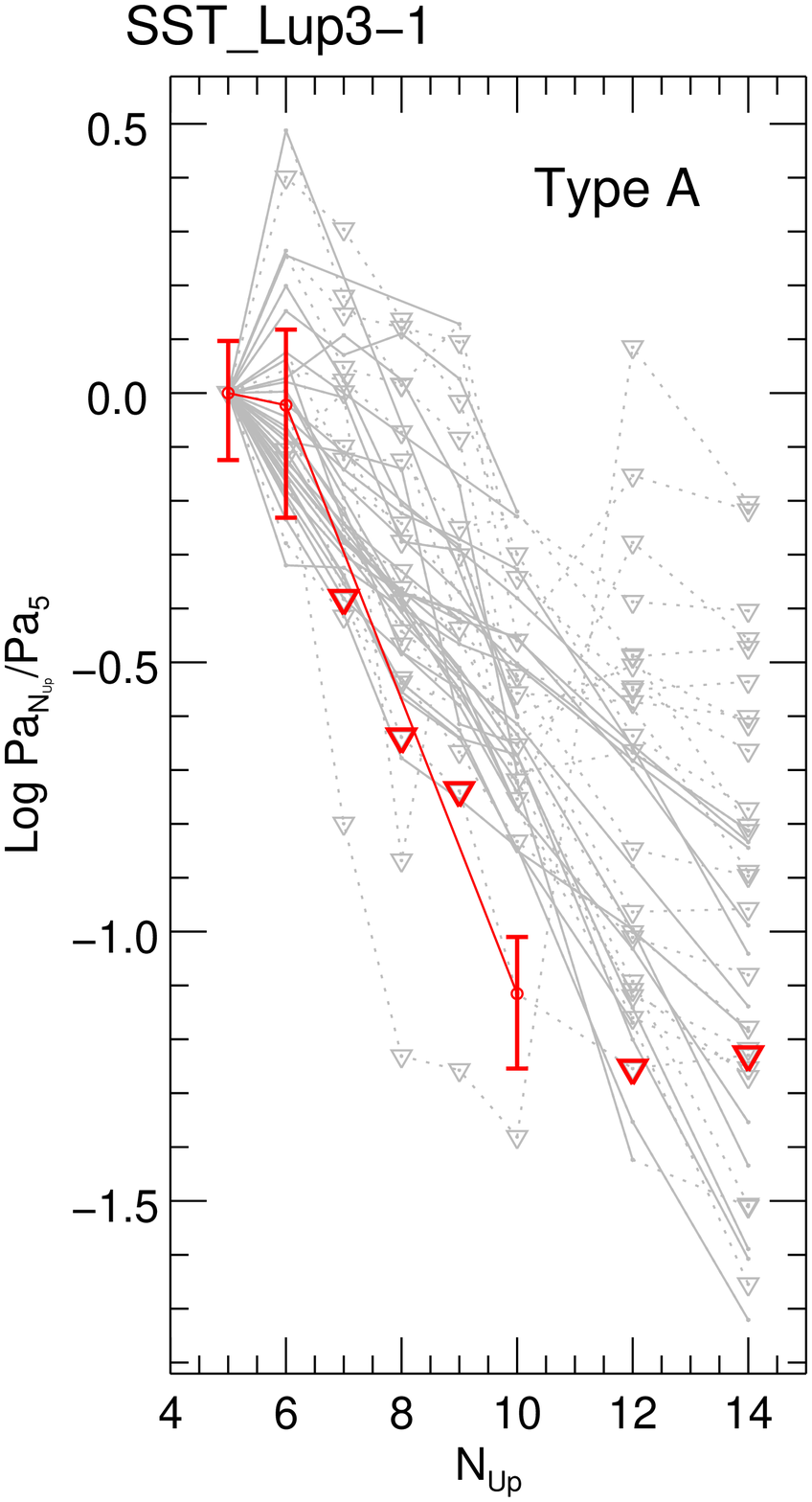}
\includegraphics[width=4.4cm]{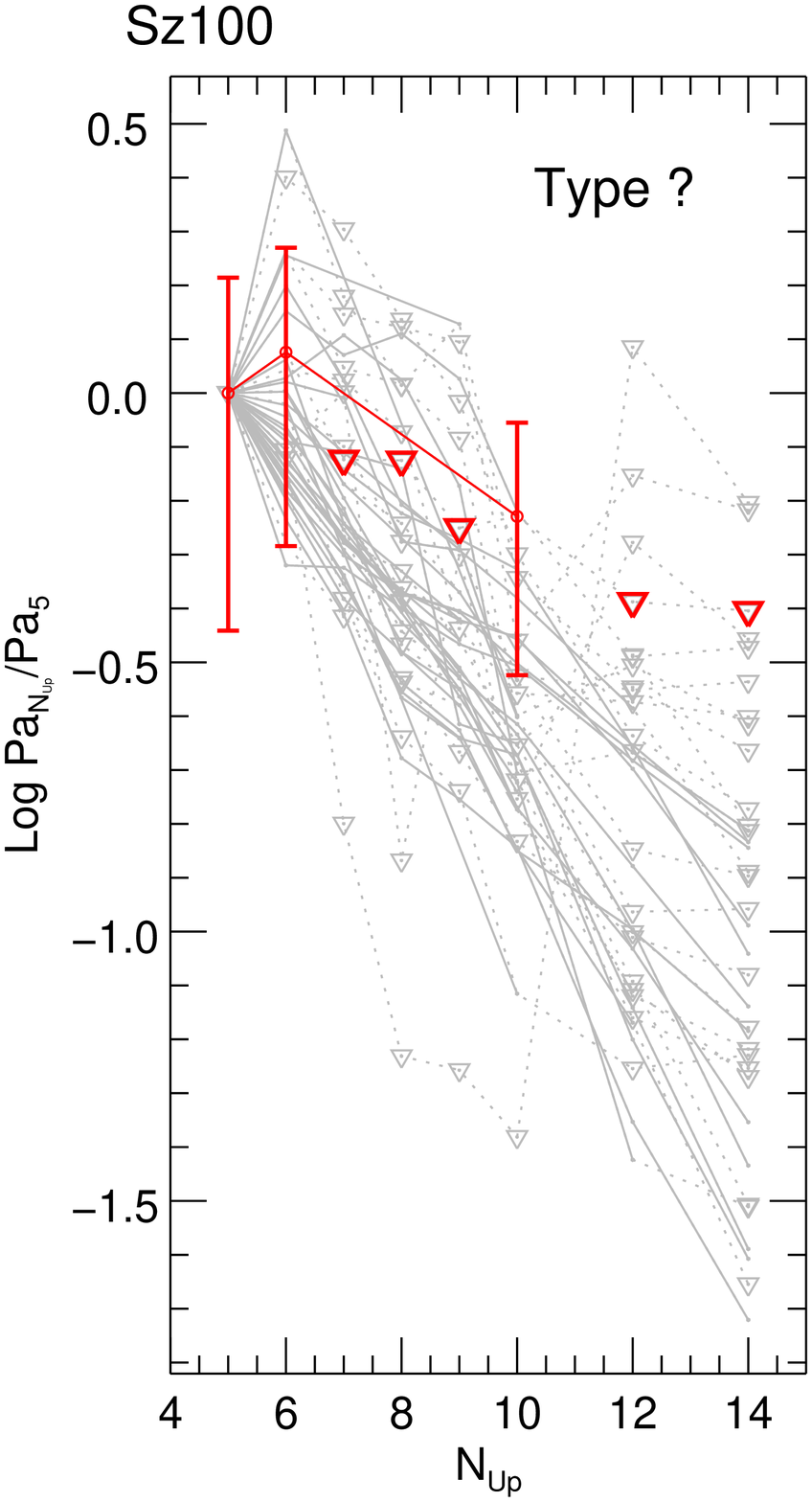}
\includegraphics[width=4.4cm]{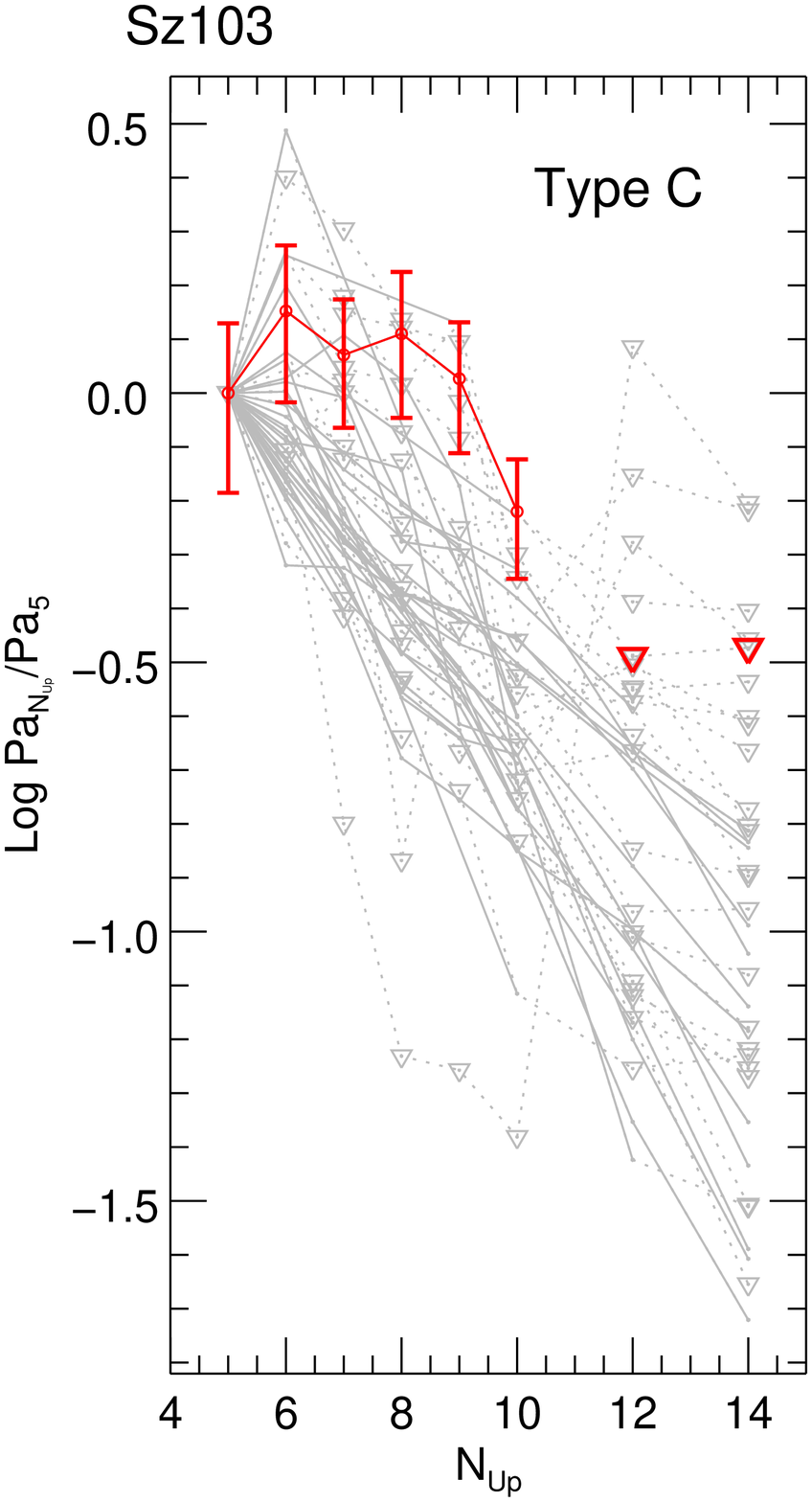}\\
\caption{\label{fig:decs:pa} Paschen decrements computed with respect to Pa$\beta$. For each object, the decrement shape is highlighted in red against all observed decrements, which are plotted as grey curves. Triangles indicate upper limits.} 
\end{figure*}

\setcounter{figure}{0}
\begin{figure*}[t]
\centering
\includegraphics[width=4.4cm]{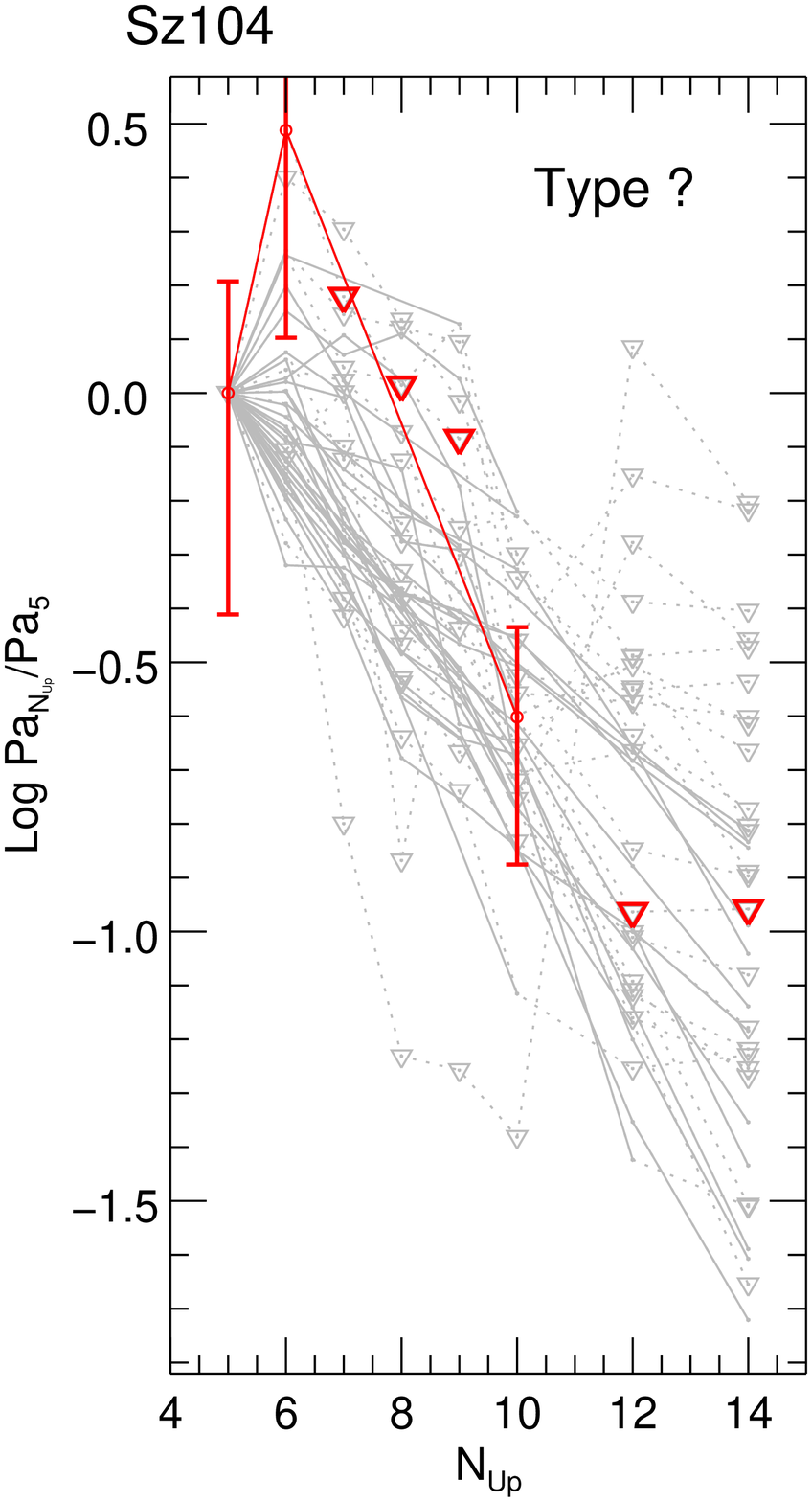}
\includegraphics[width=4.4cm]{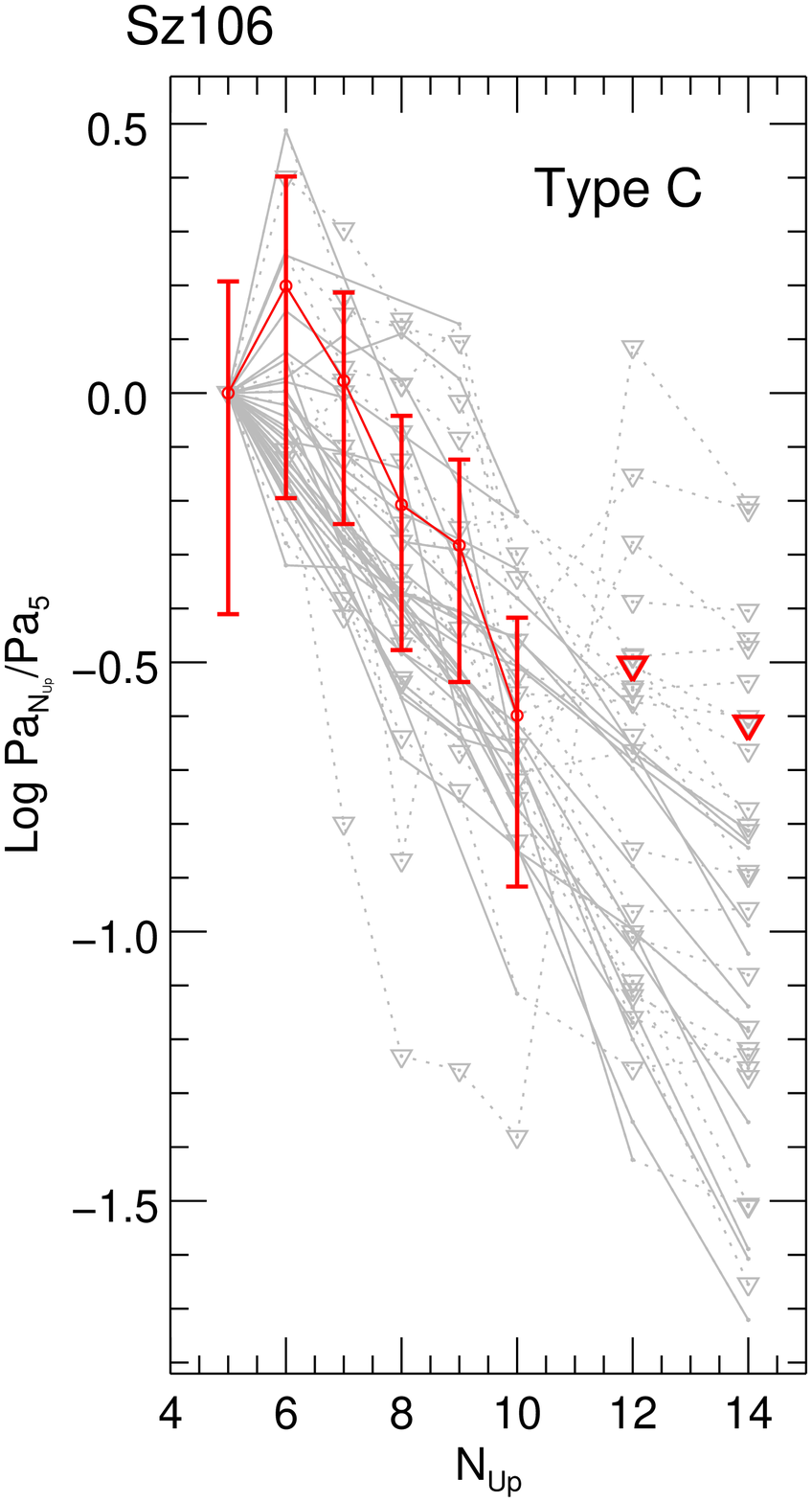}
\includegraphics[width=4.4cm]{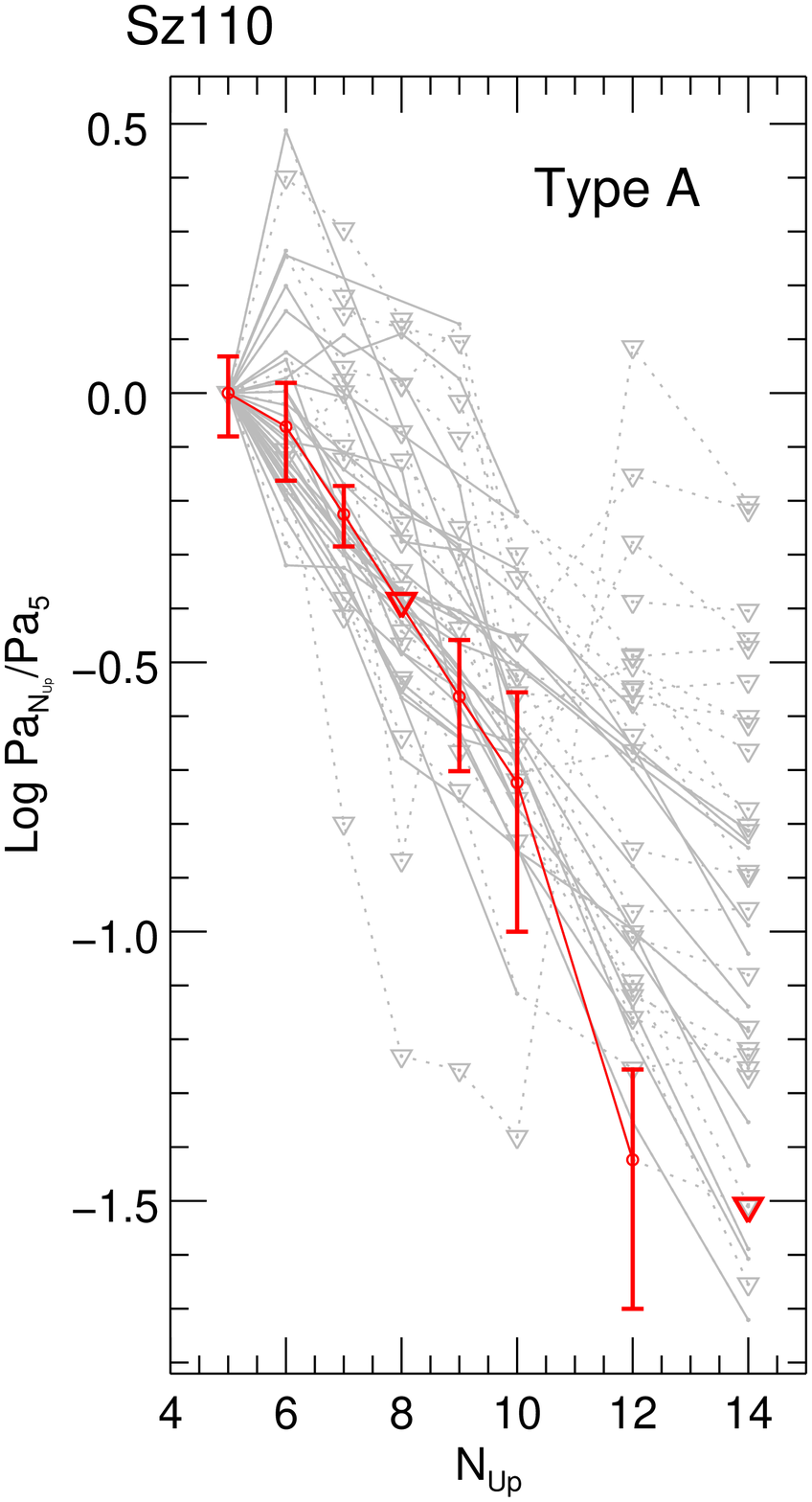}
\includegraphics[width=4.4cm]{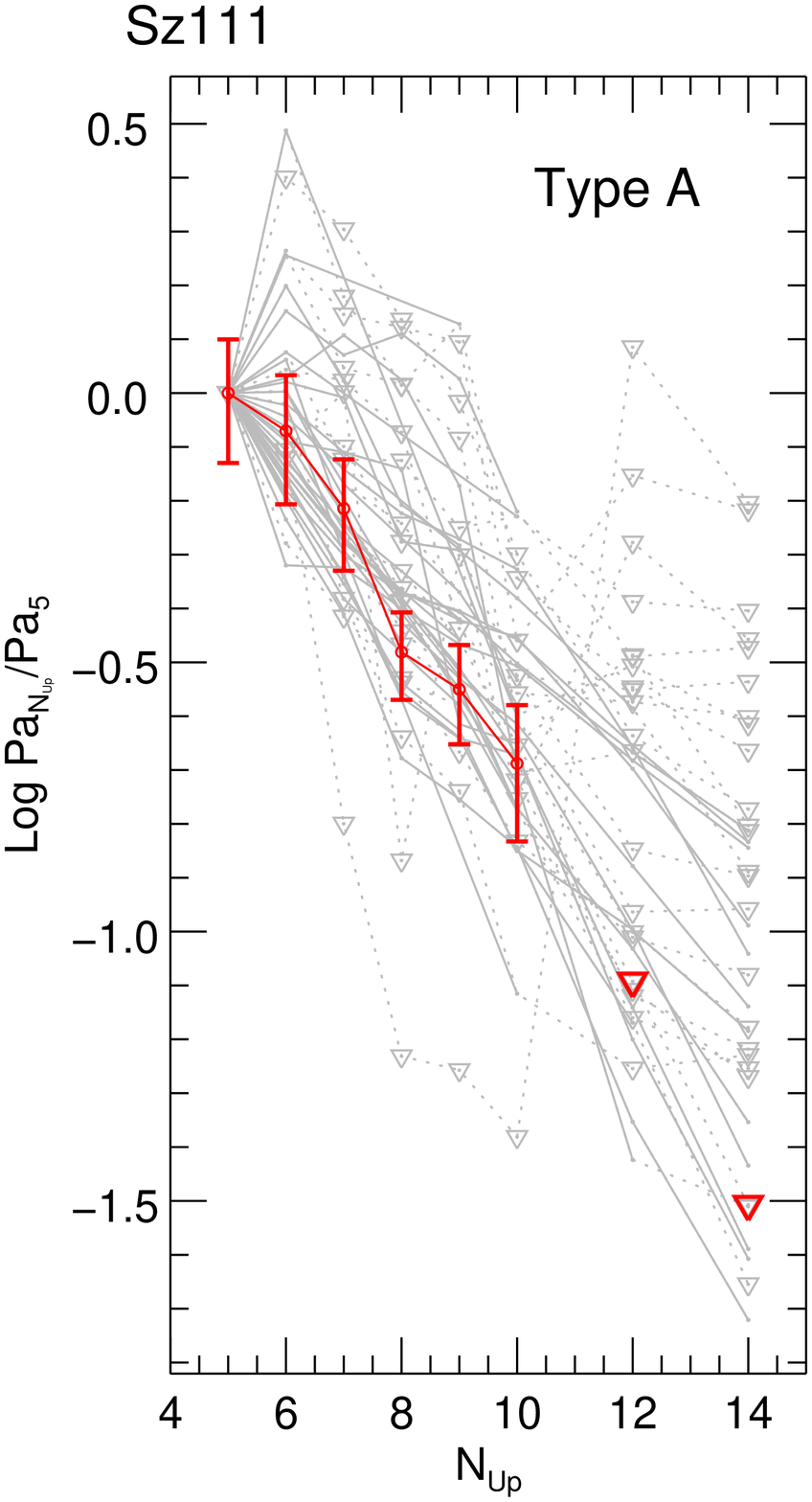}\\
\includegraphics[width=4.4cm]{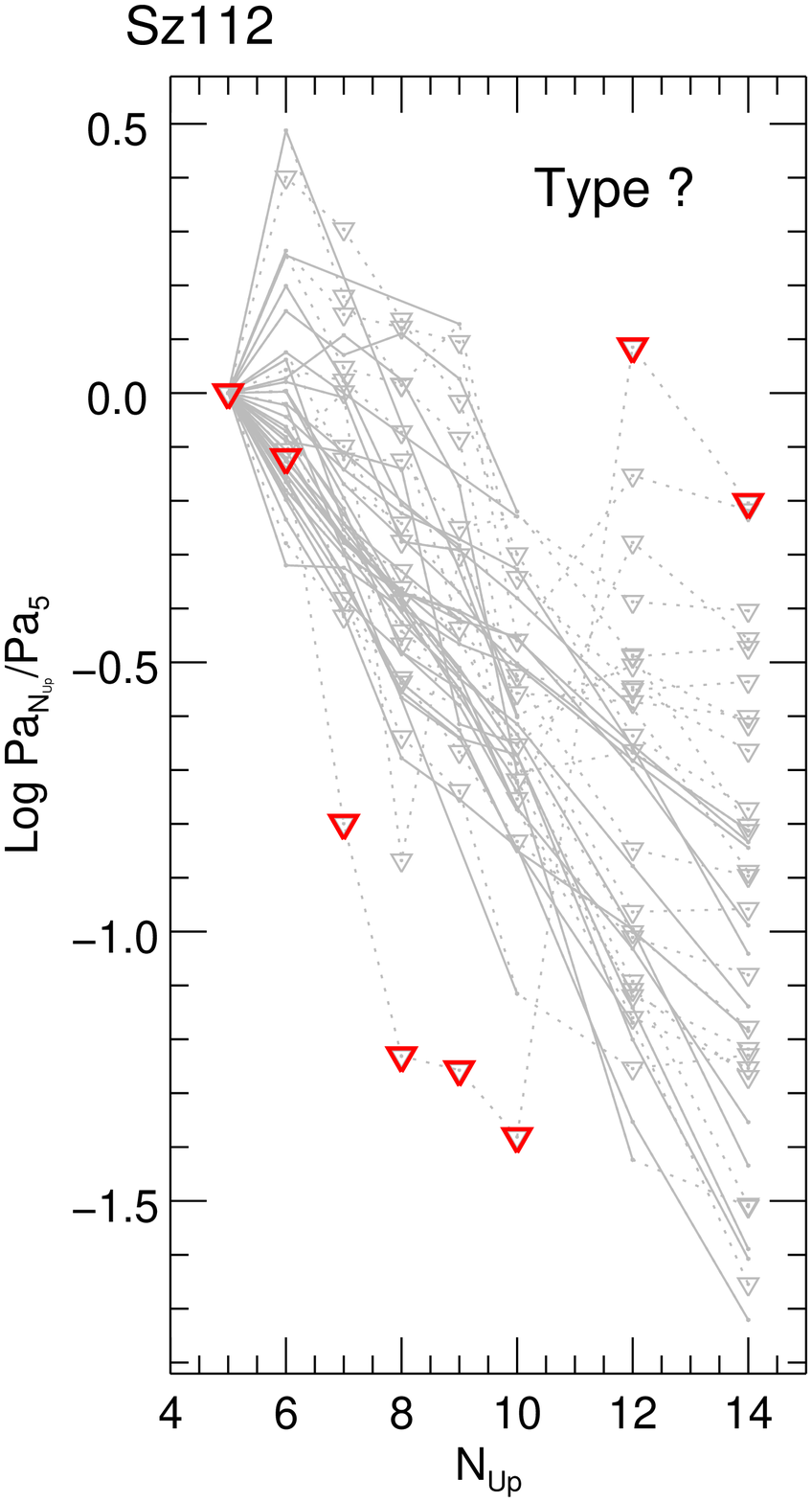}
\includegraphics[width=4.4cm]{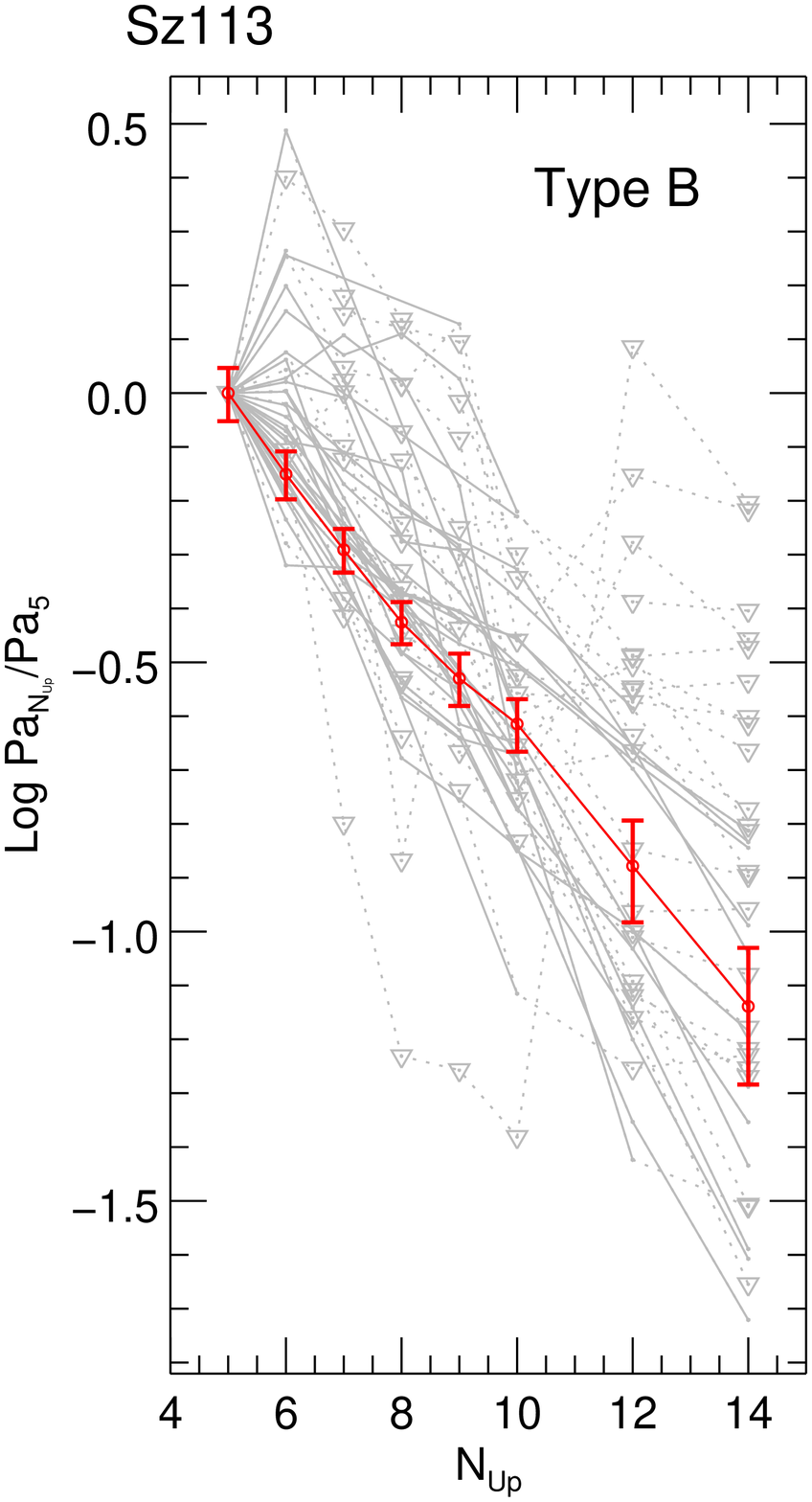}
\includegraphics[width=4.4cm]{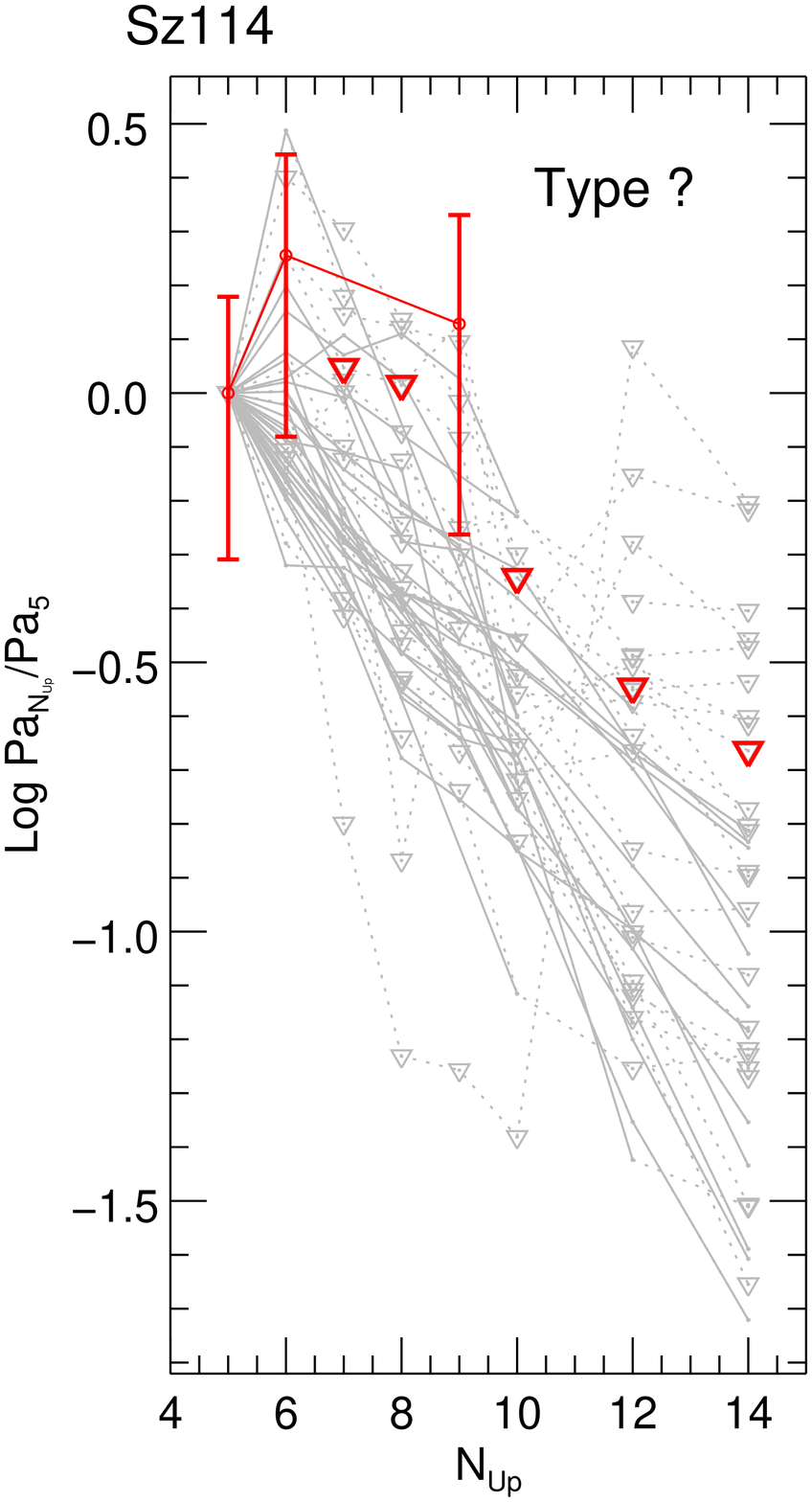}
\includegraphics[width=4.4cm]{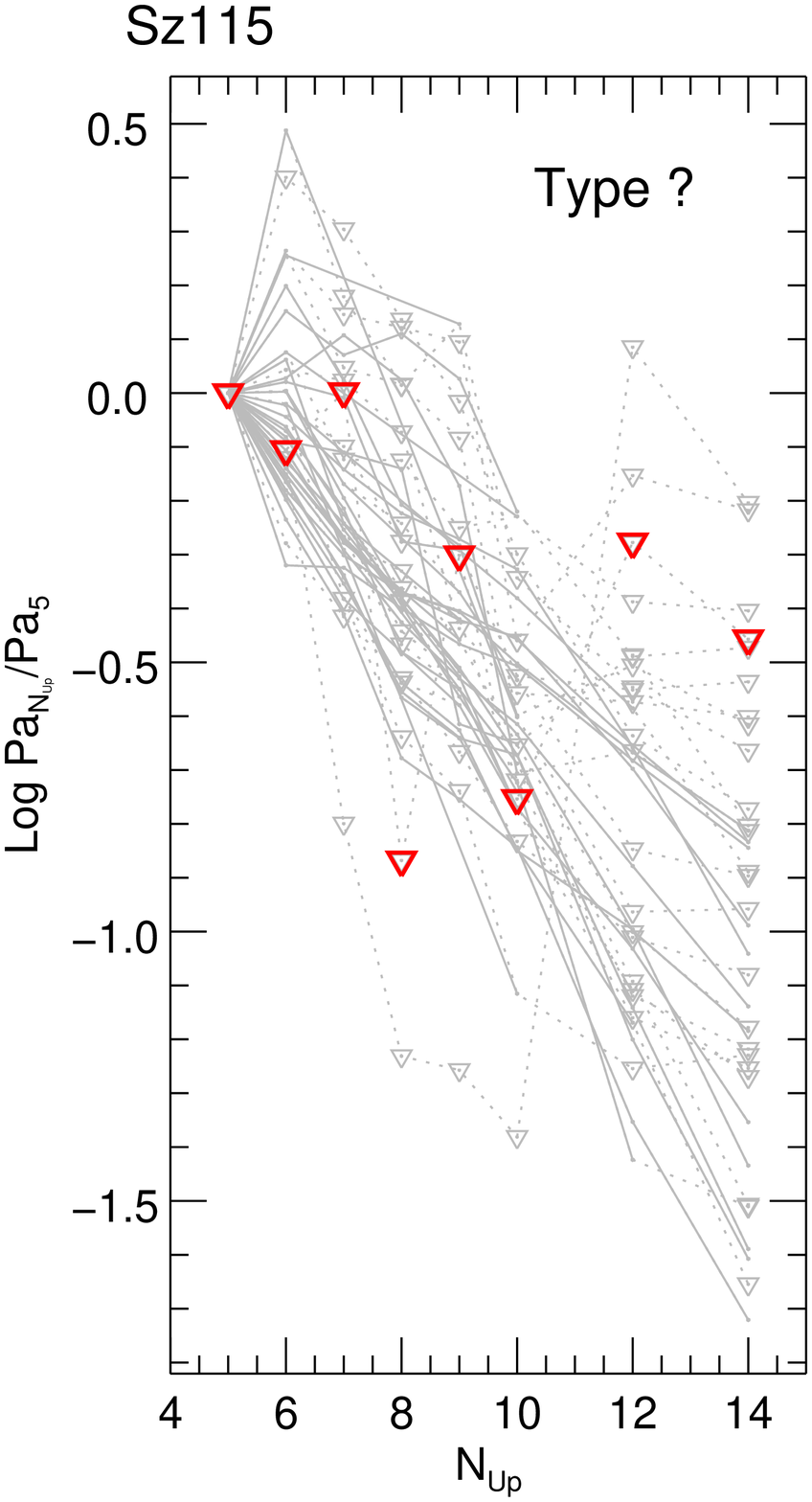}\\
\includegraphics[width=4.4cm]{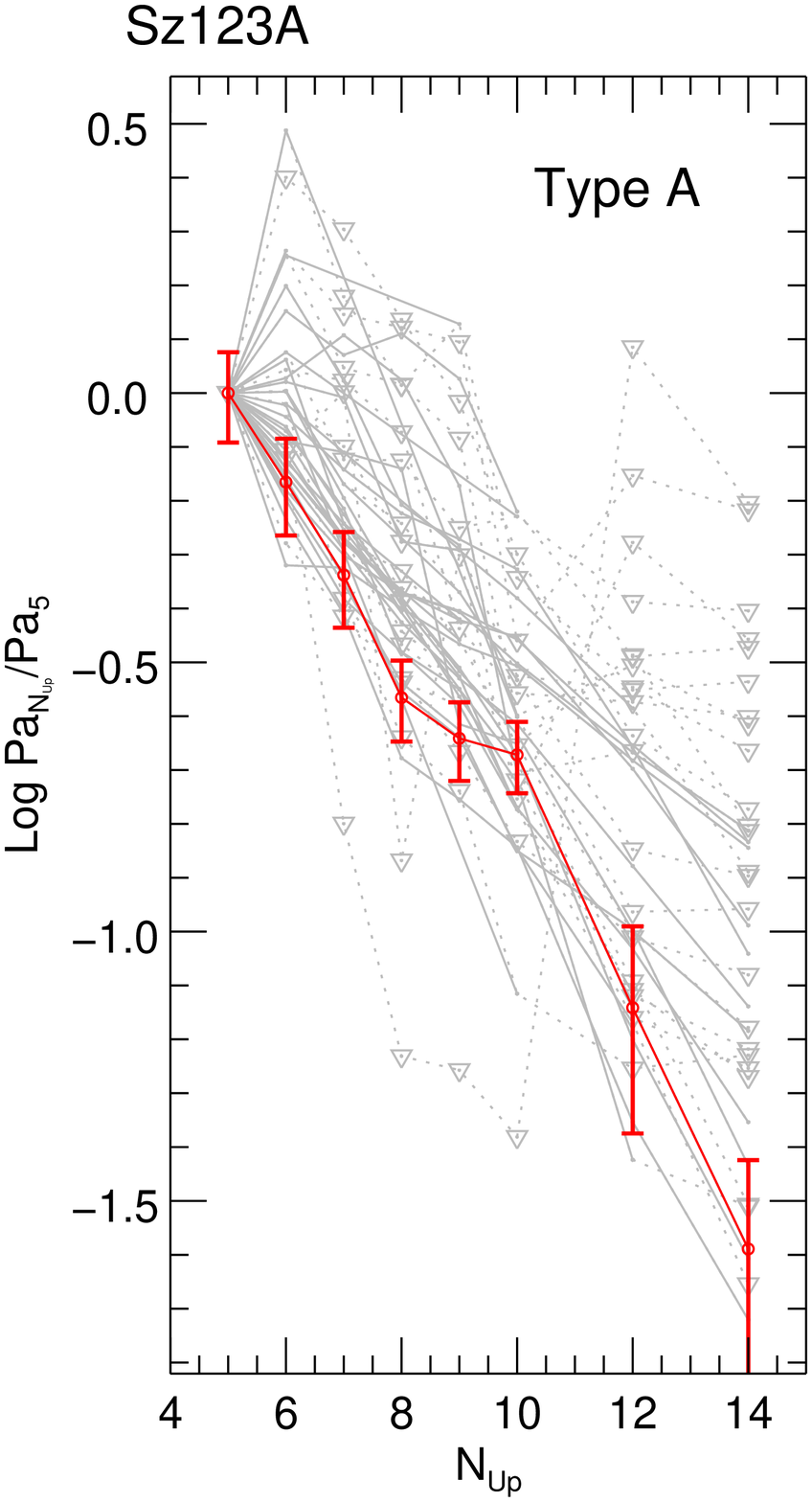}
\includegraphics[width=4.4cm]{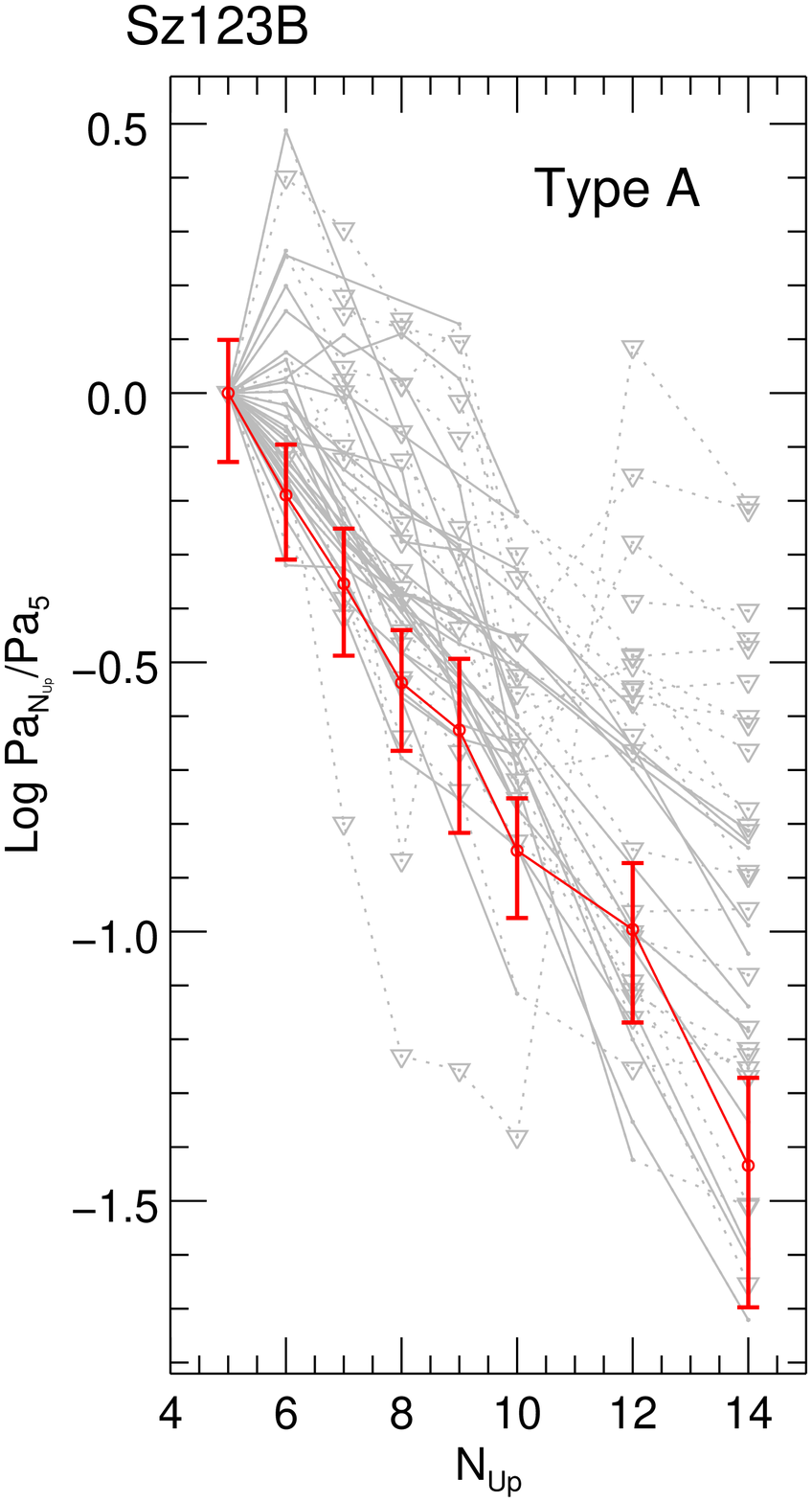}
\includegraphics[width=4.4cm]{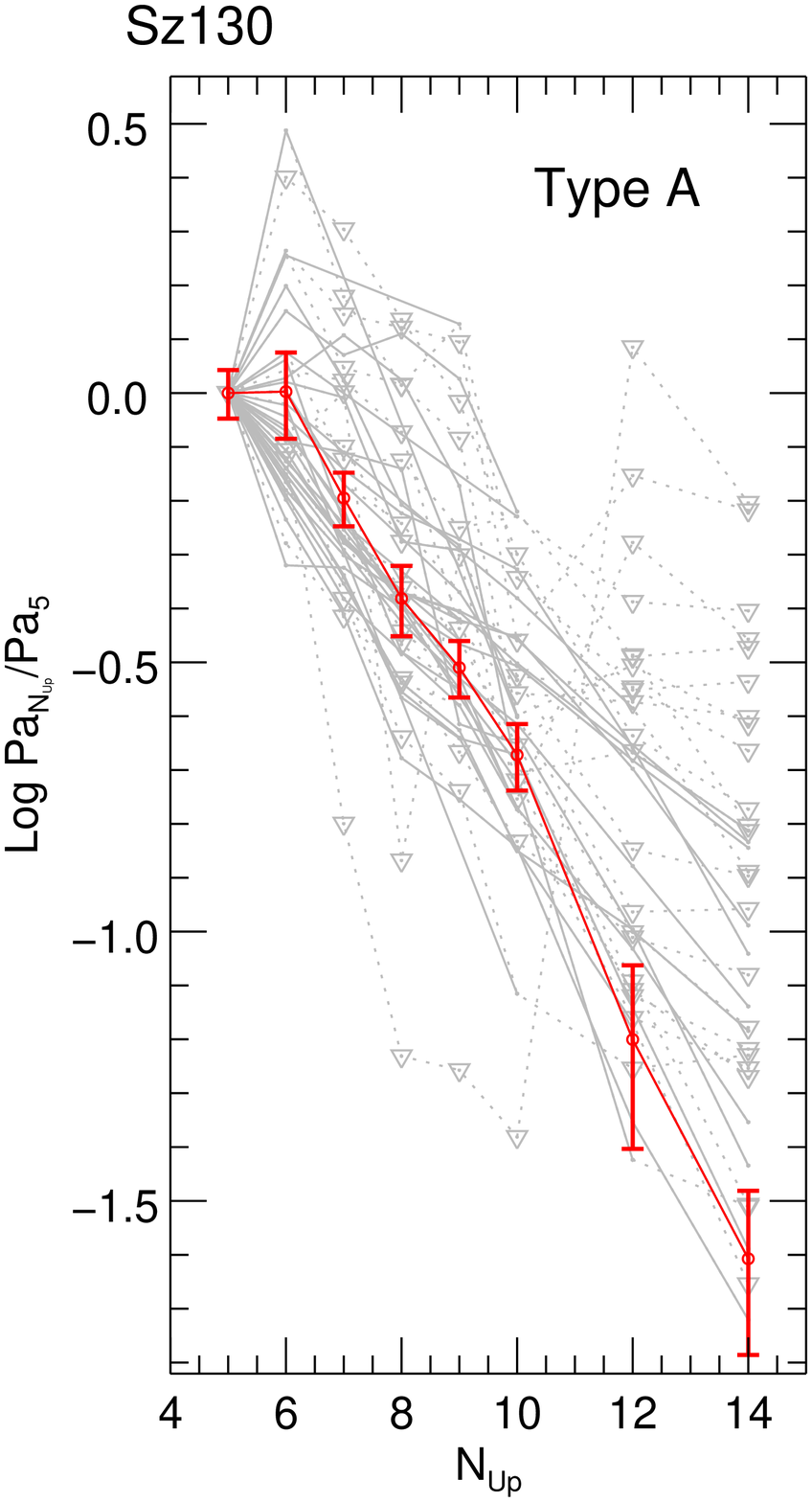}
\includegraphics[width=4.4cm]{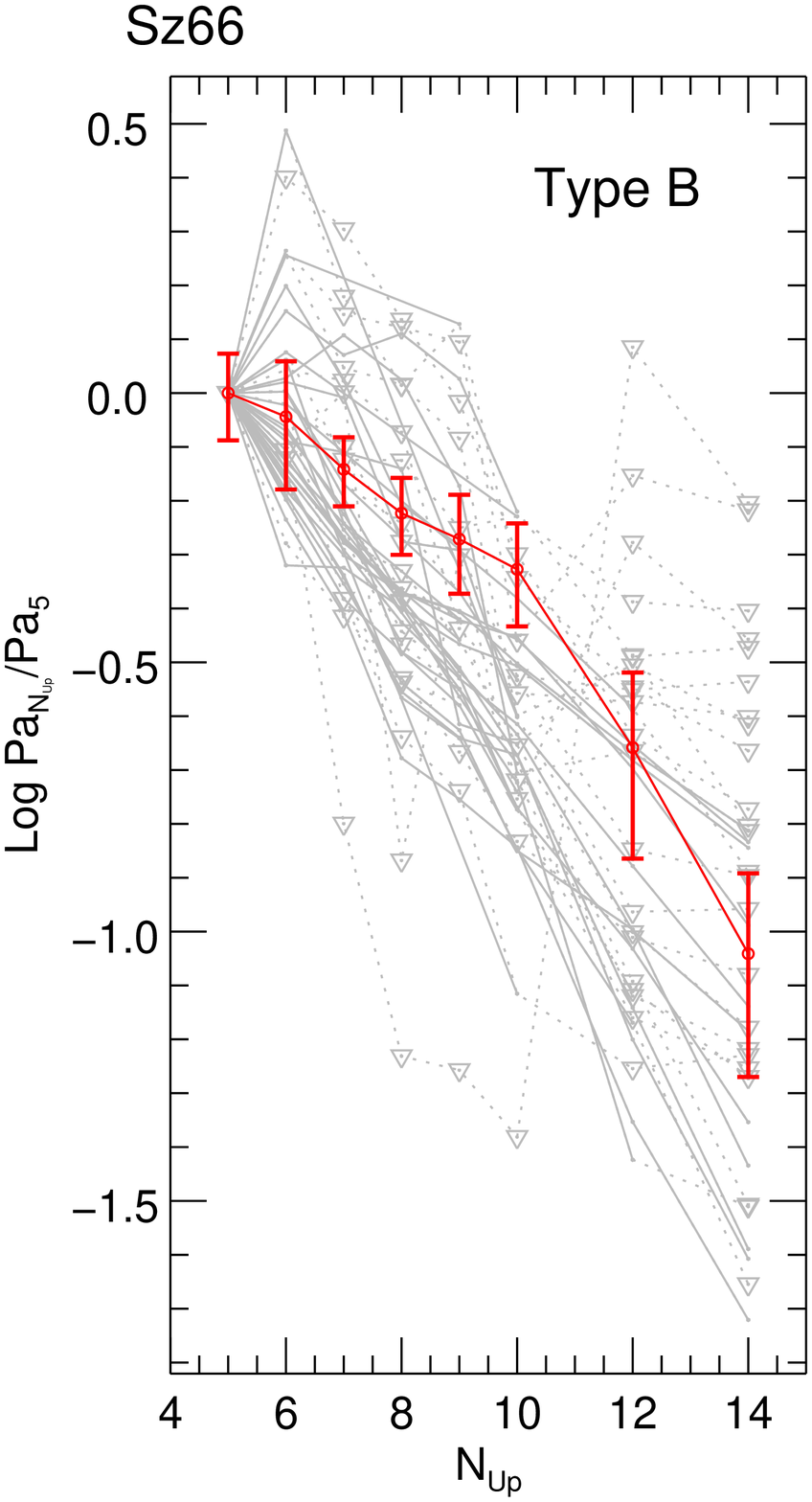}\\
\caption{Continued.} 
\end{figure*}

\setcounter{figure}{0}
\begin{figure*}[t]
\centering
\includegraphics[width=4.4cm]{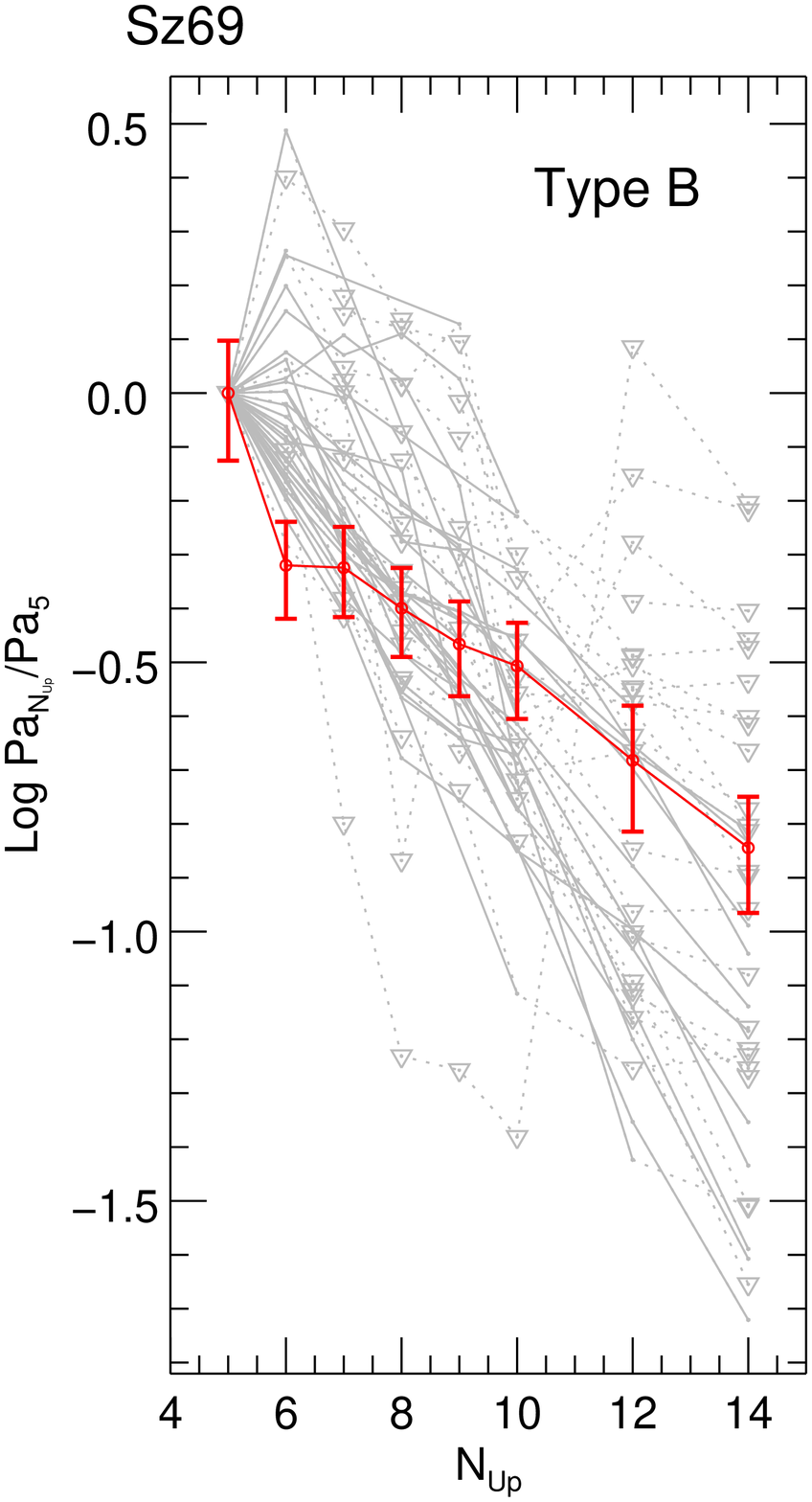}
\includegraphics[width=4.4cm]{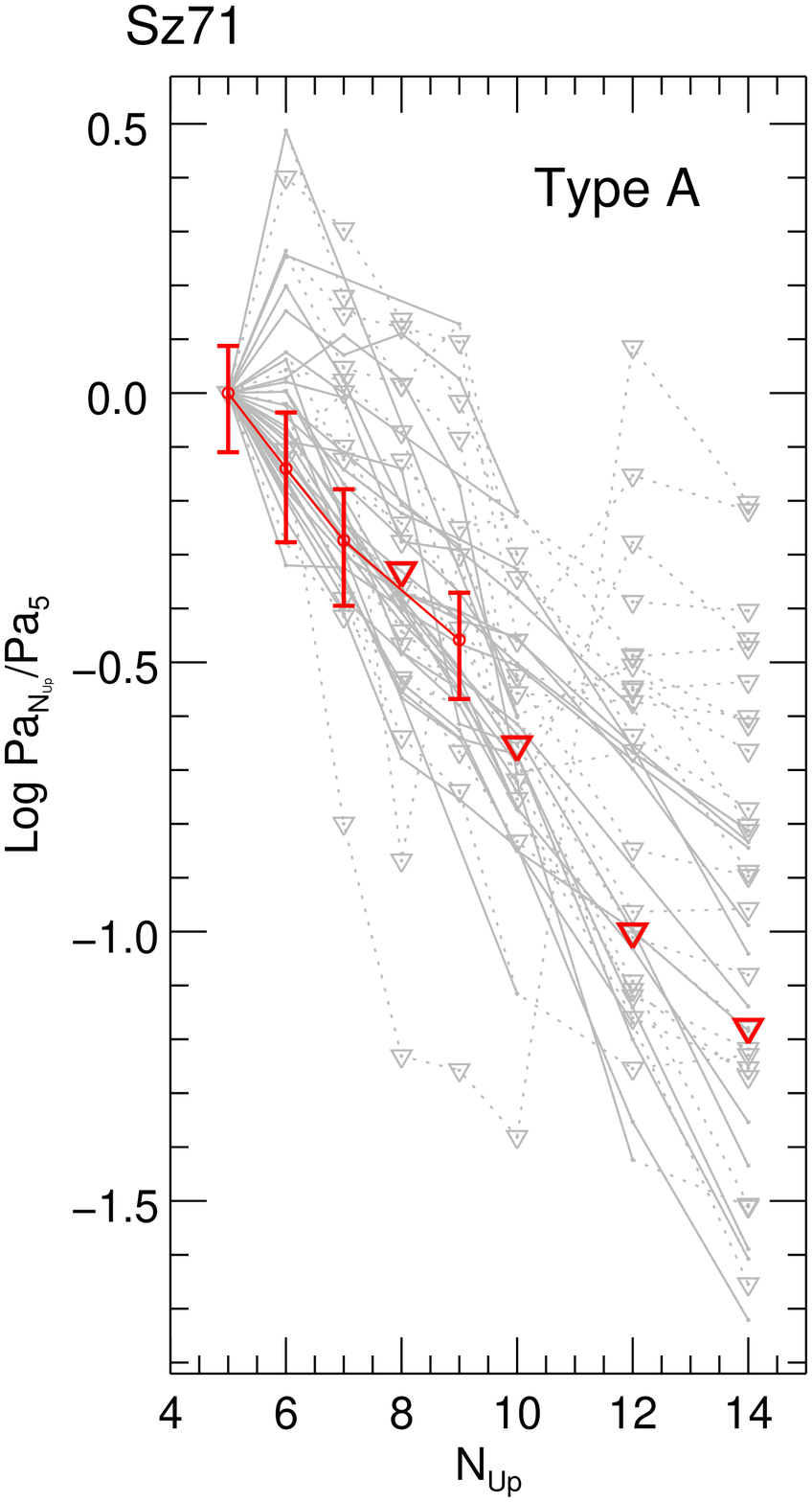}
\includegraphics[width=4.4cm]{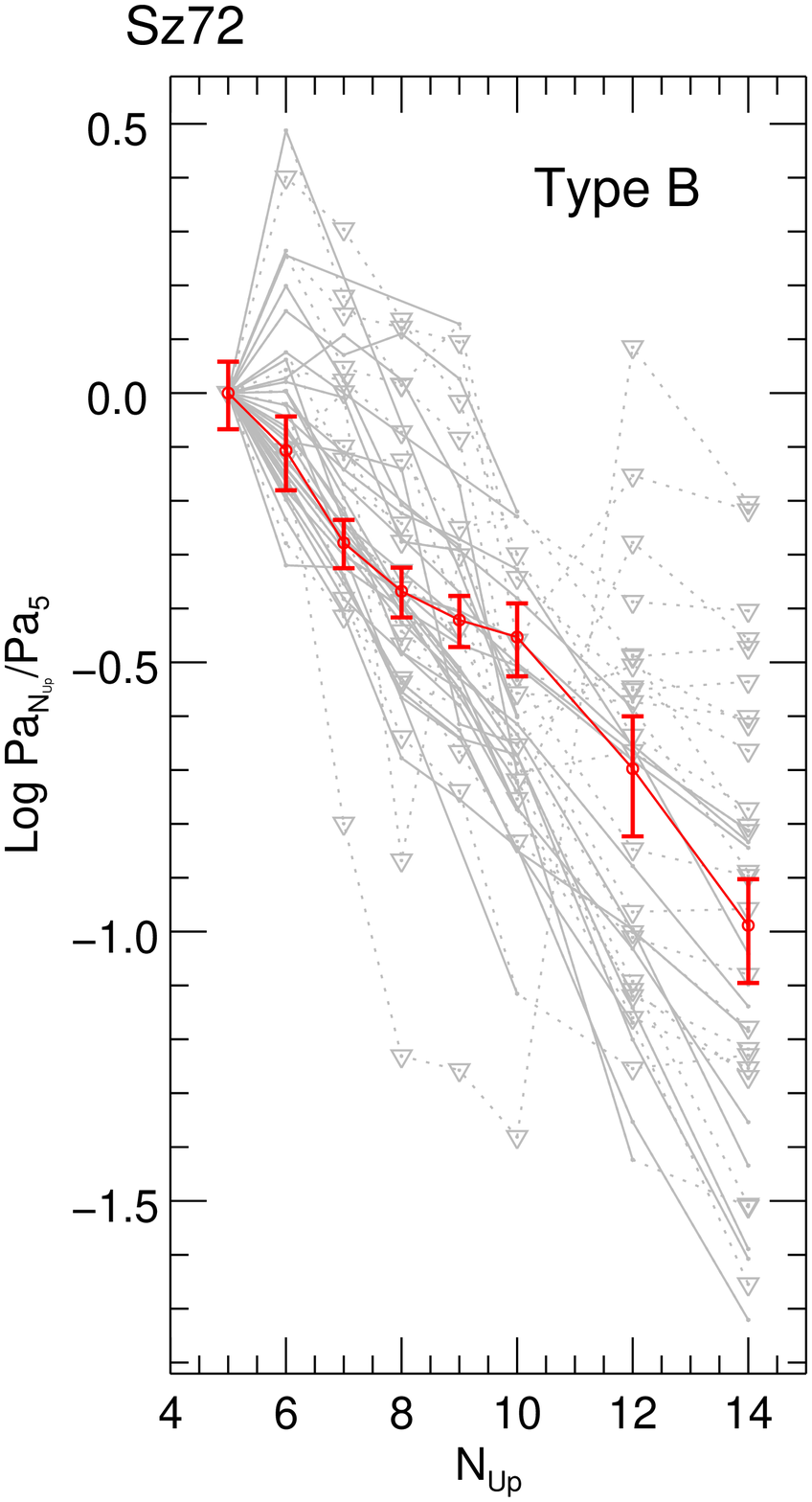}
\includegraphics[width=4.4cm]{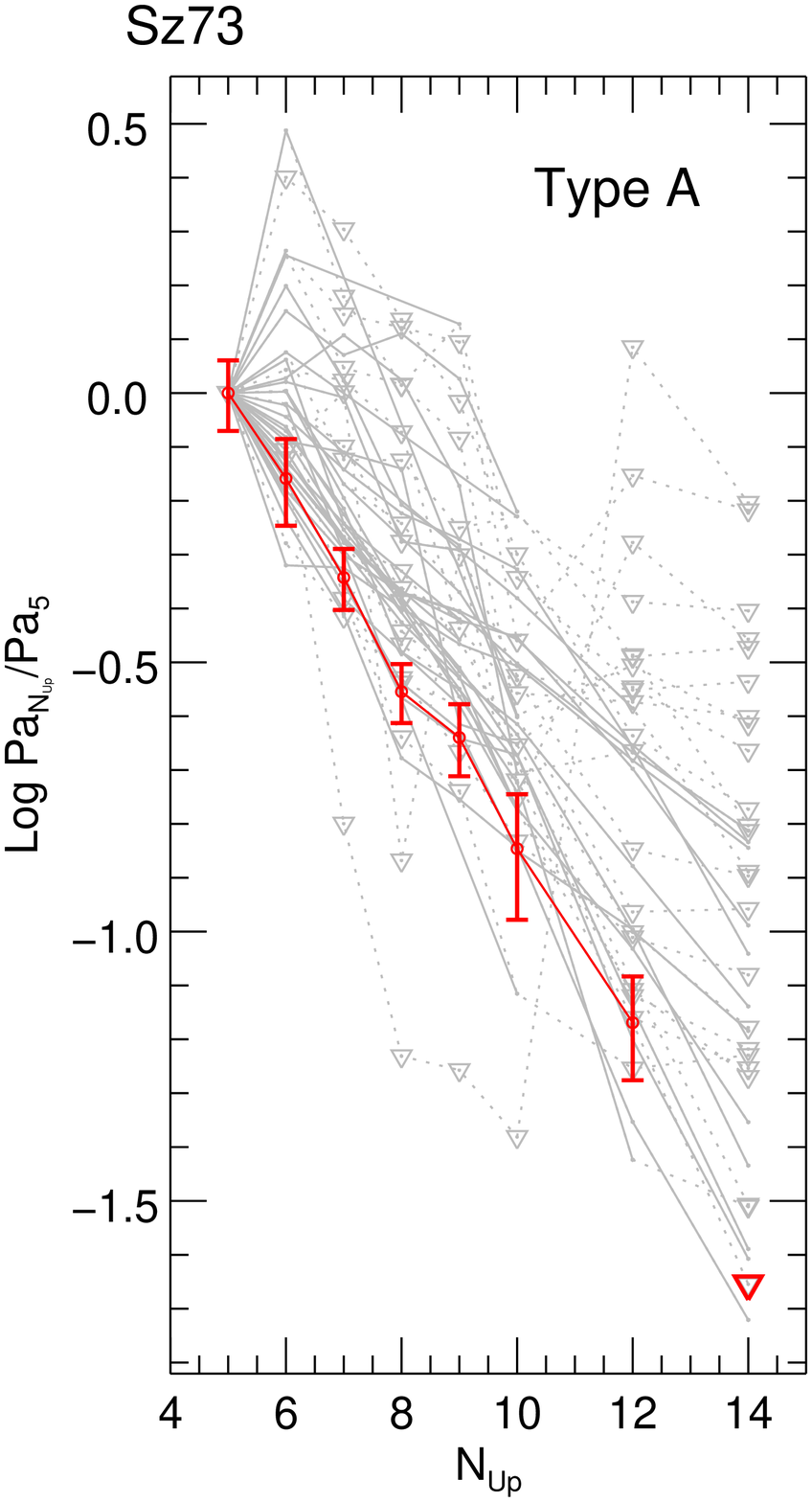}\\
\includegraphics[width=4.4cm]{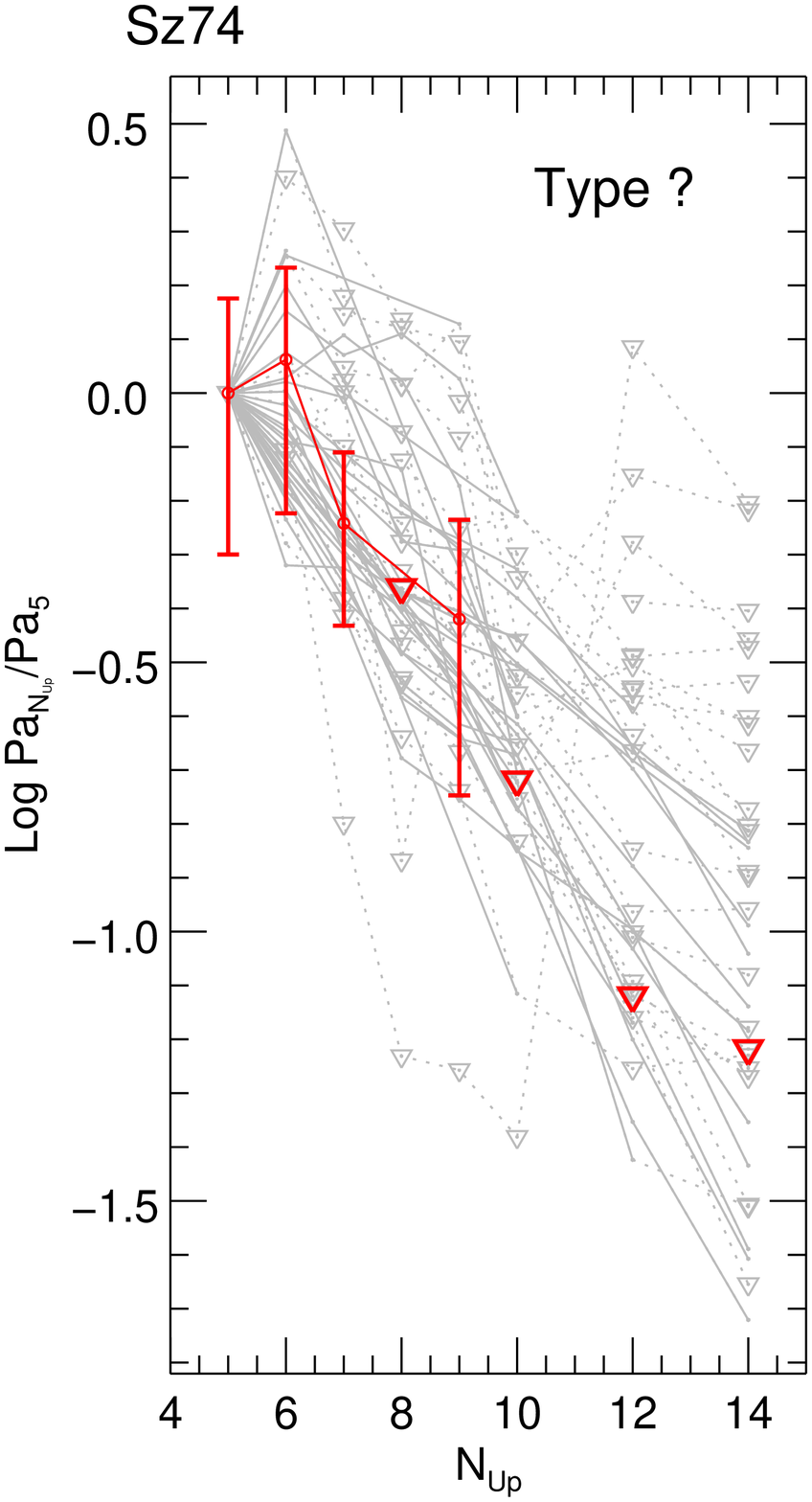}
\includegraphics[width=4.4cm]{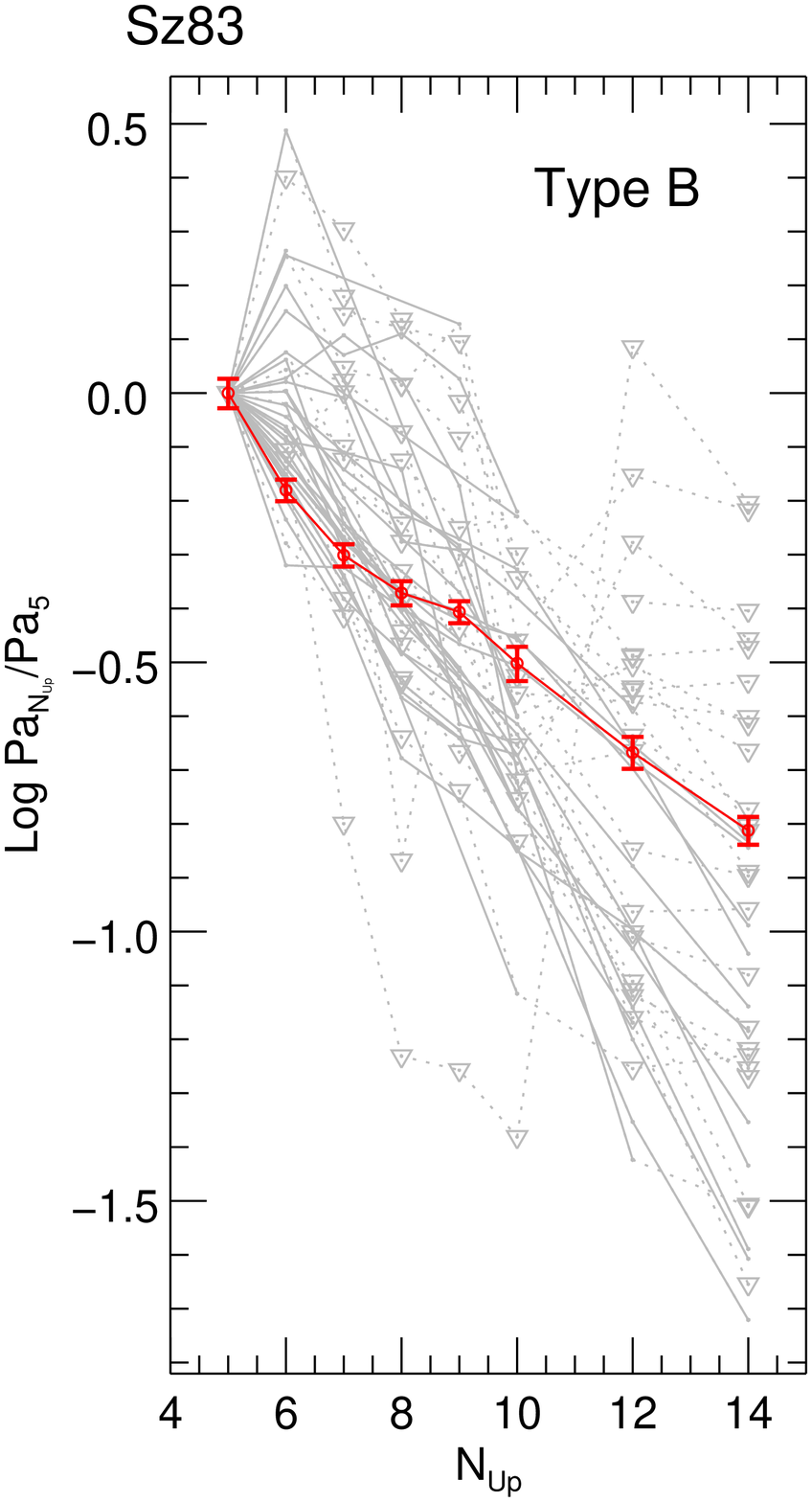}
\includegraphics[width=4.4cm]{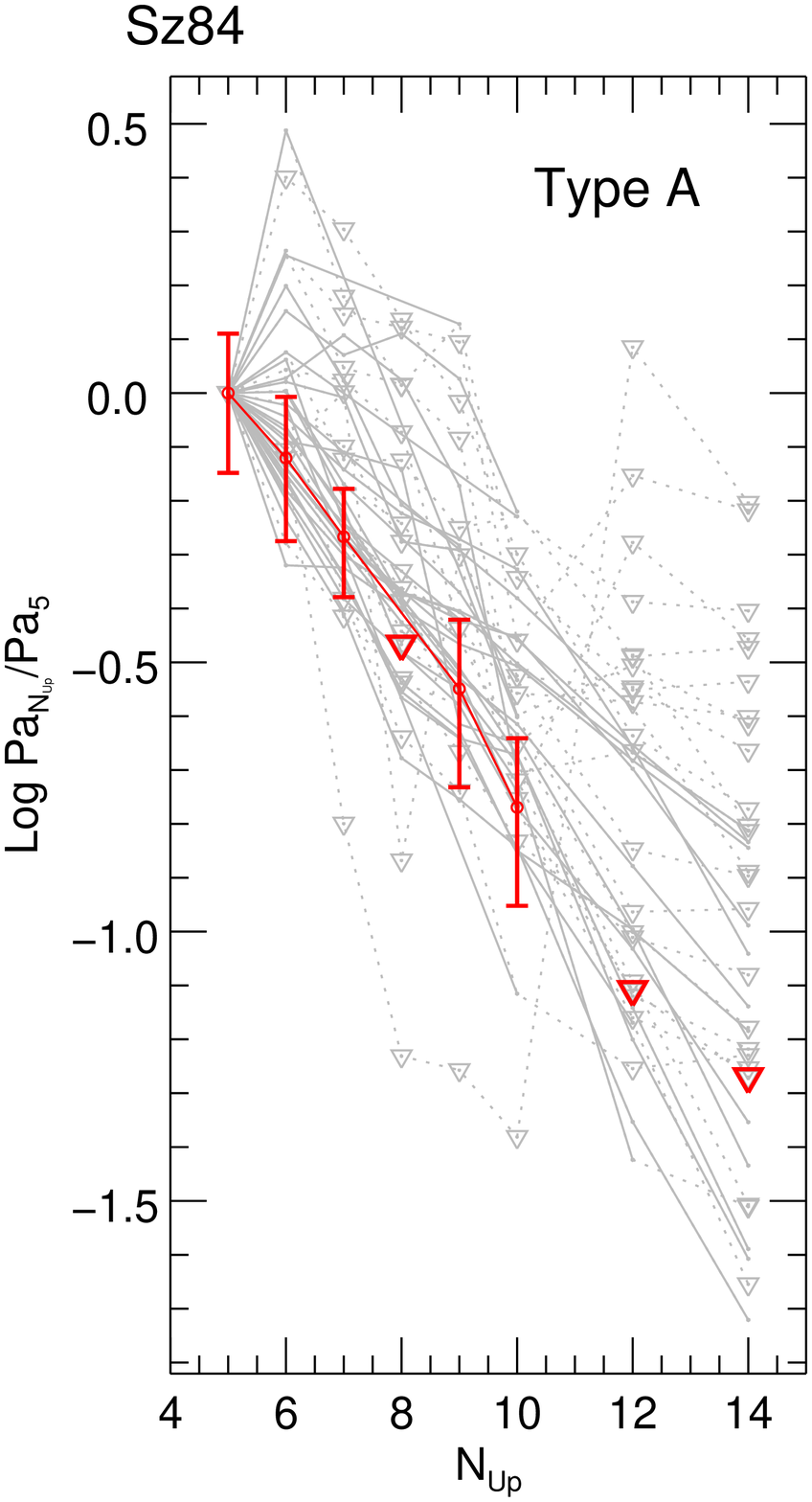}
\includegraphics[width=4.4cm]{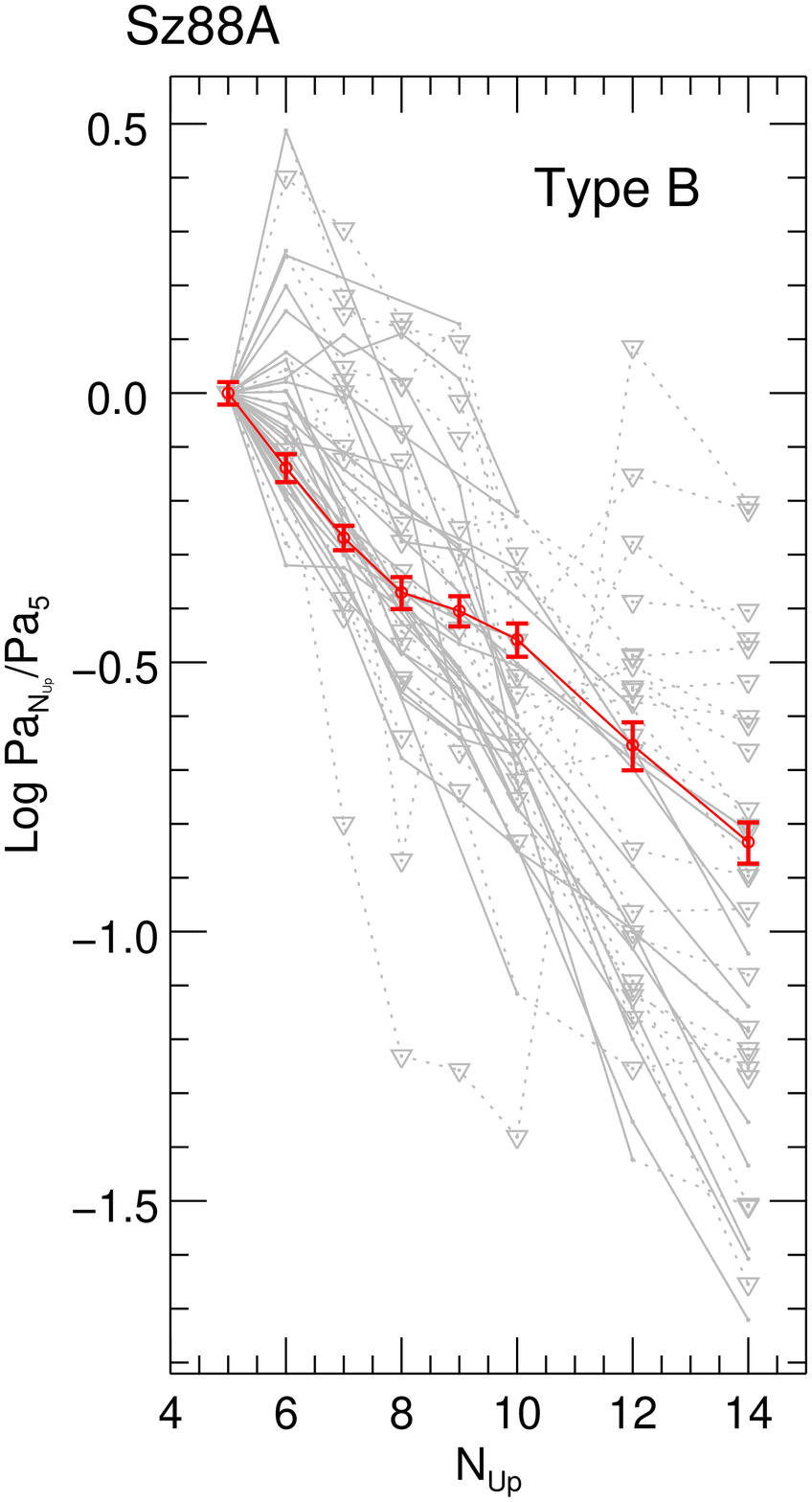}\\
\includegraphics[width=4.4cm]{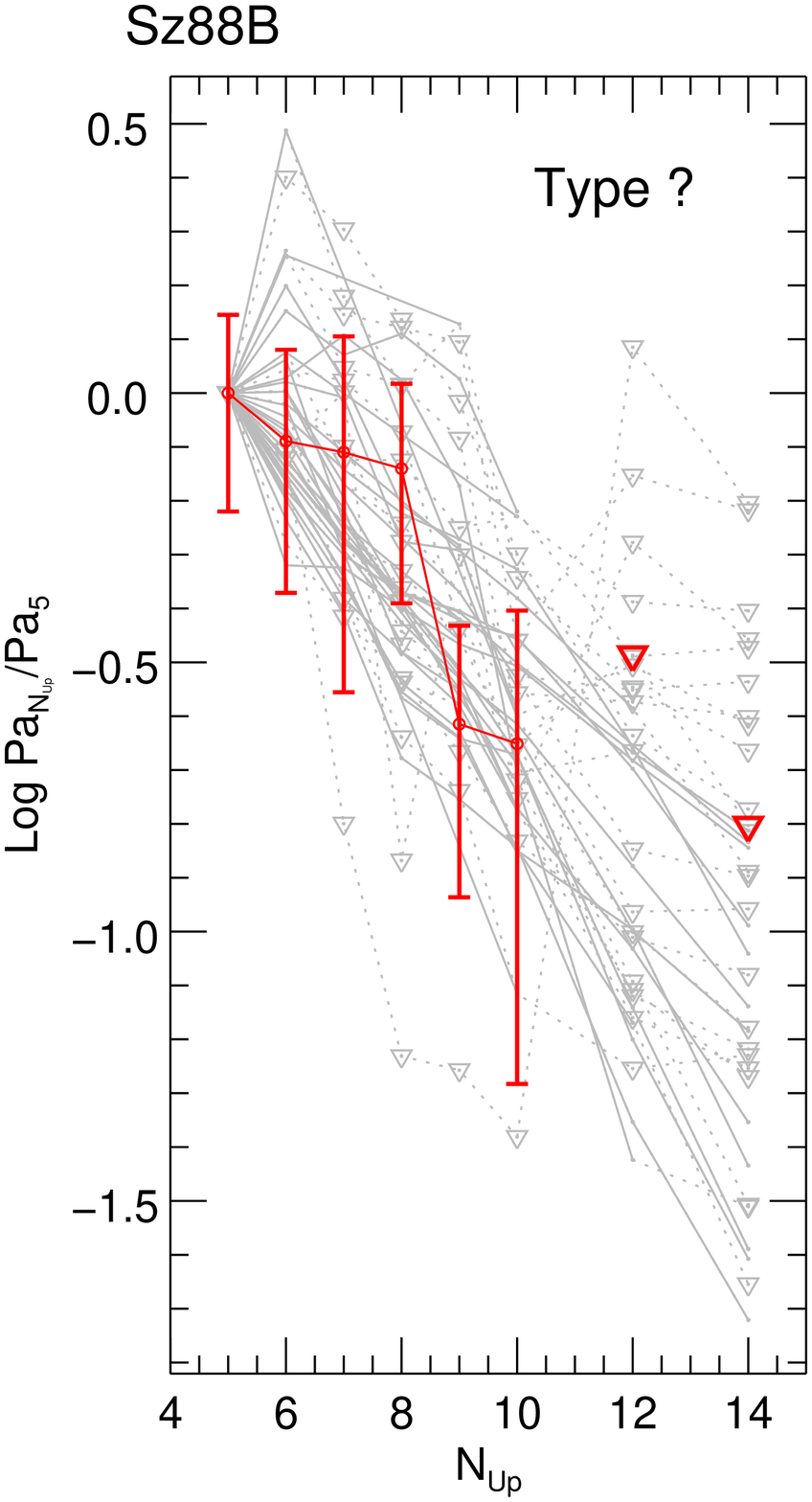}
\includegraphics[width=4.4cm]{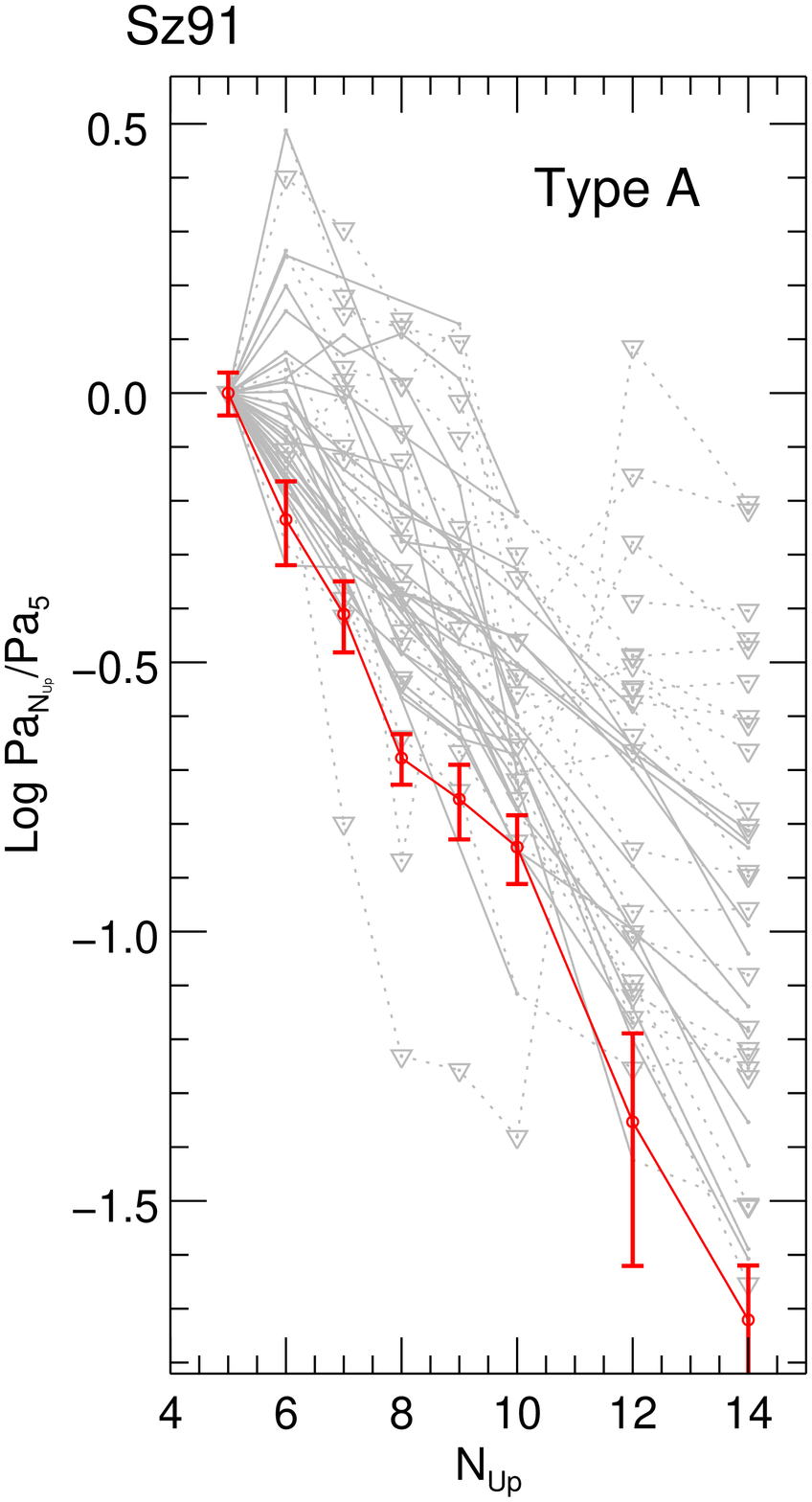}
\includegraphics[width=4.4cm]{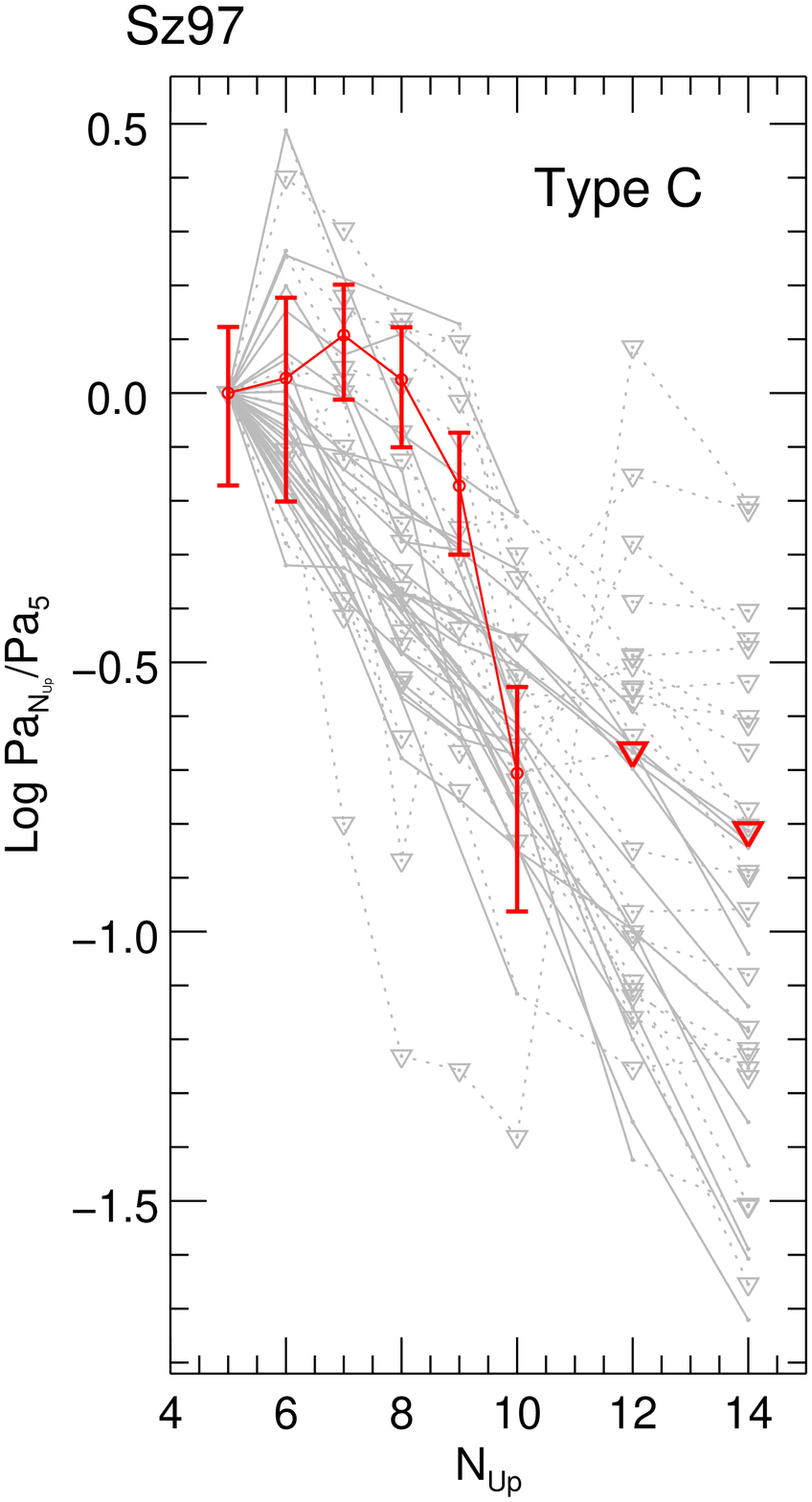}
\includegraphics[width=4.4cm]{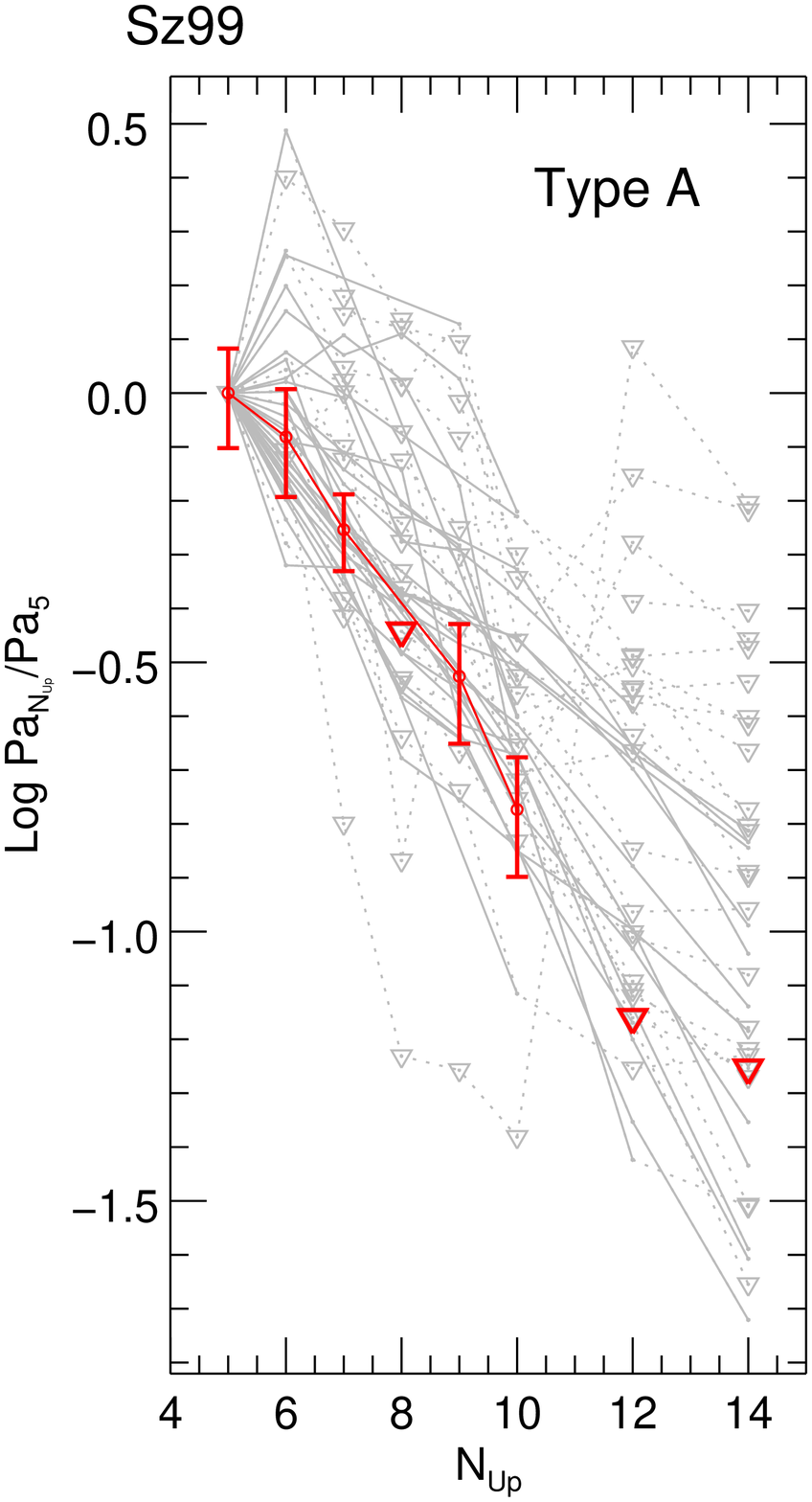}
\caption{Continued.} 
\end{figure*}

\end{document}